\documentclass[article]{aa} 
\usepackage{epsfig,deluxe} 
\begin{document}

\newcommand{\gsim}{\hbox{\rlap{$^>$}$_\sim$}} 
% A&A Section 6: Form. struct. and  evolut. of stars} 
%  \thesaurus{06     % A&A Section 6: Form. struct. and evolut. of stars   
\authorrunning{S. Dado, A. Dar \& A. De R\'ujula} 
\titlerunning{Radio Afterglows of GRBs} 
\title{On the Radio Afterglow of Gamma Ray Bursts} 
\author{Shlomo Dado$^{^1}$, Arnon Dar$^{^1}$ and 
A. De R\'ujula$^{^2}$} 
\institute{1. Physics Department and Space Research Institute, Technion\\   
               Haifa 32000, Israel\\ 
           2. Theory Division, CERN, CH-1211 Geneva 23, Switzerland} 
 
\maketitle 
 
\begin{abstract} 
 
We use the cannonball (CB) model of gamma ray bursts (GRBs) to predict the
spectral and temporal behaviour of their radio afterglows (AGs). A single
simple expression describes the AGs at all times and frequencies; its
high-frequency limit reproduces the successful CB model predictions for
optical and X-ray AGs. We analyze all of the observed radio AGs of GRBs
with known redshifts, including those of the exceptionally close-by GRB
980425.  We also study in detail the time-evolution of the AGs' spectral
index. The agreement between theory and observations is excellent, even
though the CB model is extremely frugal in the number of parameters
required to explain the radio observations. We propose to use the
scintillations in the radio AGs of GRBs to verify and measure the
hyperluminal speed of their jetted CBs, whose apparent angular velocity is
of the same order of magnitude as that of galactic pulsars, consistently
measured directly, or via scintillations. 
 
\end{abstract}

\section{Introduction} 
 
The Cannonball Model is based on the hypothesis that GRBs and their 
afterglows are made in supernova explosions by the jetted ejection of 
relativistic plasmoids: ``cannonballs'' made of ordinary baryonic matter   
(Dar and De R\'ujula 2000a), similar to the ones observed in quasars and   
microquasars (e.g., Mirabel and Rodriguez 1994; 1999 and references 
therein). The name cannonball (CB) originates in the contention that 
---due to a mechanism that we have explicitly discussed in Dado et al.   
2001--- the ejected plasmoids stop expanding very early in the afterglow   
(AG) phase. 
 
The CB paradigm gives a good description of the properties of 
the $\gamma$-rays 
in a GRB, that we modelled in simple approximations in 
Dar and De R\'ujula 2000b. 
It suggests an alternative (Dar and De R\'ujula 2001a),
which is rather promising (Dado et al. 2002), to 
the ``Fe-line'' interpretation of the spectral lines observed 
in some X-ray afterglows (GRB 970508: Piro et al. 1998; 
GRB 970828: Yoshida et al. 1999, 2001; GRB 991216: Piro et al. 2000; 
GRB 000214: Antonelli et al. 2000). The model also provides 
an extremely simple and successful description of the spectrum, 
and of the shape and absolute magnitude of the light curves of the 
optical and X-ray afterglows of {\it all} GRBs of known redshift, 
at {\it all} observed times (Dado et al. 2001, hereafter called DDD 2001). 
This description is universal, it encompasses the early optical 
flash of GRB 990123, the very peculiar optical and X-ray AG of GRB 970508,   
and all of the properties of GRB 980425, associated with SN1998bw. 
 
In this paper we derive the CB model's predictions for radio afterglows,   
and compare them to {\it all} radio observations in GRBs of known   
redshift. We also study the evolution of the spectral index of AGs
as a function of time. The CB model ---in parameter-thrifty
and very simple terms--- passes these tests with 
flying colours. 
 
\section{Summary} 
 
In the CB model a GRB jet consists of $\rm n_{_{CB}}$ cannonballs, 
typically a few, 
each of them generating a prominent 
pulse in the $\gamma$-ray signal, as they 
reach the transparent outskirts of the shell of their associated supernova   
(SN).  We assume CBs to be made of ordinary matter, mainly hydrogen,
and to enclose a  magnetic field maze, as is the case for the 
observed ejections from quasars and microquasars.
The interstellar medium (ISM) the CBs traverse in the AG phase 
has been previously partially 
ionized by the GRB radiation and is fully 
ionized by Coulomb interactions as it enters the CB. 
In analogy to processes occurring in quasar and microquasar 
ejections, the ionized ISM particles are multiple 
scattered, in a ``collisionless'' way, by the CBs' turbulent magnetic   
fields. 
In the rest system of the CB the ISM swept-up nuclei are isotropically   
re-emitted, exerting 
an inwards force on the CB's surface. This allows one to compute {\it   
explicitly} 
the CB's radius as a function of time (DDD 2001). 
The radius, for typical parameters, and in minutes of observer's
time, reaches a constant  $\rm R_{max}$ of a few times $10^{14}$ cm.

The ISM nuclei (mainly protons) that a CB scatters also decelerate its
flight: its Lorentz factor, $\rm\gamma(t)$, is calculable.  Travelling at
a large $\gamma$ and viewed at a small angle $\theta$, the CB's emissions
are strongly relativistically aberrant:  in minutes of observer's time,
the CBs are parsecs away from their source.  For a constant CB radius and
an approximately constant ISM density, $\rm\gamma(t)$ has an explicit
analytical expression, as discussed in Appendix I.  Typically
$\rm\gamma=\gamma(0)/2$ at a distance of order 1 kpc from the source, and
$\gamma(0)\sim 10^3$.  Due to a limited observational sensitivity, GRBs
have been detected only up to angles $\theta$ of a few times
$1/\gamma(0)$.

The ISM electrons entering a CB are caught up and bounce off 
its enclosed magnetic domains acquiring a predictable power-law energy   
spectrum, as we argue in Section 3. In the CB's rest system
$\rm dn_e/dE\propto E^{-2}$ below an energy 
$\rm E_b(t)\simeq \gamma(t)\, m_e\, c^2$, steepening to 
$\rm dn_e/dE\propto E^{-(p+1)}$, with $\rm p\simeq 2.2$, 
above this energy\footnote{In our previous work   
$\rm p$ was called $\rm\beta_p$, referring to the proton spectrum, 
for which radiative losses are negligible.}.
The energy $\rm E_b$ does not correspond to the conventional
synchrotron ``cooling break'' but to the {\it injection bend} at the
energy at which electrons enter the CB with a Lorentz factor $\rm\gamma(t)$.
In Section 4 we discuss the observational 
support of the existence of the injection bend, which is strong.
Given the very large magnetic and radiation energy densities  
in the CB, the usual cooling break (at the energy at which 
the energy-loss rate due to synchrotron emission and inverse Compton 
scattering equals that due to bremsstrahlung, 
adiabatic losses and escape) happens only at subrelativistic energies, 
as discussed in Appendix II. 
 
The magnetic energy-density in a CB (DDD 2001) is: 
\begin{equation}\rm 
U_B= {B^2\over8\, \pi}\sim {1\over 4}\,\gamma^2\,n_p\, m_p\, c^2,
\label{emag}
\end{equation}
with $\rm n_p$ the 
ISM baryon density (seen as $\rm \gamma\, n_p$ by the CB in its rest 
system). Thus, the magnetic field is:
\begin{equation} 
\rm B(t)\sim 3 \;\left[{n_p\over 10^{-3}\,cm^3}\right]^{1/2}
\left[{\gamma(t)\over 10^3}\right]\; Gauss.
\label{mag}
\end{equation}
The pitch-angle averaged characteristic synchrotron-radiation 
frequency of electrons of energy $\rm E=E_b$ is (Rybicki and Lightman 1979):
\begin{equation}
\rm \nu_b(t)\sim 0.29\,{3\over 4}\; \gamma(t)^2\,\nu_L
\label{nub}
\end{equation} 
where $\rm\nu_L=e\,B/(2\,\pi\,m_e\,c)$ is the Larmor frequency in the  
CB enclosed magnetic field $\rm B$. 
To a good approximation, in the CB rest system and prior to 
cumulation, absorption
and limb-darkening corrections, the synchrotron radiation has a spectral 
shape: 
\begin{eqnarray} 
\rm \nu\,{dn_\gamma\over d\,\nu} &\propto& \rm 
f_{sync}(\nu,t) \equiv  
{K(p)\over \nu_b(t)}{[\nu/\nu_b(t)]^{-1/2}\over
\sqrt{1+[\nu/\nu_b(t)]^{(p-1)}}}
\nonumber \\ \rm 
 \rm K(p)&\equiv&\rm
{\sqrt{\pi}\over   \Gamma\left[{2\, p-1\over 2(p-1)}\right]
                                   \, \Gamma\left[{2\, p-3\over 2(p-1)}\right]}
\simeq{p-2\over 2\,(p-1)}\; , 
\label{sync} 
\end{eqnarray}
where we have normalized $\rm f_{sync}(\nu)$ to a unit integral 
over all frequencies and the approximation
is good to better than 8\% precision
in the range $\rm 2<p\leq 2.6$. Note that, for $\rm \nu\gg\nu_b$,
$\rm f_{syn}\propto \nu_b^{(p-2)/2}$  
i.e., it is independent of $\rm \nu_b$ for $\rm p=2$. For $\rm p\sim 2.2\, ,$  
the extremely weak dependence of $\rm f_{syn}$  
on $\rm \nu_b$ in the optical and X-ray bands was
neglected in DDD 2001. 

For the first $\sim 10^3$ seconds of observer's time,
a CB is still cooling fast and emitting
via thermal bremsstrahlung (DDD 2001), but after that
the CB emissivity integrated over frequency is 
equal to the energy deposition rate of the ISM 
electrons in the CB\footnote{The kinetic energy of a CB is mainly lost to 
the ISM protons it scatters; only a fraction $\rm\leq m_e/m_p$ is 
re-emitted by electrons, as the AG.}:
\begin{equation} \rm {dE\over dt} 
\simeq \eta\, \pi\, R_{max}^2\, n_e\, m_e\, c^3\, \gamma(t)^2 , 
\label{cbemissivity} 
\end{equation} 
where $\rm n_e\gamma$ is the ISM 
electron density in the CB rest system and $\eta$ is the fraction
of ISM electrons that enter the CB
and radiate there the bulk of their incident energy.  
In the early afterglow Eq.~(\ref{cbemissivity}) 
must be modified to account for the fact that  the bulk of the radio emission
by the incoming ISM electrons
is delayed by the time it takes them to cool down to energies
much lower than their initial one.
This implies that Eq.~(\ref{cbemissivity}) must
be modified by a multiplicative ``cumulation factor'' $\rm C(\nu,t)$,
which is $\approx 1$ at optical and X-ray wavelengths, as discussed in detail
in Section 5. Two other factors,  discussed in Sections 6 and 7, 
distinguish radio
waves from higher-frequency emissions: attenuation by self-absorption 
and limb darkening; they introduce two extra factors $\rm A_{_{CB}}[\nu]$
and $\rm L_{_{CB}}(\nu,\theta_{_{CB}})$, with $\rm \theta_{_{CB}}$ a
direction of emission relative to the CB's velocity vector.
Normalized as in Eq.~(\ref{cbemissivity}) and corrected by all these factors,
the afterglow energy flux density of a CB is:  
\begin{eqnarray} 
\rm F_{_{CB}}[\nu,t,\theta_{_{CB}}] &\simeq&  \rm 
               \eta\, \pi\,R_{max}^2\, n_e\, m_e\, c^3 \gamma(t)^2  
 \;f_{sync}(\nu,t)\nonumber \\  &\times & \rm
C(\nu,t)\, A_{_{CB}}[\nu]\, L_{_{CB}}(\nu,\theta_{_{CB}})\, , 
\label{Fnucb} 
\end{eqnarray} 
to be summed over $\rm n_{_{CB}}$ for a jet with that number of 
cannonballs. This expression, for $\rm\nu\gg\nu_b$ and the second
row set to unity, reproduces the optical and X-ray AG result 
discussed in DDD 2001.  

An observer in the GRB progenitor's rest system,
viewing a CB at an angle $\theta$ (corresponding to
$\rm\theta_{_{CB}}$ in the CB's proper frame), sees its radiation 
Doppler-boosted by a factor $\delta$:
\begin{eqnarray} 
\rm 
\delta(t)&\equiv&\rm
{1\over\gamma(t)\,(1-\beta(t)\cos\theta)}
\simeq {2\,\gamma(t)\over 1+\theta^2\gamma(t)^2}\; , 
\nonumber \\ \rm
\cos\theta_{_{CB}}&=&\rm
{\cos\theta-\beta(t)\over 1-\beta(t)\,\cos\theta}
\simeq{1-\theta^2\gamma(t)^2\over1+\theta^2\gamma(t)^2}
\label{doppler} 
\end{eqnarray} 
where the approximations are valid in the domain of interest for GRBs: large $\gamma$ and small $\theta$. Since the CB is catching-up\footnote{For
$\theta\gamma>1$, the observer sees the back of the CB as it is coming
forth towards her: $\rm \cos\theta_{_{CB}}<1$. The CB would actually
hit or pass by the observer before its back is unveiled, were it not for the 
fact that its motion is decelerated.}
with the radiation it emits, $\delta$ is also the relative time aberration:
$\rm dt_{obs}=dt_{_{CB}}/\delta$. The observed spectral energy 
density is modulated by a factor $\delta^3$, two powers of $\delta$ reflecting
the relativistic forward collimation of the radiation emitted in the CB's
rest system.  The AG spectral energy density $\rm F_{obs}$
seen by a cosmological observer at a redshift $\rm z$ 
(Dar and De R\'ujula, 2000a), is:
\begin{equation} 
\rm F_{obs}[\nu,t]\simeq 
\rm {A_{Gals}\, (1+z)\,\delta^3 
                    \over 4\, \pi\, D_L^2}\, 
F_{_{CB}}\left[{(1+z)\,\nu\over\delta(t)},{\delta(t)\,t\over 1+z} 
\right]\, , 
\label{Fnuobser} 
\end{equation} 
where $\rm A_{Gals}$ represents the absorption in the host galaxy and the
Milky Way,   $\rm F_{_{CB}}$
is as in Eq.~(\ref{Fnucb}), and $\rm D_L$ is the luminosity distance
(we use throughout a cosmology with $\Omega=1$ and 
$\Omega_\Lambda=0.7$).  In the CB model, the extinction
in the host galaxy may be time dependent: in a day or so,
CBs typically move to kiloparsec distances from their birthplace,
where the extinction should have drastically diminished.

In DDD 2001 we fit, in the CB model, the R-band AG light curves of 
GRBs. The fit involves five parameters per GRB:
the overall normalization;  $\theta$: 
the viewing angle; $\gamma_0$: the $\rm t=0$ value of the
Lorentz factor; $\rm x_\infty$: the ``deceleration''
parameter of the CBs in the ISM; and the
spectral index $\rm p$.  The value of $\rm p$, obtained
from the temporal shape of the afterglow, is in every case
very close to the expectation $\rm p=2.2$, and ---within
the often large uncertainties induced by absorption--- with
the observed spectra from optical frequencies to X-rays (DDD 2001).

In this paper we complete our previous work by making broad-band
fits to the data at all available radio and optical frequencies.
In so doing, we need to introduce {\it just one} new ``radio''
parameter:  an ``absorption frequency'' $\rm \nu_a$, corresponding
to unit CB opacity at a reference frequency. We set $\rm p=2.2$ so
that the extension to a broad-band analysis does not involve an
increase in the total number of parameters.  We have to refer  very
often to the values of the parameters that our previous experience
with the CB model made us choose as reference values.  For convenience,
these are listed in Table I.

\begin{figure}[t]  
\begin{tabular}{cc}  
\hskip 1.0truecm  
\vspace*{2cm} 
\hspace*{-2.5cm}  
\epsfig{file=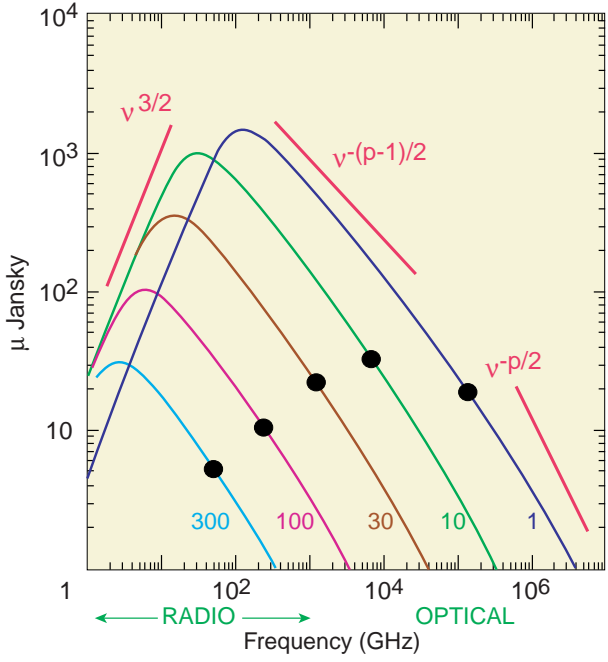, width=8cm}
\vspace*{-1.5cm}
\\ 
%\hskip 1truecm 
\hspace*{-.2cm}  
\epsfig{file=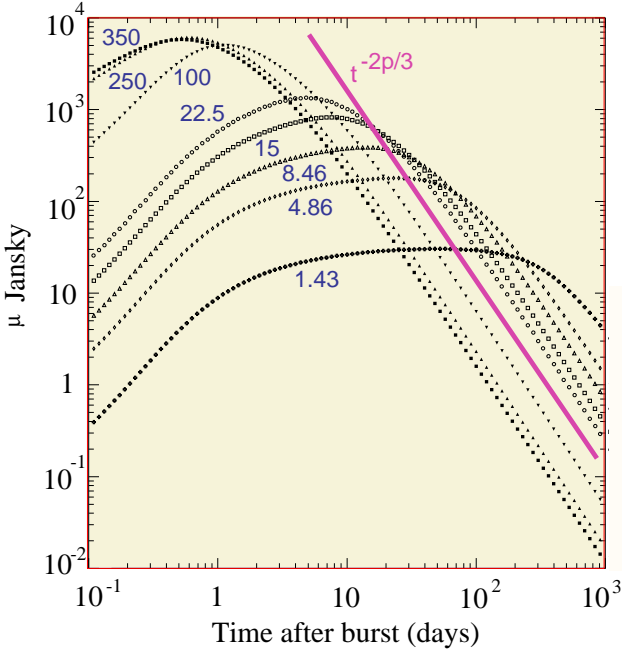, width=9.5cm} 
\end{tabular}  
\caption{Typical predictions for the CB model's radio afterglow. 
Upper panel: spectra at different times, from 1 to 300 days.
The peak frequencies correspond to CB self-opacities of $\cal{O}$(1).
The black dots are the location of the synchrotron frequency
corresponding to the injection bend.
Lower panel: Light curves at different radio frequencies,
from 350 to 1.43 GHz. The asymptotic curve is $\rm t^{-2\,(p+1)/3}$
(for $\rm \nu\gg\nu_b$, as is the case at all frequencies shown
in this figure).}
\label{figCBpreds}  
\end{figure}

The predictions of the CB model, for typical parameters,
are summarized in Fig.~(\ref{figCBpreds}). The energy density spectra 
at radio to optical frequencies are shown, at various times after
the GRB, in the upper panel. The spectral slopes before and soon after
the peak frequency are $\rm 3/2$ and $\rm -(p-1)/2$, as indicated.
The spectra peak at a frequency at which self-attenuation in the CBs
results in an opacity of $\cal{O}$(1).
At frequencies well above the frequency $\rm \nu_b$ characterizing
the injection bend, the spectrum steepens to a slope $\rm -p/2$.
In the figure's lower panel
we show light curves at various radio frequencies. At large times 
and for $\rm \nu\gg \nu_b$ ---which is the case at all frequencies
in the example of Fig.~(\ref{figCBpreds}), whose parameters
are close to those of GRB 000301c--- they
tend to $\rm t^{-2\,(p+1)/3}$, this behaviour being reached
at earlier times, the higher the frequency. 
For $\rm \nu\ll \nu_b$, the corresponding limiting behaviour
is $\rm \approx t^{-4/3}$, observable at low frequencies
in the cases of GRBs 991216, 991208 and 000418. All of the above
predictions are robust: they do not depend on the
detailed form of the attenuation, cumulation and limb-darkening
factors. The early rise
of the light curves does depend on such details, on which
we shall have to invest a disproportionate effort in Sections 5 to 7.

%  estaba la primera figura 

The CB model provides an 
excellent description of the data, as discussed in Sections 4, 8 and 9.
In the case of GRB 980425, for which the optical AG is dominated
by SN1998bw, we used the parameters that fit
its X-ray afterglow (DDD 2001) and the GRB's fluence (Dar and De R\'ujula
2000a) to argue that they are not exceptional.
The CB-model's description of the
radio data for this GRB/SN pair is excellent: there is nothing peculiar
about this GRB, nor about its associated supernova, as we discuss 
in detail in Section 9,  along with the question of the
angular separation in the sky of the SN and the associated CBs,
which may have been, or may still be, observable.

The apparent sky velocities of cosmological CBs are extremely superluminal
and their angular velocities happen to be of the same order of magnitude
as those of galactic pulsars. This implies that CB velocities can possibly
be extracted from their observed radio scintillations, as discussed in
Section 12.

\section{The  electron spectrum}

\subsection{Numerical simulations}

The acceleration of charged particles by a moving CB is not substantially
different from some of the cases already studied in the literature,
the acceleration of cosmic rays and electrons having
attracted an enormous amount of attention since Fermi's first
analysis in 1949 (for an excellent introduction, see Longair 1994).
The most efficient and thus promising mechanism is the ``first-order'' 
acceleration of particles by fast-moving shocks, extensively
studied analitically  and numerically since the pioneering
works of  Axford et al. (1977), Krymsky (1977), Bell (1978) and Blandford
and Ostriker (1978). The analysis closest to the case at hand
is that of Ballard and Heavens (1992), who studied
acceleration by relativistic shocks, with the charged particles 
deflected by highly disordered magnetic fields, rather than, as it is 
generally assumed, by small irregularities in an otherwise uniform field.
The ``relativistic'' and ``chaotic'' inputs are what make this work
particularly relevant to the case of particle acceleration by 
and within CBs.

Ballard and Heavens study numerically, for various values of a
moving discontinuity's Lorentz factor ranging up to $\rm\gamma_s=5$,
the result of its collision with an isotropic ensemble of particles
with $\rm \gamma_p=100$. They find that, for $\rm\gamma_s=5$,
the resulting particle energy distribution has a break (in this reference
system) at $\rm\gamma\sim 10\, \gamma_p$, at which point it steepens.
The particles below the break have a dominantly
very forward motion: they are the ones which have been upscattered
just once. Given this hint, it is easy to reproduce the numerical
results in an analytic approximation. In the shock's rest system,
the energy of the particles that have been scattered only once
is equal to their incoming energy: the break in the spectrum
seen in the simulations is a kinematical break occurring roughly
at the injection energy. ``Observed'' in the system in which the
shock is travelling at $\rm\gamma_s=5$, this {\it injection bend}
is very reminiscent of the familiar synchrotron-cooling ``break'',
but it has little to do with it; indeed, in the simulations of Ballard and 
Heavens (1992) cooling was entirely neglected.

\subsection{A simple analysis}

Consider the CB in its rest system and temporarily postpone the
discussion of cooling. The ISM electrons impinge on the CB in a fixed direction 
with a Lorentz factor equal to that of the CB in the GRB progenitor's rest system,  
$\rm\gamma_e=\gamma(t)$. The electrons not having ``bounced back'' 
off the CB's strong magnetic field, or having
done it only once, retain the incoming energy, $\rm E_b(t)=\gamma(t)\,m_e\, c^2$,   
so that their energy distribution is:
$\rm dn_e/dE\propto \delta[E-E_b(t)]$.
A very robust (i.e. detail independent) feature of the studies
of acceleration by relativistic shocks is that the particles
having bounced more than once acquire a spectrum 
$\rm dn_e/dE\propto E^{-p}$,
with $\rm p=2$ in analytical approximations and $\rm p\sim 2.2$
in numerical simulations. A few bounces are sufficient to attain
such a spectrum. The CB is a system of finite transverse dimensions
and the magnetic field contrast between its interior and its exterior
is very large. Thus, we do not expect the same electron to bounce
many times off the CB, as the latter catches up with it. The
acceleration should occur mainly within the CB as charged particles
bounce off its chaotically moving magnetic domains, and  it should be very
fast and efficient, since the injection is highly relativistic and
there is no distinction between ``first and second order Fermi'' processes.
The overall ``source''
spectrum of relativistic electrons is:
\begin{eqnarray}
\rm {dn_e^s\over dE}&\sim&\rm A_1(t)\,E_b(t)\,\delta[E-E_b(t)] 
\nonumber\\ &+& \rm
A_2(t)\,\Theta[E-E_b(t)] \left[{E\over E_b(t)}\right]^{-p}\!\!\! ,
\label{source}
\end{eqnarray}
with $\rm A_1$ and $\rm A_2$ of comparable magnitude and a time 
dependence which is that of the rate, 
$\rm\eta\,\pi\,R_{max}^2\,c\,n_e\,\gamma(t)$,
 at which electrons enter the CB.

\subsection{The spectrum of cooled electrons} 

The electron energy loss by synchrotron radiation is:
\begin{eqnarray}
\rm -{dE\over dt}&=&\rm A_S\,\beta^2\,E^2 ,\nonumber \\ \rm
A_S &\equiv& \rm 
{B^2\over 6\,\pi}\,{\sigma_T\,c\over (m_e\,c^2)^2}\, ,
\label{Eloss}
\end{eqnarray}
with $\beta\approx 1$ for the relativistic energies of interest and
$\rm \sigma_T=0.665$ barn the Thomson cross-section.
Let the rate at which fresh electrons are supplied by the ISM be called $\rm R$.  
The electron source distribution of Eq.~(\ref{source}) ``ages'' by cooling so that:
\begin{equation}\rm
{\partial\over\partial t}\left[{dn_e\over dE}\right]=
{d\over dE}
\left[{dE\over dt}\,{dn_e\over dE}\right]+
R\,{dn_e^s\over dE}\, .
\label{nequilib}
\end{equation}
At times longer than the synchrotron cooling time, the electron
distribution tends to a time-independent $\rm {dn_e/ dE}$,
obtained by equating to zero the l.h.s. of Eq.~(\ref{nequilib})
and integrating it with the source function of Eq.~(\ref{source}):
\begin{eqnarray}
\rm {dn_e\over dE}&\sim&\rm A_1(t)\,\Theta[E_b(t)-E]\,{1\over E^2} 
\nonumber\\
&+& \rm {A_2(t) \over p-1}\,\Theta[E-E_b(t)]\,\left[{E\over 
E_b(t)}-1\right]^{-(p+1)}\! . 
\label{result}
\end{eqnarray}

Admittedly, the process of acceleration that we have discussed is
not well understood, our derivation is heuristic and Eq.~(\ref{result})
is not even a continuous function (the step function in Eq.~(\ref{source})
should not be so abrupt, the magnetic energy in Eq.~(\ref{Eloss}) 
should not have a fixed value). All we want to conclude from this
exercise is that,  when the probability of
an electron to have been ``kicked'' only once is not negligible
($\rm A_1$ comparable to $\rm A_2$), 
the electron spectrum  has an injection bend at
$\rm E\sim E_b(t)$, around which its spectral index changes by 
$\sim 1$
from $\rm \sim 2$ to $\rm \sim p+1$. We choose to characterize this
behaviour by the function:
\begin{equation}
\rm  {dn_e\over dE} \propto {E^{-2}\over \sqrt{1 + [E/E_b(t)]^{2\,(p-1)}}}\, .
\label{approx1}
\end{equation}
Note how similar the injection bend is to a cooling break
(also a spectral steepening by roughly one unit)
even though their origins are so different.
The observational evidence for an injection bend at the
injection energy turns out to be strong, as we proceed to show.

\section{Evidence for an injection bend}

The injection bend induces the gradual transition
in the spectral energy distribution described by Eq.~(\ref{sync}), occurring
at a ``bend'' frequency:
\begin{equation}
\rm \nu_b \simeq {1.87\times 10^{15} \over 1+z}\, 
{[\gamma(t)]^3\, \delta(t)\over 10^{12}}\,
\left[{n_p\over 10^{-3}\;cm^3}\right]^{1/2}
Hz,
\label{nubend}
\end{equation}
where we have used the characteristic synchrotron frequency 
of Eq.~(\ref{nub}) for
the magnetic field of Eq.~(\ref{mag}), and transposed the result to the 
observer's frame.

%estaba la segunda figura

\begin{figure}[t]  
\begin{tabular}{cc}  
\hskip 2truecm  
\vspace*{2cm} 
\hspace*{-2cm}  
\epsfig{file=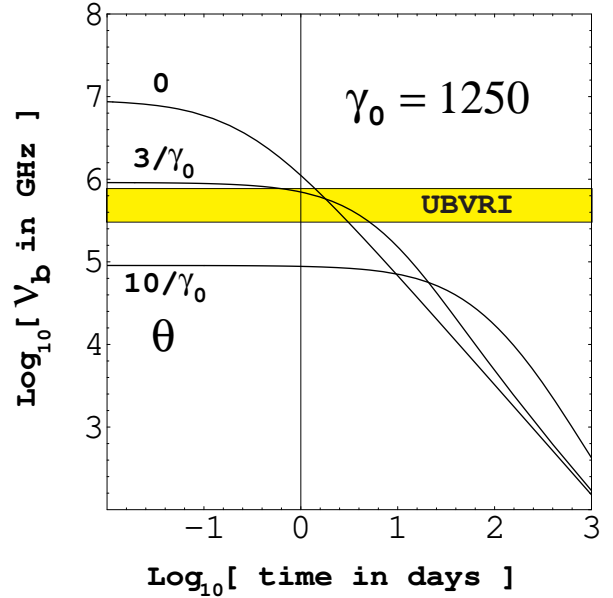, width=8cm}
\vspace*{-1.5cm}\\ 
%\hskip 1truecm 
\hspace*{.2cm}  
\epsfig{file=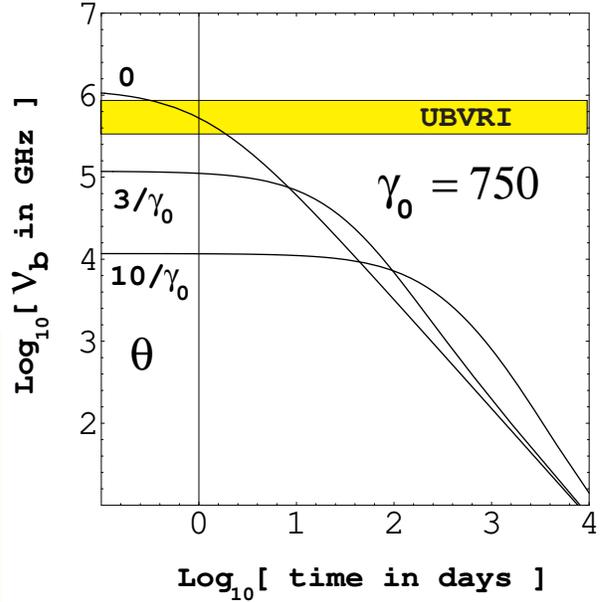, width=8cm} 
\end{tabular}  
\caption{Typical predictions for  the bend frequency in the AG spectrum
as a function of time, for $\theta=0,\,3/\gamma_0$
and $10/\gamma_0$. The ``optical'' U to I band is shown as a horizontal
band. Upper panel: for $\gamma_0=1250$. Lower panel: for $\gamma_0=750$.}  
\label{figinjection}  
\end{figure}

For the reference CB parameters and $\rm z=1$,
 $\rm\nu_b(t=0)\simeq 0.93\times 10^{15}$ Hz,
above the optical band. Since the product $\gamma^3\, \delta$ typically
declines by more than an order of magnitude within a couple of days,
the bend frequency in many GRBs
crosses the optical band into the NIR during the early afterglow.
In Fig.~(\ref{figinjection}) we present the time dependence of 
$\rm\nu_b(t)$  for $\gamma_0=1250$ and 750, characterizing the range
of the observations, for various angles $\theta$, $\rm z=1$,
and the rest of the parameters at their reference values of Table I. 
The figures show that, depending on the parameters, the bend frequency
in the early AG may be above or below the optical band, and, if it is above,
it will cross it later.

The bend frequency of the CB model is not the
break frequency of the traditional fireball model. 
The time evolution of the former is given by Eq.~(\ref{nubend}),
and is different from that of the latter, which,  
prior to the ``break'' in the AG light-curve, can be shown to be
$\rm t^{-1/2}$ (Granot and Sari, 2002).

The evolution predicted by Eq.(~\ref{sync}) from a $\nu^{-0.5\pm 0.1}$ 
to a $\nu^{-1.1\pm 0.1}$ spectral behaviour is affected by extinction.
The early behaviour corresponds to times when CBs are not yet
very far from their progenitors:  extinction in the host galaxy
may steepen the spectrum. After a day or more, when the CBs
are further away, we do not expect strong extinction in the host.
So the prediction (after extinction in the Galaxy is corrected for)
is an evolution from a behaviour close to ---or steeper than---
$\nu^{-0.5\pm 0.1}$, to a more universal $\nu^{-1.1\pm 0.1}$ at
later times.

The predicted spectral behaviour has been  observed,
with varying degrees of significance, in
the AG of several GRBs, listed in Table II. 
The first column is the bend frequency $\rm \nu_b^0$ at $\rm t=0$,
computed with Eq.~(\ref{nubend}) and the optical AG parameters of 
Table III
(the density $\rm n_p$ is extracted from the measured
$\rm x_\infty$ with use of Eq.~(\ref{range})
and our reference $\rm R_{max}$ and $\rm N_{_{CB}}$). For the listed
GRBs the bend frequency is above the visible band at $\rm t=0$ and the
early AG measurements result in effective spectral slopes, $\rm \beta(t_1)$,
not far from the expectation $-0.5\pm 0.1$, or somewhat steeper. 
A few days later, the measured
values, $\rm \beta(t_2)$, are compatible with the expectation ${-1.1\pm 0.1}$.
The second entry on GRB 990510 in Table II
(Beuermann et al. 1999)
requires an explanation. These authors argue that  
$\rm \beta(t_2)=0.55\pm 0.10$, a result that assumes a strong
extinction correction in the host galaxy. But, after a day or so,
we do not expect such an extinction. For the latest points measured by  Beuermann et al. (1999),
at day 3.85 (well after the bend), $\rm B-R=0.98\pm 0.07$
and $\rm R-I=0.49 \pm 0.06$. Converting these results ---without
extinction--- to a spectral slope yields $\rm \beta(t_1)=1.11\pm 0.12$,
in agreement with  expectation.

\begin{figure}[t]  
\hskip 0truecm   
\epsfig{file=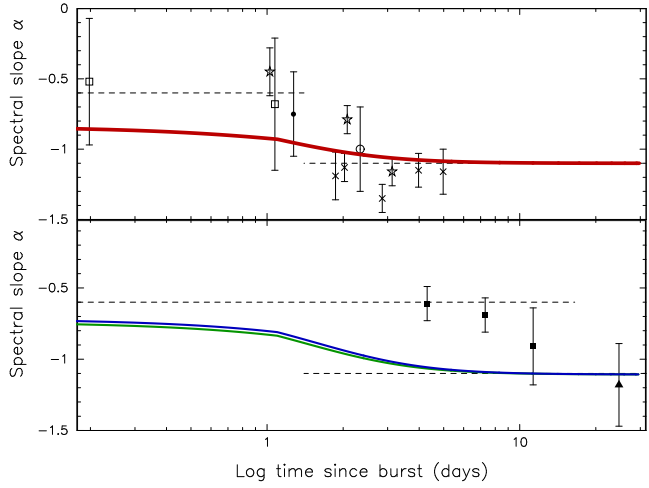, width=8.5cm}    
\vspace*{-0.1cm}
\caption{A comparison between the predicted evolution in time
of the effective spectral slope through the optical/NIR band and
the data collected by Galama et al. (1998a) for
the U, B, V, $\rm R_c$ $\rm I_c$ band of the AG of GRB 970508
(upper panel), for the K and $\rm R_c$ band  (full squares,
lower panel, Chary et al. 1998) and for the H and $\rm R_c$ band  (triangle,
lower panel, Pian et al. 1998) The three coloured lines, in the same order,
are the (parameter-less) predictions.}  
\label{index2}  
\end{figure}

The evolution from a
softer to a harder spectrum should be a gradual change in time, rather
than a sharp break, so that an AG's optical spectrum, if ``caught'' as
the injection bend is ``passing'' should have an index evolving
from $-0.5\pm 0.1$ to ${-1.1\pm 0.1}$ with the time dependence
described by  Eqs.~(\ref{sync}) and (\ref{nubend}).
In Fig.~(\ref{index2}) we test this prediction in the case of GRB 970508, for 
the time-dependent value of the ``effective'' slope
 $\rm \alpha \simeq \, \Delta\, [log\, F_\nu]/\Delta\,[log\,\nu]$, 
constructed from the theoretical expectation in the
same frequency intervals used by the observers.
The actual predicted $\rm \nu_b(t)$ in Eq.~(\ref{nubend}) is obtained by use of
the optical-AG fitted parameters ($\theta$, $\rm \gamma_0$ and
$\rm x_\infty$) that determine $\rm \gamma(t)$ and $\rm\delta(t)$, and
the density $\rm n_p$  deduced\footnote{GRB 970508 has a peculiar
AG, whose CB-model interpretation requires an ISM density
change at $\rm t\sim 1.2$ observer's days (DDD 2001).} from
$\rm x_\infty$ and the reference $\rm N_{_{CB}}$ and $\rm R_{max}$.
The data are gathered by Galama et al. (1998a) from 
observations in the U, B, V, $\rm R_c$ and $\rm I_c$ bands
(Castro-Tirado et al. 1998, Galama et al. 1998b;
Metzger et al. 1997; Sokolov et al. 1998; Zharikov et al. 1998), 
by Chary et al. (1998)
for K band results, and by Pian et al. 1998 for the H band.

In spite of considerable uncertainties in the 
spectral slopes deduced from observations (Galama et al. 1998a), the
results shown in Fig.~(\ref{index2}) are satisfactory: the observed crossing
of the injection bend is in agreement with the theoretical {\it prediction},
based on the fit in DDD 2001 to the overall R-band light curve from which
the GRB 970508 AG parameters have been fixed; 
{\it no extra parameters have been fit.}
A couple of points in the lower panel do not agree with the prediction,
but they do not agree with the observations at very nearby frequencies reported
in the upper panel, either.

% estaba la tercera figura

\begin{figure}[t]  
%\begin{tabular}{cc}  
\hskip 0truecm   
\epsfig{file=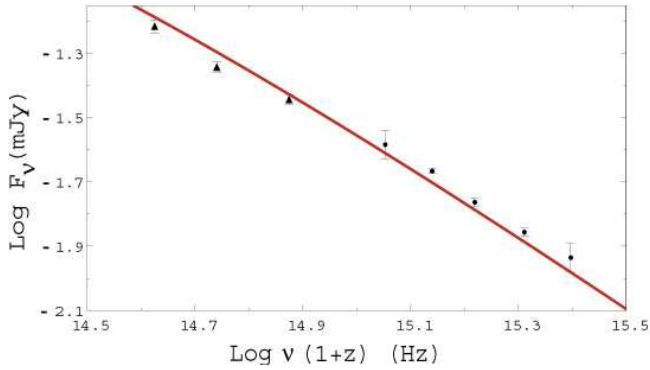, width=8.5cm}  
%\end{tabular}  
\caption{Comparison between the observations and the
(parameter-less) prediction
for the spectral {\it shape} of the optical AG of GRB 000301c, at 
$\sim 3$ days after burst. Data from Jensen et al. (2001).}  
\label{index}  
\end{figure}

A complementary analysis to that in the previous paragraph is the
study of an AG's optical spectrum at a fixed time at which
the injection bend is crossing the observed frequency range, 
or is nearby. A spectral ``snapshot'' at such time should have
the intermediate slope
and curvature described by Eq.~(\ref{sync}) for $\rm\nu\sim\nu_b$.
To test this prognosis, 
we compare in Fig.~(\ref{index}) the predicted spectral 
shape of the optical/NIR AG of GRB 000301c around March 4.45 UT 
($\sim 3$ days after burst) to its measured shape (Jensen et al. 2001).
We have selected this GRB because its extinction correction in the galactic ISM
is rather small: $\rm E(B - V)=0.05$ (Schlegel et al. 1998),
and there is no evidence for  significant extinction in the
host galaxy (Jensen et al. 2001). The theoretical line in Fig.~(\ref{index})
is given by Eq.~(\ref{sync}) with the observer's $\rm \nu_b$
of Eq.~(\ref{nubend}) ($\rm \nu_b\,(1+z)=1.75\times 10^{14}$ Hz at
$\rm t=3$ days, for the density deduced from the value of $\rm x_\infty$
of this GRB, and the reference values of $\rm N_{_{CB}}$ and $\rm R_{max}$).
In the figure the theory's normalization is arbitrary but the (slightly
evolving) slope of the theoretical curve is an absolute {\it prediction}:
it is based on the fit in DDD 2001 to the overall R-band light curve and,
once more, {\it no extra parameters have been fit}.
The result is astonishingly good, even for the curvature which 
---given the figure's aspect ratio as chosen by 
the observers--- is not easily visualized (a look at a slant angle helps).  
The late-time spectral slope deduced 
from the HST observations (Smette et al. 2001) around day 
33.5 after burst indicated a slope of $\sim -1.1$, again in agreement
with our expectation. 

% estaba la cuarta figure

We conclude that the evidence is very strong for a spectral injection bend at
the time-dependent frequency, Eq.~(\ref{nubend}), predicted in the CB model.
As illustrated in Fig.~(\ref{figCBpreds}) and contrasted with data
in Section 8, further evidence for the injection bend is provided
by the fact that it is essential to the description of the
observed broad-band spectra of GRB afterglows.

\section{The cumulation factor}

Three factors that are irrelevant in the optical and X-ray domains
play a role in the description of the longer radio wavelengths
and the early radio AG. In this section we discuss the first one of them.

Electrons that enter a CB with an injection Lorentz factor  $\rm \gamma(t)$
are rapidly Fermi accelerated to a distribution that we have argued
to be roughly that of Eq.~(\ref{source}). On a longer time scale, they
lose energy by synchrotron radiation, and their
energy distribution evolves as in Eq.~(\ref{nequilib}). 
Electrons with a large $\rm \gamma\sim {\cal{O}}\,[\gamma(t)]$ emit
synchrotron radiation, with no significant time-delay, 
at the observer's optical and X-ray wavelengths.
But the emission of radio is delayed by the time it takes the electrons
to ``descend'' to an energy at which their characteristic emission
is in the observer's radio band. At the start of the afterglow, when 
equilibrium conditions have not yet been reached, this implies a
dearth of radio emission relative to the higher-frequency bands.
This introduces a ``cumulation factor'' $\rm C(\nu,t)$ in Eq.~(\ref{Fnucb}).

Consider a fixed observed radio frequency $\rm \nu_{obs}$. 
It corresponds to a time changing frequency 
$\rm \nu=(1+z)\nu_{obs}/\delta(t)$ in the CB system.
The CB electrons preferentially emitting at this frequency (over an
unconstrained range of pitch angles) are those
whose Lorentz factor $\rm\gamma_e$ satisfies the relation
$\rm \nu \sim 0.22\,\gamma_e^2\,\nu_L$, in analogy to Eq.~(\ref{nub}). To
estimate\footnote{We can solve Eq.~(\ref{nequilib}) exactly for a given
source spectrum by the Mellin transform
methods so familiar in Quantum Chromodynamics, but this would 
be unjustified: the acceleration mechanism is not understood well
enough for the study of exact cooling solutions to be currently justifiable.} 
the time $\rm \Delta t$ it takes an electron to decelerate from
$\rm\gamma\sim \gamma(t)$ to $\rm\gamma=\gamma_e$, substitute the magnetic
energy density of Eq.~(\ref{emag}) into the electron energy loss of Eq.(\ref{Eloss})
and integrate, to obtain
\begin{eqnarray}\rm
\Delta t&=&\rm {3\,m_e\over n_p\,m_p\,\sigma_T\,c}\,{1\over\gamma^2}\,
\left({1\over \gamma_e}-{1\over \gamma}\right) \nonumber \\
&=&\rm [8.27\times 10^7\,s]\left[{10^{-3}\,cm^{-3}\over n_p}\right]
{10^6\over \gamma^2}\,
\left({1\over \gamma_e}-{1\over \gamma}\right)\, .
\label{Deltat}
\end{eqnarray}
The function $\gamma$ is given by Eq.~(\ref{gammacb}) of Appendix I,
which we may rewrite as:
\begin{eqnarray}\rm
{1\over \gamma^2}&=&\rm{1\over \gamma_0^2}\left[1+{t\over t_0}\right]
\nonumber\\ \rm
t_0&\equiv&\rm [5.14\times 10^7\, s]
    \left[{x_\infty\over 1\;Mpc}\right]
\left[{10^3\over \gamma_0}\right]^2
\label{1overg2}
\end{eqnarray}

The electrons emitting the observed radio frequencies have 
$\rm\gamma_e\!\sim\! {\cal{O}}(1)$, so that the proper CB times 
$\rm\Delta t$ and $\rm t_0$ are of ${\cal{O}}(1)$ year,
corresponding to observer's times ---foreshortened
by a factor $\rm(1+z)/\delta$--- of ${\cal{O}}(1)$ day. 
For optical and radio observations $\rm\gamma_e\!\sim\! {\cal{O}}(\gamma)$
there is no significant delay in their emission. Moreover, the 
electron accumulation rate ($\rm \eta \pi\,R_{max}^2\,n_e\,c\,\gamma$ in the
CB system)  is orders of magnitude larger than the characteristic
synchrotron cooling time $\rm E/(dE/dt)$ of Eq.~(\ref{Eloss}), even for
$\gamma\sim 10^3$. Thus, the optical and X-ray AG emission
starts as soon as the CB is transparent to its enclosed radiation: for 
each CB, a few observer's seconds after the corresponding $\gamma$-ray
pulse (DDD 2001). The radio signal, on the other hand, must await 
a time $\rm \Delta t$ for the
cumulated electrons to cool down. 

The simple way to parametrize the frequency-dependent
``cumulation effect'' is to use the expression for the total number
of electrons $\rm N(t)$ incorporated by the CB up to time $\rm t$ 
(Eq.~(\ref{acctot}) of Appendix I) and to posit\footnote{The rapid onset
of the radio signals from SNe is not understood, see e.g.
Weiler et al. 2001. Perhaps electron cumulation also plays a role there.}:
\begin{equation}\rm
C(\nu,t)={N(t-\Delta t)\over N(t)}\,\Theta(t-\Delta t),
\label{cumul}
\end{equation}
where the frequency dependence is via $\rm \Delta t=\Delta t(\nu,t)$,
and the sharp start at $\rm t>\Delta t$ is an artifact of our simplifications.
For optical and X-ray frequencies, $\rm \Delta t=0$ and $\rm C(\nu,t)=1$.
In practice we find that, except for GRB 980425 whose viewing angle
is exceptionally large, one may also use {\it within} the radio band an
approximation to Eq.~(\ref{cumul}): 
\begin{equation}\rm
C(\nu,t)\sim C(t)=\left[1-{\gamma(t)\over\gamma_0}\right]^{1/2},
\label{cumulapp}
\end{equation}
which is also frequency independent.

\section{The attenuation factor}

At optical and X-ray frequencies the CB is transparent and, for the spectrum
of Eq.~(\ref{sync}), the bulk of the radiation's energy is emitted
around the bend frequency $\rm\nu_b$. At such relatively high
frequencies, as illustrated in Fig.~(\ref{figCBpreds}), absorption is 
unimportant. Thus, for optical and
X-ray afterglows (DDD 2001) it suffices to know that all of the incoming
electron's energy is reradiated, the spatial distribution of the
radiating electrons within the CB is irrelevant. But in the radio,
where absorption is important, the location of these electrons
inevitably plays a role. In the next sections
we argue that it is plausible
that the radiating electrons be close to the surface ``illuminated''
by the ISM (\S 6.1), and that the values of the CB's plasma frequency (\S 6.2)
and free-free absorption coefficient (\S 6.3) actually suggest that they may 
be relatively close to that surface. In \S 6.4 we deduce the final form
of the attenuation factor in the CB model, characterized by a single parameter.

\subsection{Electron penetration}

Using numerical simulations, Achterberg al. (2001) have shown 
that for simple geometries  the bulk of highly relativistic particles 
encountering a collisionless shock escape before
they undergo diffusive shock acceleration.  In reality, the
geometry of the CB, its density distribution and its magnetic field 
distribution are very complicated, making  
the fraction of the ISM electrons that penetrate inside the CB, and
their distribution there, very uncertain.

Several length scales play a role in discussing the fate of an electron that 
enters the CB with $\rm\gamma_e=\gamma(t)$. The Larmor radius  
is $\rm R_L=m_e\,c^2\,\gamma/(e\,B)$,
which is independent of $\gamma$ for $\rm B$ scaling as in
Eq.~(\ref{mag}). For our reference parameters, $\rm R_L\sim 6$ km is many
orders of magnitude smaller than the CB's radius and does not play
a crucial role. The length of an electron's curled-up trajectory as 
it radiatively loses energy  is $\rm c\,E/(dE/dt)$ or $\sim 2.6\times 
10^{15}$ cm for the cooling rate of Eq.~(\ref{Eloss}) and an initial 
$\gamma=10^3$. This is only an order of 
magnitude larger than the reference CB's radius $\rm R_{max}$.
We have no way to estimate the typical coherence size of a CB's 
magnetic domain $\rm L_B$, but the depth 
$\rm D\sim (c\,\tau_\gamma\;L_B)^{1/2}$ to which an electron penetrates, 
even for a relatively simple magnetic mess ($\rm L_B$ not much smaller than
$\rm R_{max}$) is smaller than the CB's radius.
For $\rm L_B$ as small as $\rm R_L$,
$\rm D\sim 4\times 10^{10}$ cm, some four orders of magnitude smaller than 
$\rm R_{max}$. Even this concrete value is uncertain, for it depends on the
surface magnetic field as $\rm B^{-3/4}$, and  the surface $\rm B$-value
may be different from that of Eq.~(\ref{mag}), which is a volume average.

In addition to all of the above uncertainties, it is possible that a CB's
illuminated working surface be turbulent, and harbour fast plasma motions, 
if only to establish local charge neutrality, which is disrupted as electrons
and protons penetrate the CB to different depths. 
We conclude that the the fraction of ISM electrons that enter inside
the CB {\it may} be small  and the synchrotron-radiating
electrons {\it may} be concentrated close to the CB's surface, as opposed
to be acquiring a uniform distribution over the CB's 
volume. 

\subsection{The plasma frequency}

The plasma frequency in a CB with an average free inner electron density
$\rm \bar n_e^{free}$ is: 
\begin{equation} 
\rm \nu_p = \left[{\bar
n_e^{free}\, e^2\over\pi\, m_e}\right]^{1/2}\simeq 28\, \left[{\bar
n_e^{free}\over 10^7\;cm^{-3}}\right]^{1/2}\, MHz. 
\label{plasma}
\end{equation} 
For a fully ionized CB $\rm \bar n_e^{free}=\bar n_e$, to
whose reference value we normalized the above result (the fraction of
electrons swept up from the ISM is small, relative to the total number in
the CB --to whose free fraction Eq.~(\ref{plasma}) refers--
but the CB is highly ionized, as shown in Appendix III). 

For $\rm \nu\!<\!\nu_p$
the radio emission is completely damped within a typical 
length $\rm \sim c/[\nu_p^2-\nu^2]^{1/2}$, much smaller than the CB's radius.
At very early times, $\delta\sim 10^3$ and the Doppler-boosted value 
of $\rm \nu_p$ falls in the low end of the observed range of radio signals,
where a sharp cutoff is not observed. We must conclude that the (small
fraction of)
radiating electrons is located in a CB surface layer whose total
electron density (dominated by the thermal electron constituency) is 
smaller than our reference average value, 
a one order of magnitude reduction being
comfortably sufficient to move the value
of $\rm \nu_p$ to a position below the currently observed frequencies.
We have explicitly checked that our 
fits do not improve significantly with the inclusion of $\rm \nu_p$ 
as a free parameter: the minimization procedure always ``gets rid''
of the fit $\rm \nu_p$ by choosing it somewhat below the reference value of
Eq.~(\ref{plasma}), and below the lowest measured frequencies.

\subsection{Free-free attenuation}

At the MHz frequencies in the CB system corresponding to
the observed radio frequencies, the 
synchrotron emission is strongly attenuated by free-free absorption 
(inverse bremsstrahlung) in the CB; free-free absorption dominates over self-synchrotron absorption, as shown in Appendix 4. For a 
hydrogenic plasma, the free-free
absorption coefficient at radio frequencies is: 
\begin{equation} 
\rm \chi_\nu \simeq 0.018\, g_{ff}\,\bar n_e\, \bar n_i\, T^{-3/2}\,   
\nu^{-2}\, cm^{-1}\, , 
\label{chifree} 
\end{equation} 
where $\rm \bar n_i\simeq \bar n_e$ is the free ion 
density in the CB, in units of $\rm 
cm^{-3}$, $\rm T$ is the plasma temperature in degrees Kelvin,
$\nu$ is in Hertz
and $\rm g_{ff}$ is a Gaunt factor for free-free emission, of $\cal{O}$(10)
at the relevant frequencies.

The opacity $\tau_\nu$ of a surface layer of depth $\rm D$ is:
\begin{equation} 
\rm \tau_\nu(D,t) =\int_{R_{max}-D}^{R_{max}}\chi_\nu \, dr\equiv 
\bar \chi_\nu[t]\,D\, . 
\label{taufree} 
\end{equation}
Equilibrium between photoionization of atomic hydrogen in the CB by 
synchrotron radiation and 
recombination of free electrons and protons to hydrogen 
keeps the CB plasma partially ionized during the observed AG. 
The Coulomb relaxation rate in the CB is very fast because of its 
high density. Consequently, the CB plasma is 
approximately in quasi thermal equilibrium. Because of the exponential 
dependence of the Saha equation 
on temperature, and the high ionization rate, the CB's temperature is kept 
practically constant around a few eV, and the ion density and free 
electron density become proportional to $\rm \gamma(t)$ (Appendix III).  
Using $\rm T_0= 10^5\, K$  and the reference
average densities $\rm \bar n_e=\bar n_i\simeq 10^7\, cm^{-3}$ 
in Eqs.~(\ref{chifree}) and (\ref{taufree}), we obtain:
\begin{equation}
\rm
\tau_\nu(D,t)\sim 1.4\times 10^2 \left[{D\over R_{max}}\right]
\left[{1\;GHz\over \nu}\right]^2 \,
\left[{T\over 10^5}\right]^{-3/2}\!\! ,
\label{taubis}
\end{equation}
which is very large for $\rm D\sim {\cal{O}}(R_{max})$. A reduction in surface
or average CB
density of one order of magnitude or more ---which, as we have seen, renders
unobservable the unobserved plasma-frequency cutoff---
reduces $\tau_\nu$ by two orders of magnitude, or more. 
For $\rm D\sim R_{max}/10$, this would make $\rm \tau(D)$,
as required, of order unity at the peak frequency 
$\sim 10^2$ GHz of the early-time spectrum of Fig.~(\ref{figCBpreds}) 
(at which time the observed and CB frequencies differ by a factor 
$\delta/(1+z)\sim 10^3$).

The conclusion is that a reasonable deviation of the properties
of the CB from their reference bulk average values (a reduction
of the total number-density of free electrons in 
a synchrotron-emitting surface layer) implies, not only that the
plasma-frequency break is not observable in the current data, but also
that the magnitude of the free-free attenuation is the required one.
 Our ignorance of the depth, temperature and density of ions and
electrons in the radio-emitting surface of a CB can be absorbed into
a single parameter: a characteristic absorption frequency,
$\rm \nu_a$, in the opacity of 
Eqs.~(\ref{chifree},  \ref{taufree}, \ref{taubis}):
\begin{equation}\rm
\tau_\nu\equiv\left[{\nu\over\nu_a}\right]^{-2}\,\left[{\gamma(t)\over
\gamma_0}\right]^{2}.
\label{absfreq}
\end{equation}
The frequency dependence of the free-free attenuation, 
$\chi_\nu\propto\nu^{-2}$, is fairly well supported by the observed
radio spectra at their lowest frequencies, as our comparisons
with observations in Sections 8 and 9 demonstrate.

\subsection{Attenuation in slabs and spheres}

We do not know a priori  the geometry  of the working surface
from which a CB's synchrotron radiation is emitted. In the case of
optical AGs this is immaterial, for the CB is transparent to radiation
at the corresponding CB-system wavelengths: the bulk of the radiation
energy is emitted at these frequencies.
For the case of radio AGs, attenuation is important
and the shape of the emitting surface layers plays a role: the expression
for attenuation as a function of opacity is geometry-dependent.

For a planar-slab geometry, the familiar expression for the attenuation is:
\begin{equation} 
\rm 
A[\nu]={1-e^{-\tau_\nu}\over \tau_\nu}\, . 
\label{Attslab} 
\end{equation} 
For the emission from a sphere of constant properties, we obtain:
\begin{equation} 
\rm 
A[\nu]= {3\, \tau_\nu^2-6\, \tau_\nu+6\, [1-e^{-\tau_\nu}]\over 
\tau_\nu^3}\, , 
\label{Attsphere1} 
\end{equation} 
while for the emission from a thin spherical surface, the result
of Eq.~(\ref{Attslab}) is recovered.

For the sake of definiteness, we adhere to CBs that are spherical
in their rest system. This means that, as the frequencies increase
and the CB evolves from being opaque to being transparent, we
should use an attenuation evolving from Eq.~(\ref{Attsphere1})
to Eq.~(\ref{Attslab}). Rather than doing that, we have checked
explicitly that our results are insensitive to the use of one
or the other form, and used the simpler one.

\section{The illumination and limb-darkening factor}

Consider a spherical CB in its rest system. It is ``illuminated'' by  
incoming ISM electrons only in its ``front'' hemisphere. If observed
at an angle $\rm\theta_{_{CB}}\neq 0$, a fraction of the ``dark'' CB
is also exposed to the observer, like the Moon in phases other than
totality. For radio waves ---to which the CB is not transparent--- these
geometrical facts play a non-trivial role.

Place the direction of the CB motion, or of its illumination, at
$(\theta,\phi)=(0,0)$; at a direction $\rm \vec n_i=(0,0,1)$ in Cartesian
coordinates. The normal to a sphere's surface point at $(\theta,\phi)$
is $\rm \vec n_s=(\cos\theta\sin\phi,\sin\theta,\cos\theta\cos\phi)$. 
The observer is in the direction $\rm (0,\theta_{_{CB}})$, where we have taken
the liberty to label ``$\theta$'' what in this parametrization is 
an azimuthal angle; the corresponding
unit vector is $\rm \vec n_{_{CB}}=(\sin\theta_{_{CB}},0,\cos\theta_{_{CB}})$.
The relation between $\rm \theta_{_{CB}}$ and the terrestrial observer's
viewing angle is that of Eq.~(\ref{doppler}).

When attenuation plays a significant role, an element of a CB's 
surface reemits an amount of energy proportional to the cosine of
the illumination angle: $\rm \vec n_i \cdot \vec n_s$.
Because of the limb-darkening effect, the reemitted radiation
depends on the cosine of the angle between the surface element
and the observer: $\rm \vec n_s \cdot \vec n_{_{CB}}\equiv \mu$.
A simple characterization of the functional form of the limb darkening effect
(see e.g. Shu, 1991) is:
\begin{equation}\rm
F(\mu)=\left({2\over 5}+{3\over 5}\,\mu\right)\,\Theta[\mu]\, .
\label{limb}
\end{equation}
The combined effect of illumination and limb darkening is an emitted
radiation proportional to:
\begin{eqnarray}\rm
E(\cos\theta_{_{CB}})&=&\rm
\int^1_{-1} d\cos\theta\int^{\pi/2}_{\theta_{_{CB}}-{\pi/ 2}}
d\phi\,F(\vec n_s \cdot \vec n_{_{CB}})\, \vec n_i \cdot \vec n_s\nonumber\\
&=&\rm {1\over 5}\,[2+(2+\pi-\theta_{_{CB}})\cos\theta_{_{CB}}
+\sin\theta_{_{CB}}].
\label{ild}
\end{eqnarray}
An excellent and simple approximation to Eq.~(\ref{ild}) is:
\begin{equation}\rm
E(x)=E(1)\,{1\over 10}\,
[4+x][1+x]\, ,
\label{cosapp}
\end{equation}
with $\rm x=\cos\theta_{_{CB}}$.

For negligible self-attenuation $\rm A[\nu]=1$, as in the optical, 
there is no limb darkening and illumination effect. As absorption
becomes increasingly important for longer wavelengths, the
effect becomes fully relevant. We interpolate between these two
extremes by writing:
\begin{equation}\rm
L_{_{CB}}(\nu,\cos\theta_{_{CB}})\simeq A[\nu]+(1-A[\nu])\,
{E(\cos\theta_{_{CB}})\over E(1)}
\label{illlumdar}
\end{equation}
 to obtain the overall illumination and limb-darkening
correction factor to the energy flux density of Eq.~(\ref{Fnucb}).

\section{Broad band spectra: radio and optical results}

In practice, it is not an effortless task to test
a prediction for an AG's spectrum extending, as in the
upper panel of Fig.~(\ref{figCBpreds}),
to all measured wavelengths from radio to X-rays. The problem
is not related to the model, but to the data. First, the
corrections due to absorption, particularly in the host
galaxy, are frequency-dependent and notoriously difficult
to ascertain with confidence. Second, the integration
times employed in the radio observations are long,
so that the theoretical prediction varies within the time
window, and so do the optical energy flux densities,
measured over much shorter periods, as well as some of
the radio observations themselves. Unavoidably, this
will make our spectral figures look a bit peculiar,
with two theoretical curves bracketing the expectations,
and various observational points at the same frequency.

We study the AG light-curves and broad-band spectra of all
GRBs with known-redshift whose AG was measured both in the
radio and optical bands\footnote{The domain extending from the optical to
the X-ray regime ---but for the early injection bend discussed in
Section 4--- is compatible  with the expected behaviour $\rm
\nu^{-p/2}\sim\nu^{-1.1}$ (DDD 2001).}. 

Our predictions are given by Eq.~(\ref{Fnuobser}), 
fit to the optical and radio observations.
The fitted parameters are the overall 
normalization, $\gamma(0)$, $\theta$, the
deceleration parameter $\rm x_\infty$ (whose meaning and role are
reviewed in Appendix I) and the CB self absorption frequency $\rm \nu_a$ of
Eq.~(\ref{absfreq}).  We found in DDD 2001 that $\rm p$ is very narrowly
distributed around its theoretical value $\rm p=2.2$, and we
fix it to that value for all GRBs in the current analysis.
Thus, {\it the total number of parameters in our broad-band fits
is the same as we used in
DDD 2001 to describe just the R-band light-curve}.

The values of the parameters, listed in Table III, are very similar 
to those deduced in DDD
2001 by fitting only the R-band optical data with the high-$\nu$ limit of
Eq.~(\ref{Fnuobser}). The small differences are due 
not only to the use of radio data and optical bands other than R,
but also to the inclusion of the effects of the injection bend in the CB
synchrotron AG global formula, Eq.~(\ref{Fnuobser}), and (to a small
extent) to the use of a fixed $\rm p=2.2$. The results show that the theory 
agrees with observations both at radio and optical wavelengths. 
For some GRBs a slightly 
better fit to the radio data is obtained if the absorption frequency 
$\rm\nu_a$ is best fitted to the radio data alone or if a fitted power-law
dependence on time is used for the CB opacity instead of 
Eq.~(\ref{absfreq}), with all other parameters taken from the global fit.
Because of scintillations, and of the very detail-dependent character of our
prediction for the time dependence of a CB's opacity,
it is difficult to assess whether or not
the slightly improved $\chi^2$ values are significant or not.

Notice in Table III that the distributions of parameters are fairly
narrow, in particular for $\gamma_0$. Of particular interest, since
it can be predicted, is the distribution in $\gamma \times\theta$.
Since AGs are discovered at optical and X-ray frequencies,
the angular distribution is that of the high frequency limit of
Eq.~(\ref{Fnuobser}). For small $\theta$, 
$\rm d\,N/d\theta\propto \theta/(1+\gamma^2\theta^2)^{4.1}$.
This distribution has a maximum at $\gamma\theta\sim 0.37$
and a median at $\sim 0.5$. In Table III there are four cases
with $\gamma\theta$ below the median and five above. The
worst ``outlier'' in $\theta$ is much less so in $\gamma\theta$.
The conclusion that this distribution is perfectly compatible
with the expectation can also be reached from Fig.~(39) of
DDD 2001, whose results were obtained from only
optical data, but for which the statistics is a bit better.

We first discuss the broad-band spectra and light curves of three
representative GRBs: 000301c, 000926 and 991216.
The optical AG of GRB 000301c is practically unextinct, 
that of GRB 000926 has strong extinction in the host galaxy 
(e.g., Fynbo et al. 2001) and that of GRB
991216 has strong extinction both in the host galaxy and in
ours (e.g., Halpern et al. 2000). We discuss GRBs 991208, 000418,
000510, 990123 and 970508 in slightly less detail. The apparently
special case of GRB 980425 is discussed separately in the next chapter.
   
\subsection{GRB 000301c}

\begin{figure}[t]  
%\begin{tabular}{cc}  
\hskip 0truecm   
\epsfig{file=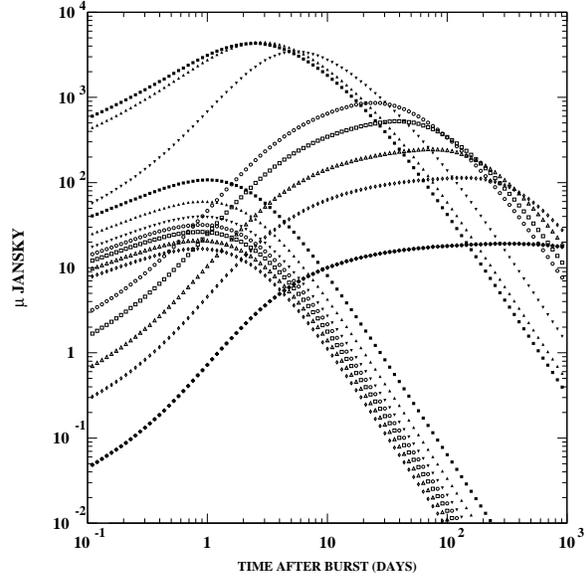, width=8.5cm}  
%\end{tabular}  
\caption{Results of a fit to radio and optical observations of 
the light curves of GRB 000301c. The narrowly spaced lines refer
---from top to bottom--- to the K, J, I, R, V, B and U bands.
The more widely spaced lines refer
---from top to bottom at the figure's left side--- to 
frequencies of 1.43, 4.86, 8.46, 15, 22.5, 100, 250 and 350 GHz.
The comparison with data is shown in Figs.~(\ref{K301}) to (\ref{rad-opt301b}).
}  
\label{all301}  
\end{figure}

\begin{figure}[t]  
%\begin{tabular}{cc}  
\hskip 0truecm   
\epsfig{file=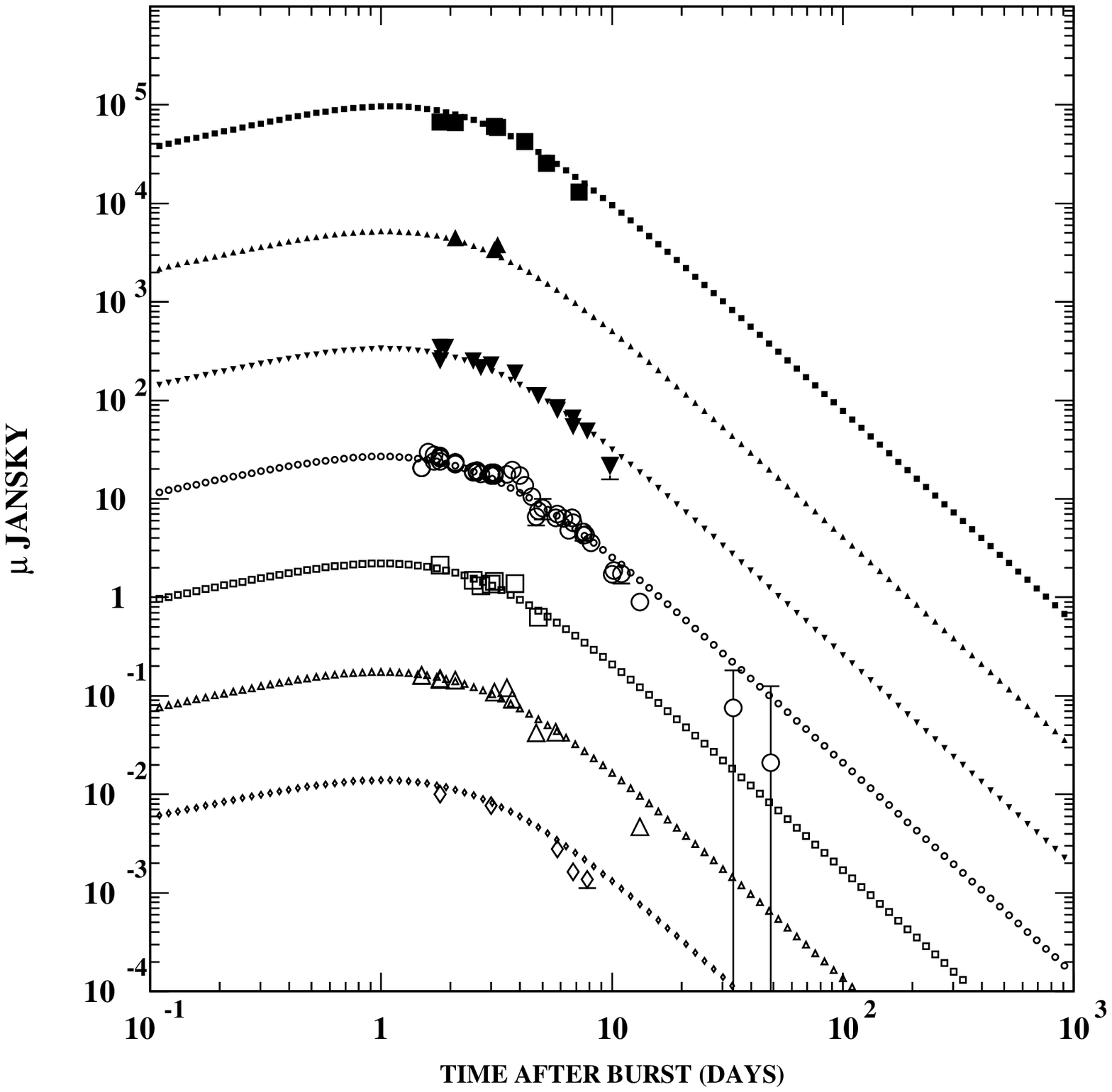, width=8.5cm}  
%\end{tabular}  
\caption{Comparisons between our fitted CB model AG of GRB 000301c, 
at $\rm z=2.033$, 
Eq.~(\ref{Fnuobser}) and Fig.~(\ref{all301}), with the observed optical
data. 
The figure shows (from top to bottom) 1000 times the K-band results,
100 times the J-band, 10 times the I-band, the R-band, 1/10 of the V-band,
1/100 of the B-band and 1/1000 of the U-band.
The contributions of the underlying galaxy and 
an expected (but,  in this case, unobservable) SN1998bw-like 
SN have been subtracted. }  
\label{K301}  
\end{figure}

\begin{figure}[t]  
\begin{tabular}{cc}  
\hskip 2truecm  
\vspace*{2cm} 
\hspace*{-1.7cm}  
\epsfig{file=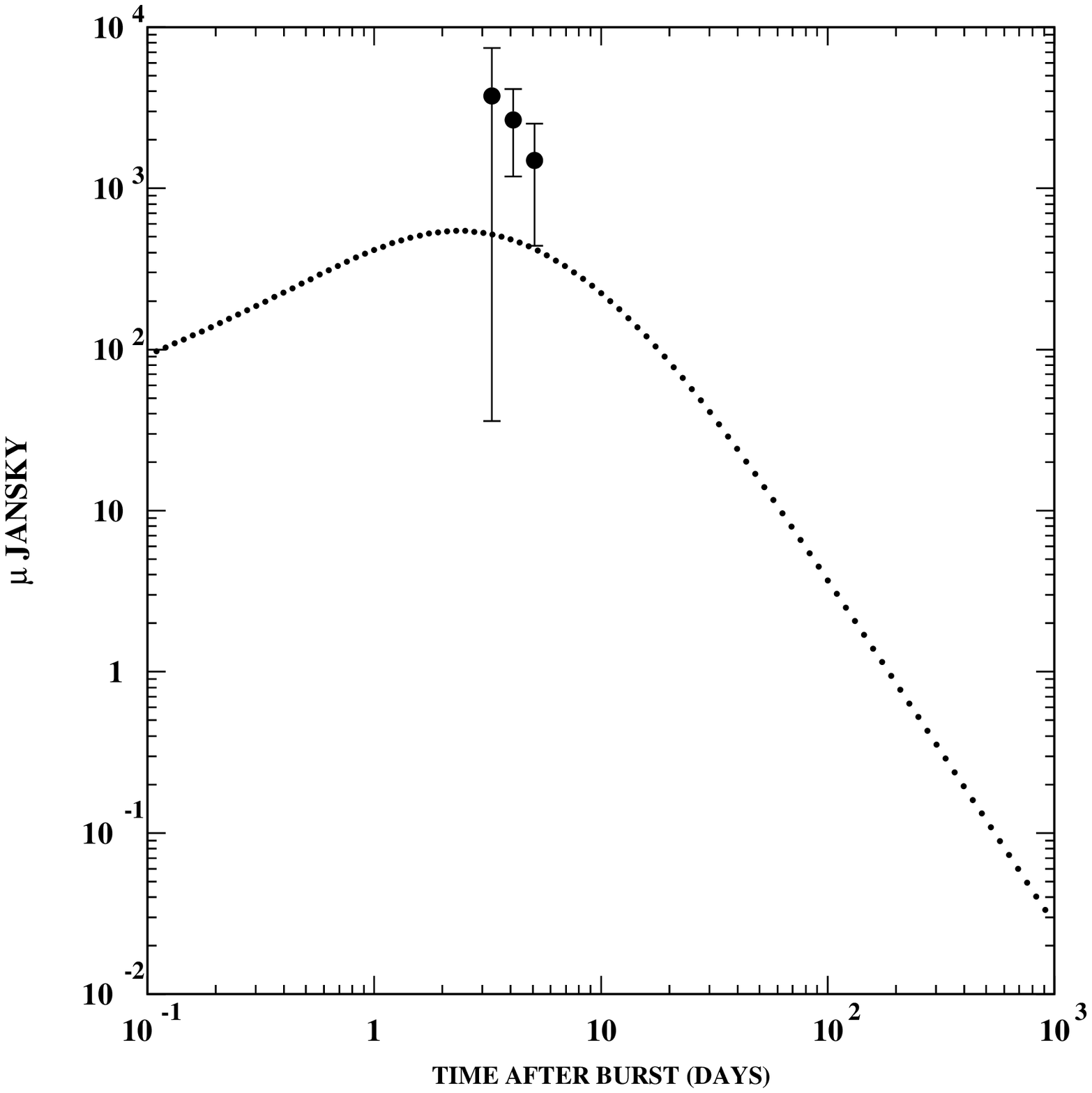, width=8cm} \\ 
%\hskip 1truecm  
\hspace*{.5cm}  
\epsfig{file=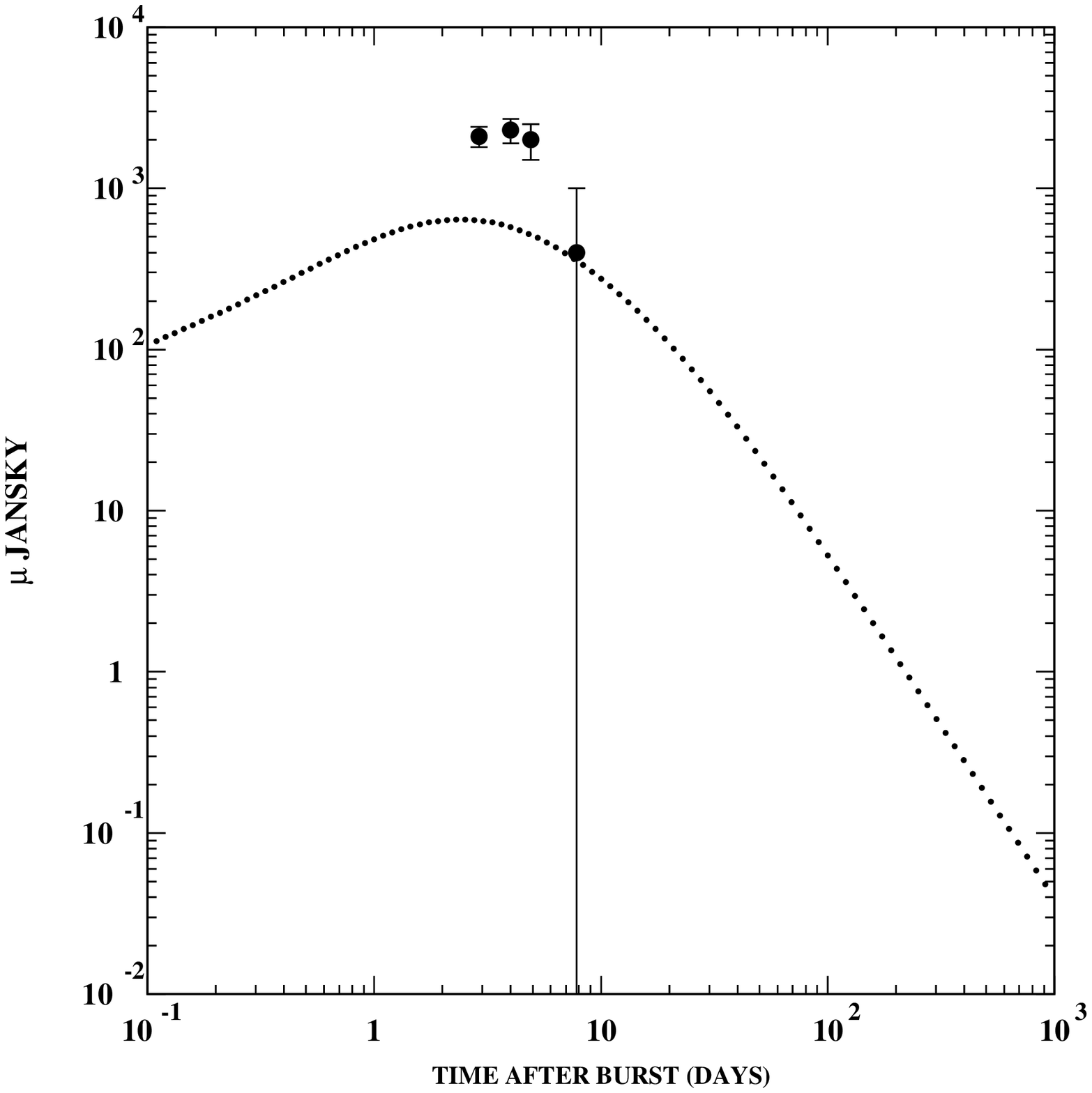, width=8cm} 
\end{tabular}  
\caption{Comparisons between our fitted CB model afterglow, 
Eq.~(\ref{Fnuobser}), and the observed radio afterglow of GRB 000301c. 
Upper panel: the light curve at 350 
GHz. Lower panel: the light curve at 250 GHz.}  
\label{figr030101}  
\end{figure}

\begin{figure}[t]  
\begin{tabular}{cc}  
\hskip 2truecm  
\vspace*{2cm} 
\hspace*{-1.7cm}  
\epsfig{file=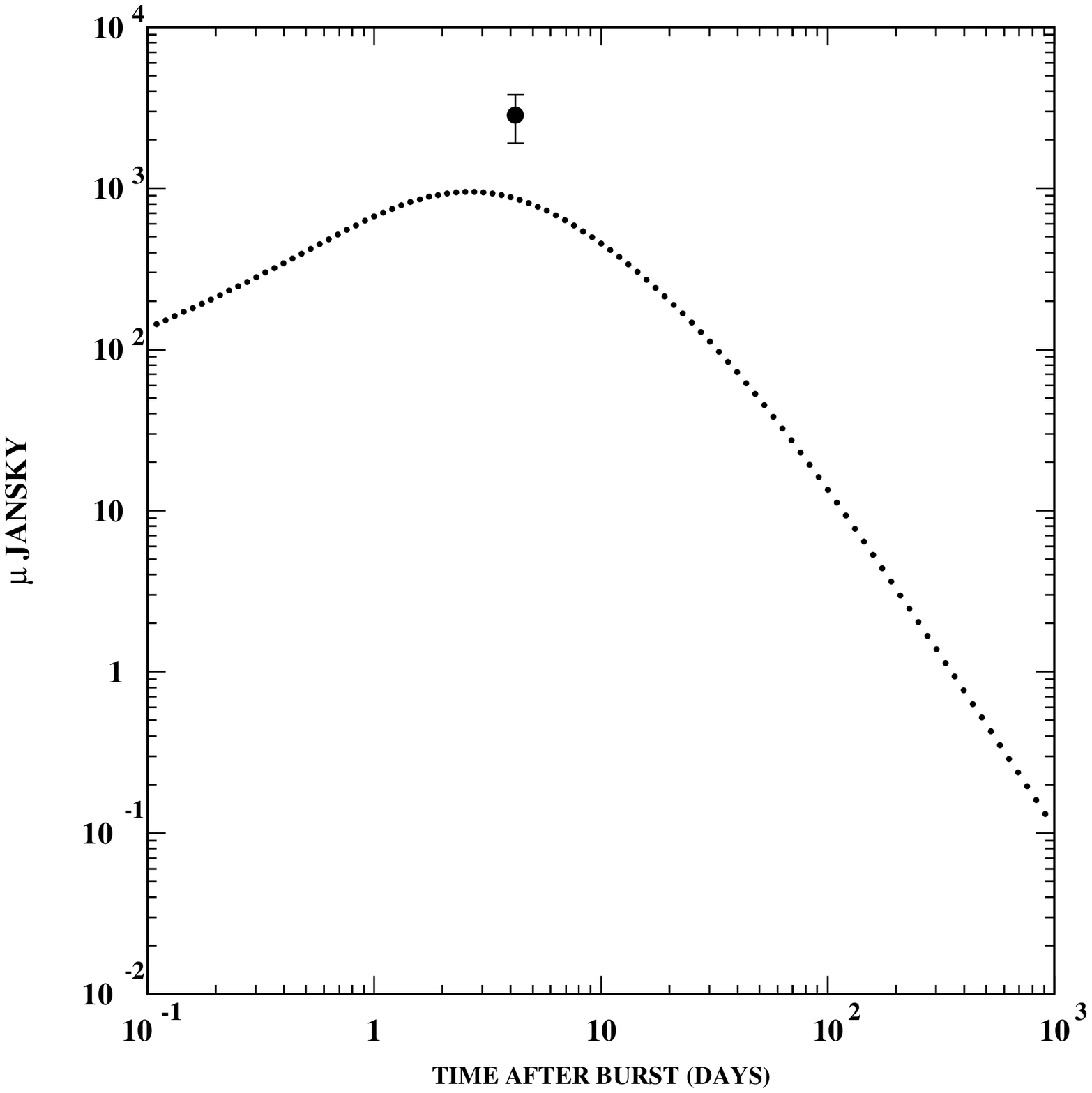, width=8cm} \\ 
%\hskip 1truecm  
\hspace*{.5cm}  
\epsfig{file=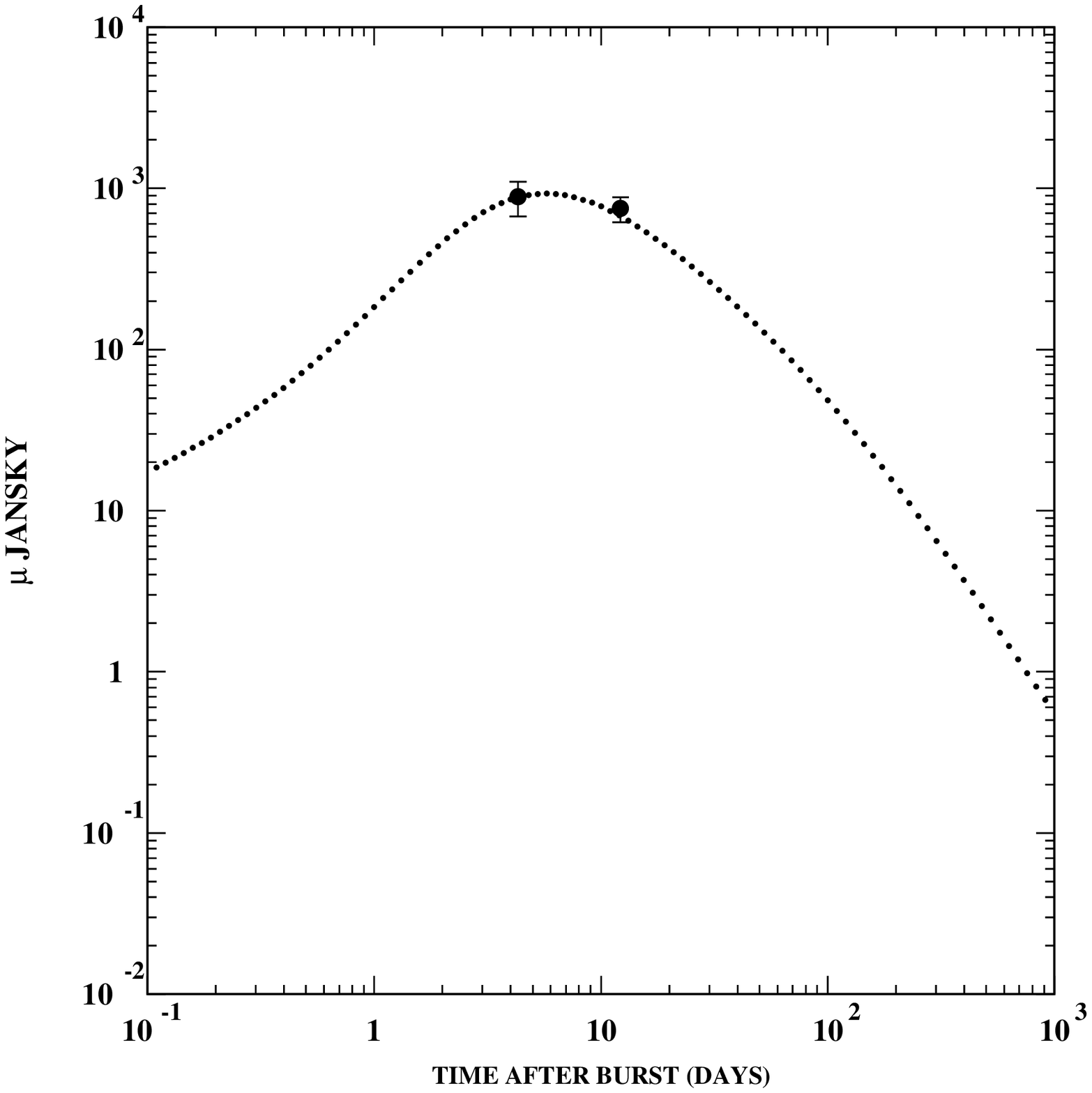, width=8cm} 
\end{tabular}  
\caption{Comparisons between our fitted CB model afterglow, 
Eq.~(\ref{Fnuobser}), and the observed radio afterglow of GRB 000301c.
Upper panel: the light curve at 100 
GHz. Lower panel: the light curve at 22.5 GHz.}  
\label{figr030102}  
\end{figure}

\begin{figure}[t]  
\begin{tabular}{cc}  
\hskip 2truecm  
\vspace*{2cm} 
\hspace*{-1.7cm}  
\epsfig{file=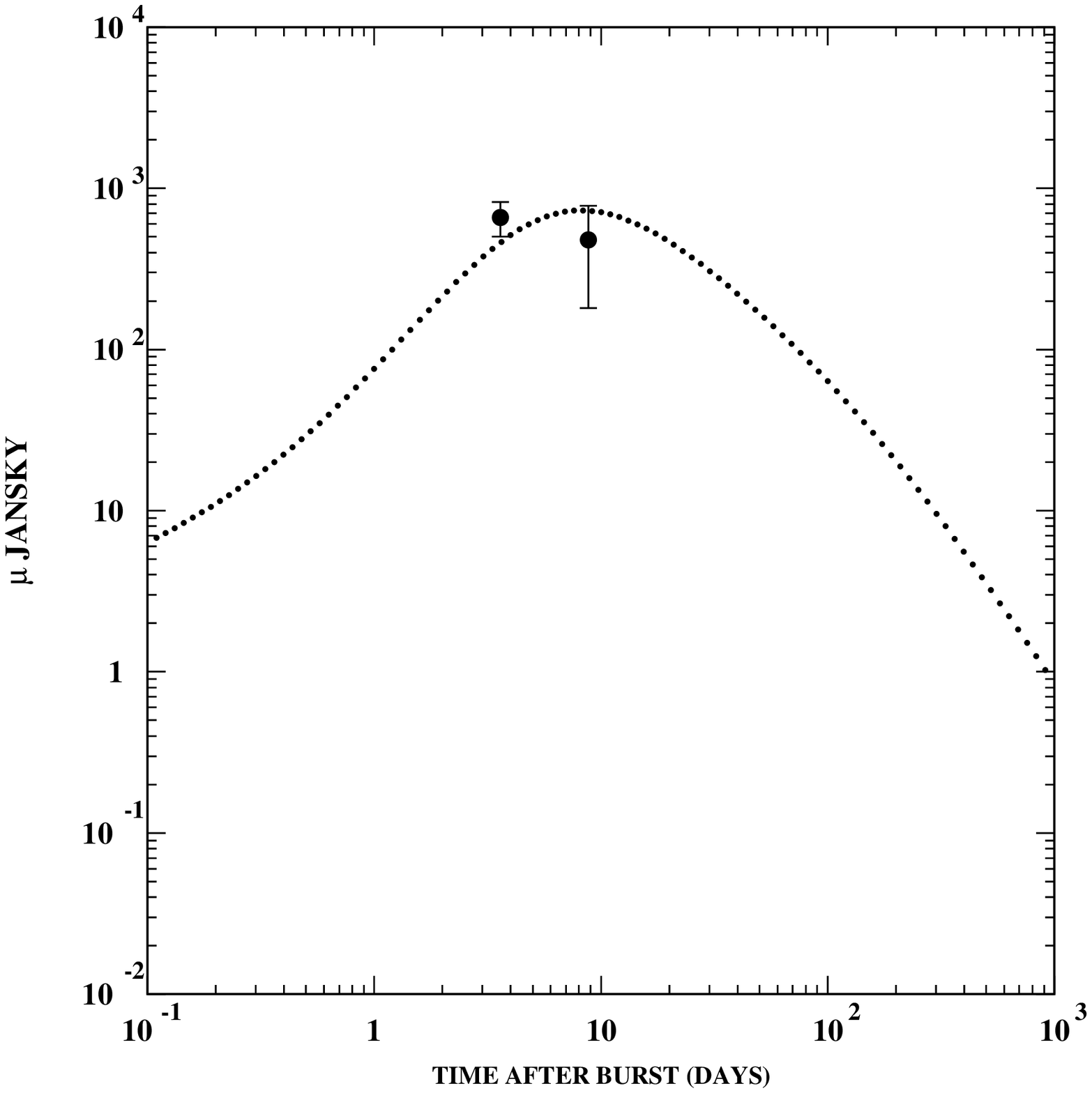, width=8cm} \\ 
%\hskip 1truecm  
\hspace*{.5cm}  
\epsfig{file=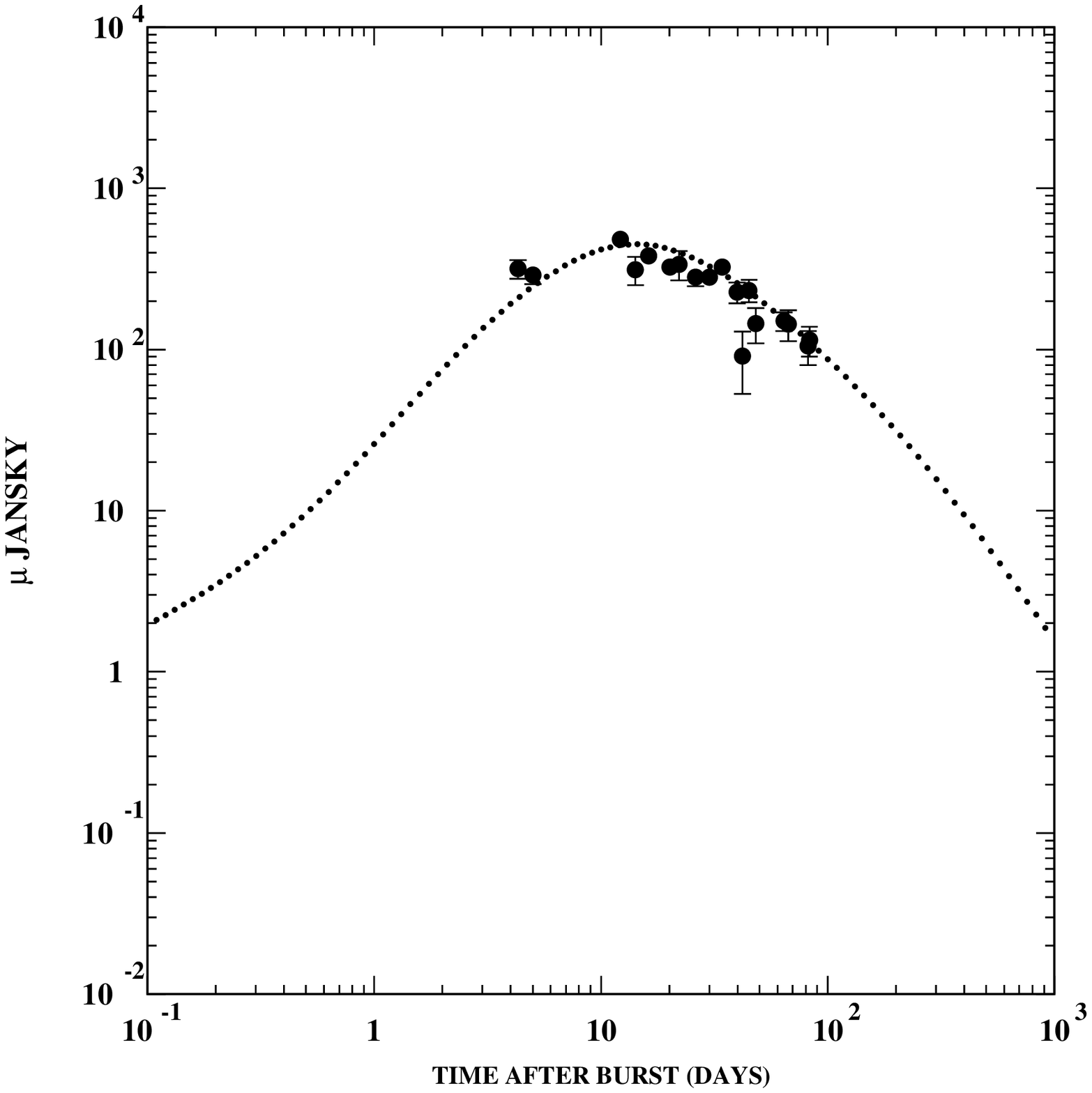, width=8cm} 
\end{tabular}  
\caption{Comparisons between our fitted CB model afterglow, 
Eq.~(\ref{Fnuobser}), and the observed radio afterglow of GRB 000301c.
Upper panel: the light curve at 15 
GHz. Lower panel: the light curve at 8.46 GHz.}  
\label{figr030103}  
\end{figure}

\begin{figure}[t]  
\begin{tabular}{cc}  
\hskip 2truecm  
\vspace*{2cm} 
\hspace*{-1.7cm}  
\epsfig{file=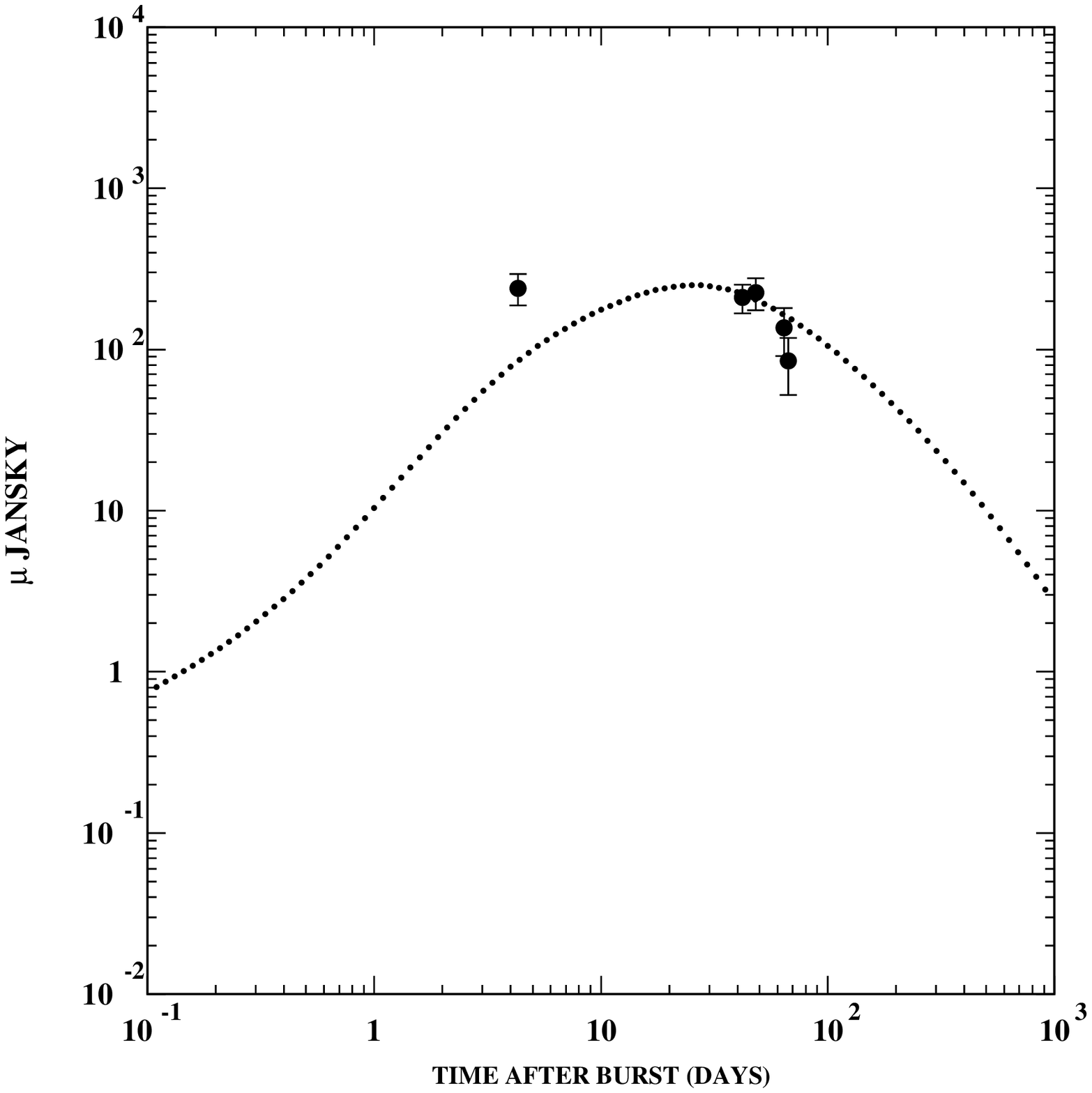, width=8cm} \\ 
%\hskip 1truecm  
\hspace*{.5cm}  
\epsfig{file=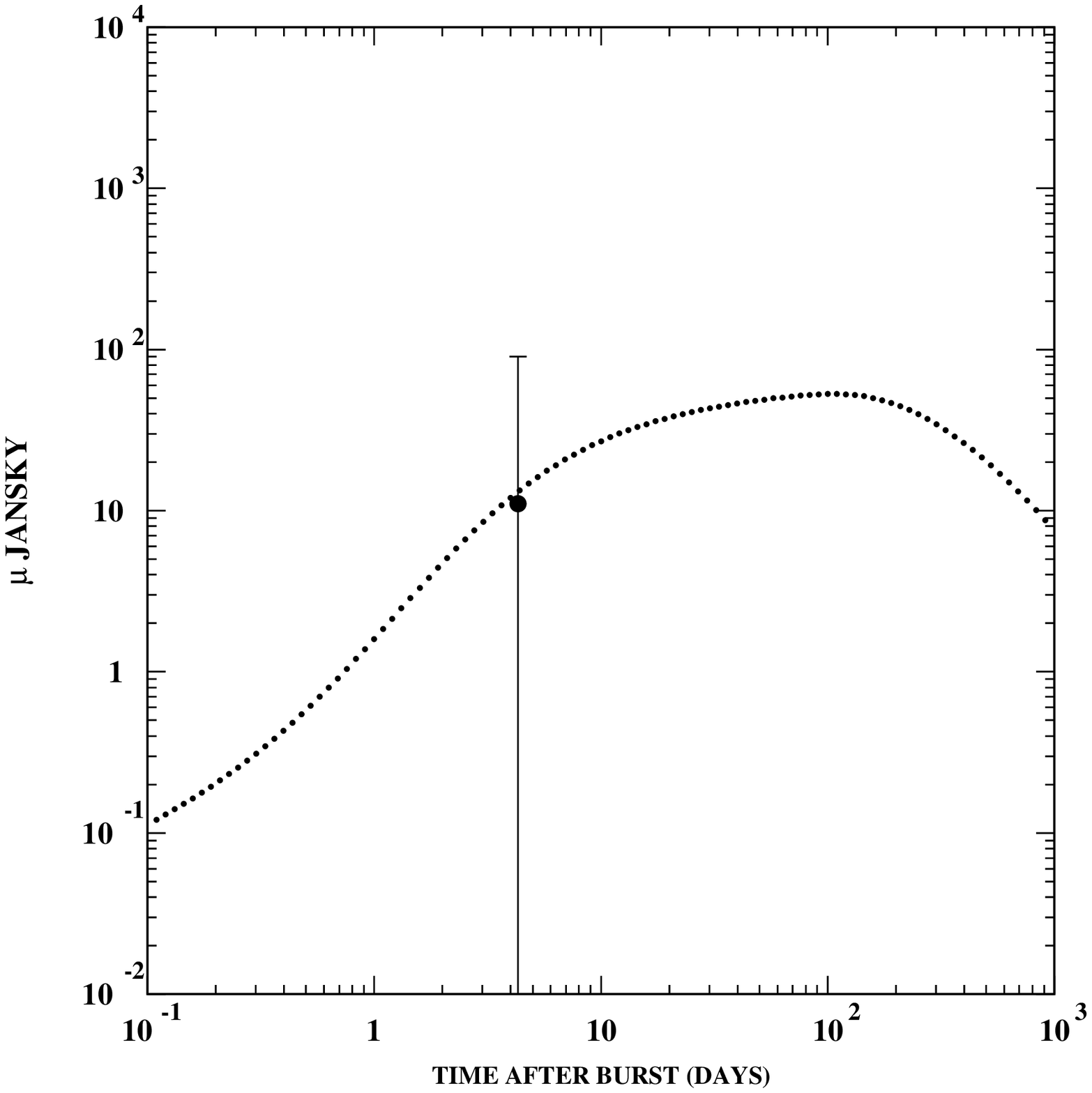, width=8cm} 
\end{tabular}  
\caption{Comparisons between our fitted CB model afterglow, 
Eq.~(\ref{Fnuobser}), and the observed radio afterglow of GRB 000301c.
Upper panel: the light curve at 4.86 
GHz. Lower panel: the light curve at 1.43 GHz.}  
\label{figr030104}  
\end{figure}

\begin{figure}[t]  
\begin{tabular}{cc}  
\hskip 2.5truecm  
\vspace*{2cm} 
\hspace*{-2.7cm}  
\epsfig{file=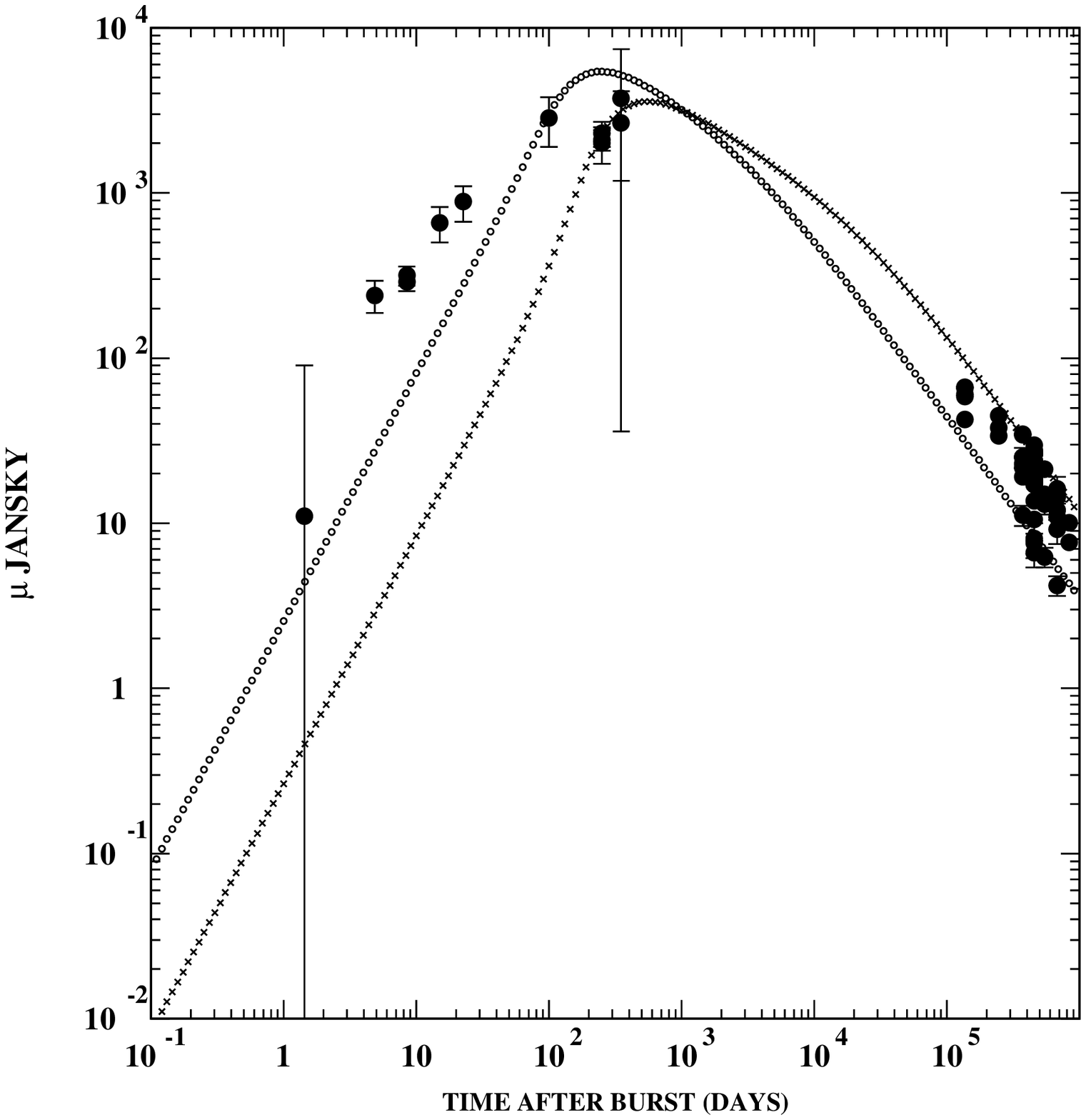, width=8cm}
\vspace*{-1.5cm}
\\ 
%\hskip 1truecm 
\hspace*{-.2cm}  
\epsfig{file=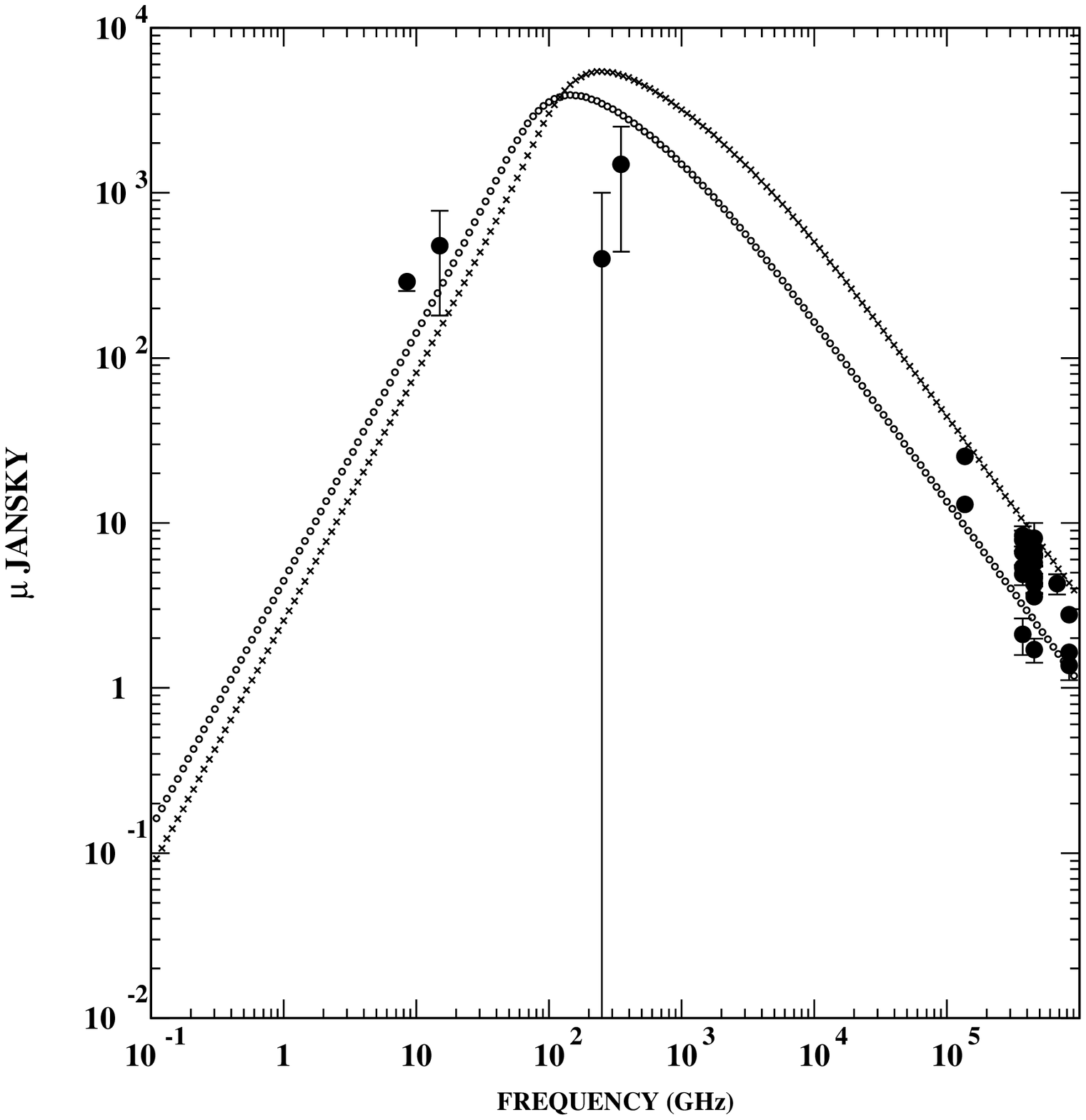, width=8cm} 
\end{tabular}  
\caption{The spectrum of the AG of GRB 000301c from radio to optical
frequencies.
Upper panel: in the time interval between 1 and 5 days after burst.
Lower panel:  in the time interval between 5 and 10 days after burst. 
The highest peaking curve in the upper pannel corresponds to the later 
time and in the lower panel to the earlier time.} 
\label{rad-opt301}  
\end{figure}

\begin{figure}[t]  
\begin{tabular}{cc}  
\hskip 2.5truecm  
\vspace*{2cm} 
\hspace*{-2.7cm}  
\epsfig{file=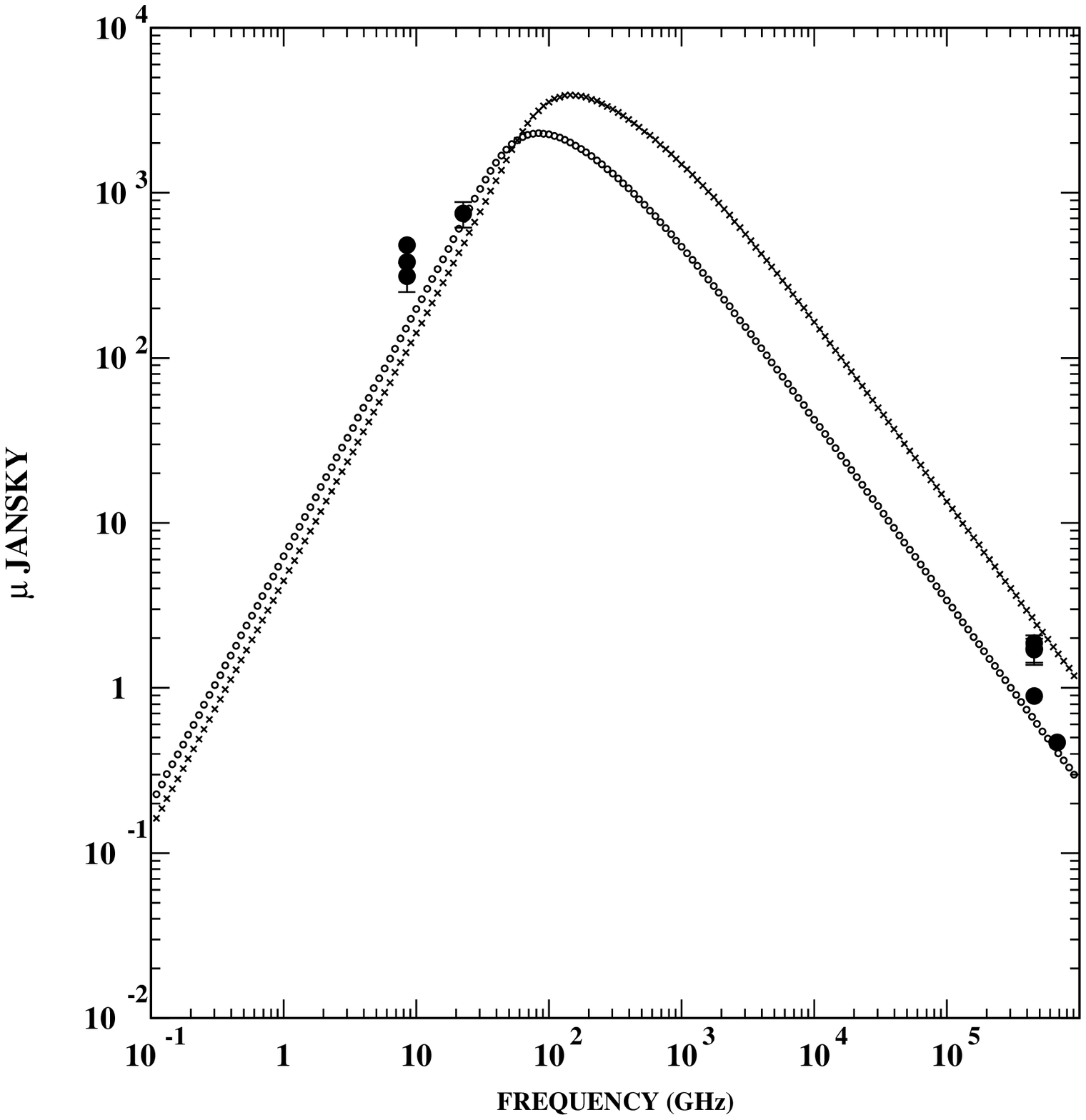, width=8cm}
\vspace*{-1.5cm}
\\ 
%\hskip 1truecm 
\hspace*{-.2cm}  
\epsfig{file=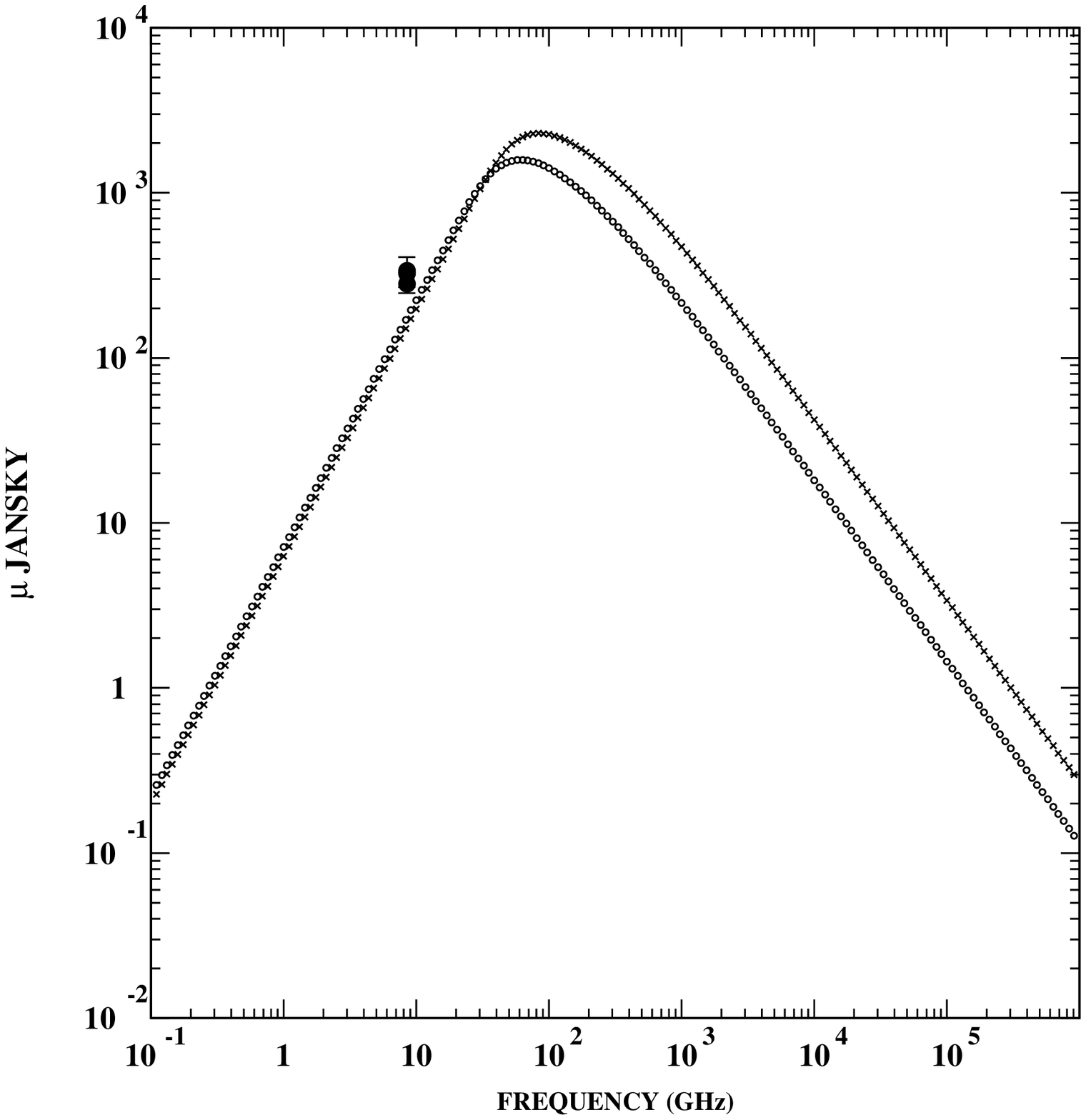, width=8cm} 
\end{tabular}  
\caption{The spectrum of the AG of GRB 000301c from radio to optical
frequencies.
Upper panel: in the time interval between 10 and 20 days after burst.
Lower panel:  in the time interval between 20 and 30 days after burst. 
In both cases the highest peaking curve
corresponds to the earlier time.}  
\label{rad-opt301b}  
\end{figure}

For this GRB we fit the radio data of Berger et al. (2000)
and the optical data of Garnavich et al. (2000b), Jensen et al.
(2001), Masetti et al., (2000), Rhoads and Fruchter (2001) and
Sagar et al. (2001). Our results for the light curves at all observed
optical and radio frequencies are gathered in Fig.~(\ref{all301}),
which is representative of the trends seen in all GRBs.
The narrowly spaced lines in the figure are the optical light curves
for ---from top to bottom--- the K, J, I, R, V, B and U bands.
Their very satisfactory comparison with data is reported in 
Fig.~(\ref{K301}). The results for the
radio AG are the more spaced lines in Fig.~(\ref{all301}), which
correspond ---from top to bottom at the figure's left side--- to 
frequencies of 1.43, 4.86, 8.46, 15, 22.5, 100, 250 and 350 GHz.
Their very satisfactory comparison with observations is reported
in Figs.~(\ref{figr030101}) to (\ref{figr030104}). Notice that all 
features of the
data have precisely the trends summarized in Fig.~(\ref{all301}).
In Figs.~(\ref{rad-opt301}) and (\ref{rad-opt301b}) we present the
complementary information, by comparing
our fits with the observations for the radio-to-optical spectra
of GRB 000301c in four radio time-integration brackets;
1 to 5, 5 to 10, 10 to 20, and 20 to 30 days. 
The pronounced peaks are at  (observer's) frequencies 
for which the opacity of Eq.~(\ref{absfreq}) is 
$\rm \tau_\nu\sim{\cal{O}}(1)$. The injection bend at a higher
frequency is clearly visible, it is responsible for the agreement
between the radio and optical magnitudes and frequency trends.
The two curves in these figures, and many later ones, refer to the
expectation at the two times which bracket the actual radio observation.
The results are quite satisfactory.

\subsection{GRB 000926}

\begin{figure}[t]  
%\begin{tabular}{cc}  
\hskip 0truecm   
\epsfig{file=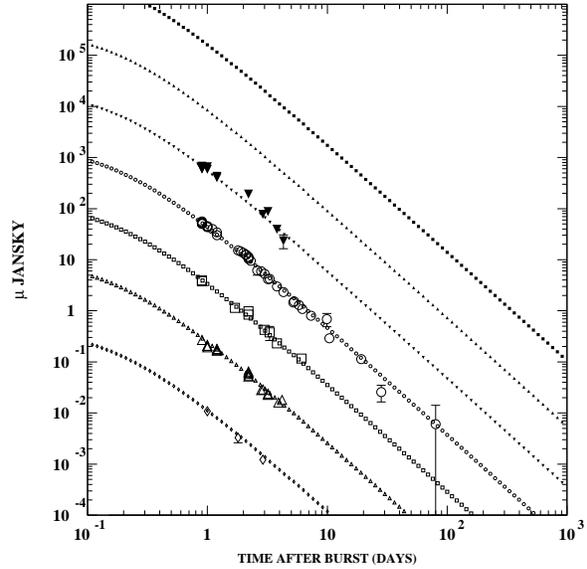, width=8.5cm}  
%\end{tabular}  
\caption{Comparisons between our fitted CB model afterglow, 
Eq.~(\ref{Fnuobser}), 
and the observed optical afterglow of GRB 000926 
at $\rm z=2.037$. 
The figure shows (from top to bottom) 1000 times the K-band results,
100 times the J-band, 10 times the I-band, the R-band, 1/10 of the V-band,
1/100 of the B-band and 1/1000 of the U-band.
The contributions of the underlying galaxy and 
an expected (but,  in this case, unobservable) SN1998bw-like 
SN have been subtracted. }  
\label{opt926}  
\end{figure}

\begin{figure}[t] 
\begin{tabular}{cc} 
\hskip 2truecm 
\vspace*{2cm} 
\hspace*{-1.7cm} 
\epsfig{file=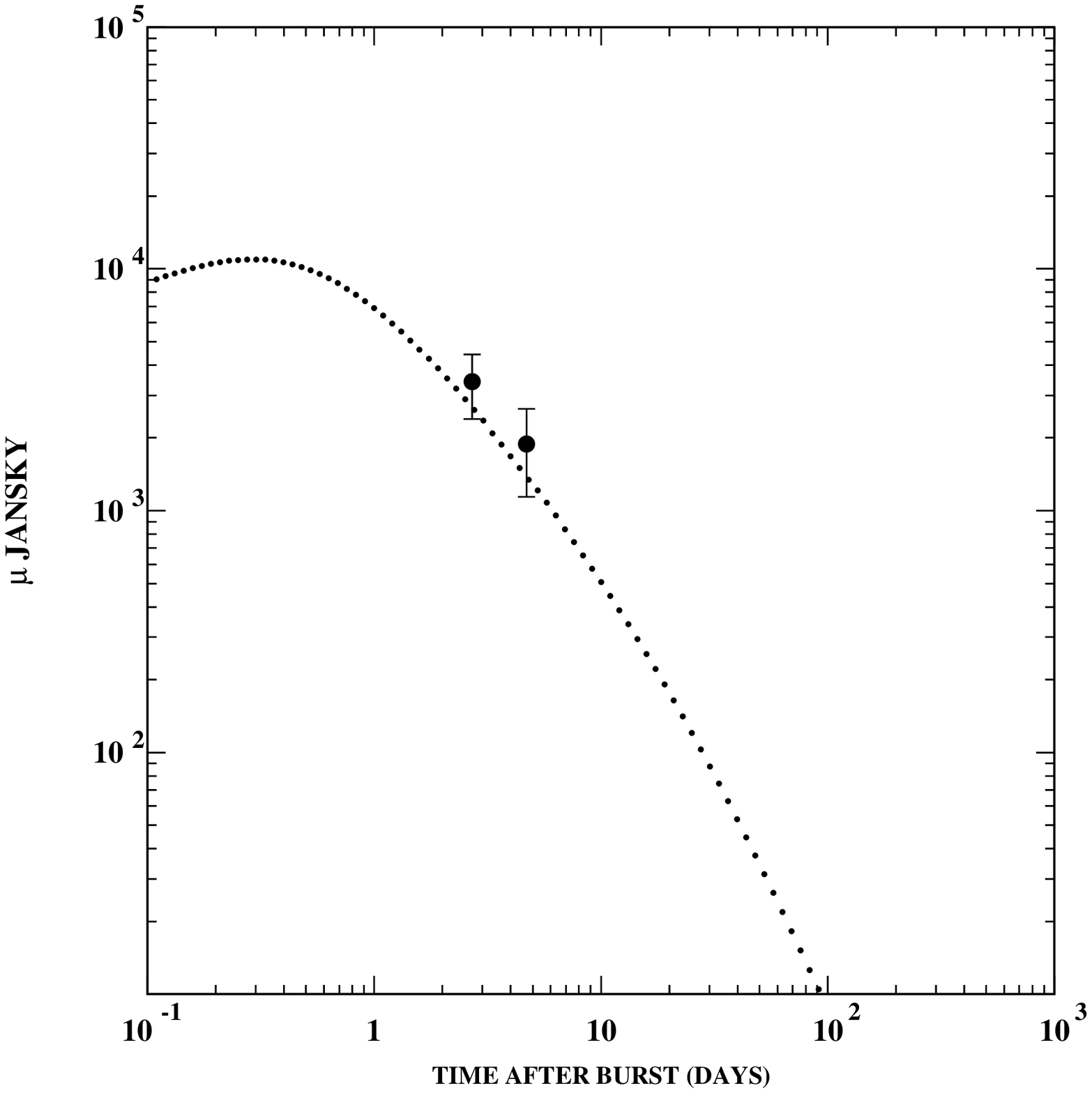, width=8cm} \\ 
%\hskip 1truecm 
\hspace*{.5cm} 
\epsfig{file=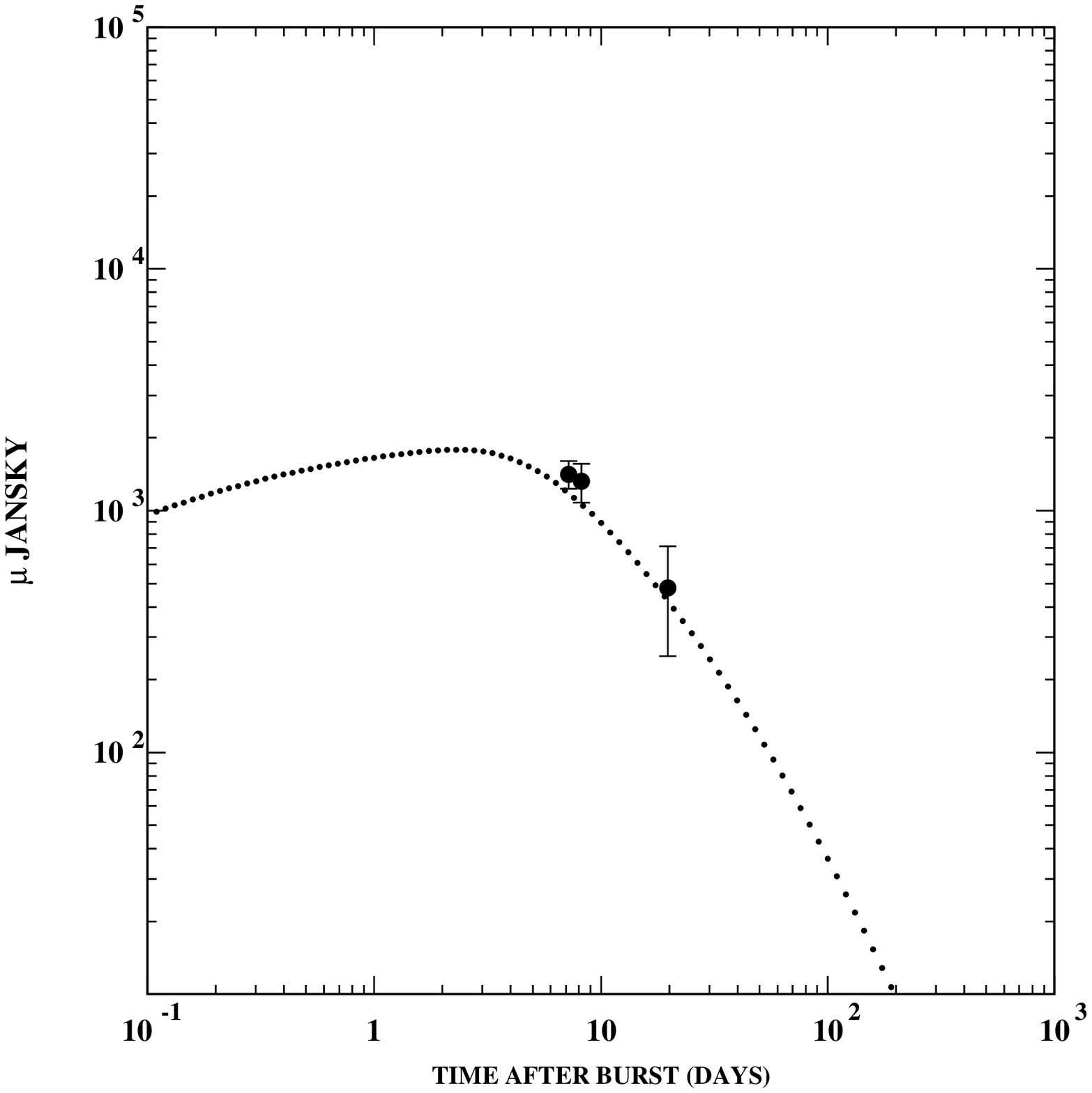, width=8cm} 
\end{tabular} 
\caption{Comparisons between our fitted CB model afterglow, 
Eq.~(\ref{Fnuobser}), 
and the observed radio afterglow of GRB 000926. 
Upper panel: the light curve at 98.48 GHz. 
Lower panel: the light curve at 22.5 GHz.} 
\label{figr092601} 
\end{figure}

\begin{figure}[t] 
\begin{tabular}{cc} 
\hskip 2truecm 
\vspace*{2cm} 
\hspace*{-1.7cm} 
\epsfig{file=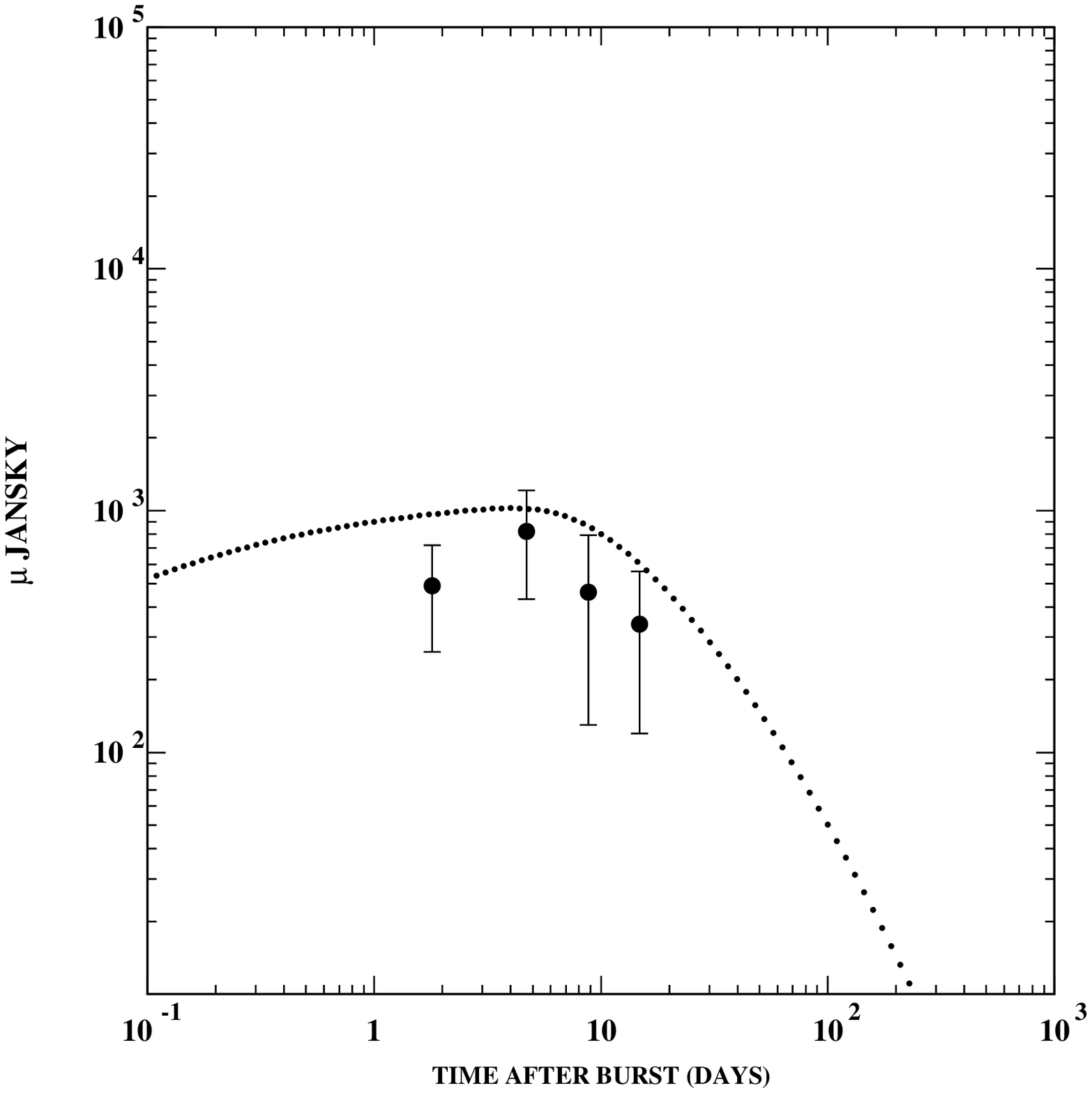, width=8cm} \\ 
%\hskip 1truecm 
\hspace*{.5cm} 
\epsfig{file=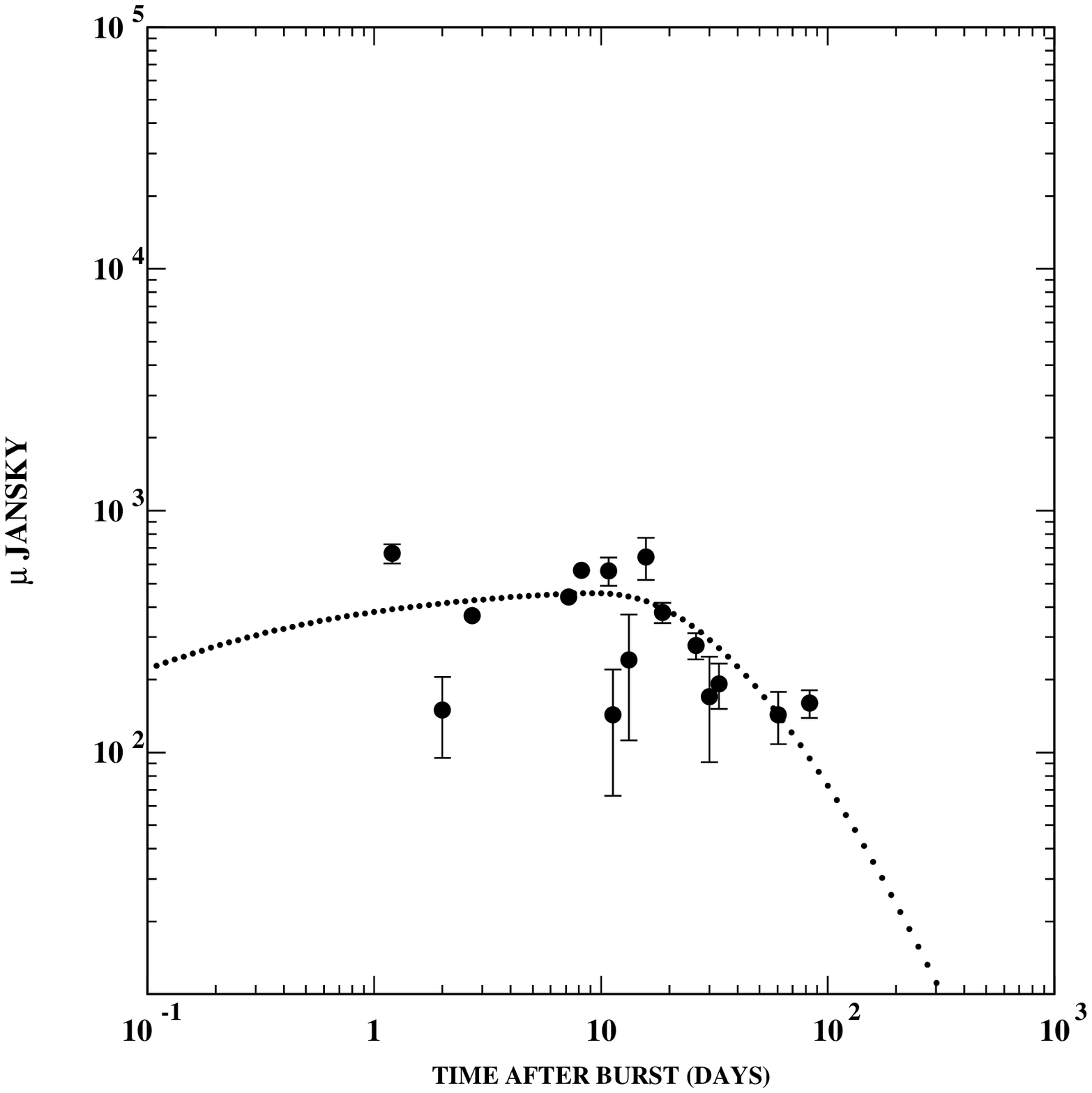, width=8cm} 
\end{tabular} 
\caption{Comparisons between our fitted CB model afterglow, 
Eq.~(\ref{Fnuobser}), 
and the observed radio afterglow of GRB 000926. 
Upper panel: the light curve at 15 GHz. 
Lower panel: the light curve at 8.46 GHz.} 
\label{figr092602} 
\end{figure}

\begin{figure}[t] 
\begin{tabular}{cc} 
\hskip 2truecm 
\vspace*{2cm} 
\hspace*{-1.7cm} 
\epsfig{file=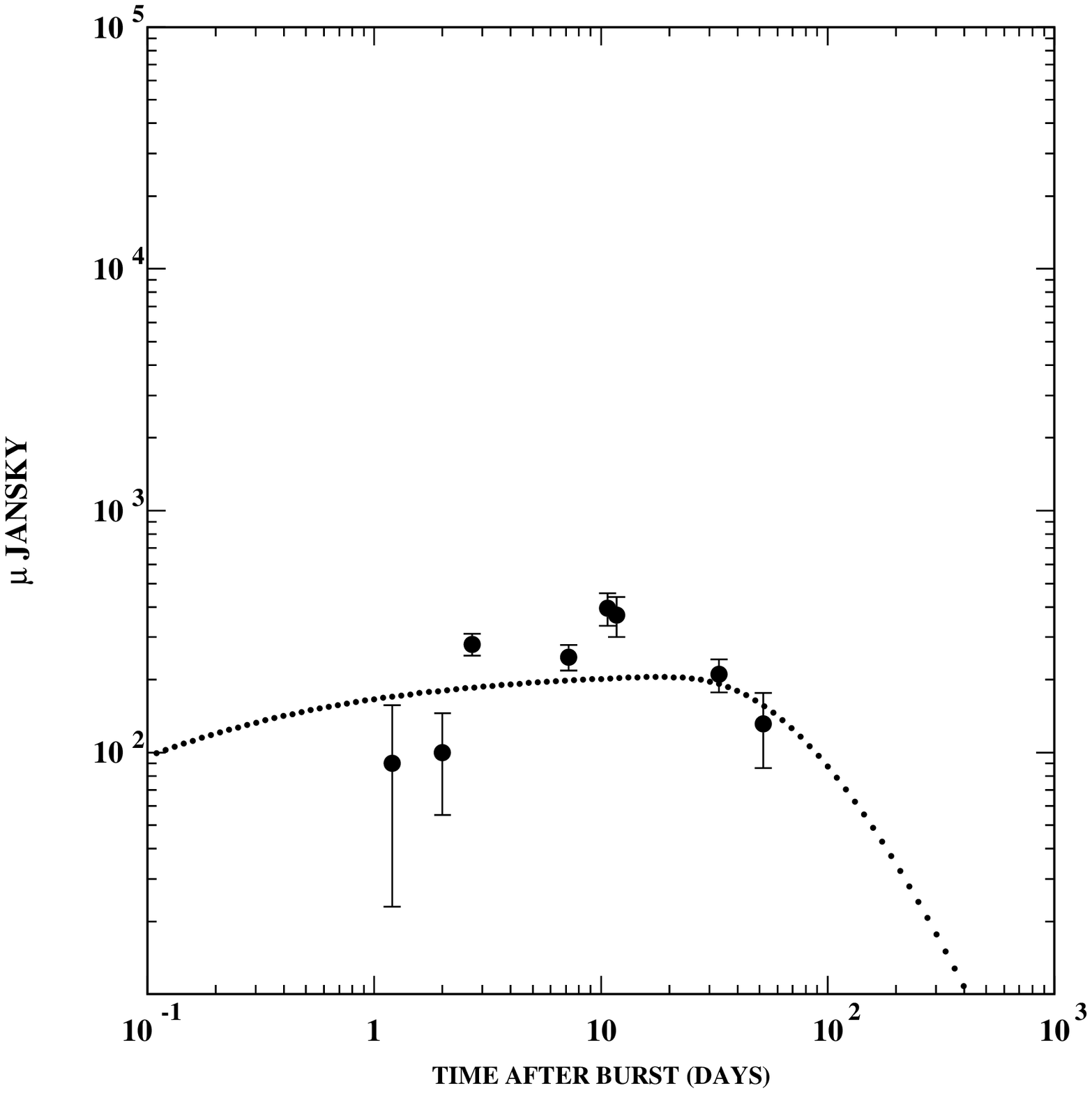, width=8cm} \\ 
%\hskip 1truecm 
\hspace*{.5cm} 
\epsfig{file=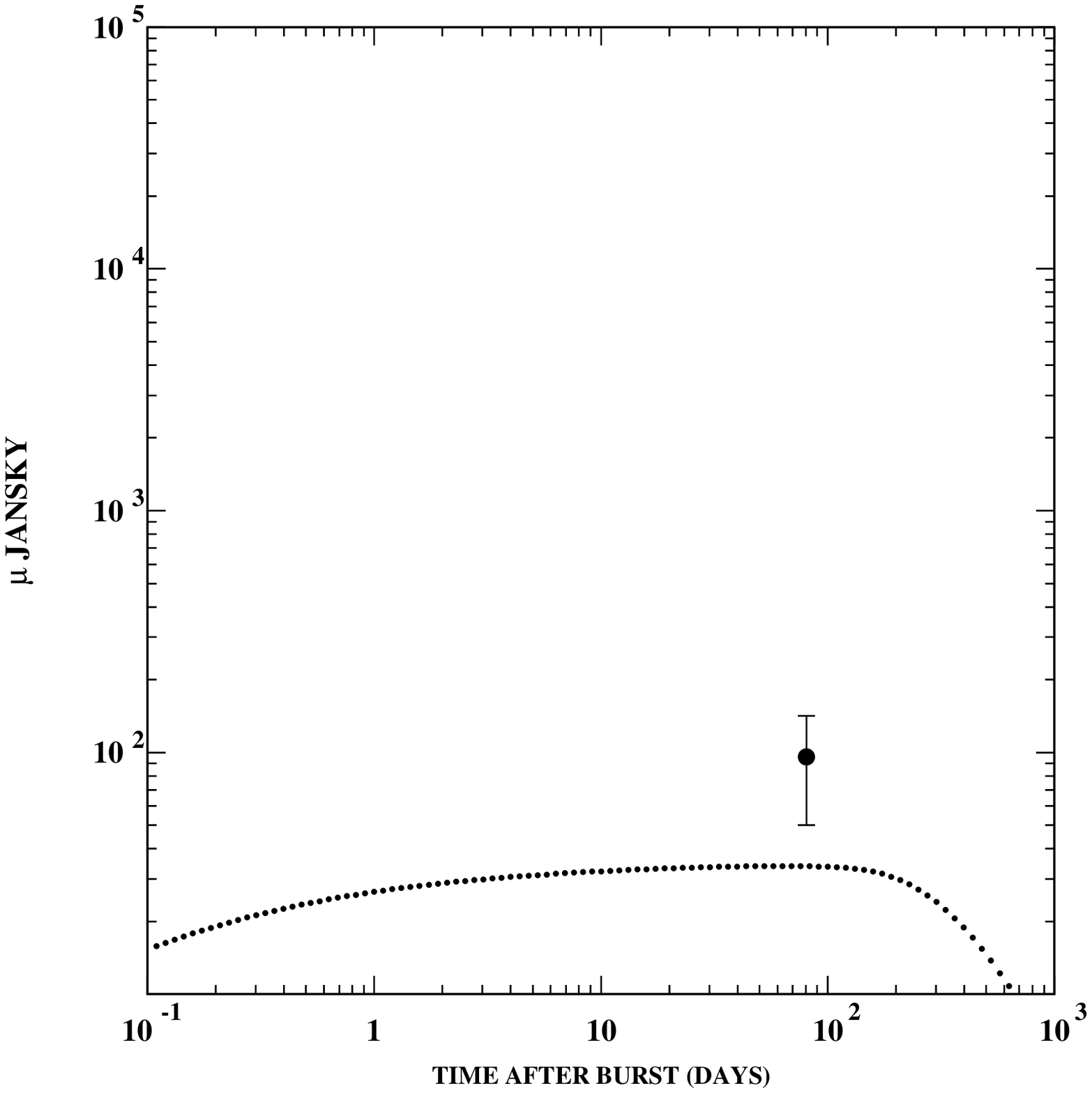, width=8cm} 
\end{tabular} 
\caption{Comparisons between our fitted CB model afterglow, 
Eq.~(\ref{Fnuobser}), 
and the observed radio afterglow of GRB 000926. 
Upper panel: the light curve at 4.86 GHz. 
Lower panel: the light curve at 1.43 GHz.} 
\label{figr092603} 
\end{figure}

We have made a global fit to the NIR/optical data (Di Paola et al. 2000; 
Fynbo et al. 2001, Harrison et al. 2001; Price et al. 2001; Sagar et al.  
2001) and the radio data (Harrison et al. 2001) on this GRB. In 
Fig.~(\ref{opt926}) 
we compare the fitted CB-model predictions with the measured light
curves for the I, R, V, B and U bands, after subtraction of the
host galaxy and SN contributions (DDD 2001).
The theoretical predictions were corrected for galactic 
extinction E(B - V)=0.0235 (Schlegel et al. 1998) and for the estimated
extinction in the host galaxy, E(B - V)=0.40 (Harrison et al. 2001).
In Figs.~(\ref{figr092601}) to (\ref{figr092603})
we present the radio light curves for
six frequencies ranging from 98.48 to 1.43 GHz.

\begin{figure}[t]  
\begin{tabular}{cc}  
\hskip 2.5truecm  
\vspace*{2cm} 
\hspace*{-2.7cm}  
\epsfig{file=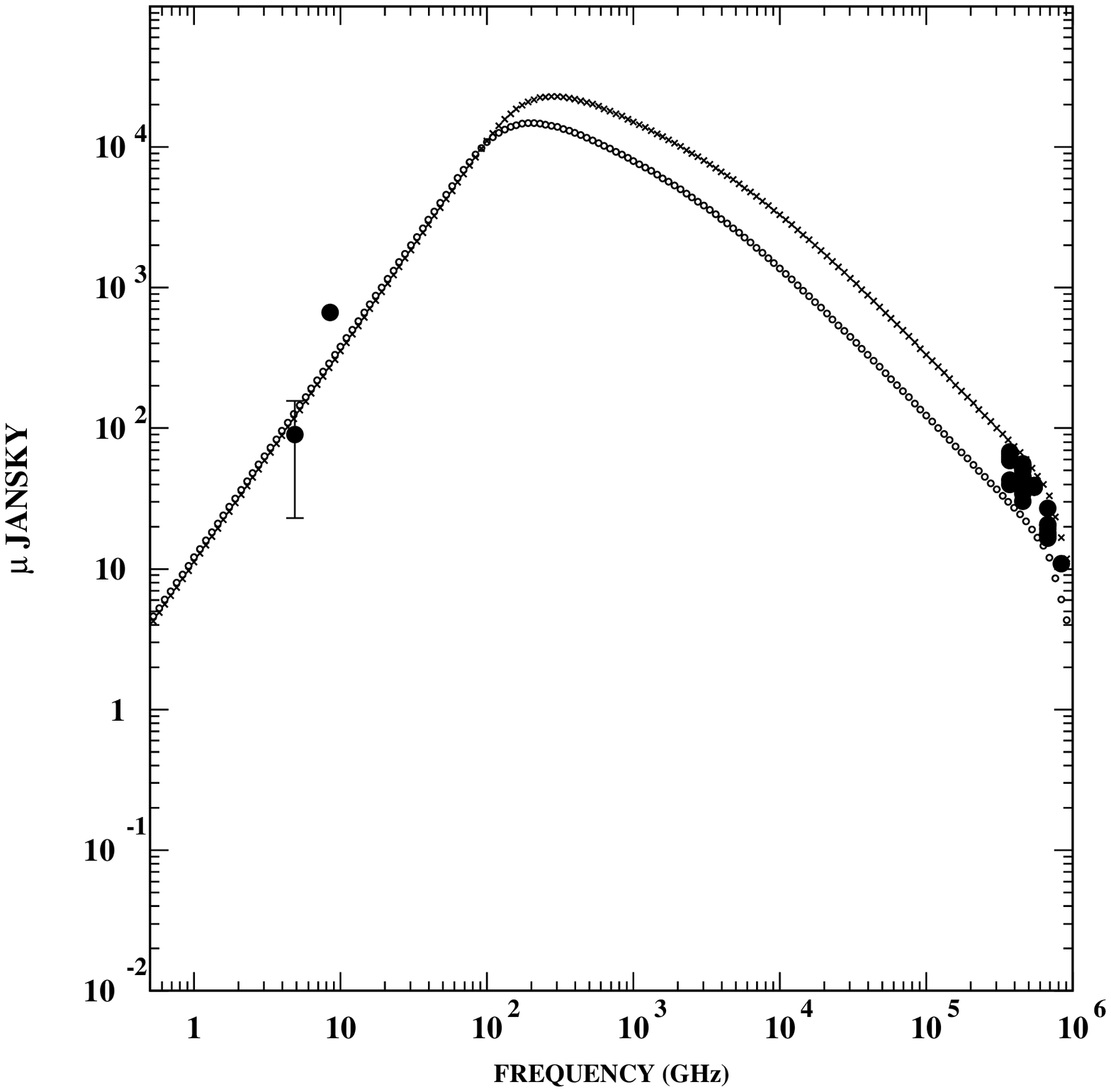, width=8cm}
\vspace*{-1.5cm}
\\ 
%\hskip 1truecm 
\hspace*{-.2cm}  
\epsfig{file=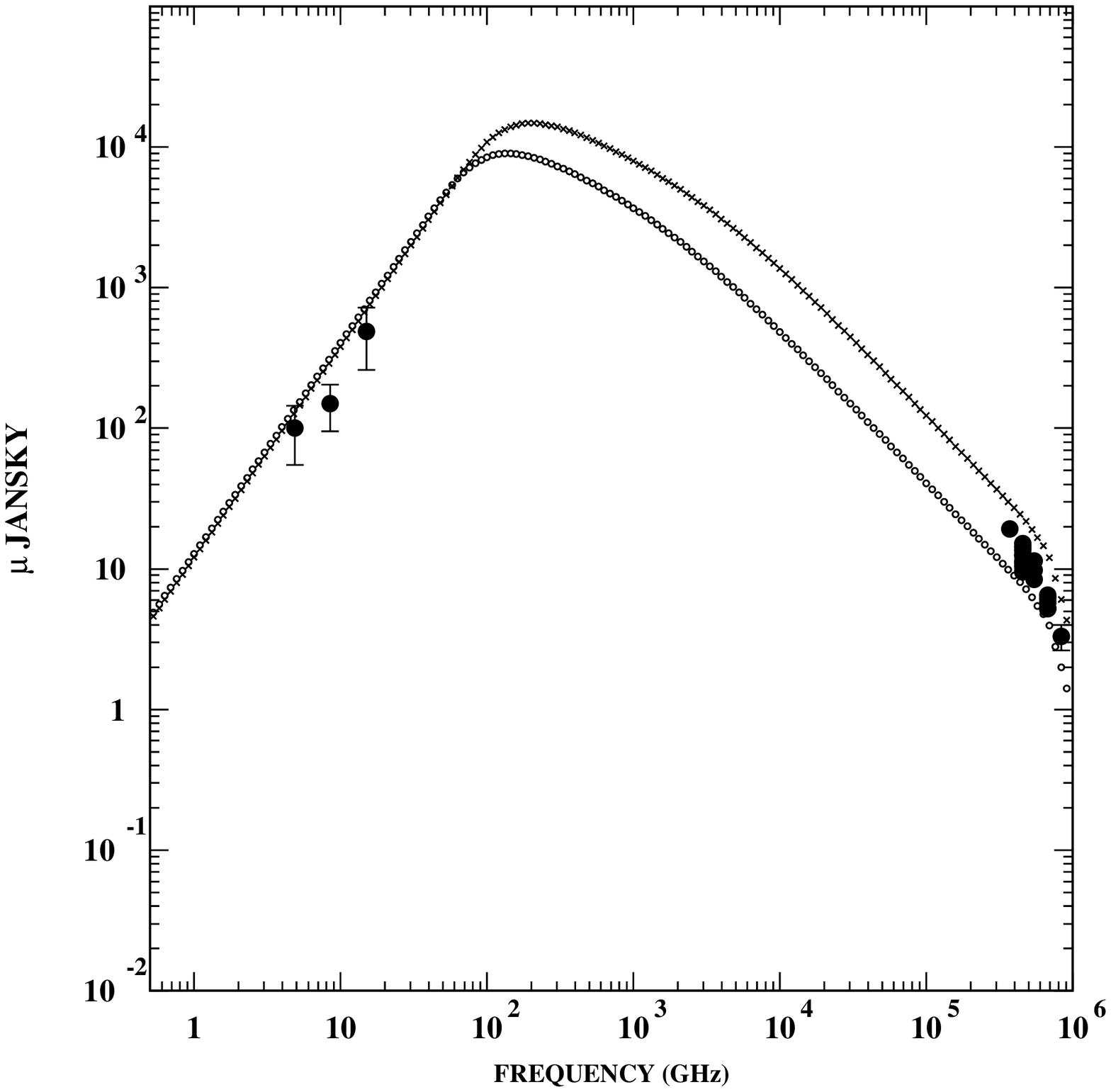, width=8cm} 
\end{tabular}  
\caption{The spectrum of the AG of GRB 000926 
from radio to optical frequencies.
Upper panel: in the time interval between 0.8 and 1.4 days after burst.
Lower panel:  in the time interval between 1.4 and 2.5 days after burst.
In both cases the highest peaking curve
corresponds to the earlier time. }  
\label{rad-opt926}  
\end{figure}

\begin{figure}[t]  
\begin{tabular}{cc}  
\hskip 2.5truecm  
\vspace*{2cm} 
\hspace*{-2.7cm}  
\epsfig{file=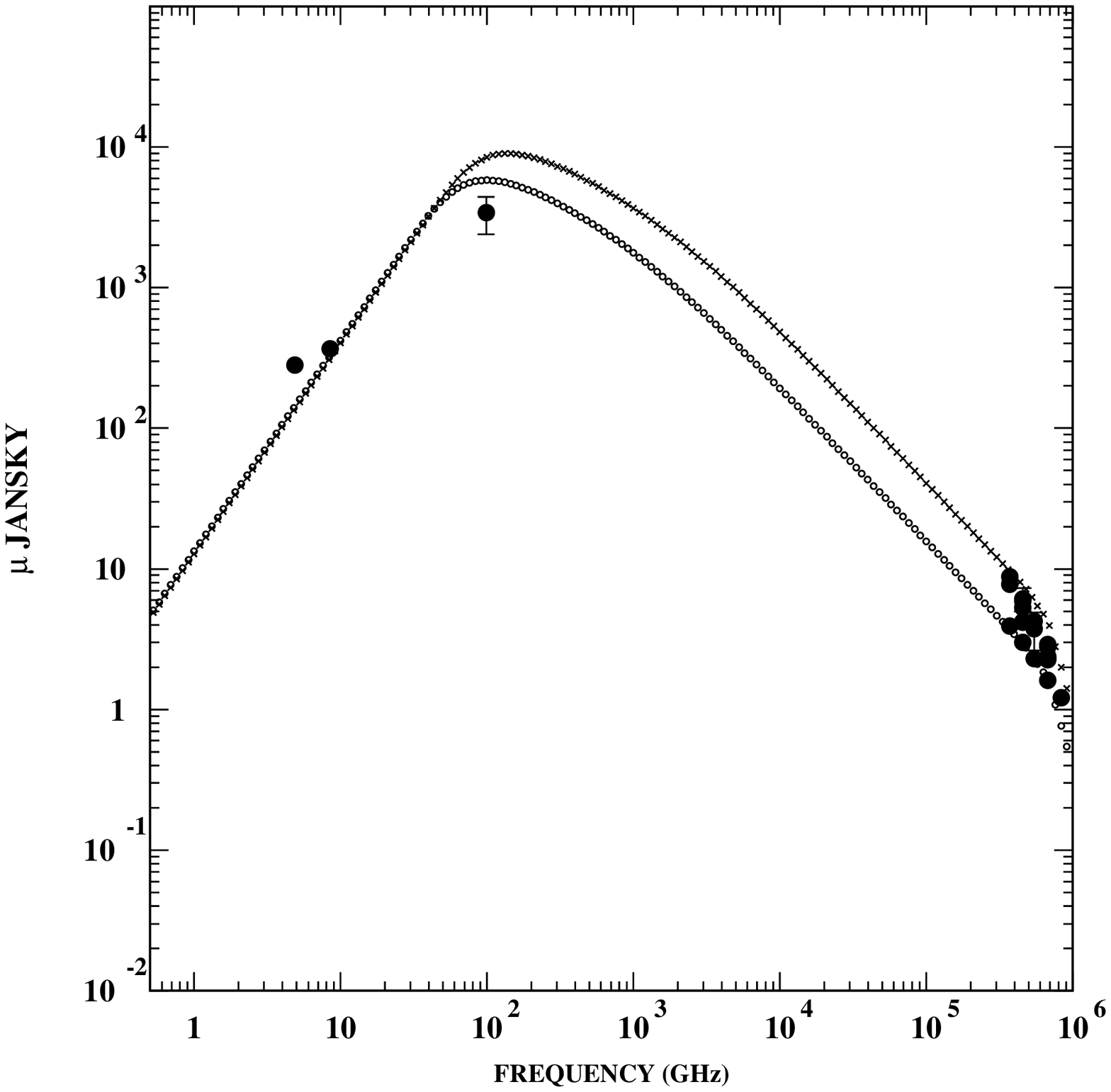, width=8cm}
\vspace*{-1.5cm}
\\ 
%\hskip 1truecm 
\hspace*{-.2cm}  
\epsfig{file=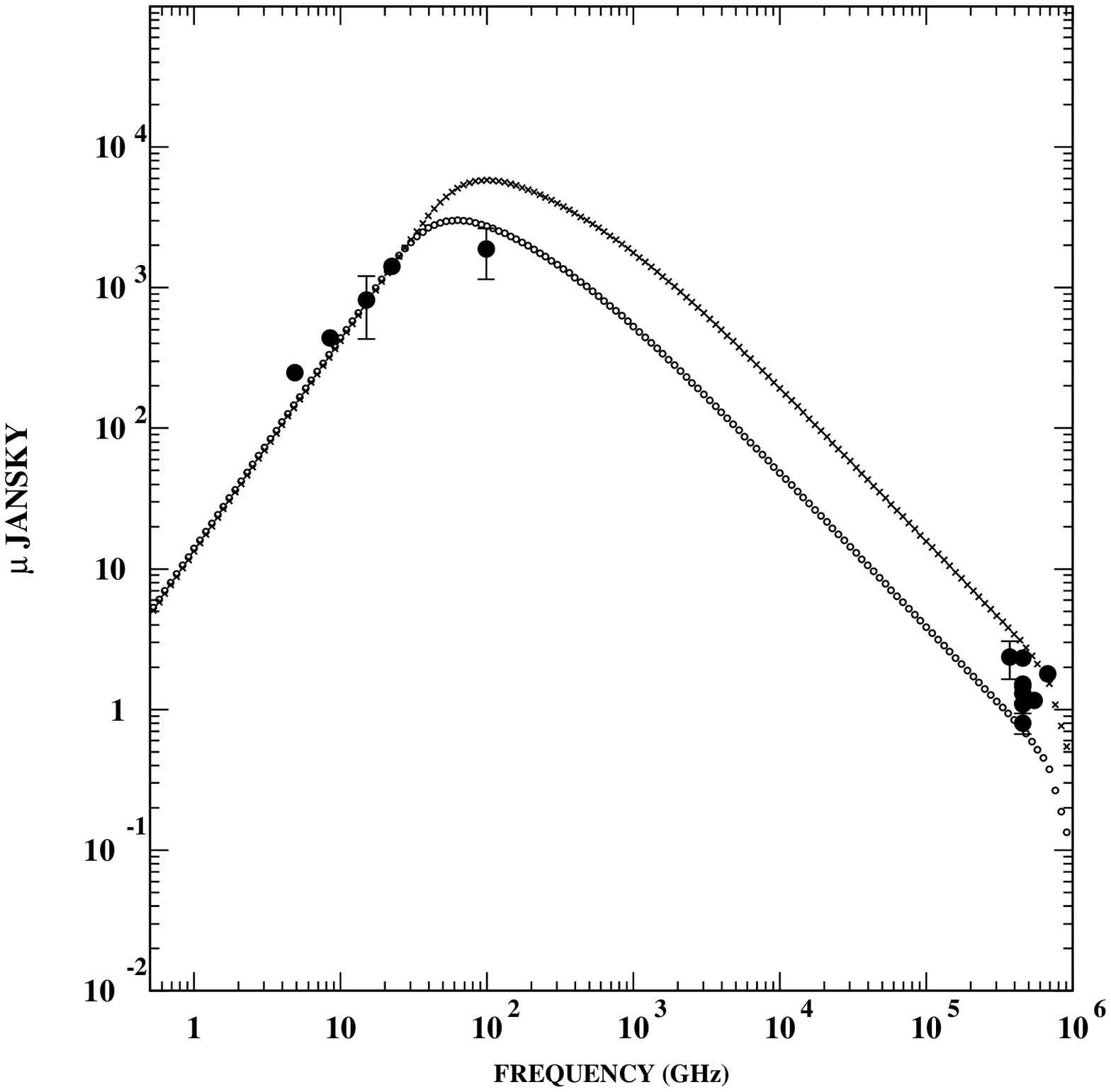, width=8cm} 
\end{tabular}  
\caption{The spectrum of the AG of GRB 000926 
from radio to optical frequencies.
Upper panel: in the time interval between 2.5 and 4 days after burst.
Lower panel:  in the time interval between 4 and 8 days after burst.
In both cases the highest peaking curve
corresponds to the earlier time.}  
\label{rad-opt926b}  
\end{figure}

\begin{figure}[t]  
\begin{tabular}{cc}  
\hskip 2.5truecm  
\vspace*{2cm} 
\hspace*{-2.7cm}  
\epsfig{file=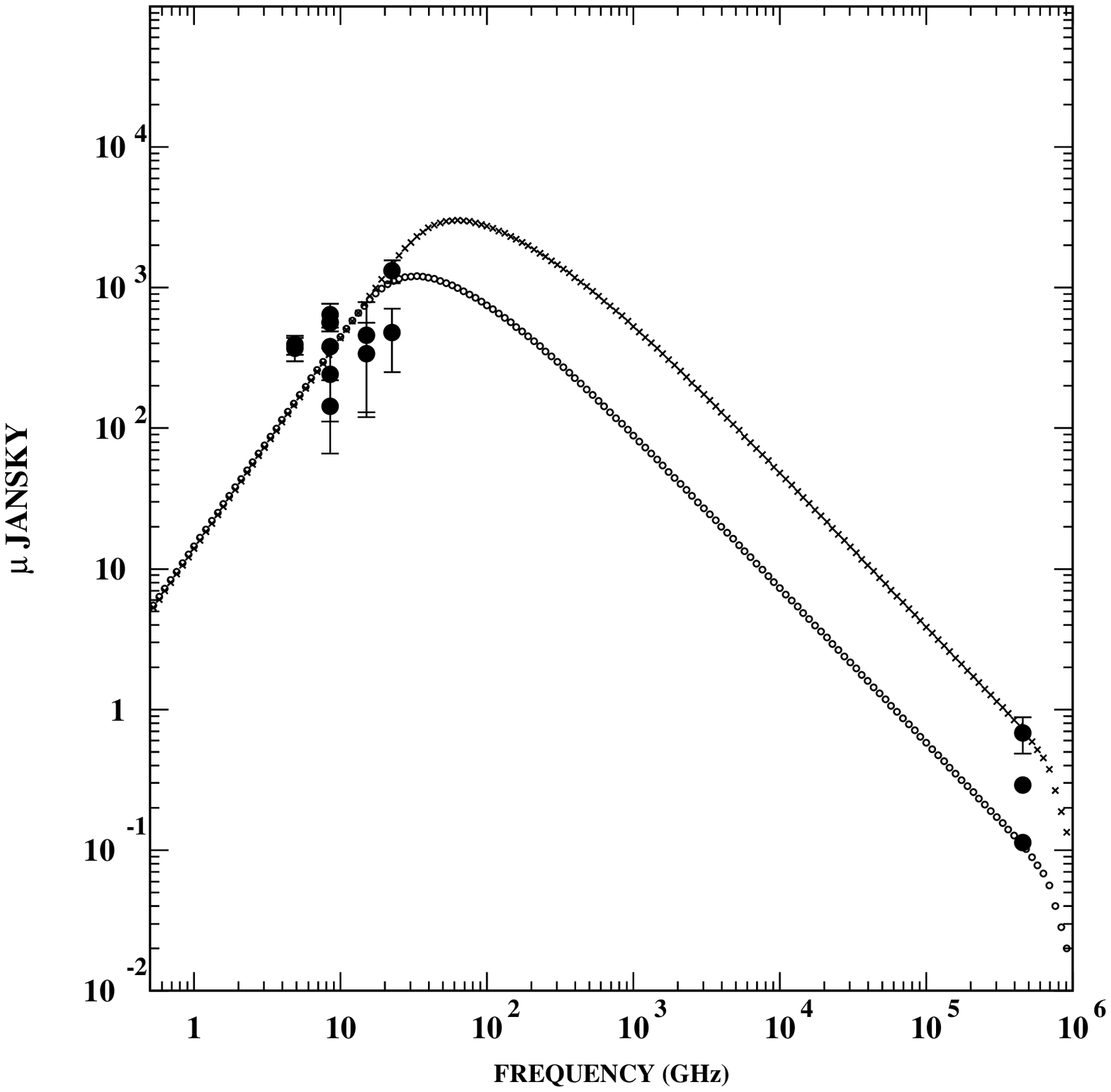, width=8cm}
\vspace*{-1.5cm}
\\ 
%\hskip 1truecm 
\hspace*{-.2cm}  
\epsfig{file=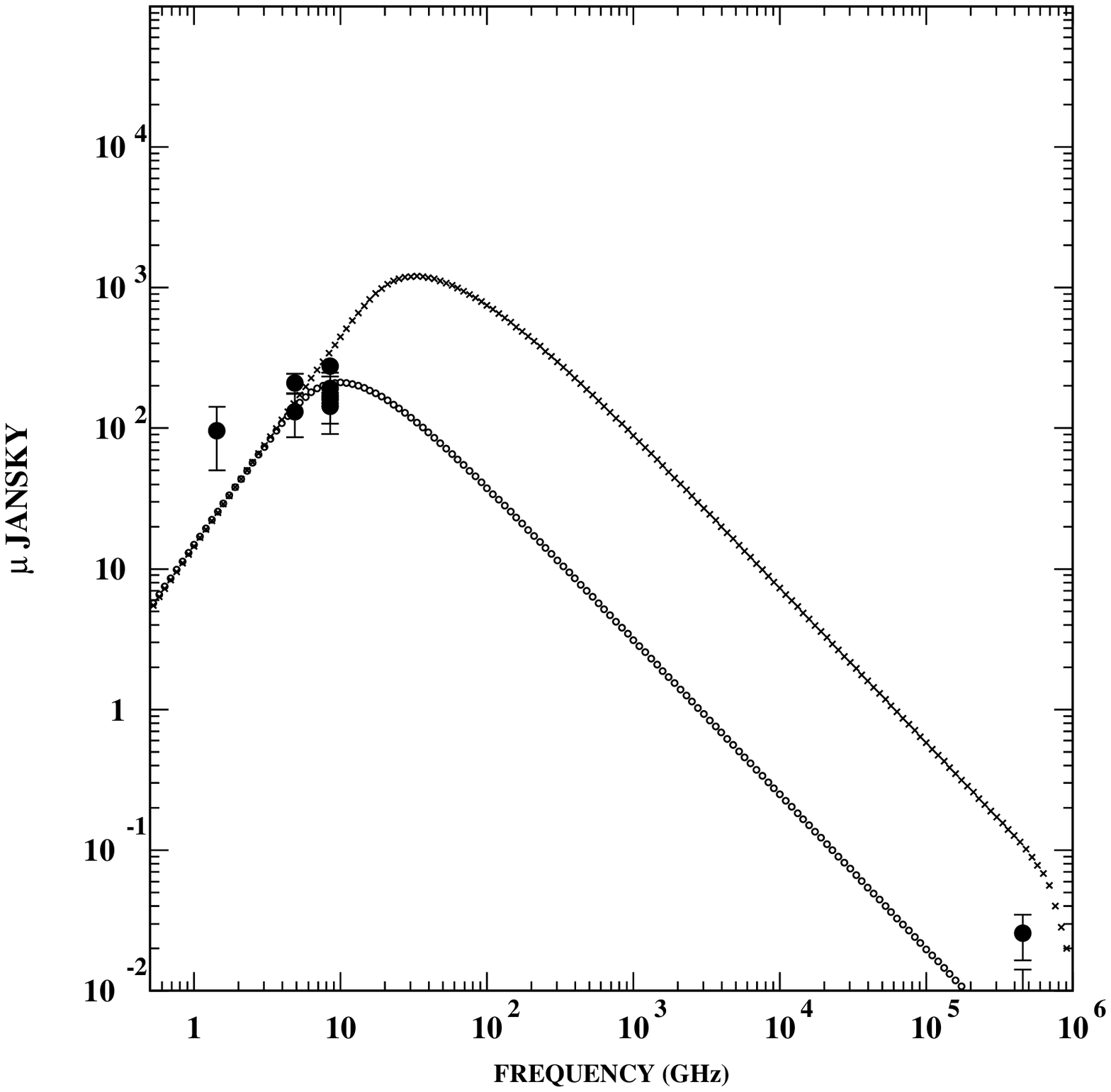, width=8cm} 
\end{tabular}  
\caption{The spectrum of the AG of GRB 000926 
from radio to optical frequencies.
Upper panel: in the time interval between 8 and 20 days after burst.
Lower panel:  in the time interval between 20 and 100 days after burst.
In both cases the highest peaking curve
corresponds to the earlier time.}  
\label{rad-opt926c}  
\end{figure}

In Figs.~(\ref{rad-opt926}) to (\ref{rad-opt926c}) we
make the complementary comparison of theory and
observations for the radio-to-optical spectra,
in six time intervals extending from 0.8 to 100 days. The results, in spite
of the crude estimate of extinction in
the host galaxy and the scintillations so clearly visible in the radio
light curves, are satisfactory.

\subsection{GRB 991216} 

\begin{figure}[t]  
%\begin{tabular}{cc}  
\hskip 0truecm   
\epsfig{file=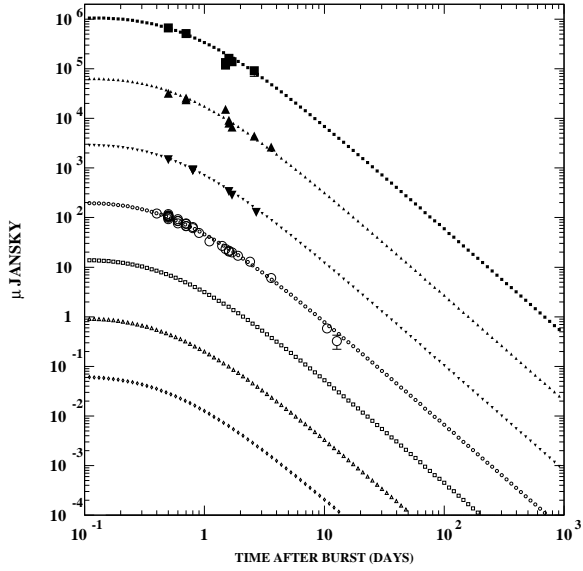, width=8.5cm}  
%\end{tabular}  
\caption{Comparisons between our fitted CB model afterglow, 
Eq.~(\ref{Fnuobser}), 
and the observed optical afterglow of GRB 991216 
at $\rm z=1.02\, .$ 
The figure shows (from top to bottom) 1000 times the K-band results,
100 times the J-band, 10 times the I-band, the R-band, 1/10 of the V-band,
1/100 of the B-band and 1/1000 of the U-band.
The contributions of the underlying galaxy and the
expected SN1998bw-like 
SN have been subtracted. In a CB-model fit, there is in this case
some evidence for such a SN (DDD 2001).}
\label{opt216}  
\end{figure}

\begin{figure}[t]  
\begin{tabular}{cc}  
\hskip 2truecm  
\vspace*{2cm} 
\hspace*{-1.7cm}  
\epsfig{file=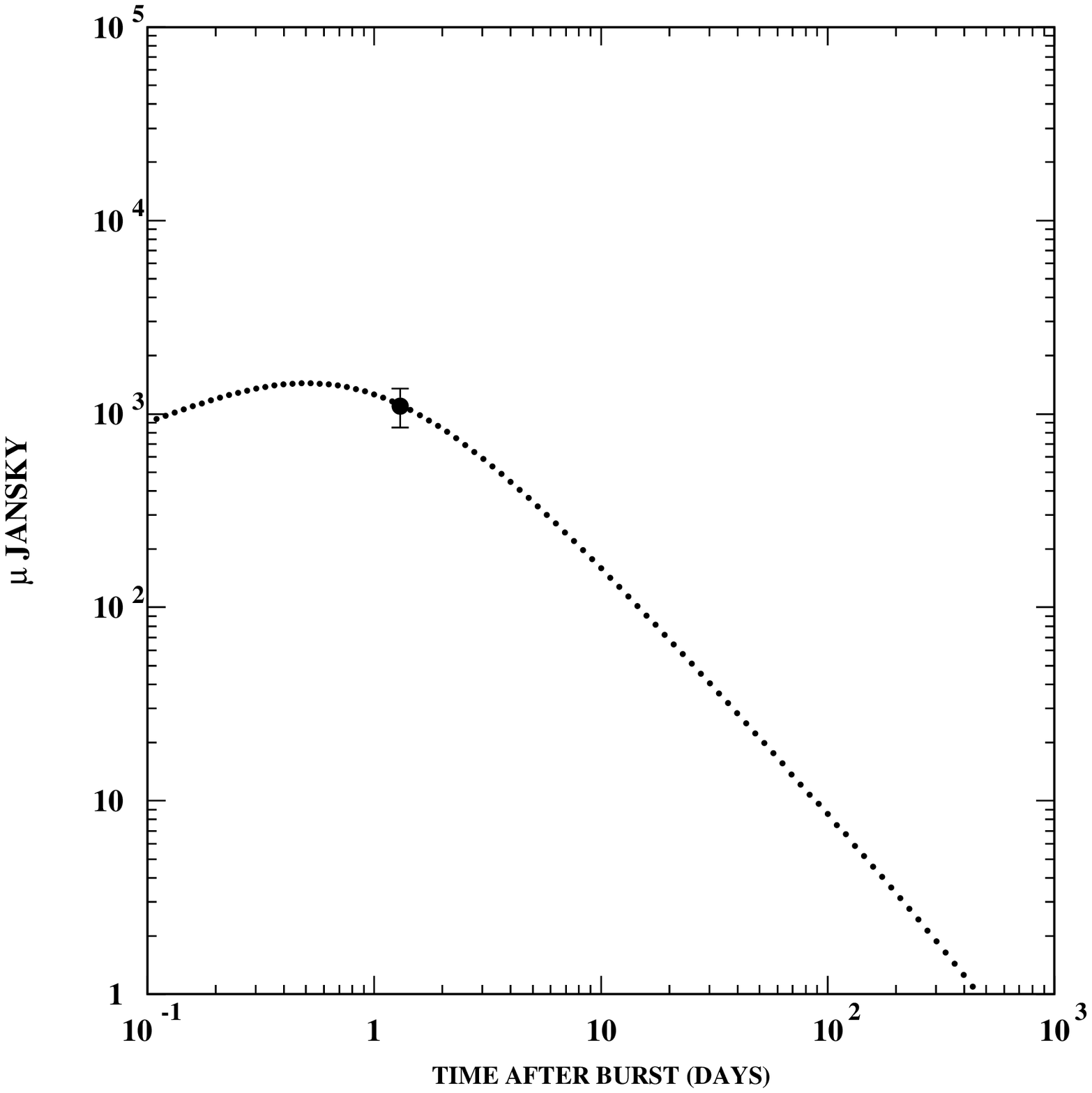, width=8cm} \\ 
%\hskip 1truecm  
\hspace*{.5cm}  
\epsfig{file=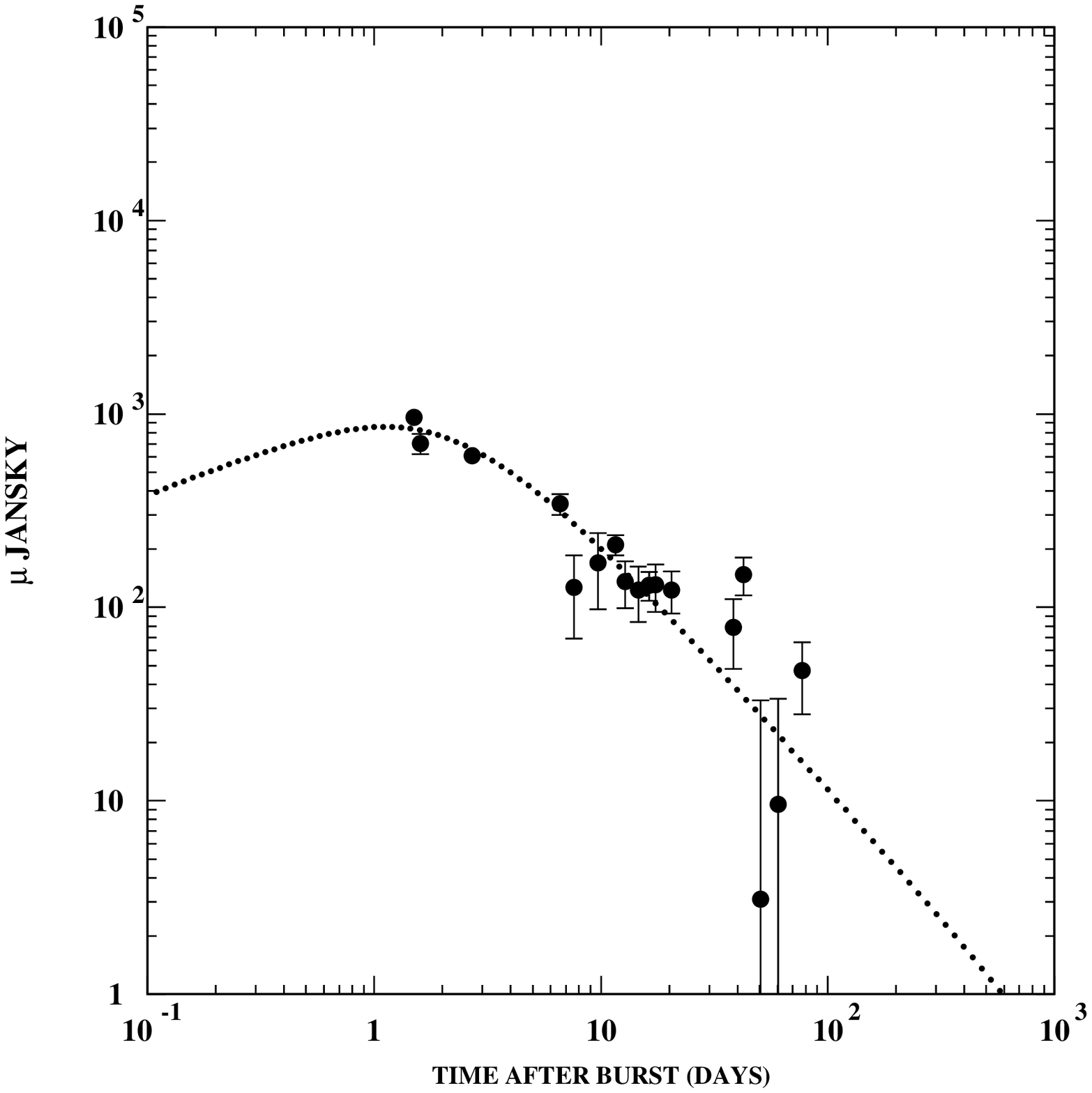, width=8cm} 
\end{tabular}  
\caption{Comparisons between our fitted CB model afterglow, 
Eq.~(\ref{Fnuobser}), and the observed radio afterglow of GRB 991216.
Upper panel: the light curve at 15 
GHz. Lower panel: the light curve at 8.46 GHz.}  
\label{figr121602}  
\end{figure}

\begin{figure}[t]  
\begin{tabular}{cc}  
\hskip 2truecm  
\vspace*{2cm} 
\hspace*{-1.7cm}  
\epsfig{file=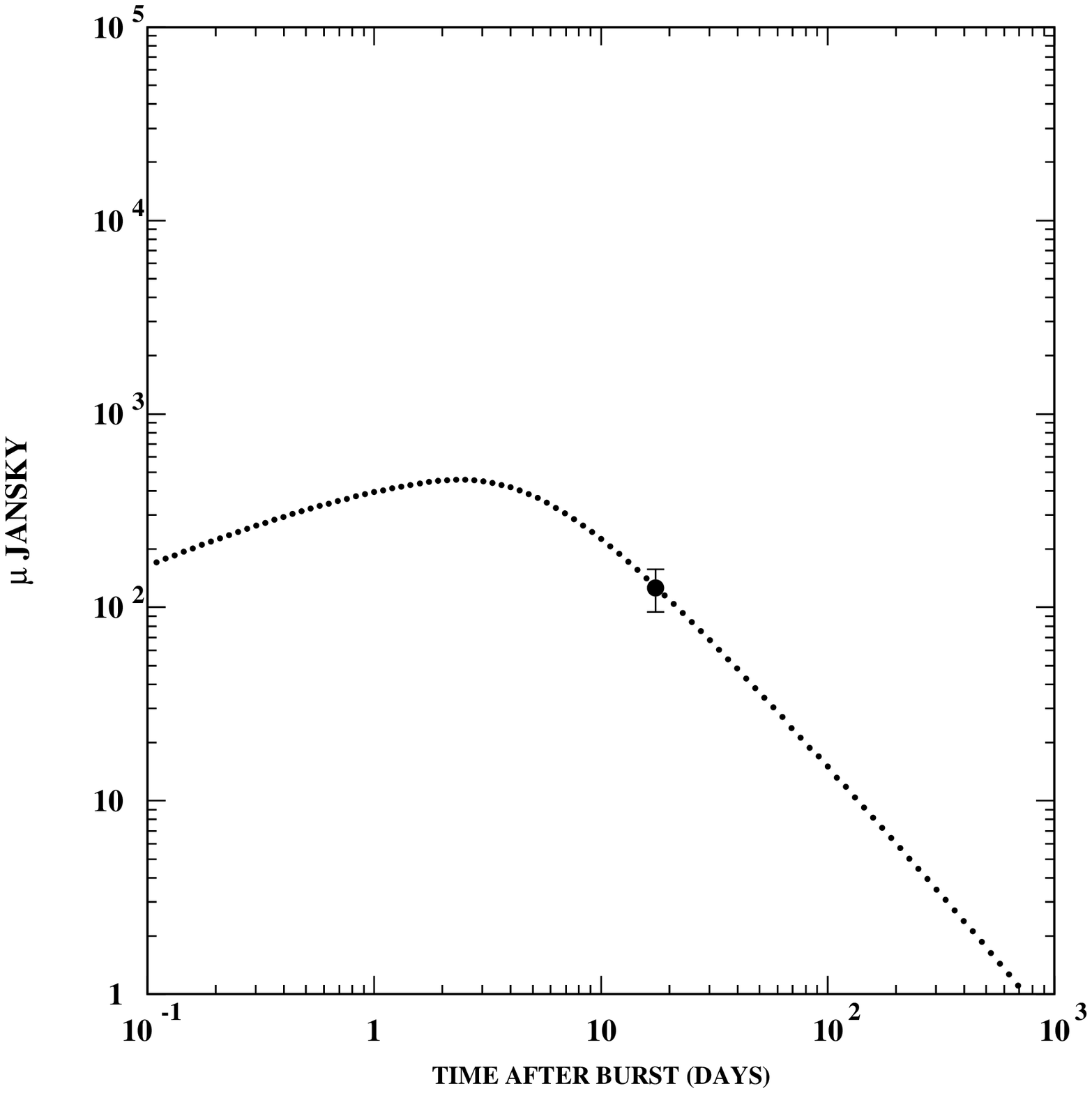, width=8cm} \\ 
%\hskip 1truecm  
\hspace*{.5cm}  
\epsfig{file=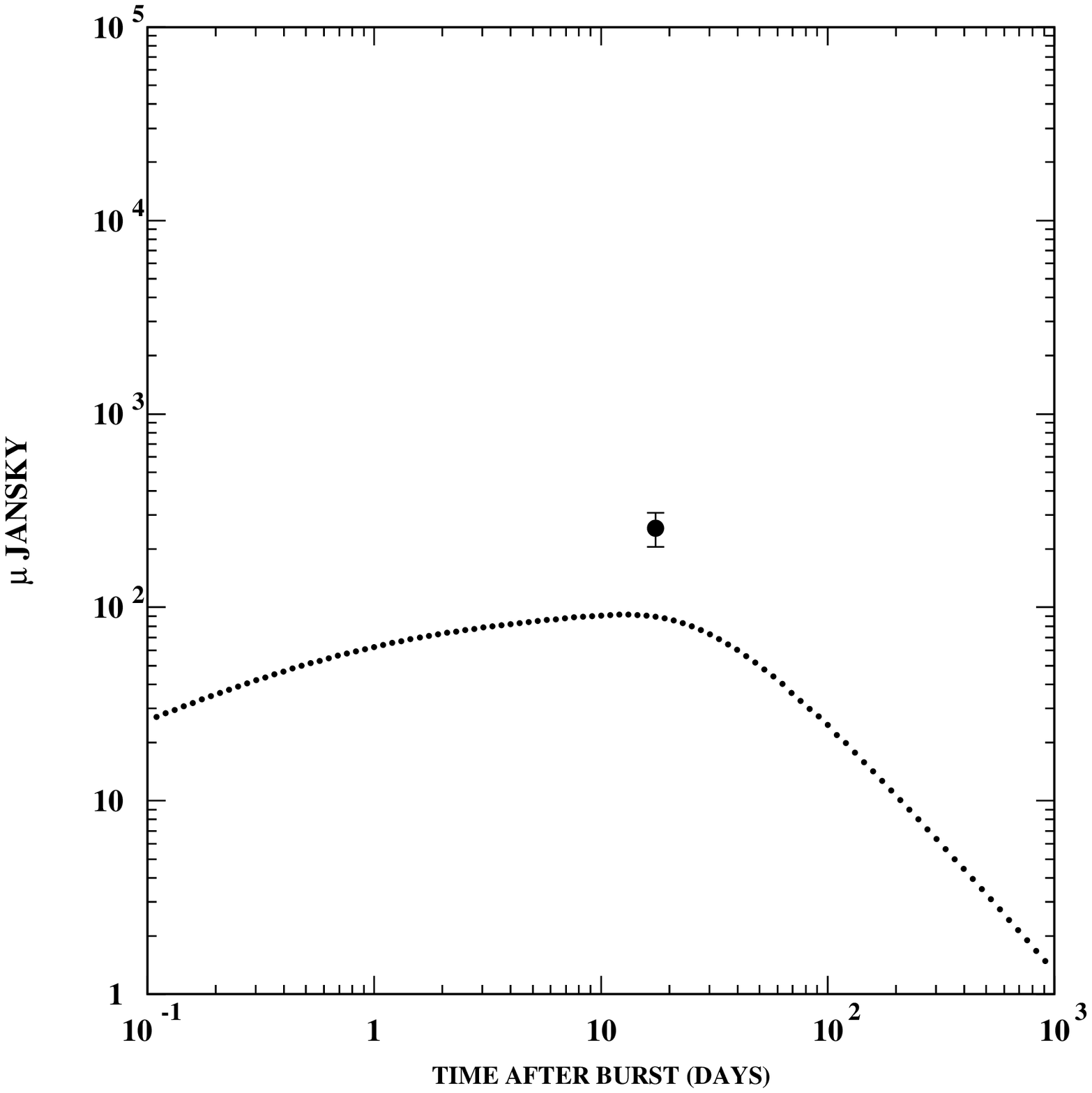, width=8cm} 
\end{tabular}  
\caption{Comparisons between our fitted CB model afterglow, 
Eq.~(\ref{Fnuobser}), and the observed radio afterglow of GRB 991216.
Upper panel: the light curve at 4.86 
GHz. Lower panel: the light curve at 1.43 GHz.}  
\label{figr121603}  
\end{figure}

\begin{figure}[t]  
\begin{tabular}{cc}  
\hskip 2.5truecm  
\vspace*{2cm} 
\hspace*{-2.7cm}  
\epsfig{file=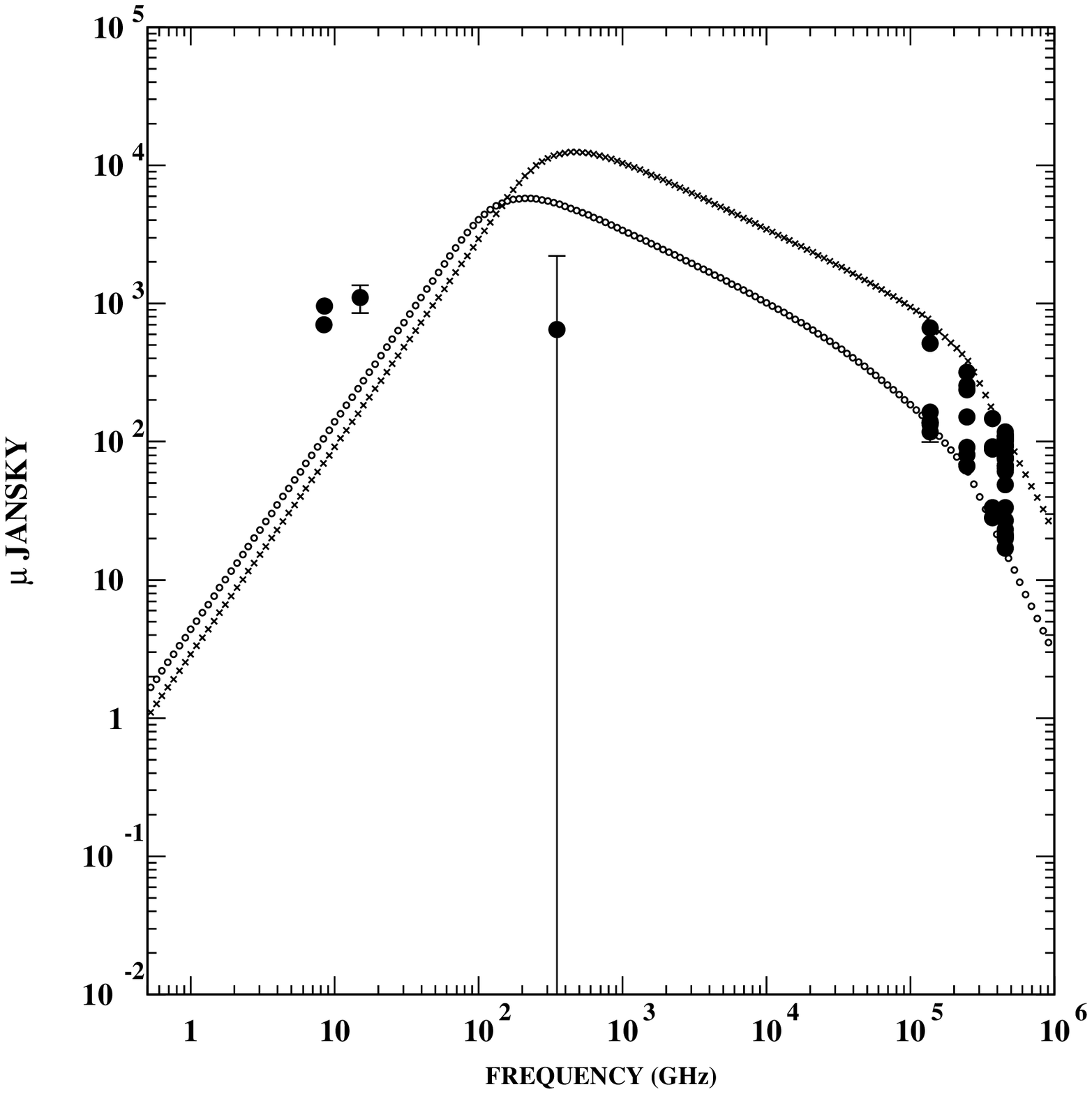, width=8cm}
\vspace*{-1.5cm}
\\ 
%\hskip 1truecm 
\hspace*{-.2cm}  
\epsfig{file=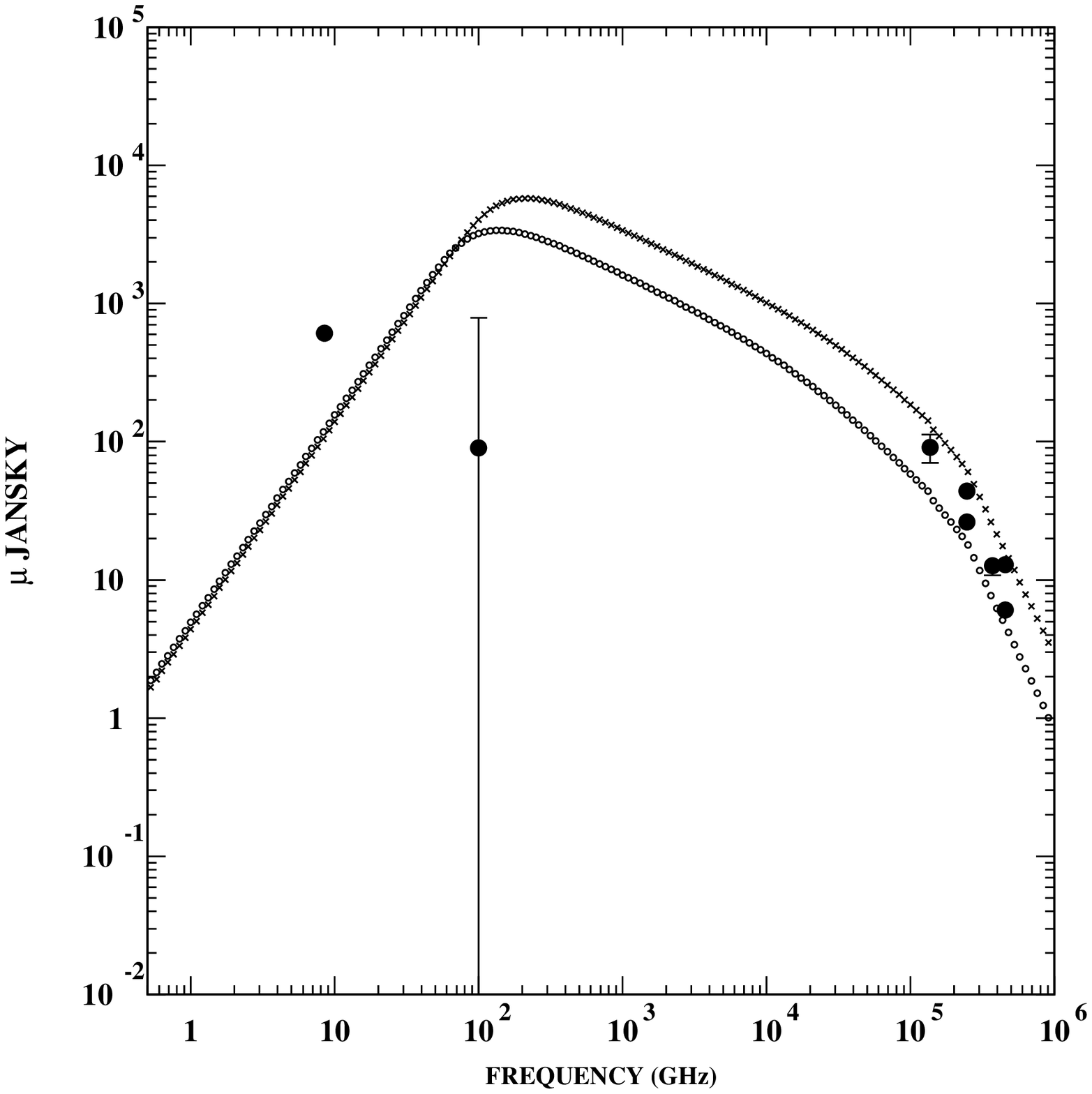, width=8cm} 
\end{tabular}  
\caption{The spectrum of the AG of GRB 991216 from radio to optical
frequencies. 
Upper panel: in the time interval between 0.44 and 2 days after burst.
Lower panel:  in the time interval between 2 and 4 days after burst. 
In both cases the highest peaking curve
corresponds to the earlier time.}  
\label{rad-opt216}  
\end{figure}

\begin{figure}[t]  
\begin{tabular}{cc}  
\hskip 2.5truecm  
\vspace*{2cm} 
\hspace*{-2.7cm}  
\epsfig{file=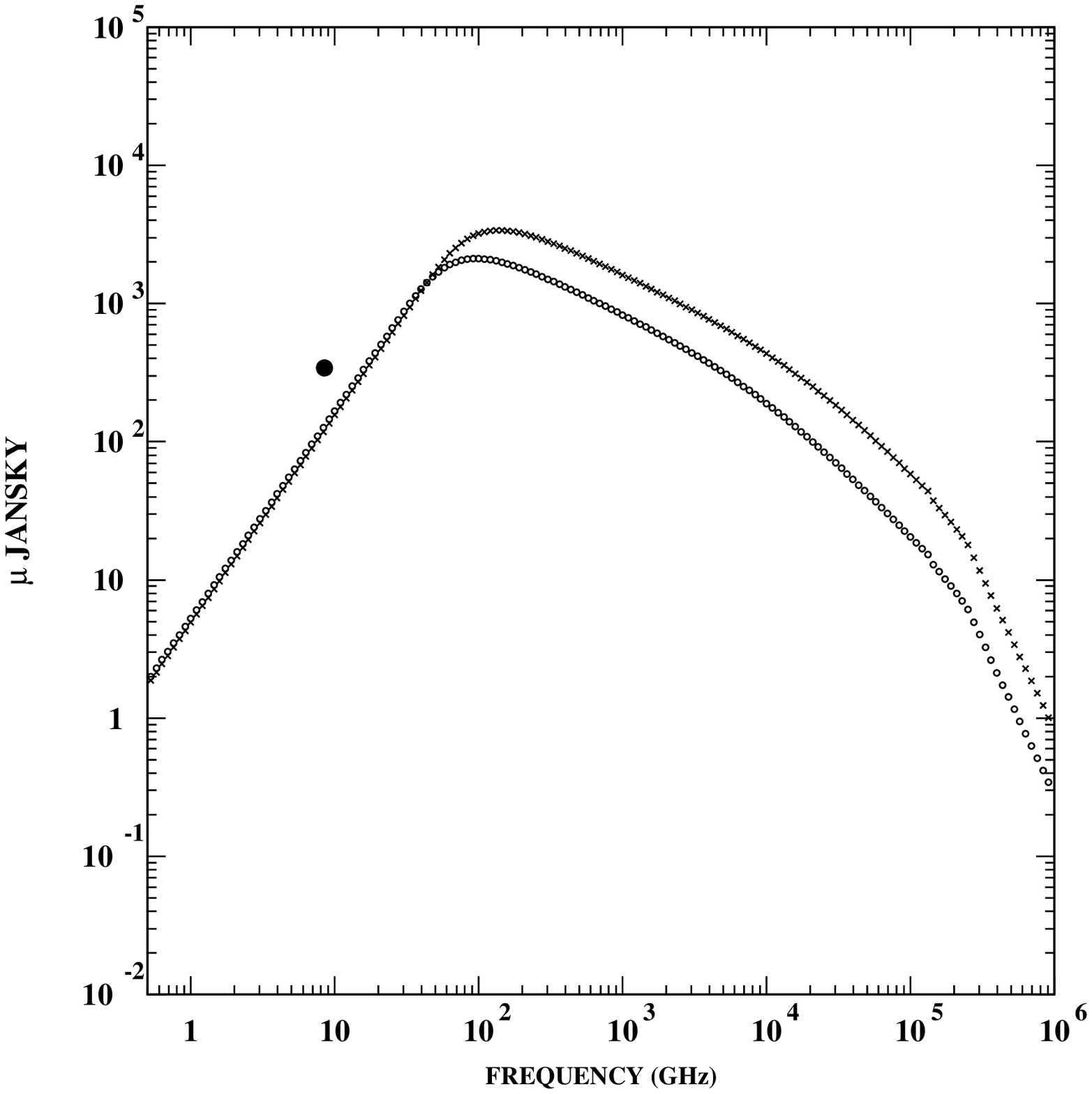, width=8cm}
\vspace*{-1.5cm}
\\ 
%\hskip 1truecm 
\hspace*{-.2cm}  
\epsfig{file=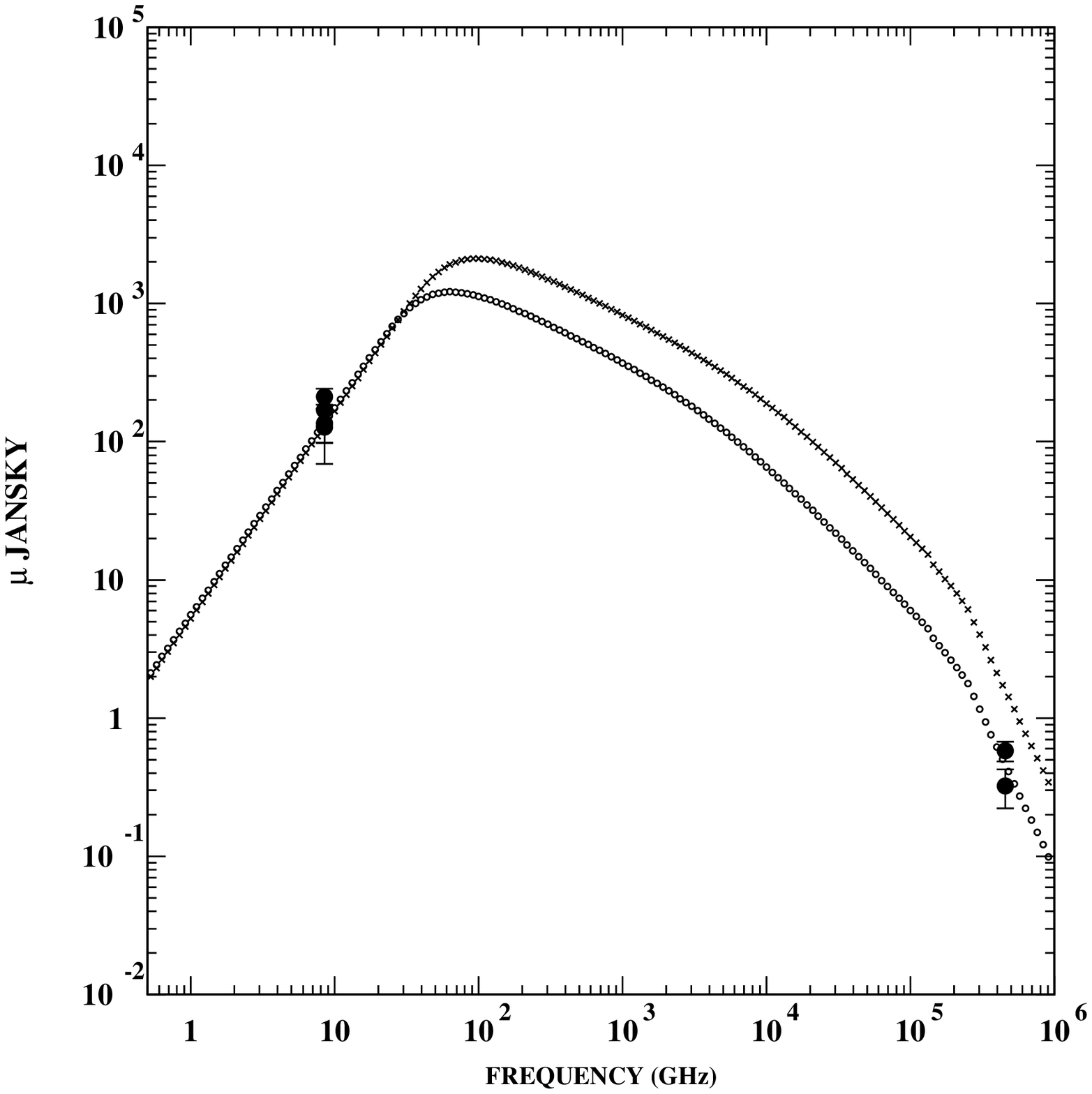, width=8cm} 
\end{tabular}  
\caption{The spectrum of the AG of GRB 991216 from radio to optical
frequencies. 
Upper panel: in the time interval between 4 and 7 days after burst.
Lower panel:  in the time interval between 7 and 13 days after burst. 
In both cases the highest peaking curve
corresponds to the earlier time.}  
\label{rad-opt216b}  
\end{figure}

\begin{figure}[t]  
\begin{tabular}{cc}  
\hskip 2.5truecm  
\vspace*{2cm} 
\hspace*{-2.7cm}  
\epsfig{file=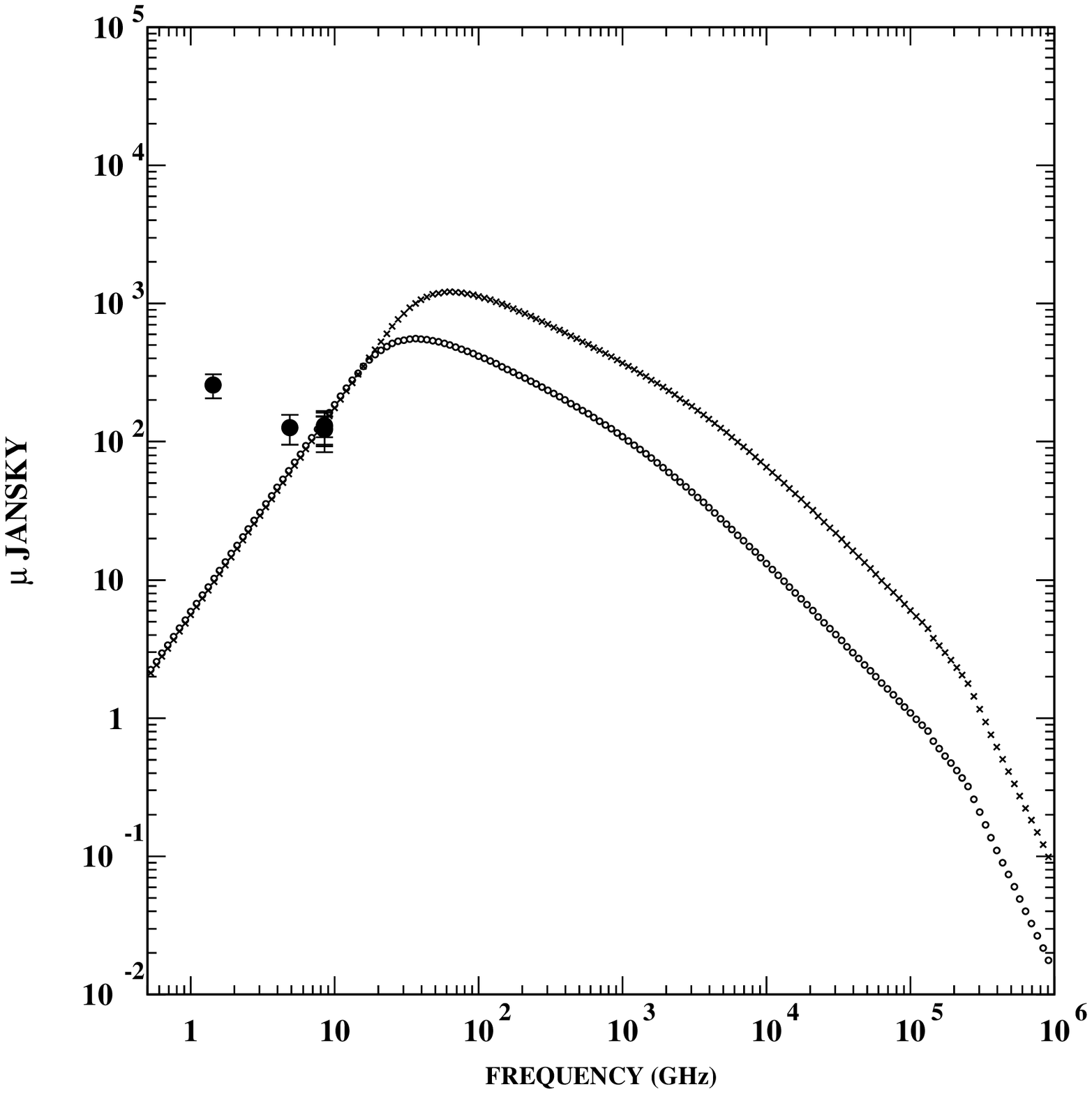, width=8cm}
\vspace*{-1.5cm}
\\ 
%\hskip 1truecm 
\hspace*{-.2cm}  
\epsfig{file=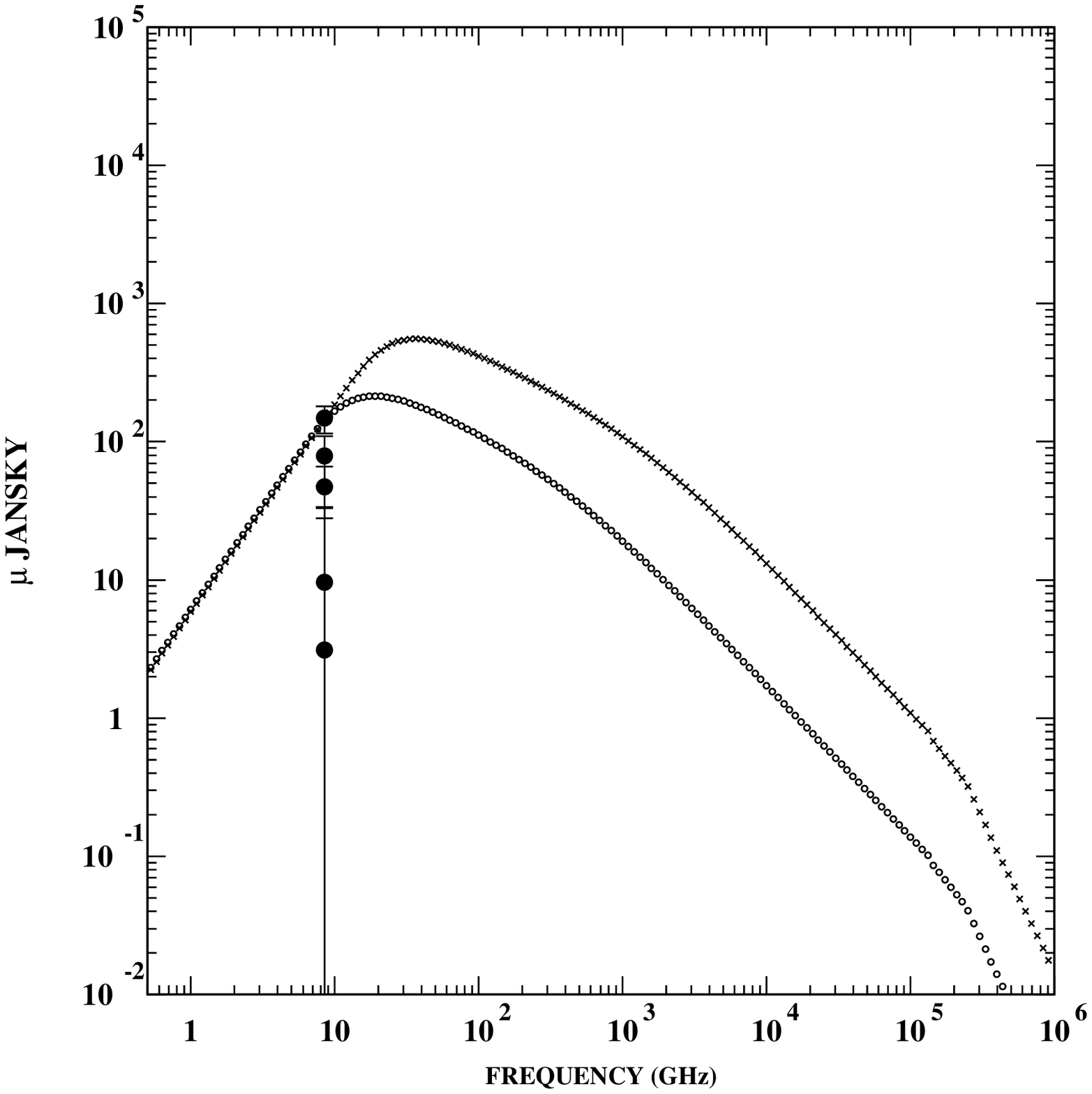, width=8cm} 
\end{tabular}  
\caption{The spectrum of the AG of GRB 991216 from radio to optical
frequencies. 
Upper panel: in the time interval between 13 and 30 days after burst.
Lower panel:  in the time interval between  30 and 80 days after burst. 
In both cases the highest peaking curve
corresponds to the earlier time.}  
\label{rad-opt216c}  
\end{figure}

The NIR/optical data for this GRB are from
Halpern et al. (2000) and Garnavich et al. (2000a); the
radio data from Frail et al. (2000b). 
In Fig~(\ref{opt216}) we present the comparison between
the measured light curves for the K, J, I, R bands, after
subtraction of the host galaxy and SN contributions (DDD 2001), and the
fitted CB model predictions. The predictions were corrected for
extinction in the host galaxy and ours, as estimated by Halpern
et al. (2000): E(B - V)=0.40. 
In Figs.~(\ref{figr121602}) to (\ref{figr121603}) 
we present the radio light curves at six frequencies from
350 to 1.43 GHz.
In Figs.~(\ref{rad-opt216}) to
(\ref{rad-opt216c}) we make the complementary comparison of
theory and observations 
for the radio to optical spectra,
in six time intervals extending
from 0.44 to 80 days. The results are once again satisfactory.

\subsection{GRB 991208}

\begin{figure}[t]  
%\begin{tabular}{cc}  
\hskip 0truecm   
\epsfig{file=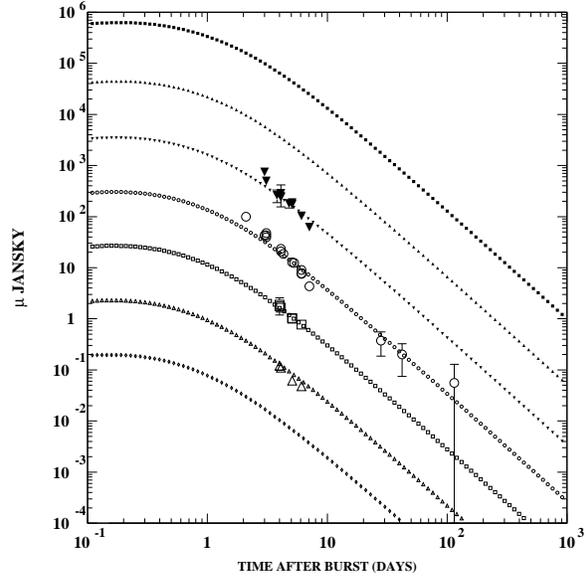, width=8.5cm}  
%\end{tabular}  
\caption{Comparisons between our fitted CB model AG of GRB 991208, 
at $\rm z=0.706$,
Eq.~(\ref{Fnuobser}), with the observed optical data. 
The figure shows (from top to bottom) 1000 times the K-band results,
100 times the J-band, 10 times the I-band, the R-band, 1/10 of the V-band,
1/100 of the B-band and 1/1000 of the U-band.
The contribution of the underlying galaxy and associated
supernova has been subtracted. 
The contributions of the underlying galaxy and the
expected SN1998bw-like 
SN have been subtracted. In a CB-model fit, there is in this case
strong evidence for such a SN (DDD 2001).
}  
\label{opt208}  
\end{figure}

\begin{figure}[t] 
\begin{tabular}{cc} 
\hskip 2truecm 
\vspace*{2cm} 
\hspace*{-1.7cm} 
\epsfig{file=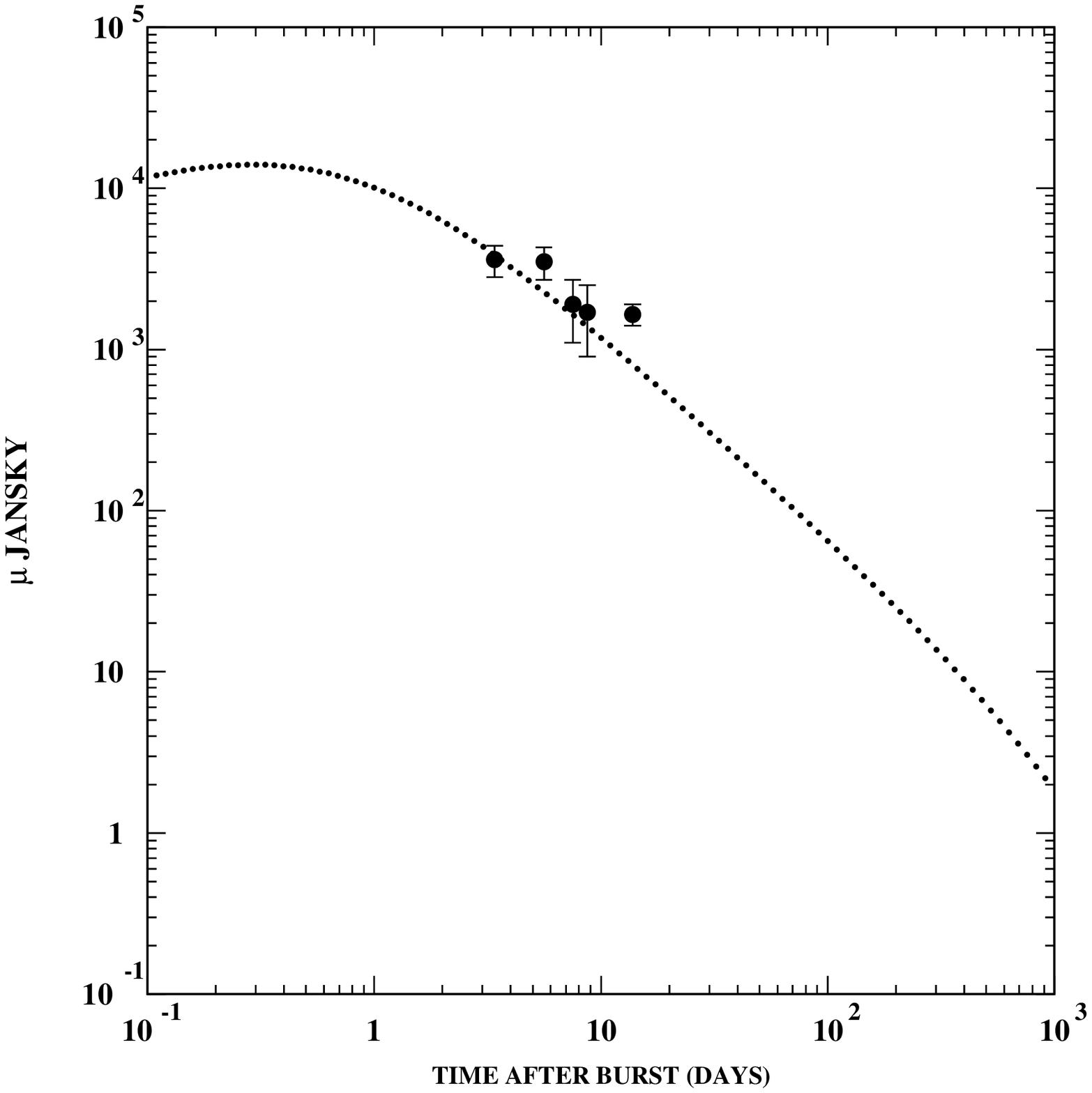, width=8cm} \\ 
%\hskip 1truecm 
\hspace*{.5cm} 
\epsfig{file=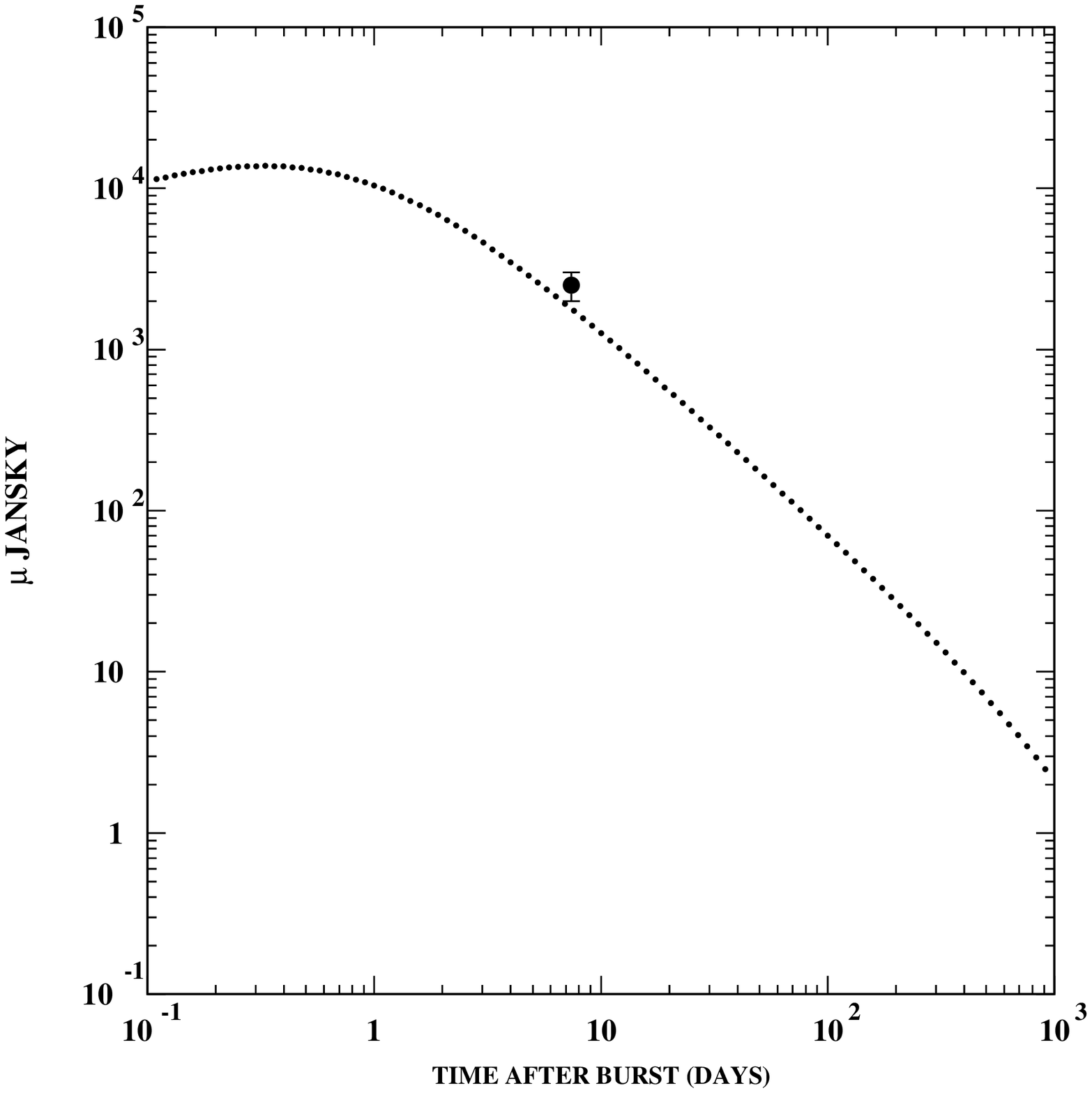, width=8cm} 
\end{tabular} 
\caption{Comparisons between our fitted CB model afterglow, 
Eq.~(\ref{Fnuobser}), 
and the observed radio afterglow of GRB 991208. 
Upper panel: the light curve at 100 GHz. 
Lower panel: the light curve at 86.2 GHz.} 
\label{figr120801} 
\end{figure}

\begin{figure}[t] 
\begin{tabular}{cc} 
\hskip 2truecm 
\vspace*{2cm} 
\hspace*{-1.7cm} 
\epsfig{file=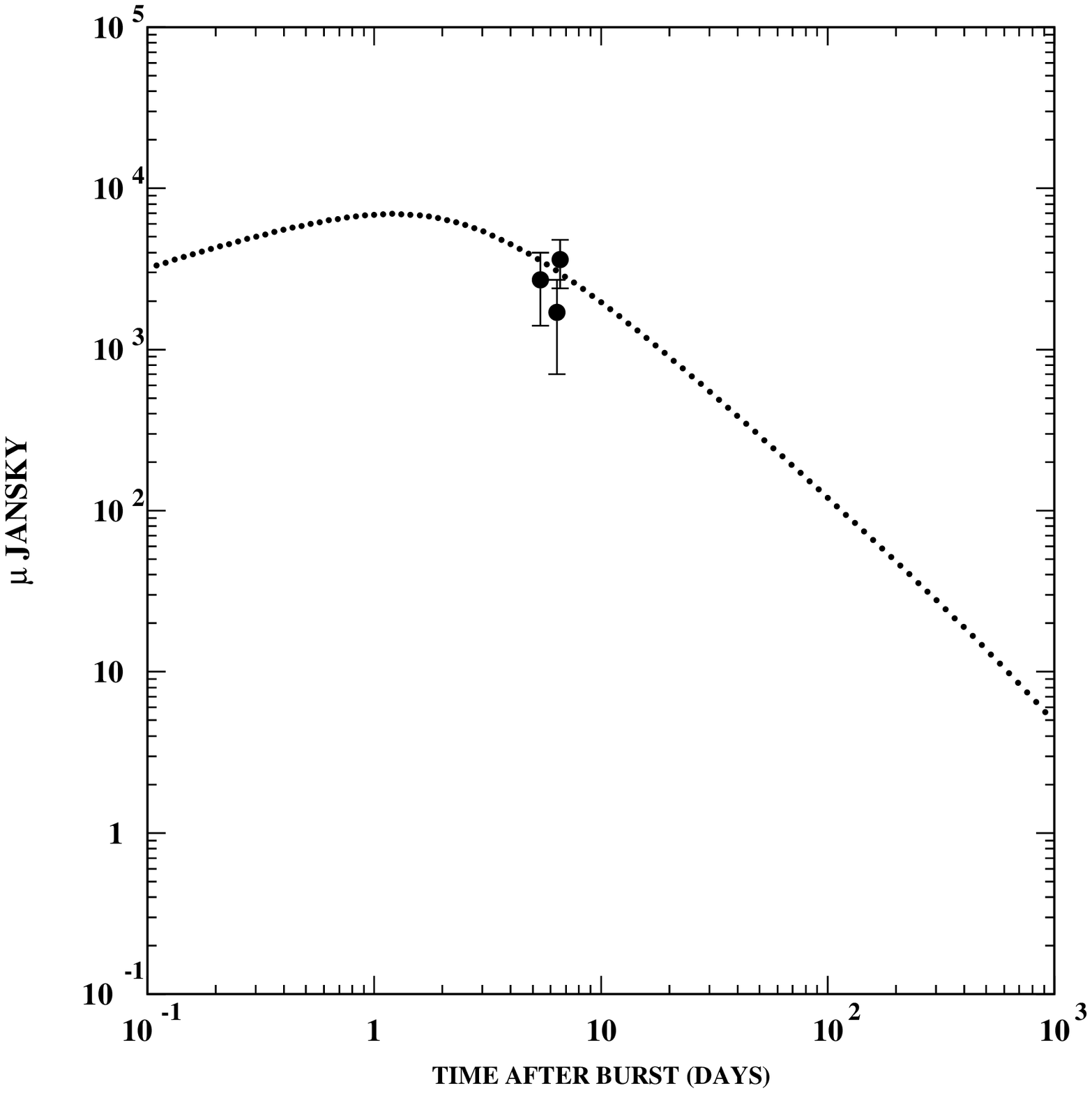, width=8cm} \\ 
%\hskip 1truecm 
\hspace*{.5cm} 
\epsfig{file=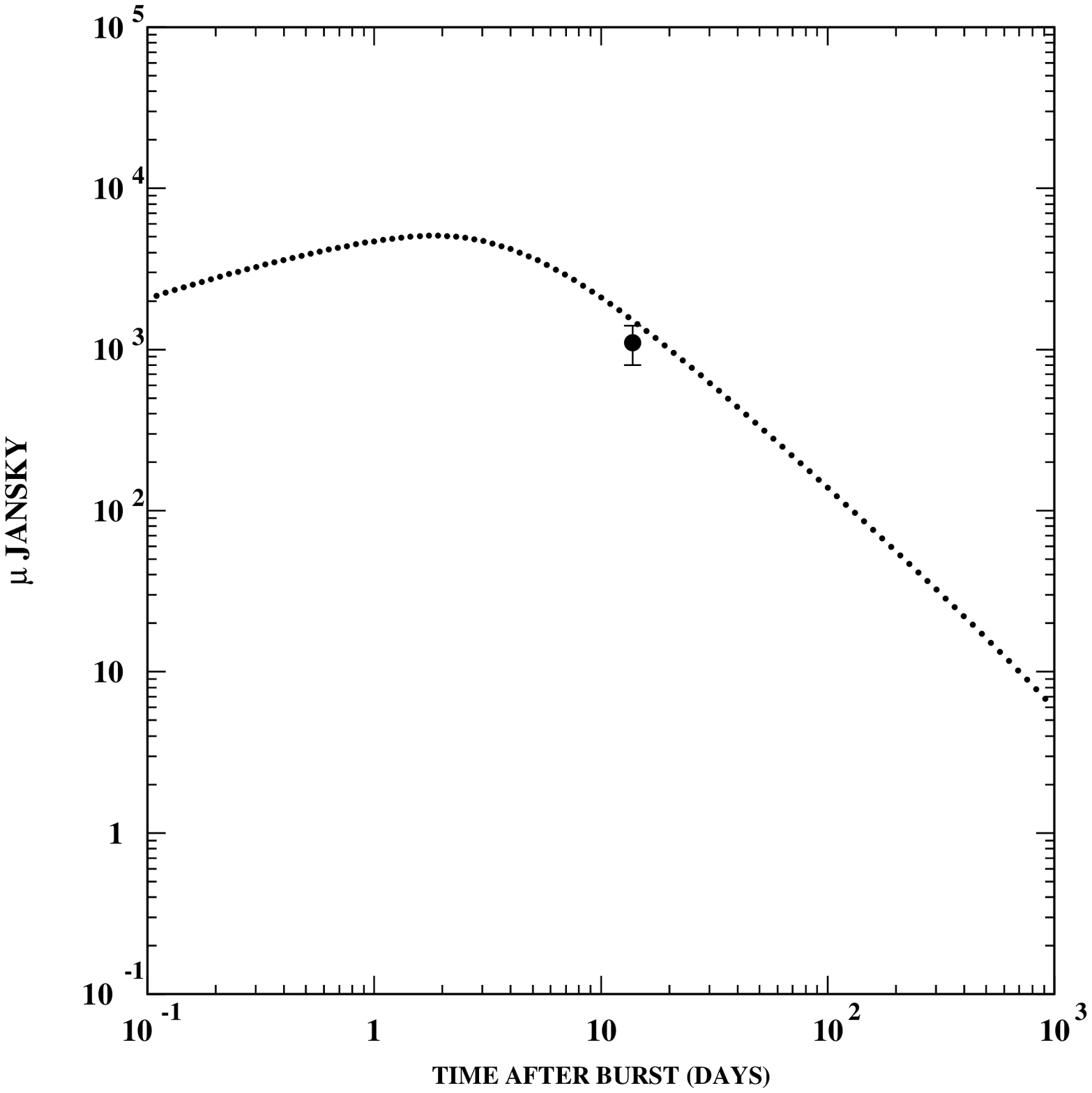, width=8cm} 
\end{tabular} 
\caption{Comparisons between our fitted CB model afterglow, 
Eq.~(\ref{Fnuobser}), 
and the observed radio afterglow of GRB 991208. 
Upper panel: the light curve at 30 GHz. 
Lower panel: the light curve at 22.5 GHz.} 
\label{figr120802} 
\end{figure}

\begin{figure}[t] 
\begin{tabular}{cc} 
\hskip 2truecm 
\vspace*{2cm} 
\hspace*{-1.7cm} 
\epsfig{file=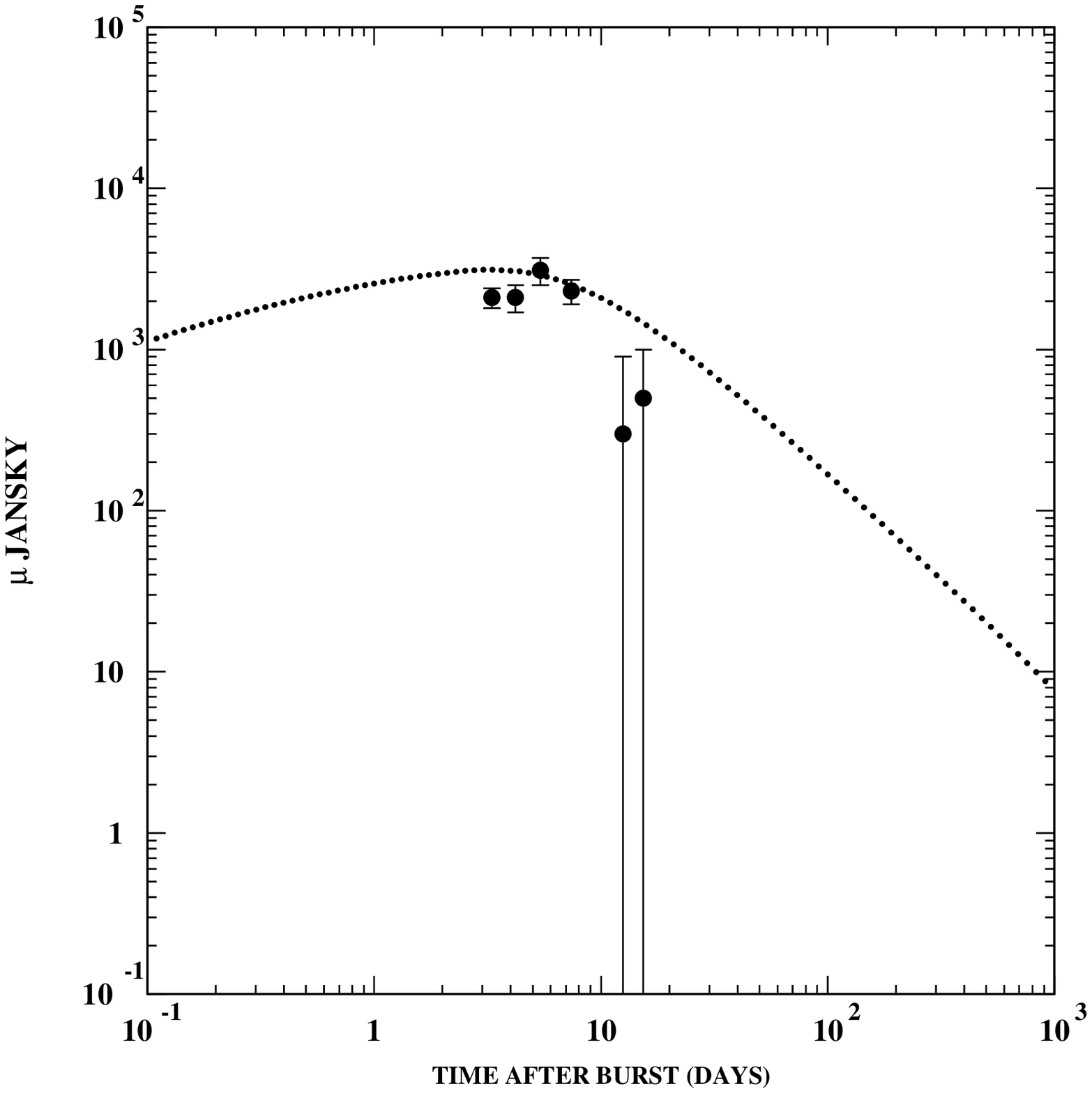, width=8cm} \\ 
%\hskip 1truecm 
\hspace*{.5cm} 
\epsfig{file=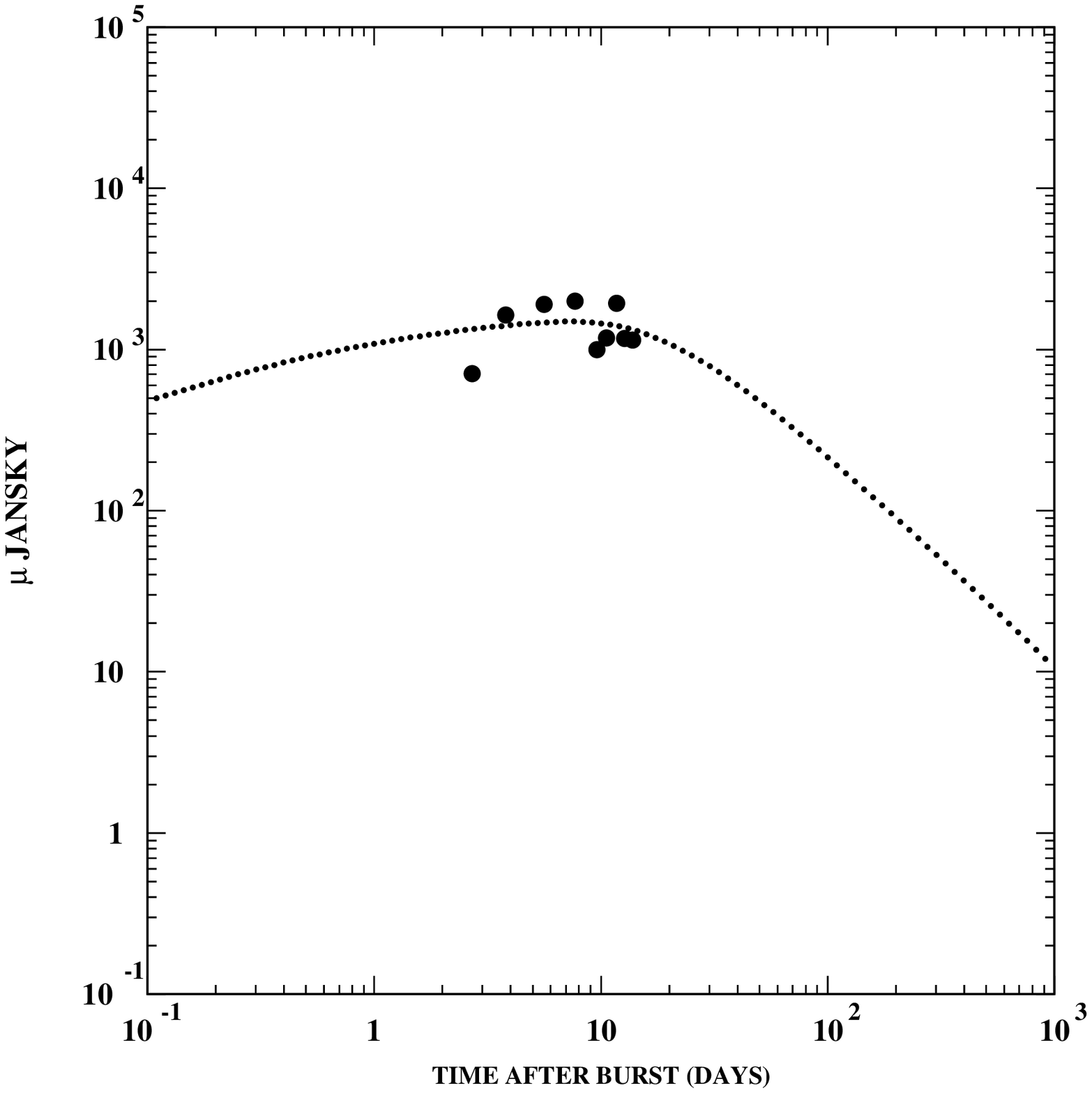, width=8cm} 
\end{tabular} 
\caption{Comparisons between our fitted CB model afterglow, 
Eq.~(\ref{Fnuobser}), 
and the observed radio afterglow of GRB 991208. 
Upper panel: the light curve at 15 GHz. 
Lower panel: the light curve at 8.46 GHz.} 
\label{figr120803} 
\end{figure}

\begin{figure}[t] 
\begin{tabular}{cc} 
\hskip 2truecm 
\vspace*{2cm} 
\hspace*{-1.7cm} 
\epsfig{file=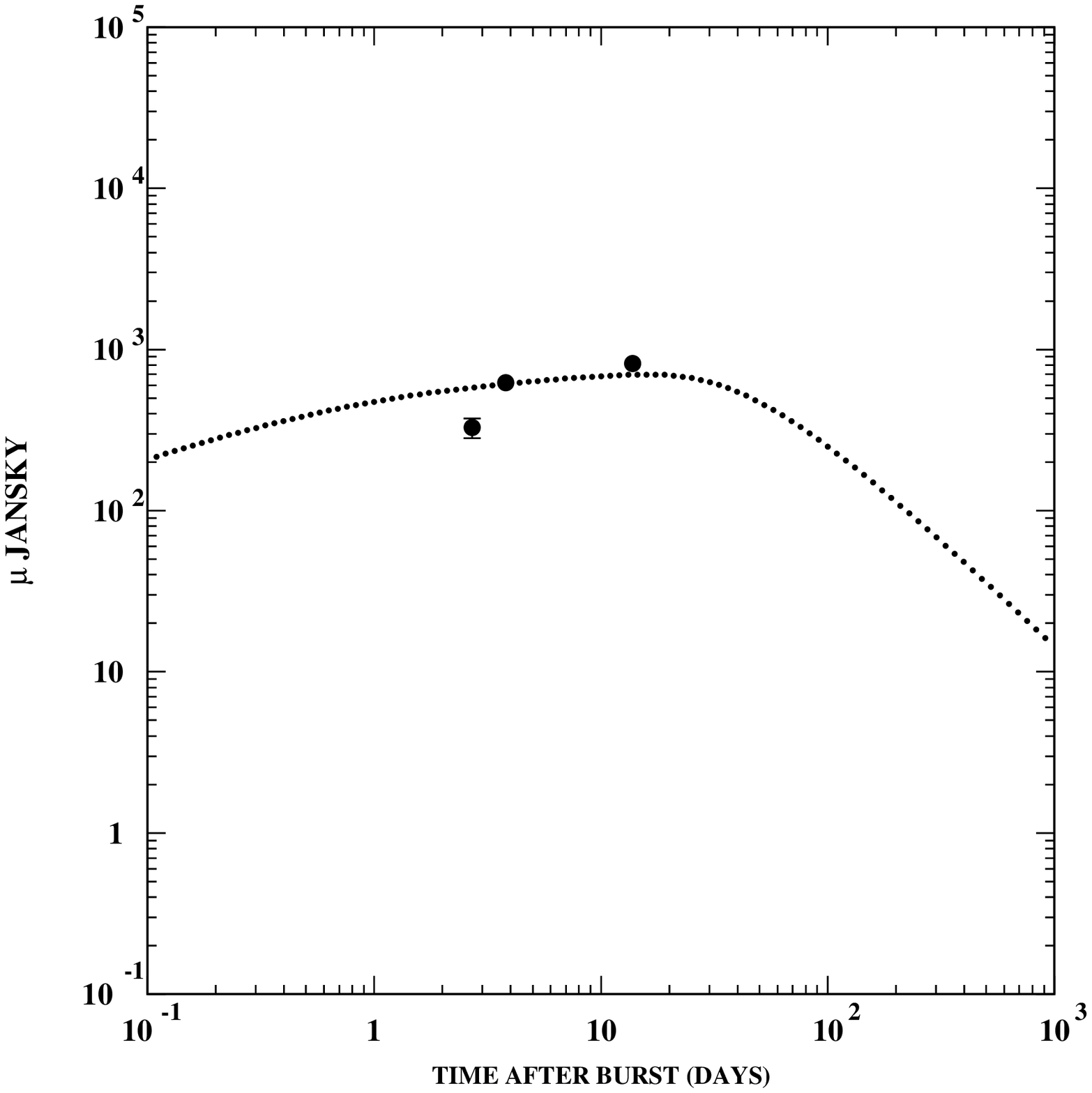, width=8cm} \\ 
%\hskip 1truecm 
\hspace*{.5cm} 
\epsfig{file=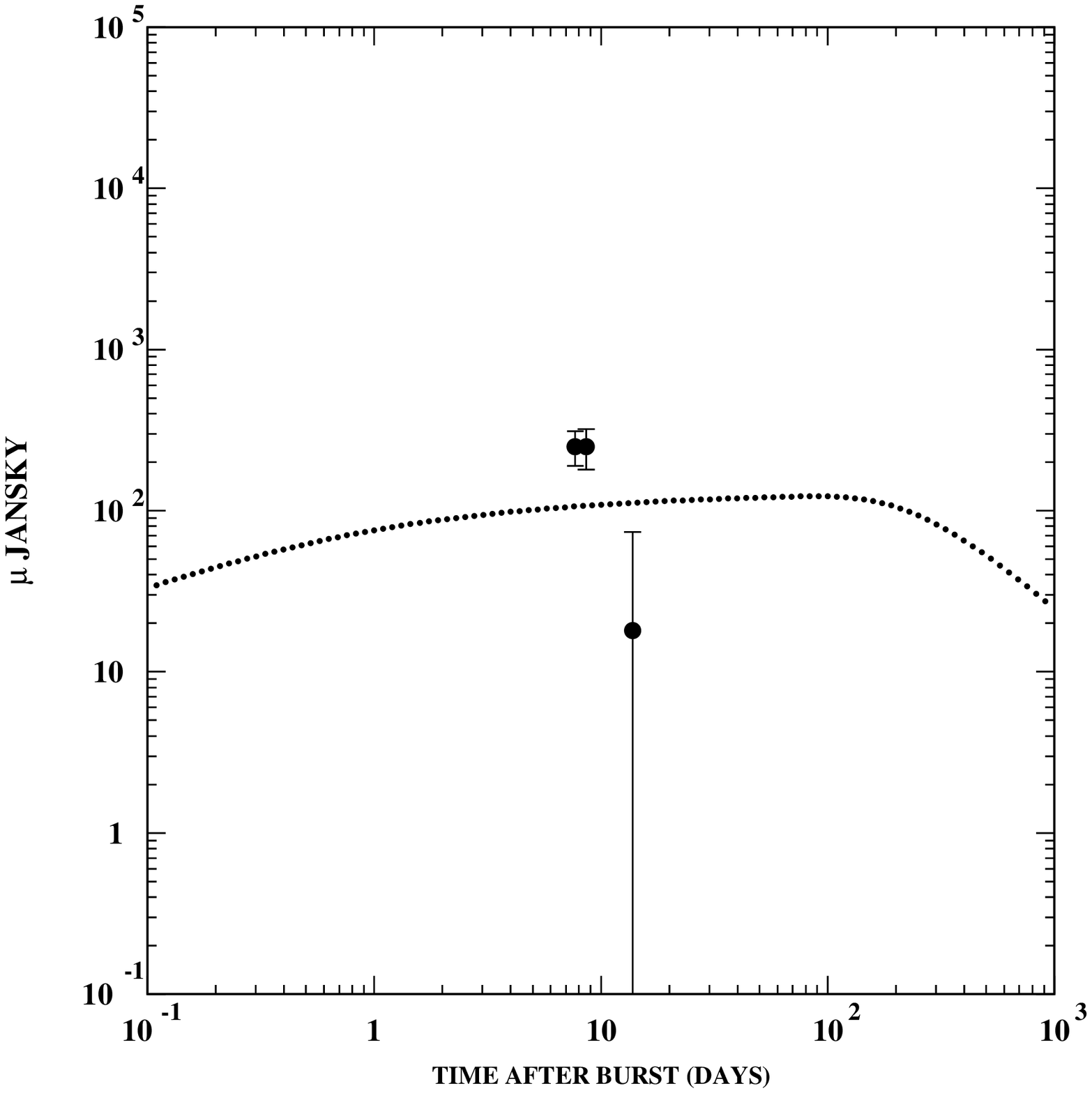, width=8cm} 
\end{tabular} 
\caption{Comparisons between our fitted CB model afterglow, 
Eq.~(\ref{Fnuobser}), 
and the observed radio afterglow of GRB 991208. 
Upper panel: the light curve at 4.86 GHz. 
Lower panel: the light curve at 1.43 GHz.} 
\label{figr120804}
\end{figure}

\begin{figure}[t] 
\begin{tabular}{cc} 
\hskip 2truecm 
\vspace*{2cm} 
\hspace*{-1.7cm} 
\epsfig{file=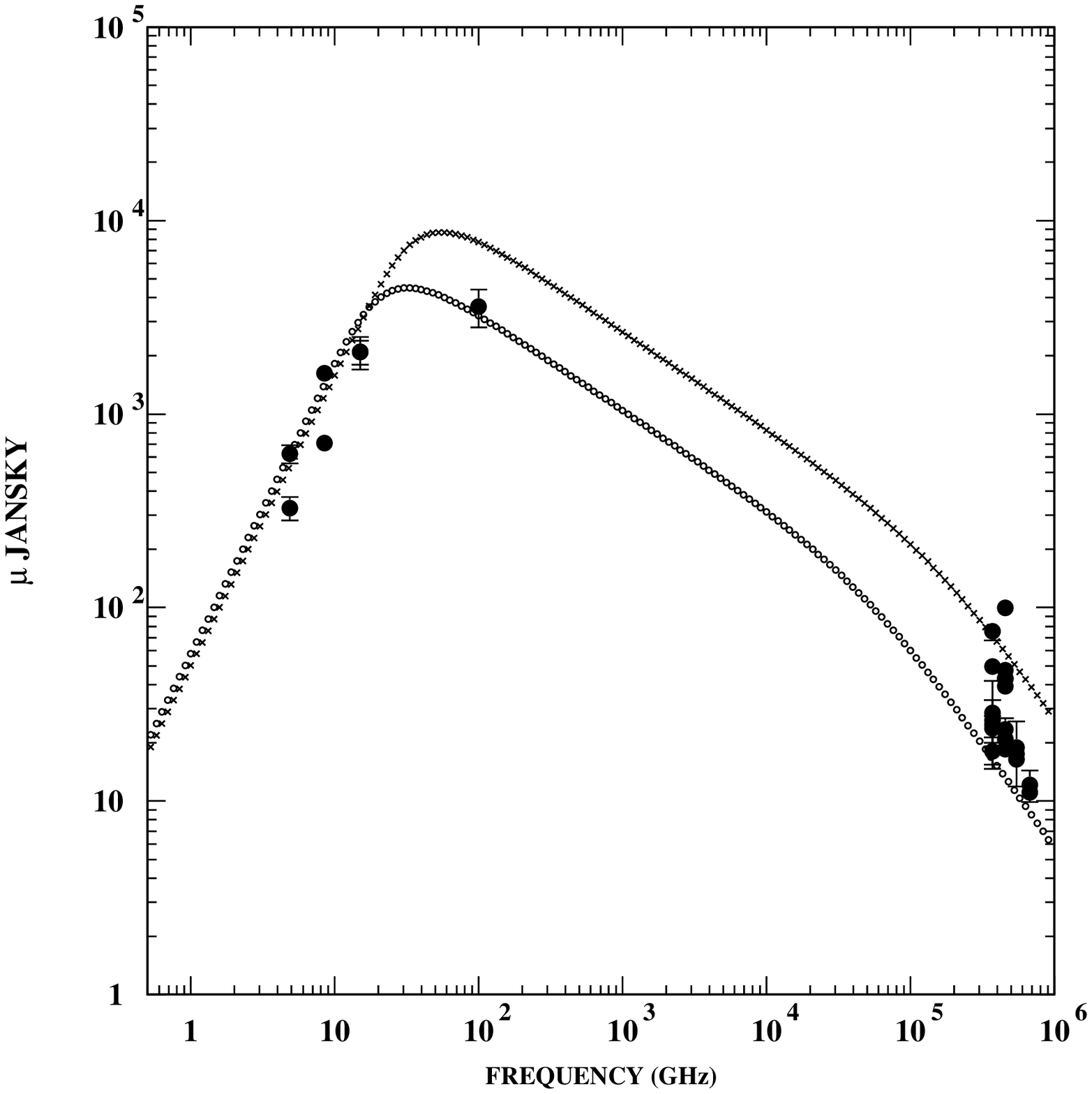, width=8cm} \\ 
%\hskip 1truecm 
\hspace*{.5cm} 
\epsfig{file=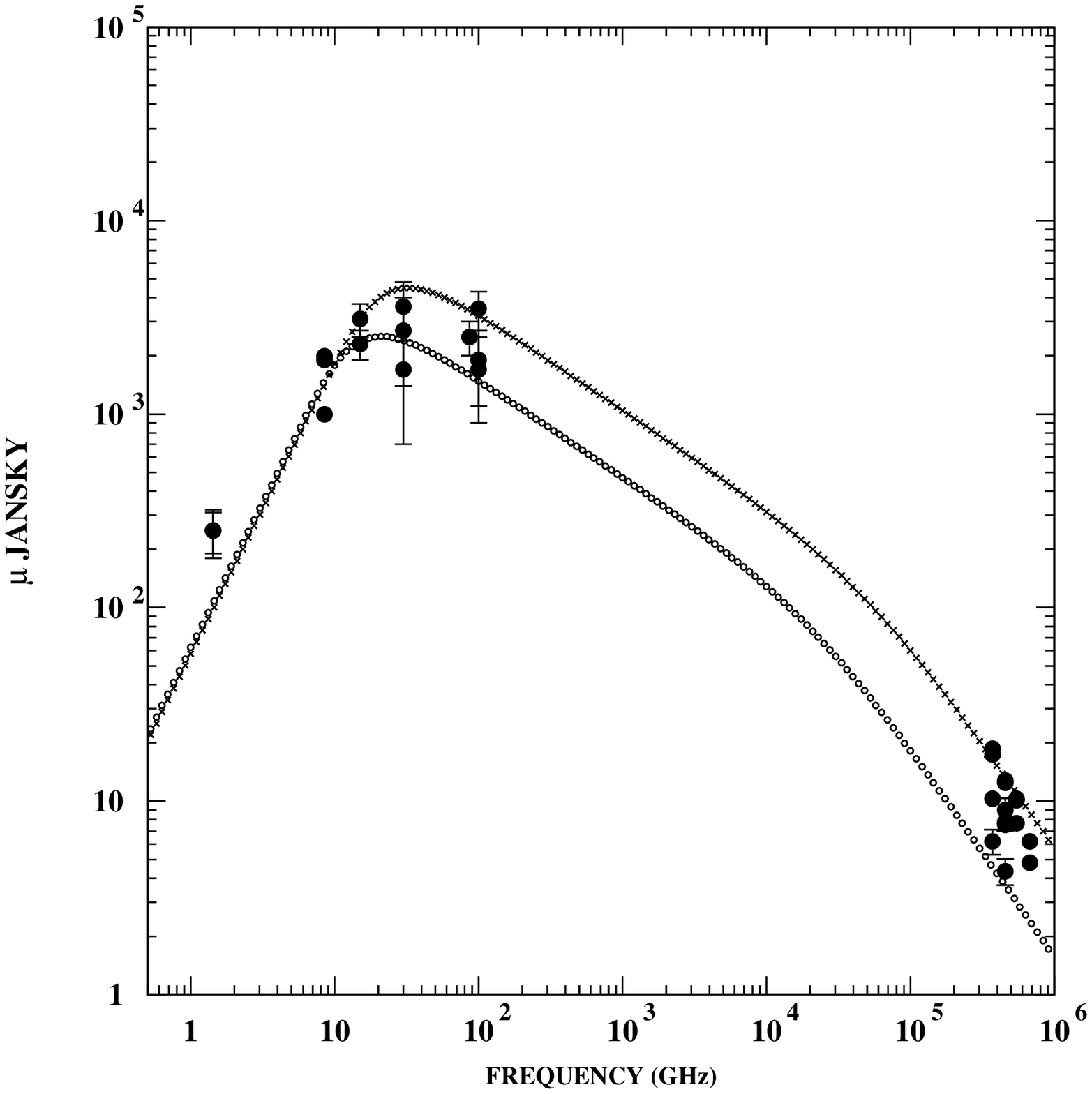, width=8cm} 
\end{tabular} 
\caption{The spectrum of the AG of GRB 991208 from radio to optical
frequencies. 
Upper panel: in the time interval between 2 and 5 days after burst.
Lower panel: in the time interval between 5 and 10 days after burst.
In both cases the highest peaking curve
corresponds to the earlier time.} 
\label{rad-opt208} 
\end{figure}

\begin{figure}[t] 
\begin{tabular}{cc} 
\hskip 2truecm 
\vspace*{2cm} 
\hspace*{-1.7cm} 
\epsfig{file=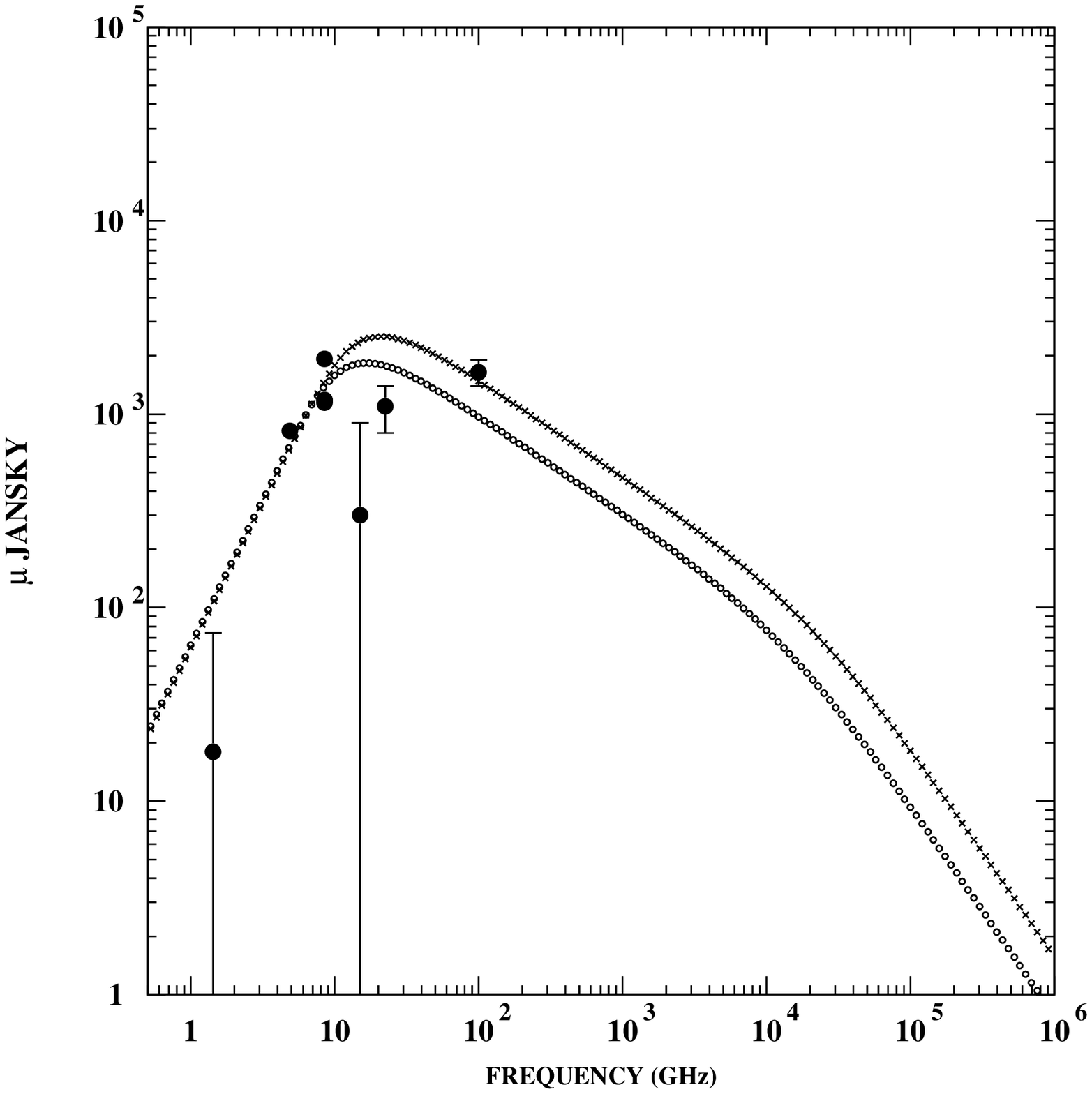, width=8cm} \\ 
%\hskip 1truecm 
\hspace*{.5cm} 
\end{tabular} 
\caption{The spectrum of the AG of GRB 991208 from radio to optical
frequencies in the time interval between 10 and 14.3 days.
The highest peaking curve corresponds to the earlier time.} 
\label{rad-opt208b} 
\end{figure}

We fit the 
NIR/optical data (Castro-Tirado et al. 2001; Sagar et al. 2000) and
the radio data (Galama et al. 2000)  on the AG of GRB 991208.
In Fig~(\ref{opt208}) 
we present the comparison between
the measured light curves for the I, R, V and B bands, and the
fitted CB model predictions, after 
subtraction of the host galaxy and SN contributions (DDD 2001).
The theoretical predictions were corrected only
for the small galactic extinction E(B - V)=0.016 (Schlegel et al. 1998) 
in the direction of this GRB, there being no spectral evidence for optical
extinction in the host galaxy. 
 In Figs.~(\ref{figr120801}) to
(\ref{figr120804}) we also present the radio light curves at 
100, 86.14, 30, 22.49, 14.97, 8.46, 4.86 and 1.43 GHz.
In Figs.~(\ref{rad-opt208}) and
(\ref{rad-opt208b}) we make the complementary comparison 
for the radio to optical spectra in three 
time intervals extending
from 2 to 14.3 days. The results are satisfactory.

\subsection{GRB 000418}

\begin{figure}[t]  
%\begin{tabular}{cc}  
\hskip 0truecm   
\epsfig{file=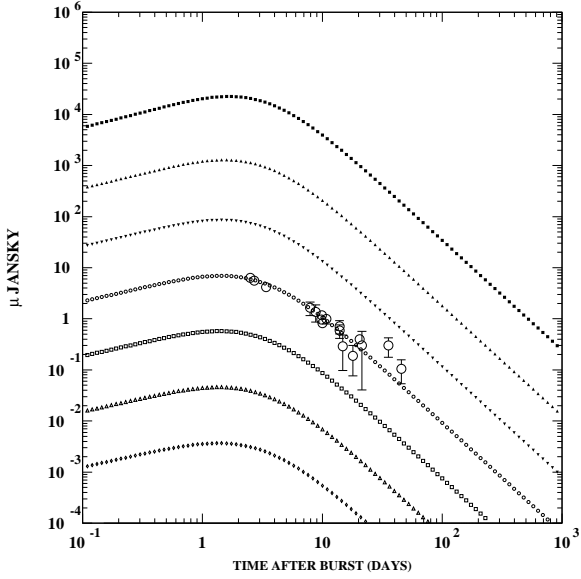, width=8.5cm}  
%\end{tabular}  
\caption{Comparisons between our fitted CB model AG of GRB 000418, 
at $\rm z=1.118$,
Eq.~(\ref{Fnuobser}), with the observed optical data. 
The figure shows (from top to bottom) 1000 times the K-band results,
100 times the J-band, 10 times the I-band, the R-band, 1/10 of the V-band,
1/100 of the B-band and 1/1000 of the U-band.
The contributions of the underlying galaxy and the
expected SN1998bw-like 
SN have been subtracted. In a CB-model fit, there is in this case
strong evidence for such a SN (DDD 2001).}

\label{opt418}  
\end{figure}

\begin{figure}[t] 
\begin{tabular}{cc} 
\hskip 2truecm 
\vspace*{2cm} 
\hspace*{-1.7cm} 
\epsfig{file=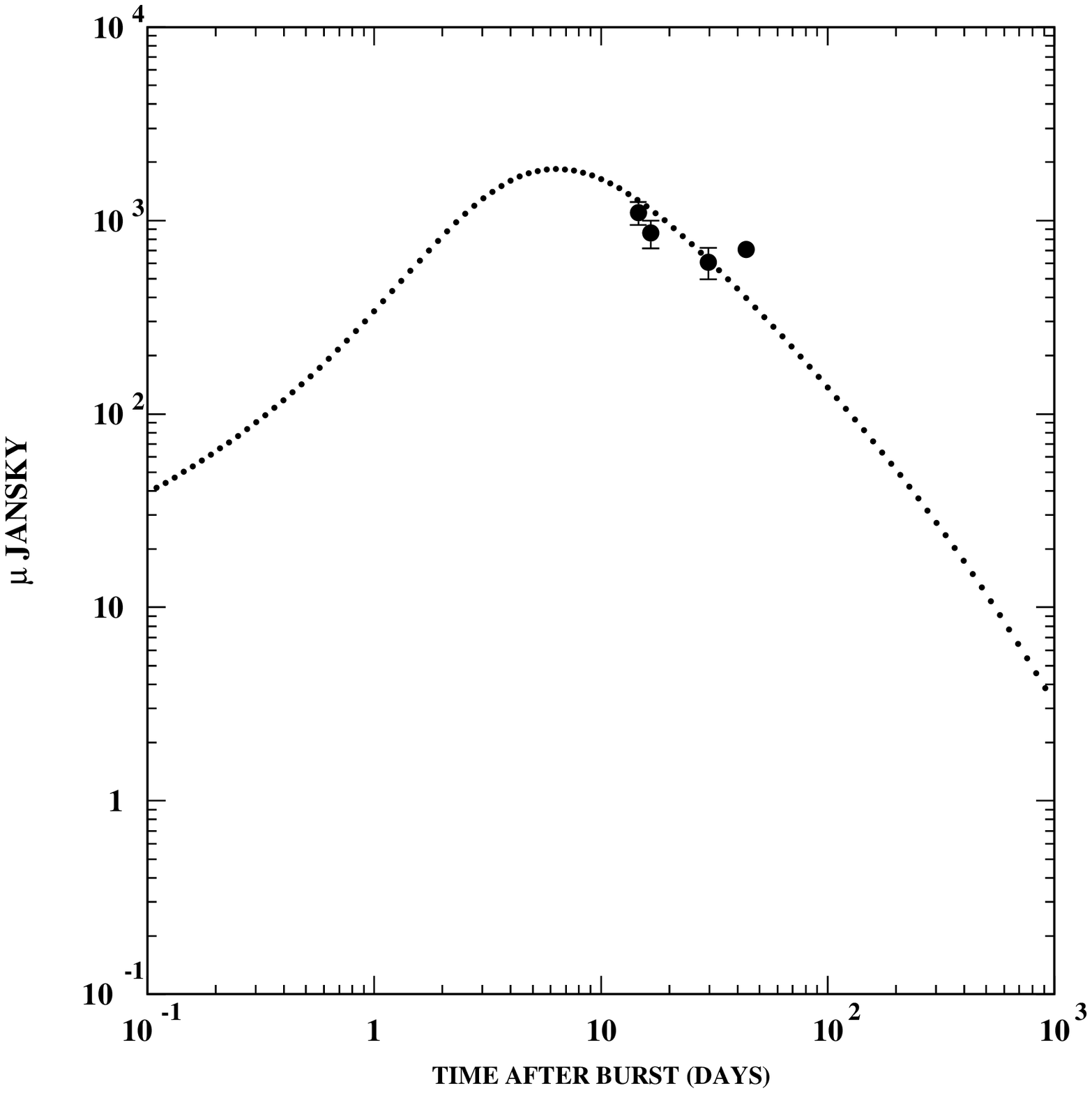, width=8cm} \\ 
%\hskip 1truecm 
\hspace*{.5cm} 
\epsfig{file=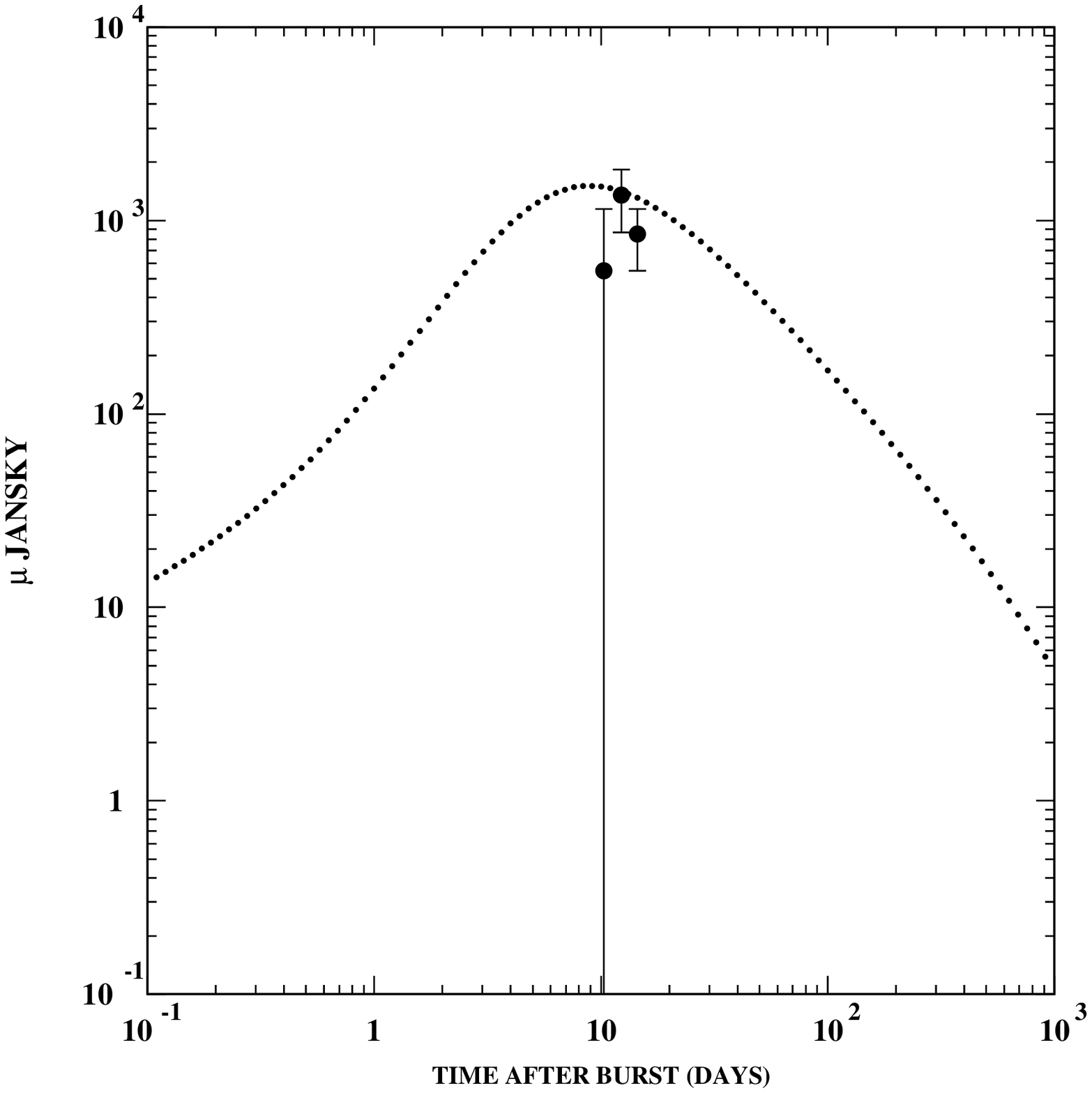, width=8cm} 
\end{tabular} 
\caption{Comparisons between our fitted CB model afterglow, 
Eq.~(\ref{Fnuobser}), 
and the observed radio afterglow of GRB 000418. 
Upper panel: the light curve at 22.46 GHz. 
Lower panel: the light curve at 15 GHz.} 
\label{figr041801} 
\end{figure}

\begin{figure}[t] 
\begin{tabular}{cc} 
\hskip 2truecm 
\vspace*{2cm} 
\hspace*{-1.7cm} 
\epsfig{file=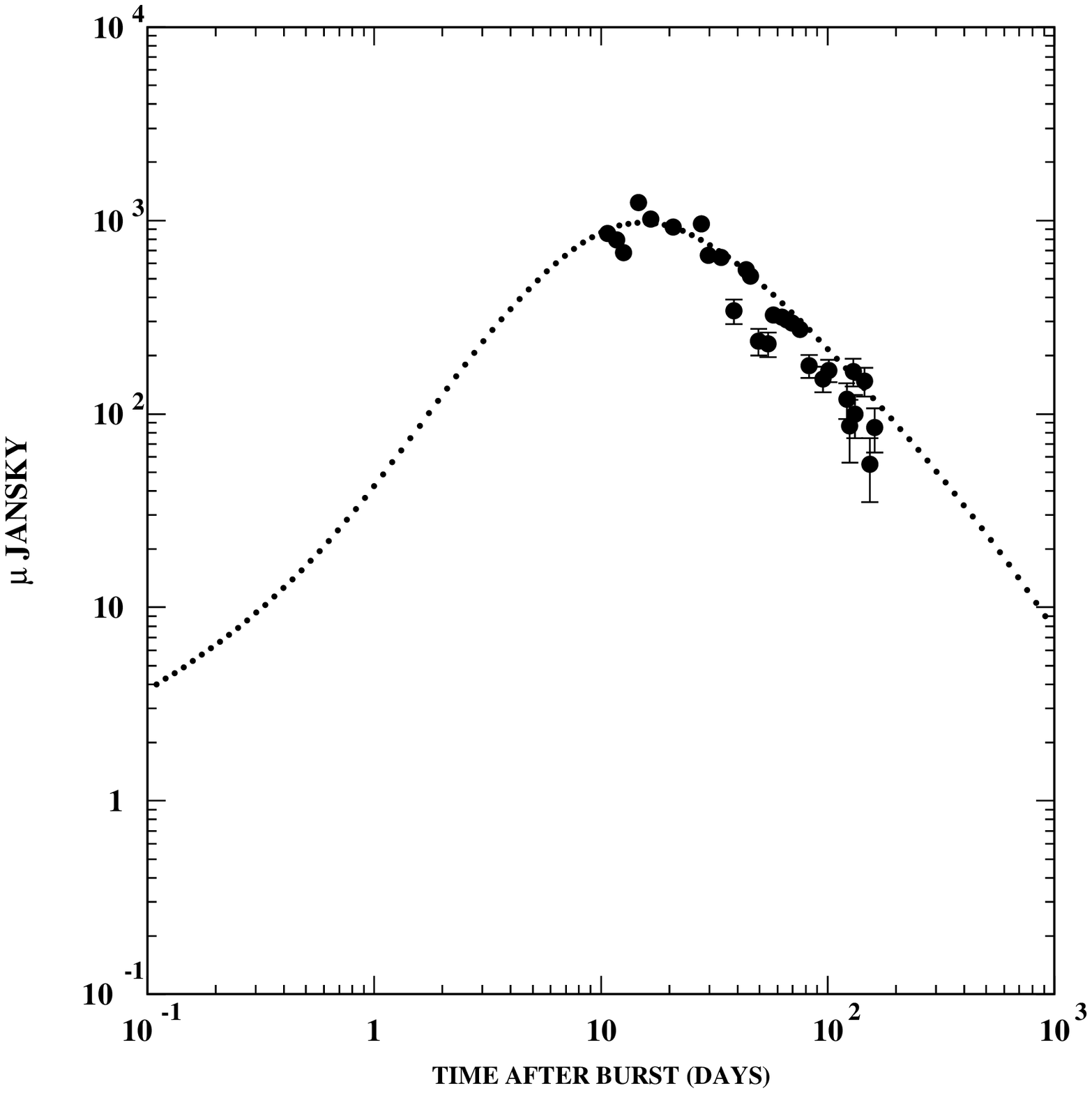, width=8cm} \\ 
%\hskip 1truecm 
\hspace*{.5cm} 
\epsfig{file=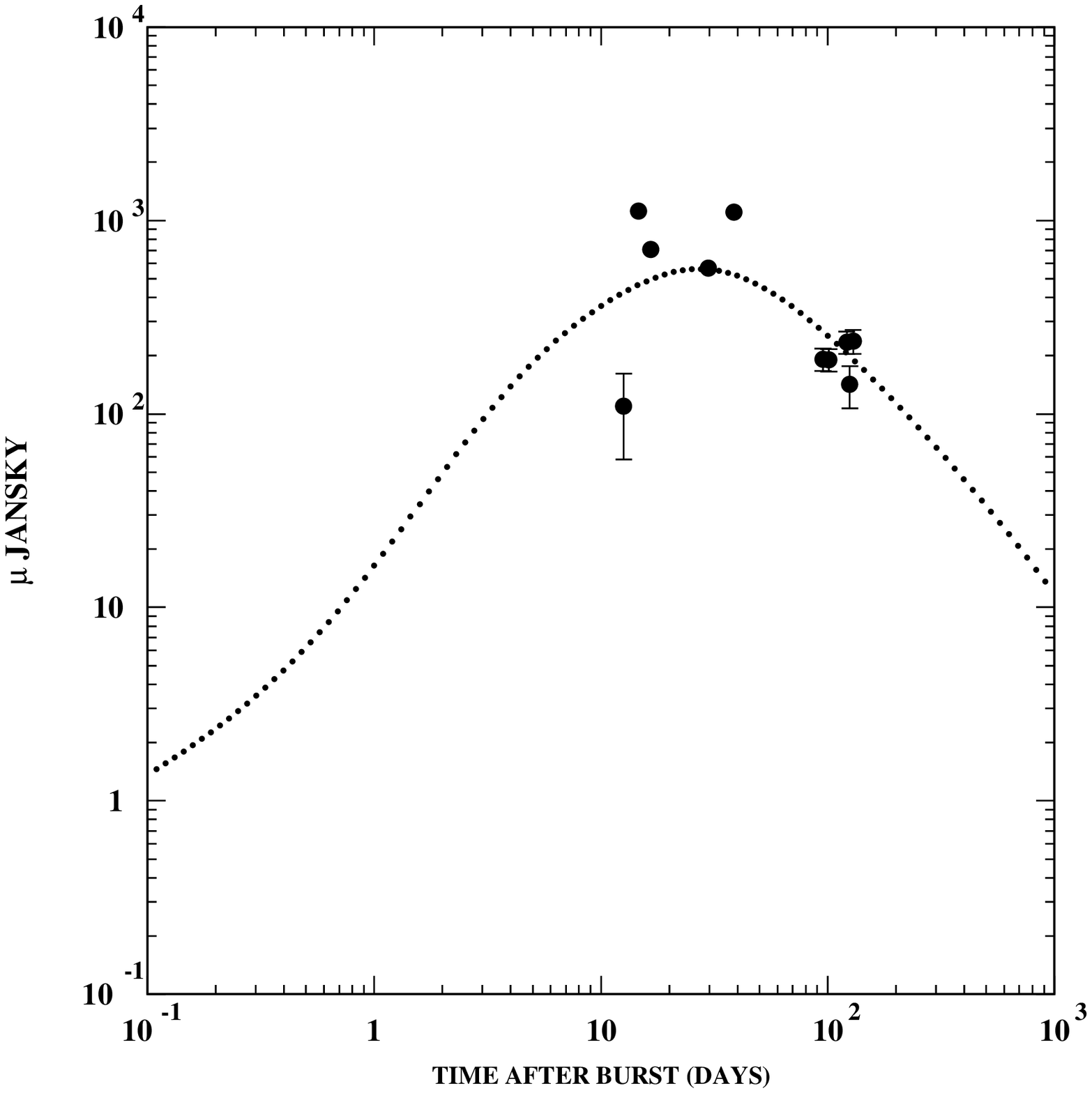, width=8cm} 
\end{tabular} 
\caption{Comparisons between our fitted CB model afterglow, 
Eq.~(\ref{Fnuobser}), 
and the observed radio afterglow of GRB 000418. 
Upper panel: the light curve at 8.46 GHz. 
Lower panel: the light curve at 4.86 GHz.
} 
\label{figr041802} 
\end{figure}

\begin{figure}[t] 
\begin{tabular}{cc} 
\hskip 2truecm 
\vspace*{2cm} 
\hspace*{-1.7cm} 
\epsfig{file=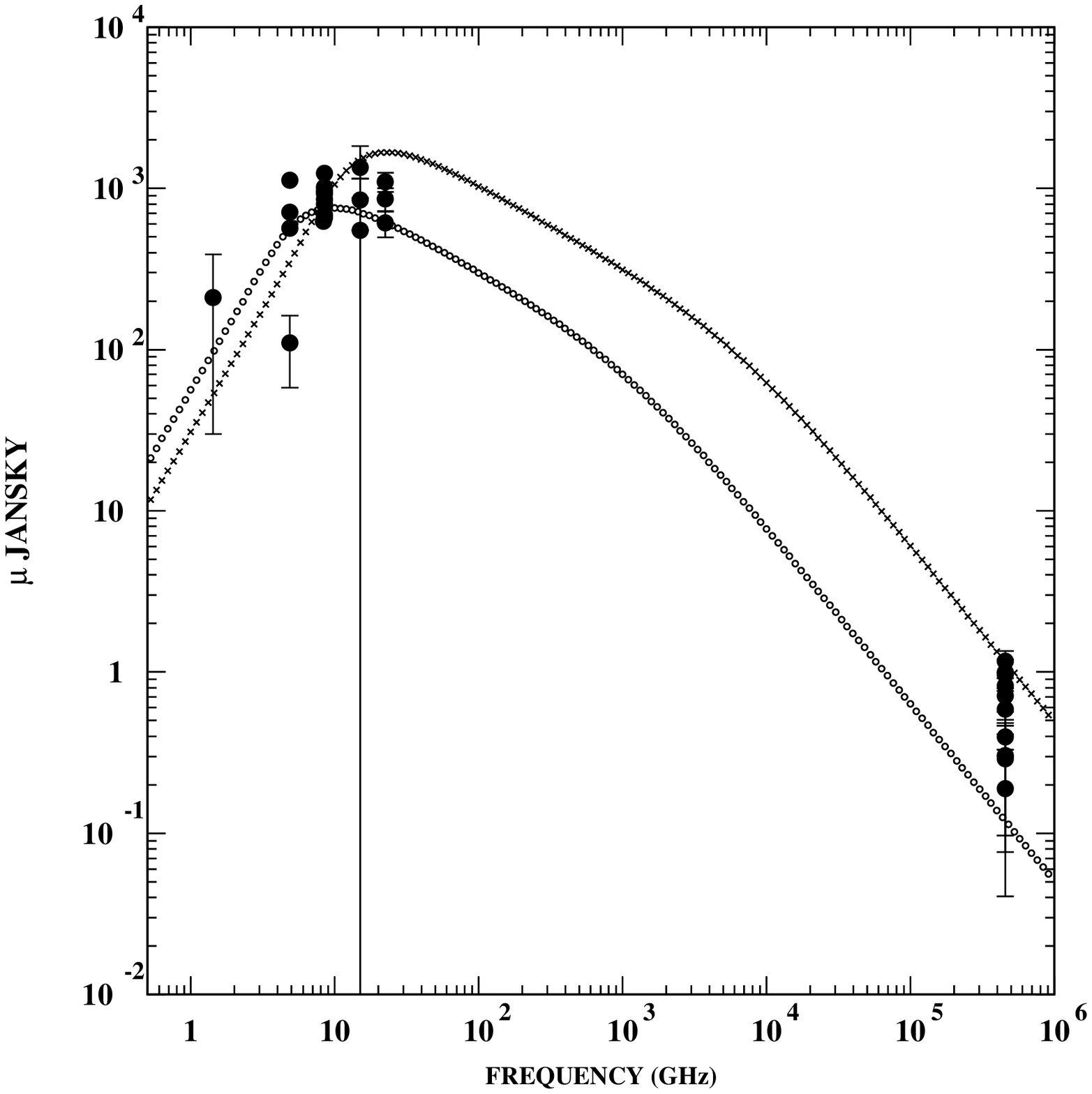, width=8cm} \\ 
%\hskip 1truecm 
\hspace*{.5cm} 
\epsfig{file=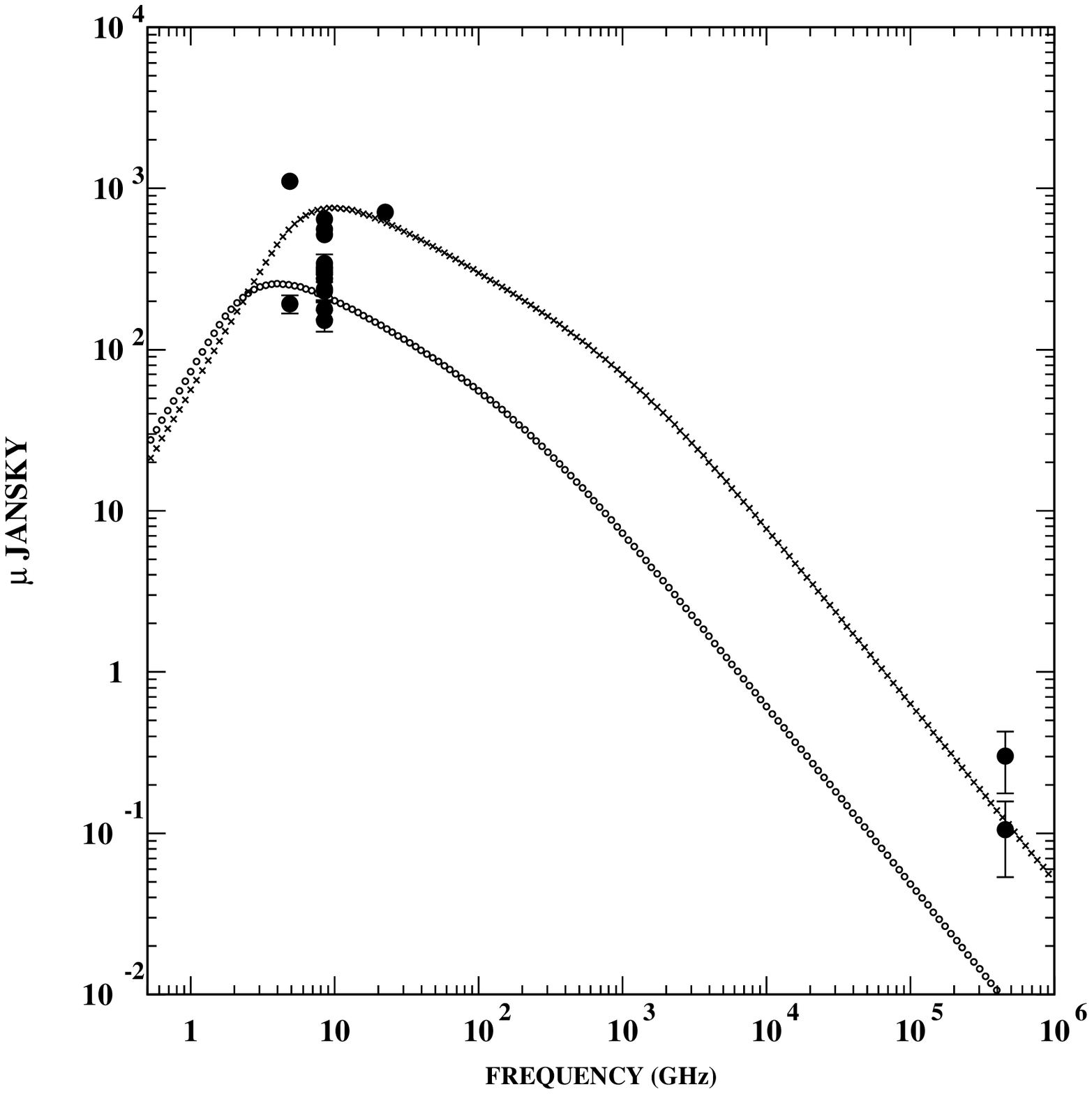, width=8cm} 
\end{tabular} 
\caption{The spectrum of the AG of GRB 000418 from radio to optical
frequencies.
Upper panel: in the time interval
between 9.5 and 30 days after burst.
Lower panel: in the time interval
between 30 and 100 days after burst.
In both cases the highest peaking curve
corresponds to the earlier time.} 
\label{rad-opt418} 
\end{figure}

The NIR/optical data are from Klose et al. (2000)
and the radio data from Berger et al. (2001a).
In Fig.~(\ref{opt418})  we compare
the fitted CB-model predictions with
the measured light curves for the R-band, after
subtraction of the host galaxy and SN contribution (DDD 2001).
The theoretical predictions were corrected for galactic extinction
and for extinction in the host galaxy as estimated by
Berger et al. (2001a): E(B - V)=0.40.
In Figs.~(\ref{figr041801}) and  (\ref{figr041802})
we also present the radio light curves at 22.5, 15, 8.46 and
4.86 GHz. In Fig.~(\ref{rad-opt418}) we
make the complementary comparison for the radio to optical spectra,
in two time
intervals extending from 9.5 to 100 days.
The results are satisfactory.

\subsection{GRB 990510}

\begin{figure}[t]  
%\begin{tabular}{cc}  
\hskip 0truecm   
\epsfig{file=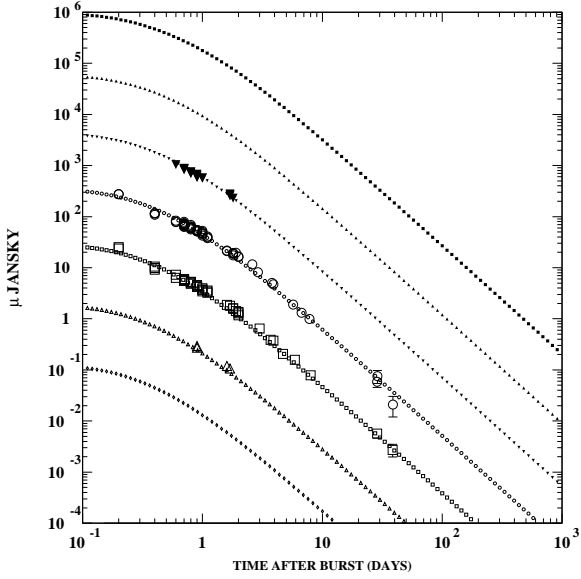, width=8.5cm}  
%\end{tabular}  
\caption{
Comparisons between our fitted CB model AG of GRB 990510,
at $\rm z=1.619$,
Eq.~(\ref{Fnuobser}), with the observed optical data.
The figure shows (from top to bottom) 1000 times the K-band results,
100 times the J-band, 10 times the I-band, the R-band, 1/10 of the V-band,
1/100 of the B-band and 1/1000 of the U-band.
The contribution of the underlying galaxy and the
(in this case unobservable) associated
supernova  has been subtracted. }  
\label{opt510}  
\end{figure}

\begin{figure}[t] 
\begin{tabular}{cc} 
\hskip 2truecm 
\vspace*{2cm} 
\hspace*{-1.7cm} 
\epsfig{file=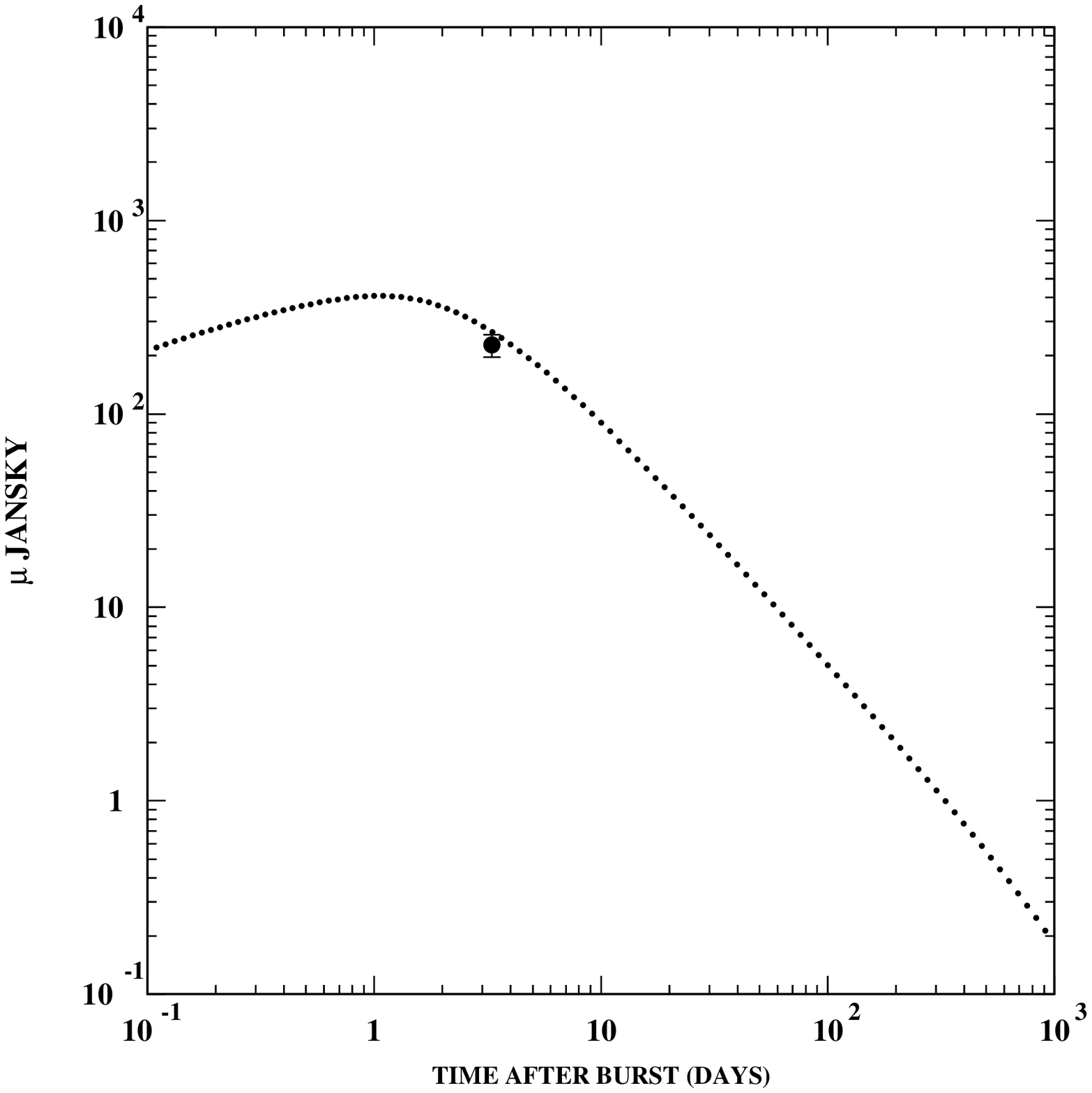, width=8cm} \\ 
%\hskip 1truecm 
\hspace*{.5cm} 
\epsfig{file=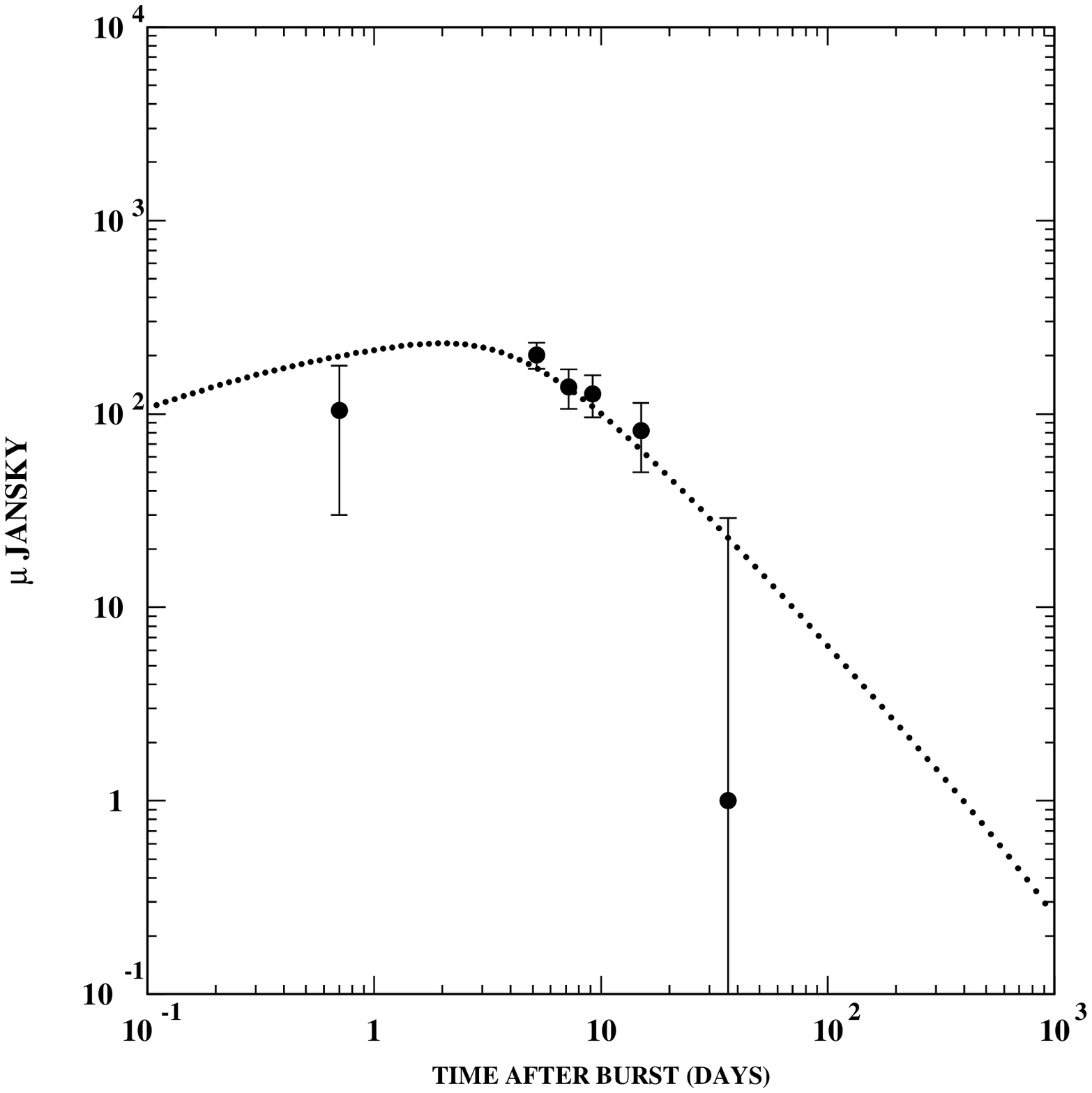, width=8cm} 
\end{tabular} 
\caption{Comparisons between our fitted CB model afterglow, 
Eq.~(\ref{Fnuobser}), 
and the observed radio afterglow of GRB 990510. 
Upper panel: the light curve at 13.68 GHz. 
Lower  panel: the light curve at 8.6-8.7 GHz.} 
\label{figr051001} 
\end{figure}

\begin{figure}[t] 
\begin{tabular}{cc} 
\hskip 2truecm 
\vspace*{2cm} 
\hspace*{-1.7cm} 
\epsfig{file=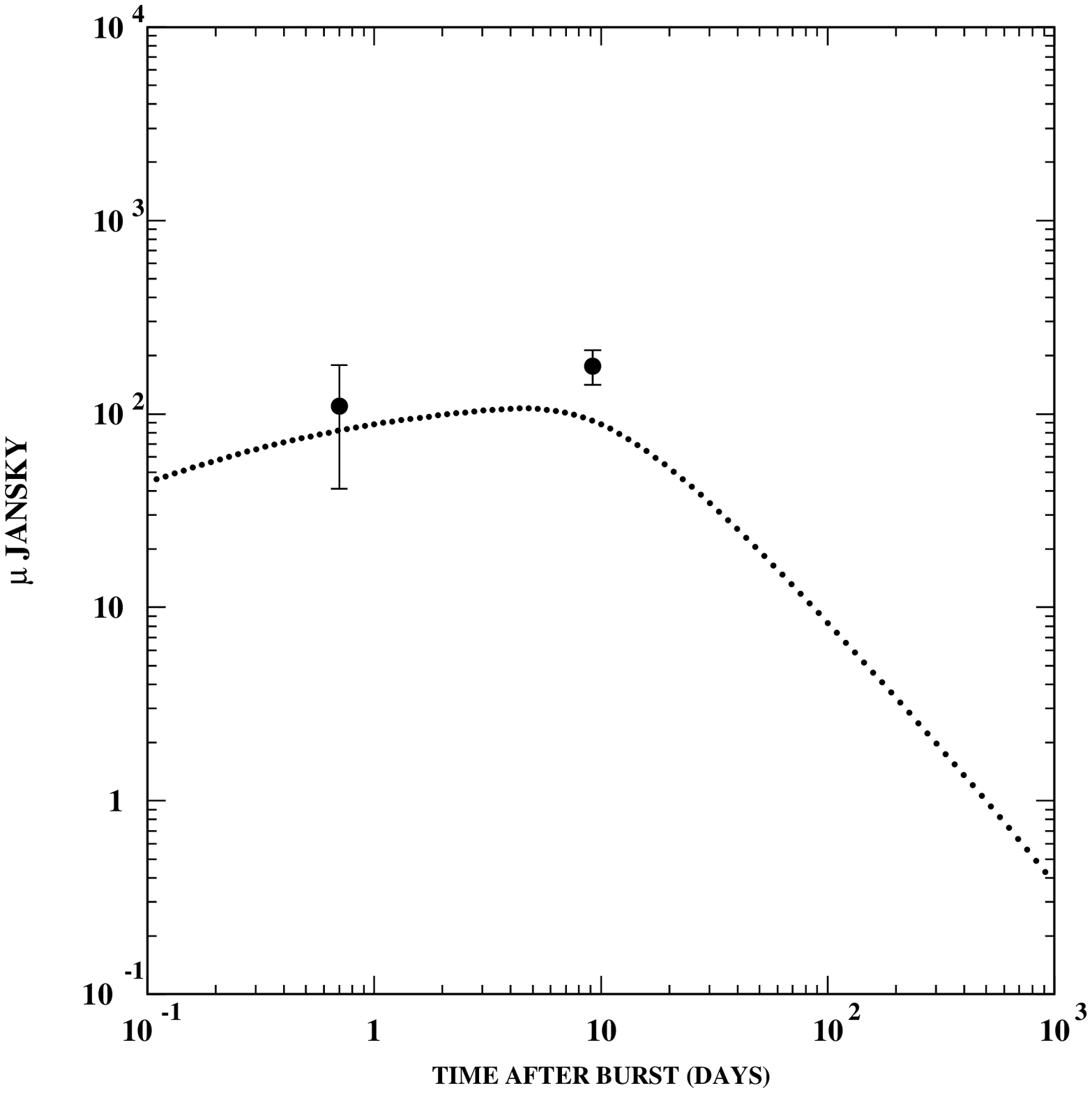, width=8cm} \\ 
%\hskip 1truecm 
\hspace*{.5cm} 
\epsfig{file=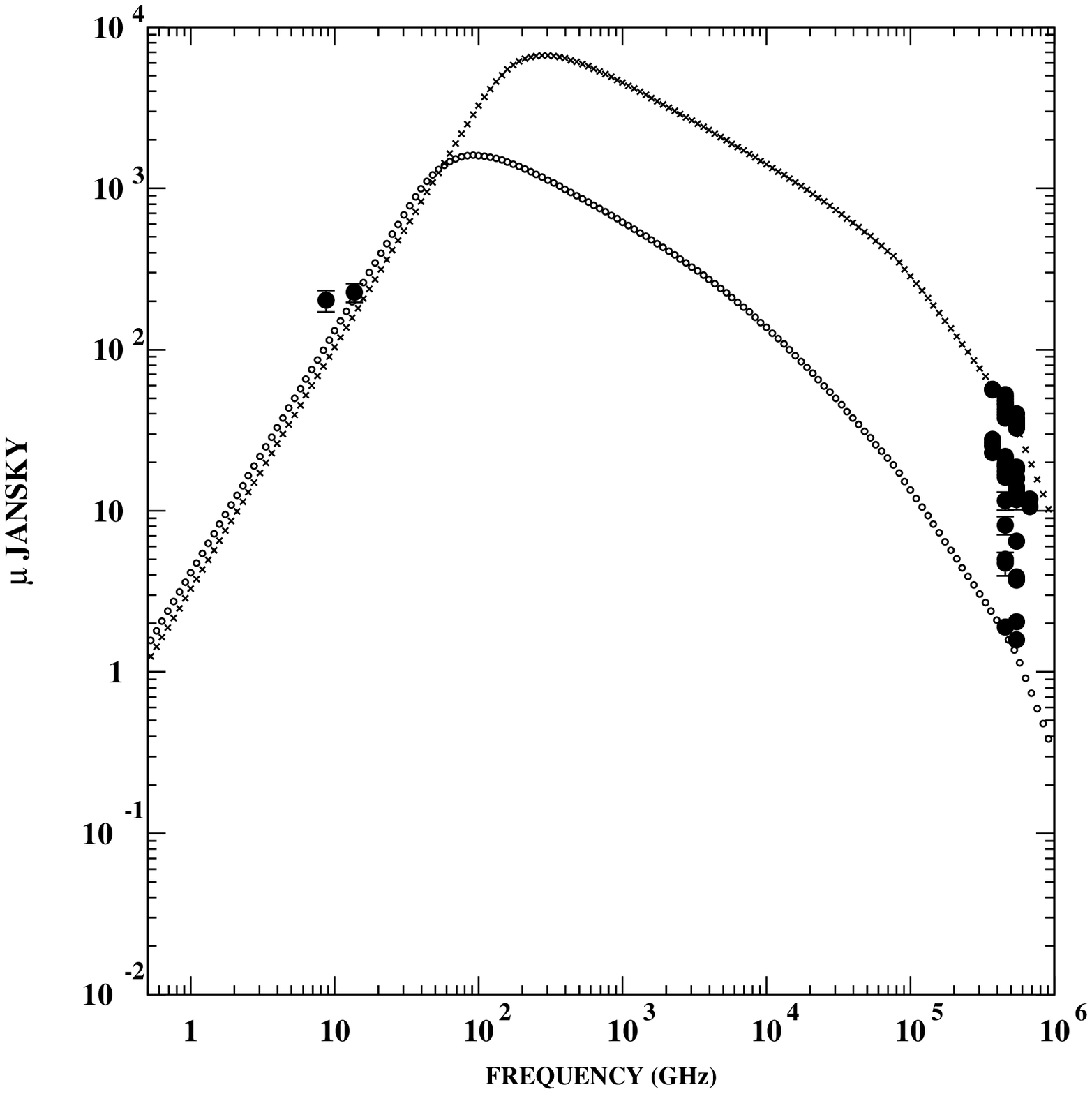, width=8cm} 
\end{tabular} 
\caption{Comparisons between our fitted CB model afterglow, 
Eq.~(\ref{Fnuobser}), 
and the observed radio afterglow of GRB 990510. 
Upper panel: the light curve at 4.86 GHz. 
Lower panel: the spectrum from radio to optical frequencies in the 
time interval between 1 and 6 days after burst. 
The highest peaking curve corresponds to the earlier time.}
\label{figr051002} 
\end{figure}

\begin{figure}[t] 
\begin{tabular}{cc} 
\hskip 2truecm 
\vspace*{2cm} 
\hspace*{-1.7cm} 
\epsfig{file=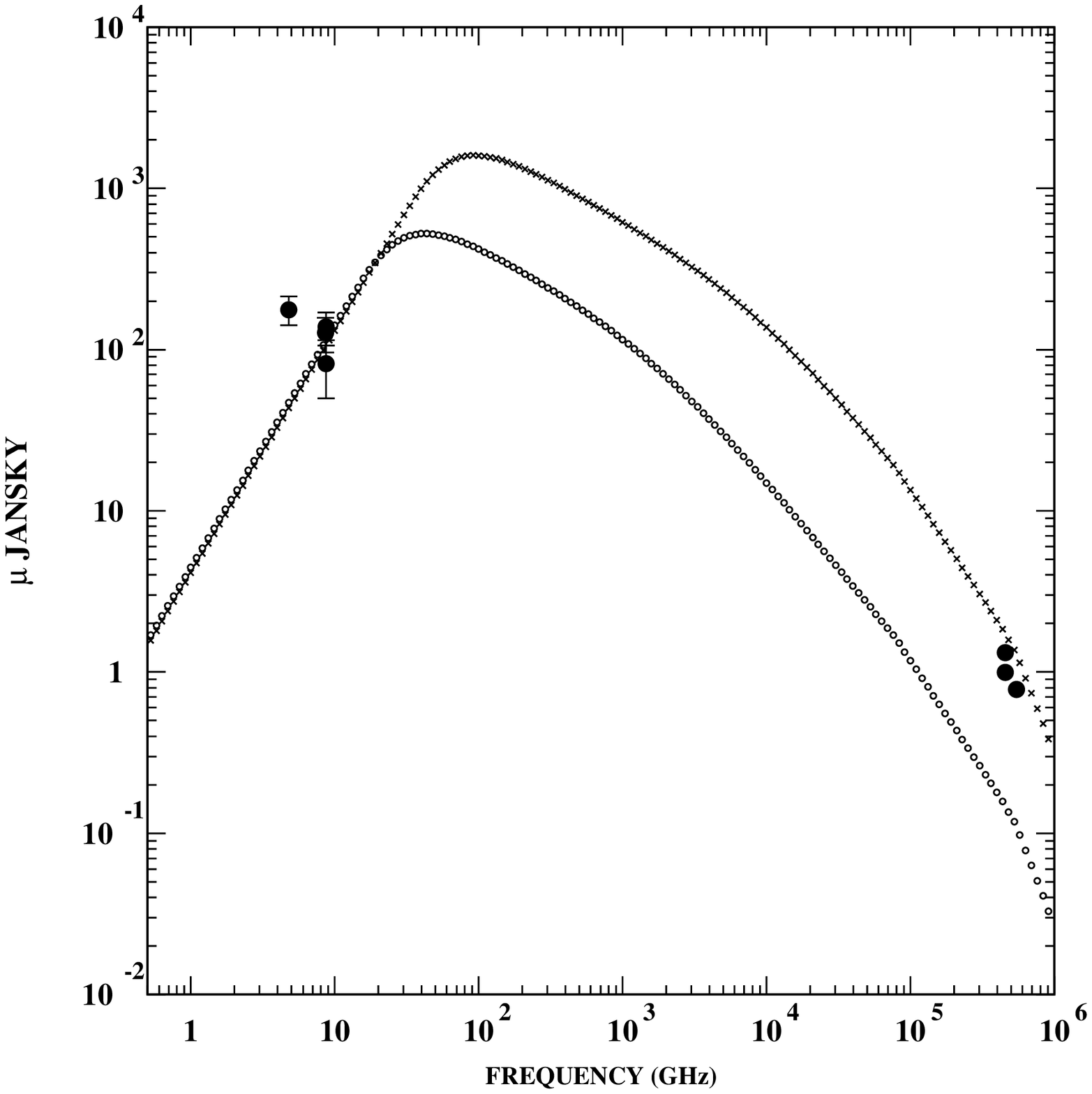, width=8cm} \\ 
%\hskip 1truecm 
\hspace*{.5cm} 
\epsfig{file=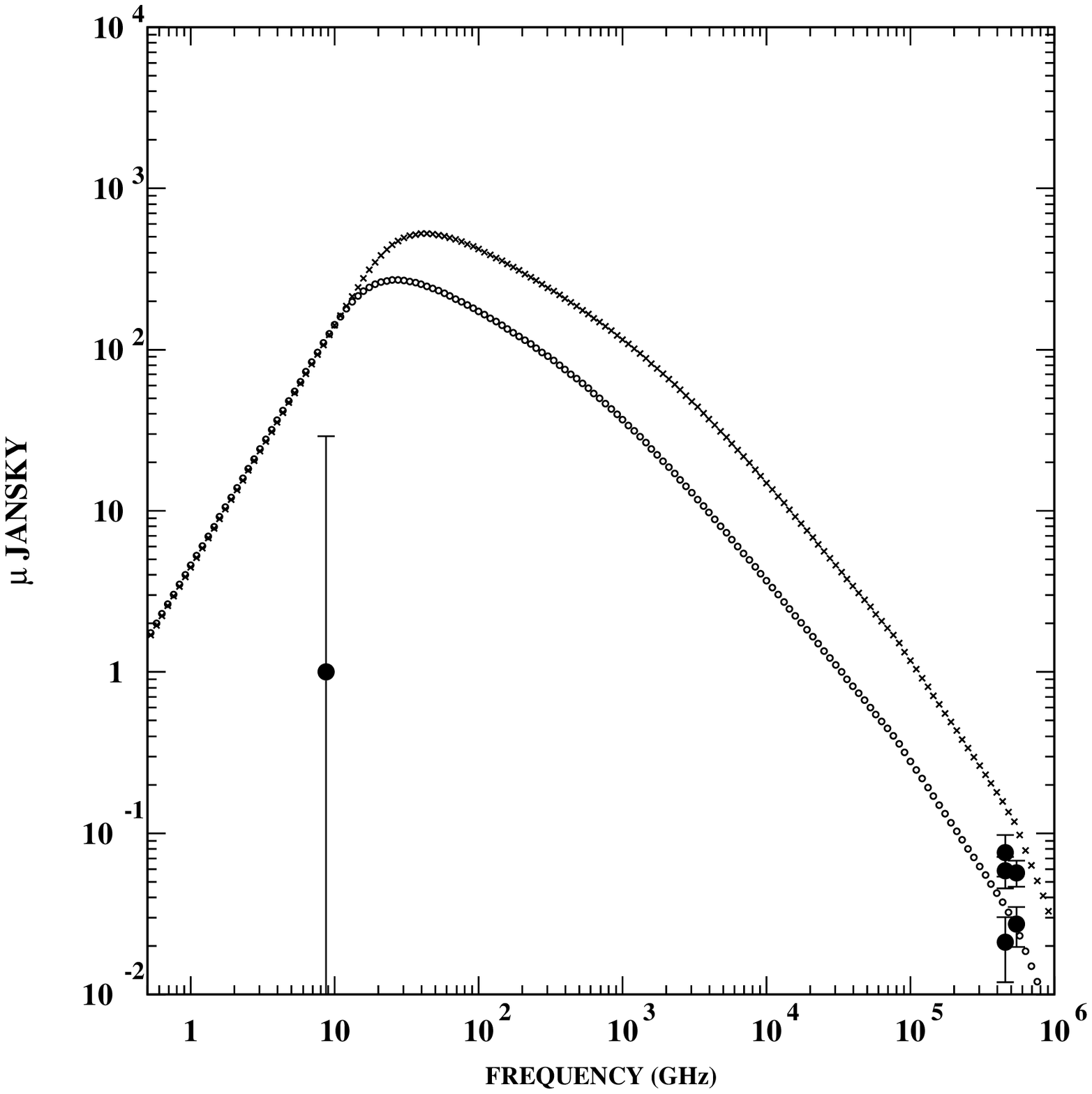, width=8cm} 
\end{tabular} 
\caption{The spectrum of the AG of GRB 990510 from radio to optical
frequencies.
Upper panel: in the time interval
between 6 and 20 days after burst. 
Lower panel: in the time interval between 20 and 40  days after burst.
In both cases the highest peaking curve
corresponds to the earlier time.}
\label{rad-opt510} 
\end{figure}

The NIR/optical data were gathered by Beuermann et al. (1999),
Harrison et al. (1999) and Stanek et al. (1999) and the radio data 
by Harrison et al. (1999). In Fig.~(\ref{opt510}) we present the comparison between
the measured light curves for the I, R, V, B bands, 
after subtraction of the
host galaxy and SN contribution (DDD 2001), and the
fitted CB model predictions, corrected for Galactic extinction
(E(B - V)=0.203, Schlegel et al. 1998) and for extinction in the host
galaxy as estimated by Stanek et al. (1999).
In Figs.~(\ref{figr051001}) and (\ref{figr051002})  
we present the radio light curves at 13.7, 8.6 and  4.8 GHz.
In Figs.~(\ref{figr051002}) and (\ref{rad-opt510})  we also 
make the complementary comparison of theory and
observations for the radio to optical spectra   
three time intervals
extending from 1 to 40 days.
The agreement between theory and observations is very good although
its significance is limited by the sparse radio data.

\subsection{GRB 990123}

\begin{figure}[t]  
%\begin{tabular}{cc}  
\hskip 0truecm   
\epsfig{file=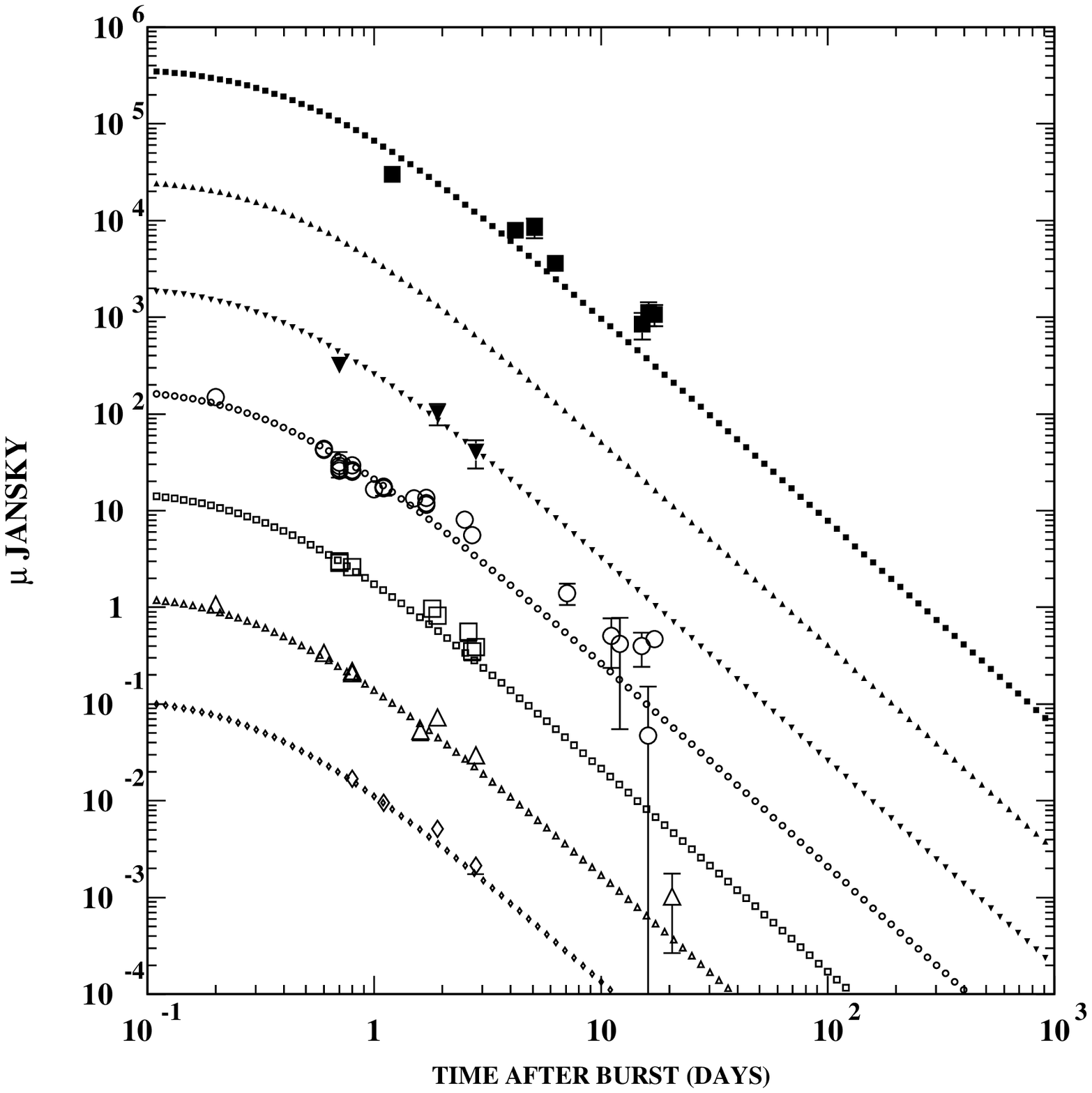, width=8.5cm}  
%\end{tabular}  
\caption{
Comparisons between our fitted CB model AG of GRB 990123, 
at $\rm z=1.600$,
Eq.~(\ref{Fnuobser}), and the observed optical data. 
The figure shows (from top to bottom) 1000 times the K-band results,
100 times the J-band, 10 times the I-band, the R-band, 1/10 of the V-band,
1/100 of the B-band and 1/1000 of the U-band.
The contributions of the underlying galaxy and 
an expected (but,  in this case, unobservable) SN1998bw-like 
SN have been subtracted. }
\label{opt123}  
\end{figure}

\begin{figure}[t] 
\begin{tabular}{cc} 
\hskip 2truecm 
\vspace*{2cm} 
\hspace*{-1.7cm} 
\epsfig{file=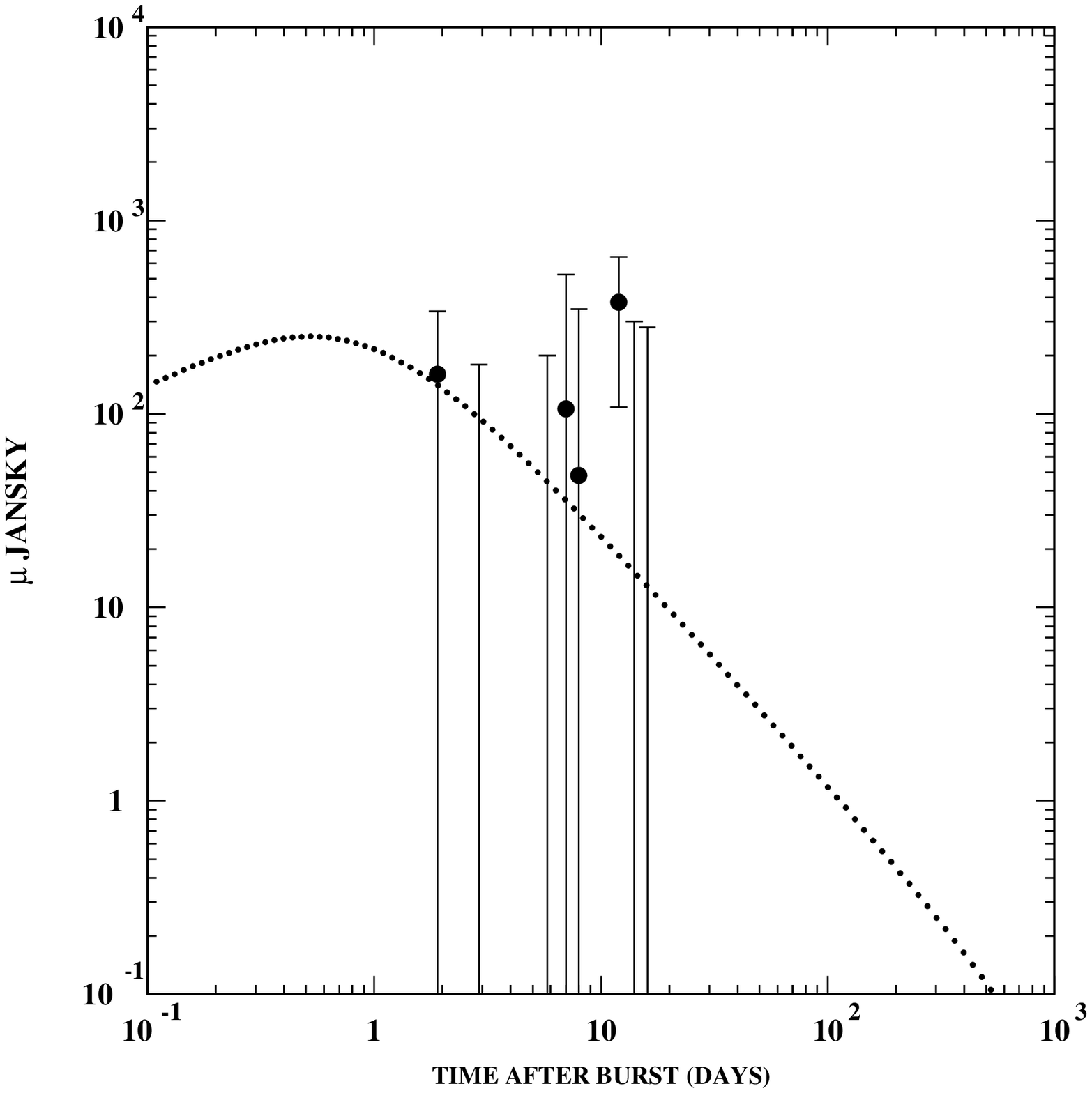, width=8cm} \\ 
%\hskip 1truecm 
\hspace*{.5cm} 
\epsfig{file=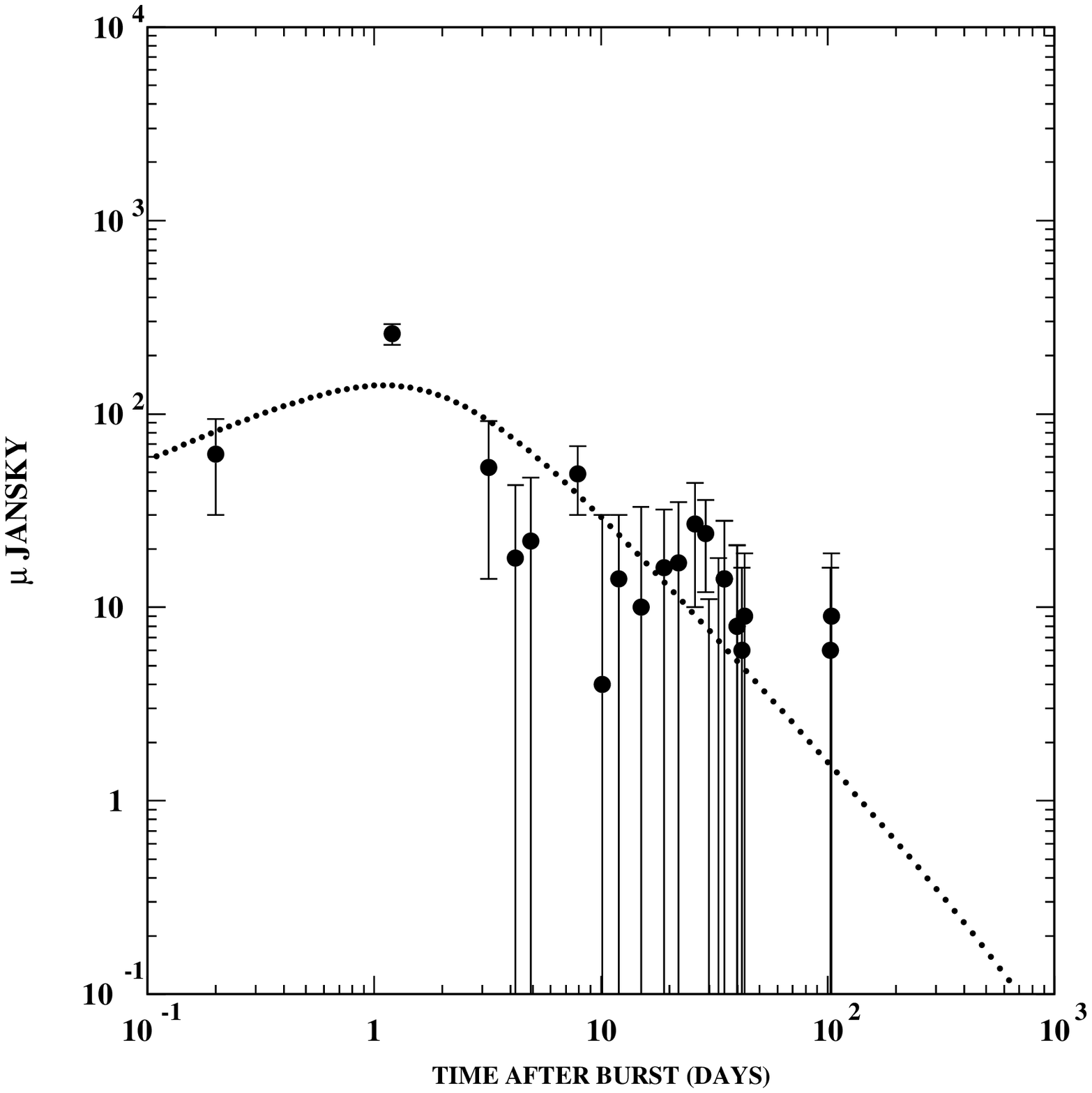, width=8cm} 
\end{tabular} 
\caption{Comparisons between our fitted CB model afterglow, 
Eq.~(\ref{Fnuobser}), 
and the observed radio afterglow of GRB 990123. 
Upper panel: the light curve at 15 GHz. 
Lower  panel: the light curve at 8.46 GHz.} 
\label{figr012301} 
\end{figure}

\begin{figure}[t] 
\begin{tabular}{cc} 
\hskip 2truecm 
\vspace*{2cm} 
\hspace*{-1.7cm} 
\epsfig{file=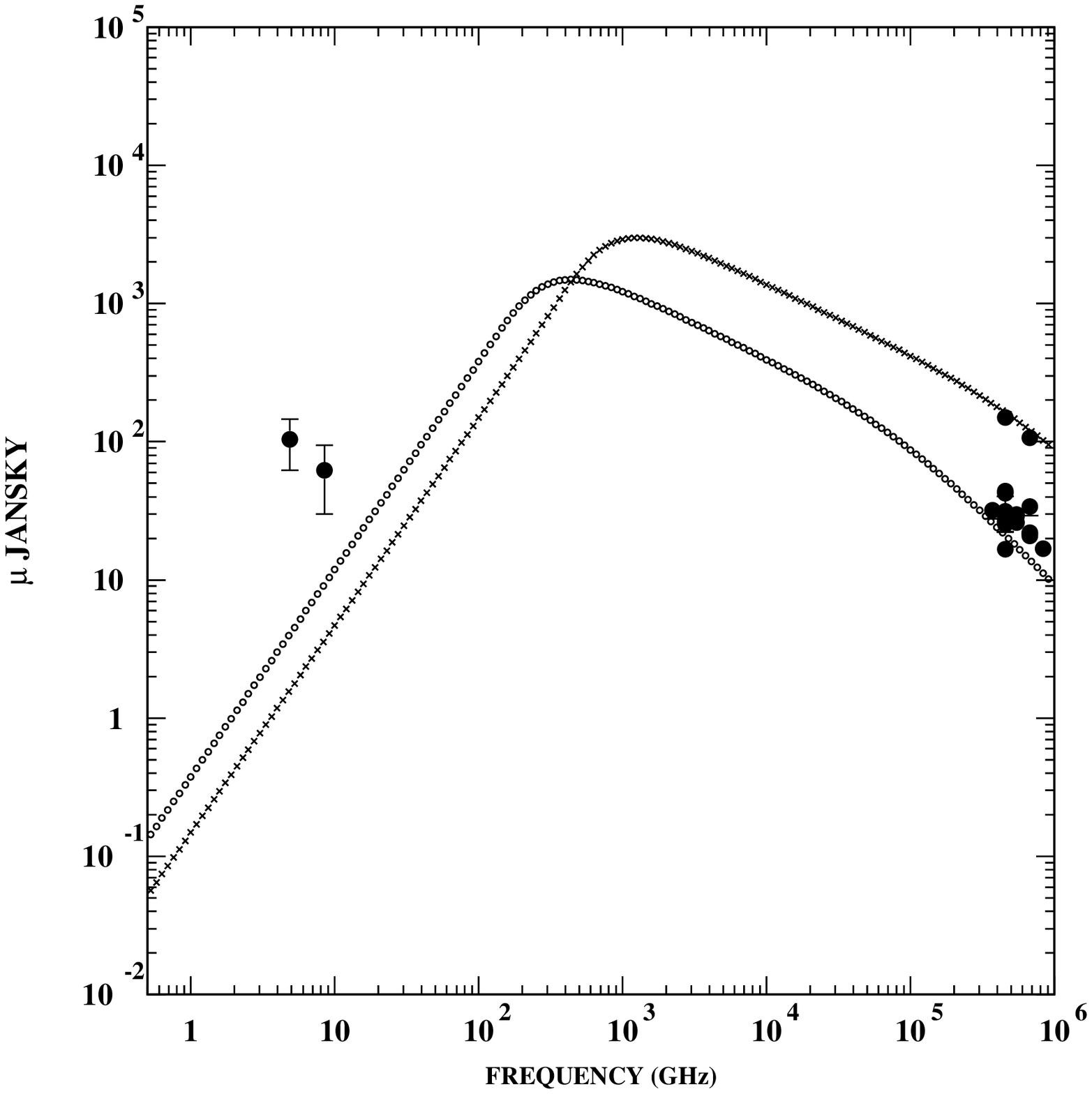, width=8cm} \\ 
%\hskip 1truecm 
\hspace*{.5cm} 
\epsfig{file=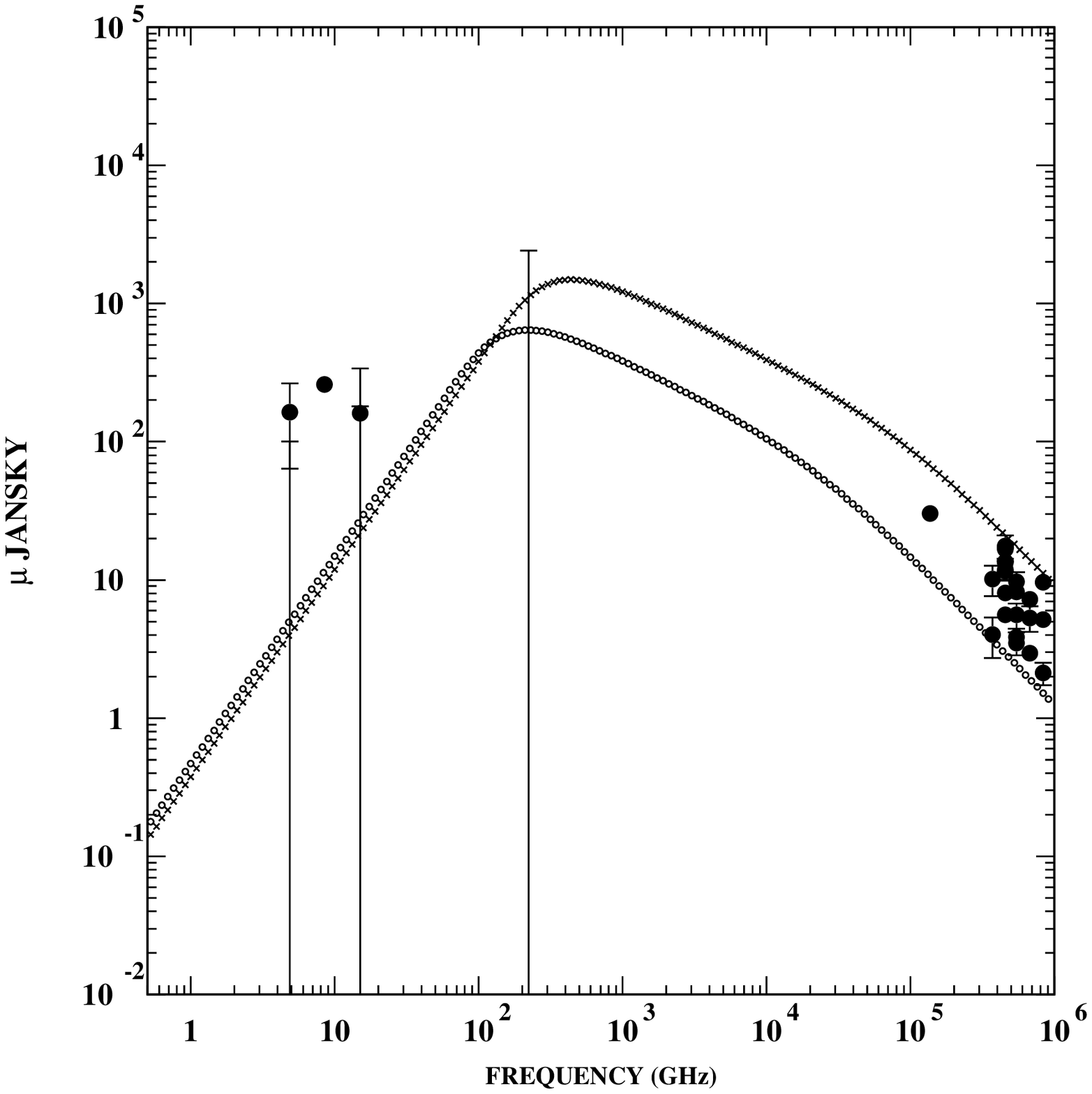, width=8cm} 
\end{tabular} 
\caption{The spectrum of the AG of GRB 990123 
from radio to optical frequencies.
Upper panel: in the time interval between 0.1 and 1 day after burst.
Lower panel: in the time interval between 1 and 3 days after burst.
In both cases the highest peaking curve
corresponds to the earlier time.} 
\label{rad-opt123a} 
\end{figure}

\begin{figure}[t] 
\begin{tabular}{cc} 
\hskip 2truecm 
\vspace*{2cm} 
\hspace*{-1.7cm} 
\epsfig{file=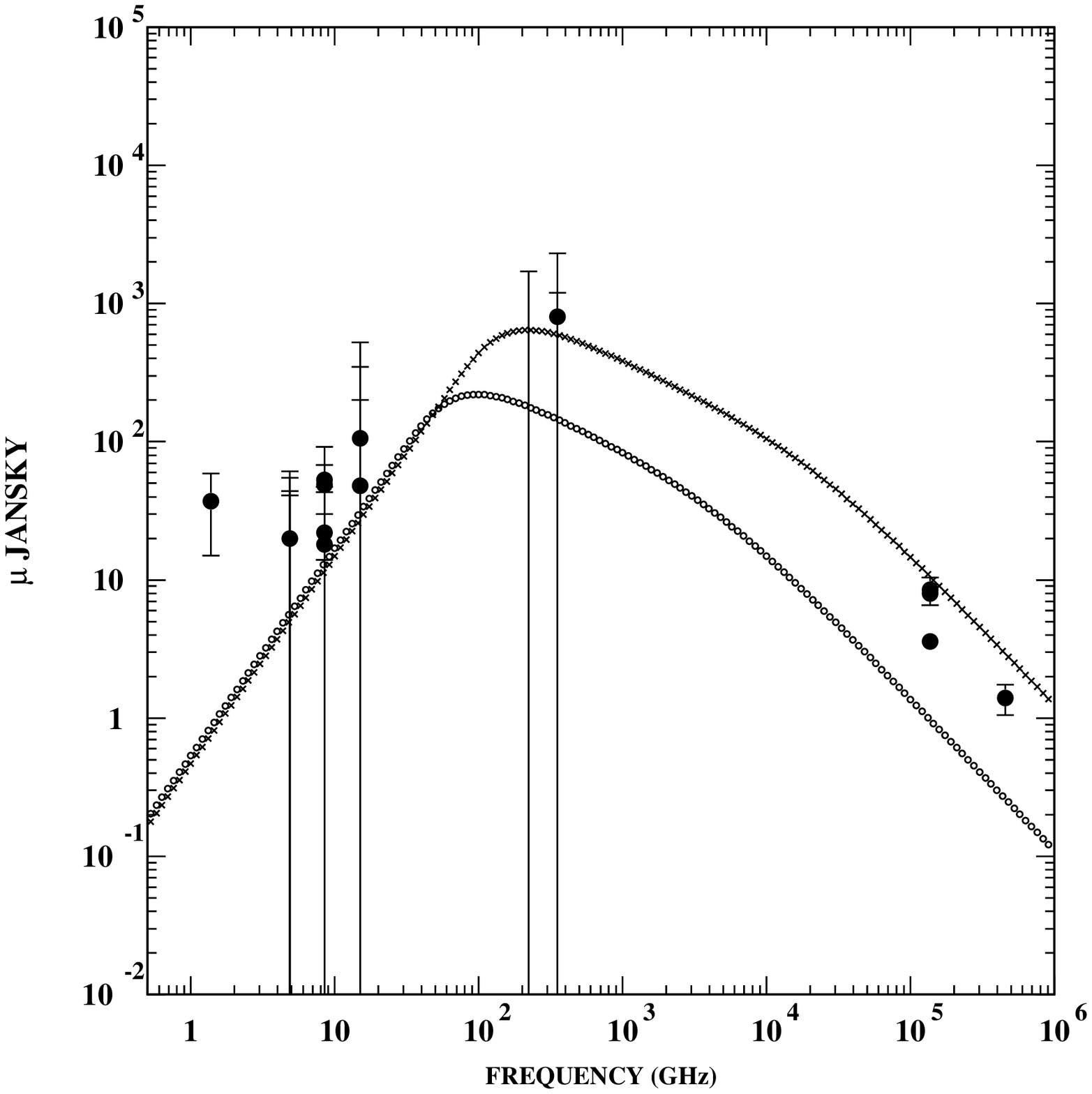, width=8cm} \\ 
%\hskip 1truecm 
\hspace*{.5cm} 
\epsfig{file=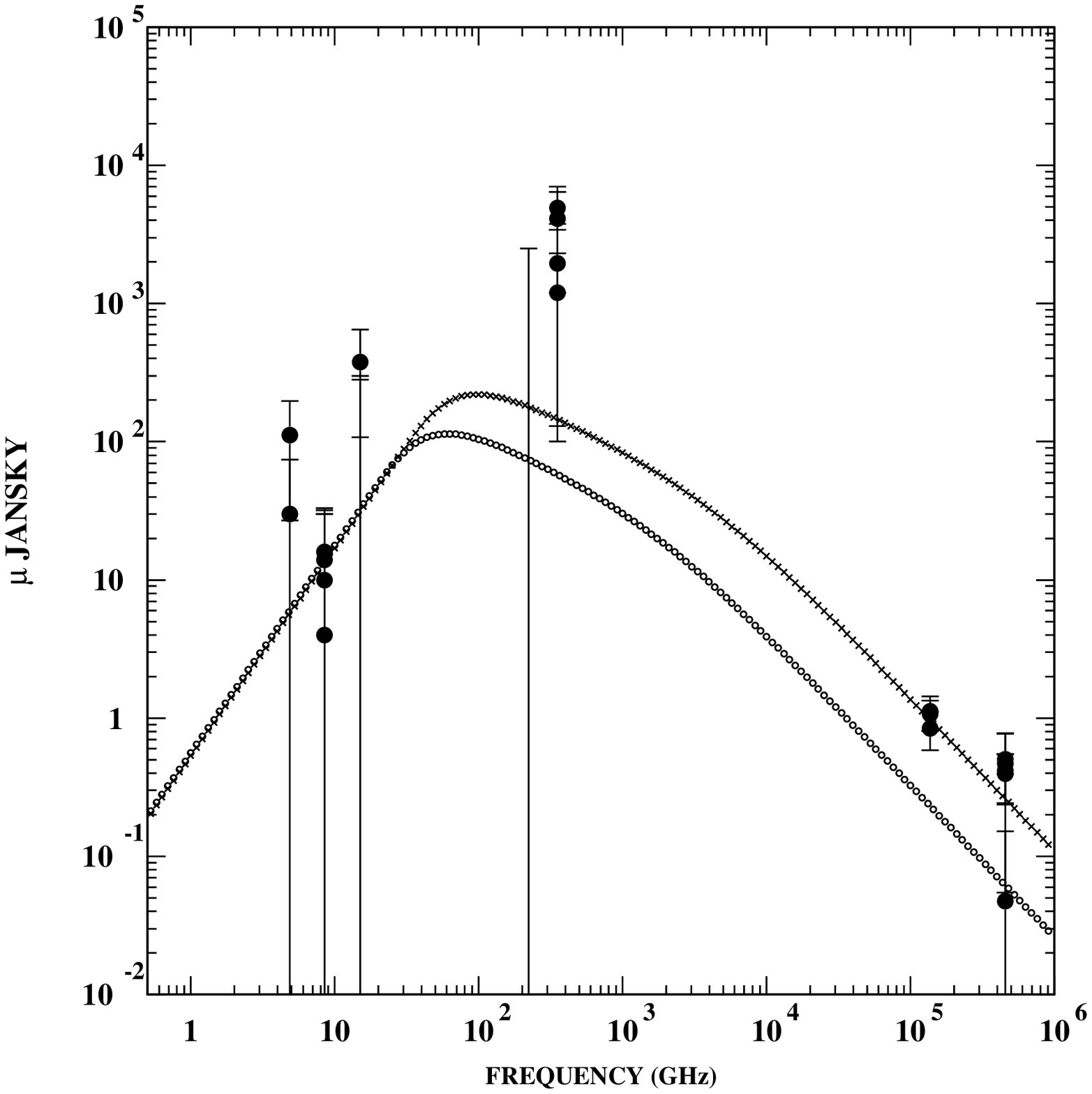, width=8cm} 
\end{tabular} 
\caption{The spectrum of the AG of GRB 990123
from radio to optical frequencies.
Upper panel: in the time interval between 3 and 10 days after burst.
Lower panel: in the time interval between 10 and 20 days after burst.
In both cases the highest peaking curve
corresponds to the earlier time.} 
\label{rad-opt123b} 
\end{figure}

We have fit the NIR/optical data (Castro Tirado 1999; Fruchter et al. 1999;
Galama et al. 1999, Holland et al. 2000; Kulkarni et al. 1999a) 
and the radio data (Galama et al. 1999; Kulkarni et al. 1999b)
for this GRB. In Fig.~(\ref{opt123}) we present the comparison between
the fitted CB model predictions ---assuming a constant ISM density after 0.1 
observer's days\footnote{the earlier optical data are discussed in 
DDD 2001.} and after subtraction of  the host galaxy and SN contributions---
with the measured light curves  for the K, I, R, V, B and U bands.
The theoretical predictions were corrected for the small Galactic extinction
in the GRB direction (E(B - V)=0.016, Schlegel et al. 1998) but not for 
extinction in the host
galaxy, since there is no spectral evidence for significant extinction 
there. In Fig.~(\ref{figr012301}) we  present the radio light curves at 
15 and 8.46 GHz. In Figs.~(\ref{rad-opt123a}) to 
(\ref{rad-opt123b}) we make the complementary comparison of theory and
observations for the radio to optical spectra,
in four time intervals extending from 0.1 to 20 days. 
The agreement between theory and observations is good despite 
the limited available data on the radio AG and its    
modulation by scintillations. 

\subsection{GRB 970508}

The optical (and X-ray) AG of GRB 970508 is the only one so far that has
been seen to rise and fall very significantly (e.g., 
Garcia et al. 1998; Galama et al. 1998b;  Pedersen et al. 1998;
Schaefer et al. 1997; Sokolov et al. 1998; Zharikov et al. 1998).  In DDD 
2001 we have shown
that a CB model fit to this AG fails, if one assumes ---like in all our other
fits---  a constant ISM density.  However, we have argued there that GRB
progenitors are presumably located in super-bubbles of 0.1 to 0.5 kpc
size. There may be instances in which the jet of CBs, after travelling for
such a distance, does not continue onwards to a similarly low-density halo
region, but encounters a higher-density domain. Indeed, we have shown that
a fairly satisfactory fit to the optical (and X-ray) AG is obtained
upon assuming an upwards jump in density by a factor
$\sim 2.2$ at $\rm t\sim 1.1$ day after burst.
This jump occurs before the first available
data points on the radio AG (Galama et al. 1998a; Frail et al. 2000b).
Therefore, we have fitted the optical data and the radio data 
with the ISM density profile that was fitted
to the R-band light curve. 

\begin{figure}[t]  
%\begin{tabular}{cc}  
\hskip 0truecm   
\epsfig{file=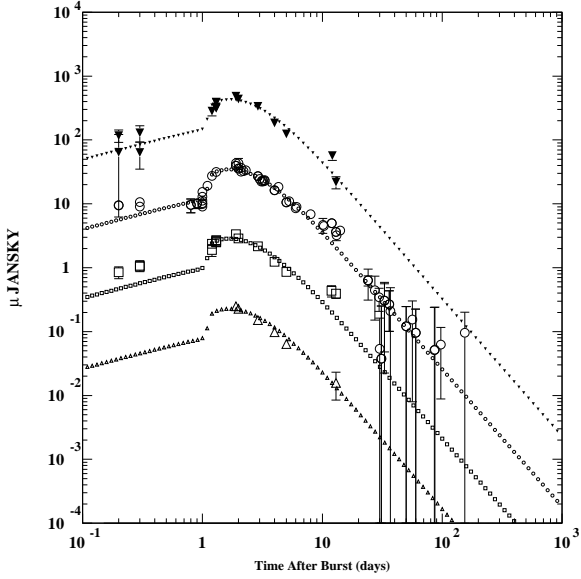, width=8.5cm}  
%\end{tabular}  
\caption{Comparisons between our fitted CB model AG of GRB 970508, 
at $\rm z=0.835$,
Eq.~(\ref{Fnuobser}), with the observed optical data. 
The figure shows (from top to bottom), 
10 times the I-band, the R-band, 1/10 of the V-band and
1/100 of the B-band.
The contributions of the underlying galaxy and the
expected SN1998bw-like 
SN have been subtracted. In a CB-model fit, there is in this case
strong evidence for such a SN (DDD 2001).}
\label{opt508}  
\end{figure}

\begin{figure}[t]  
\begin{tabular}{cc}  
\hskip 2truecm  
\vspace*{2cm} 
\hspace*{-1.7cm}  
\epsfig{file=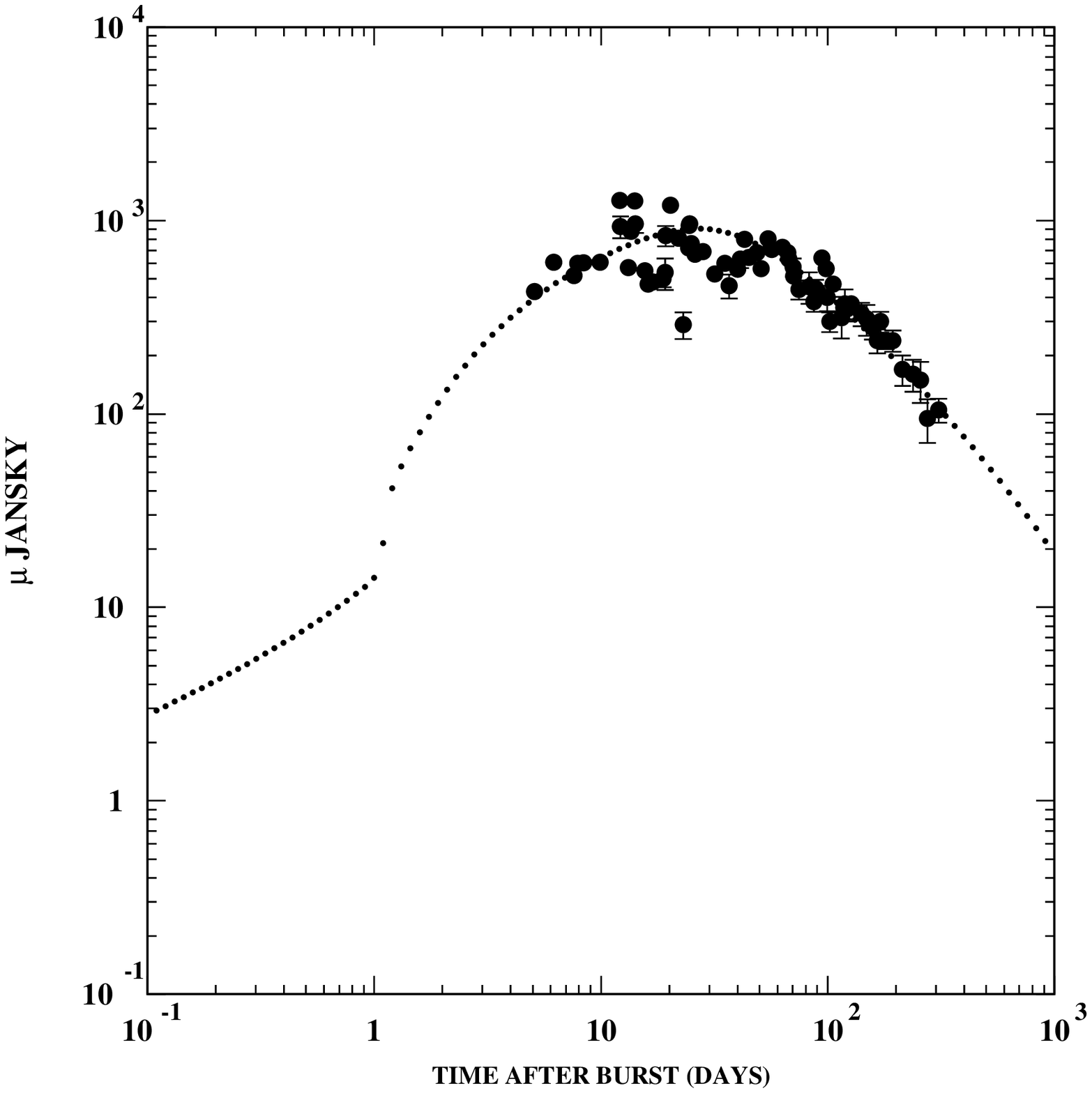, width=8cm} \\ 
%\hskip 1truecm  
\hspace*{.5cm}  
\epsfig{file=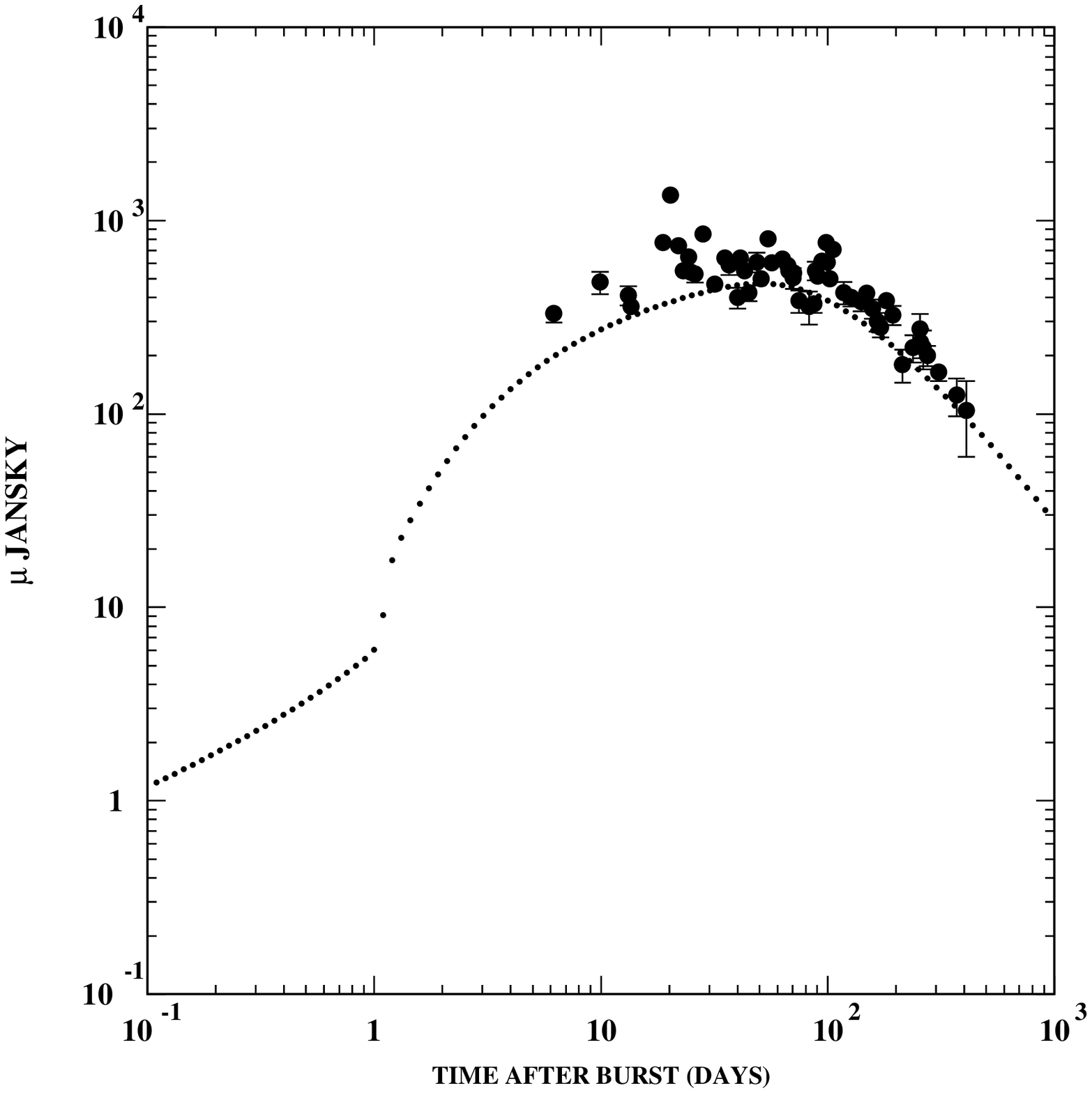, width=8cm} 
\end{tabular}  
\caption{Comparisons between our fitted CB model afterglow, 
Eq.~(\ref{Fnuobser}), and the observed radio afterglow of GRB 970508.
Upper panel: the light curve at 8.46 
GHz. Lower panel: the light curve at 4.86 GHz.}  
\label{figr050801}  
\end{figure}

\begin{figure}[t]  
\begin{tabular}{cc}  
\hskip 2truecm  
\vspace*{2cm} 
\hspace*{-1.7cm}  
\epsfig{file=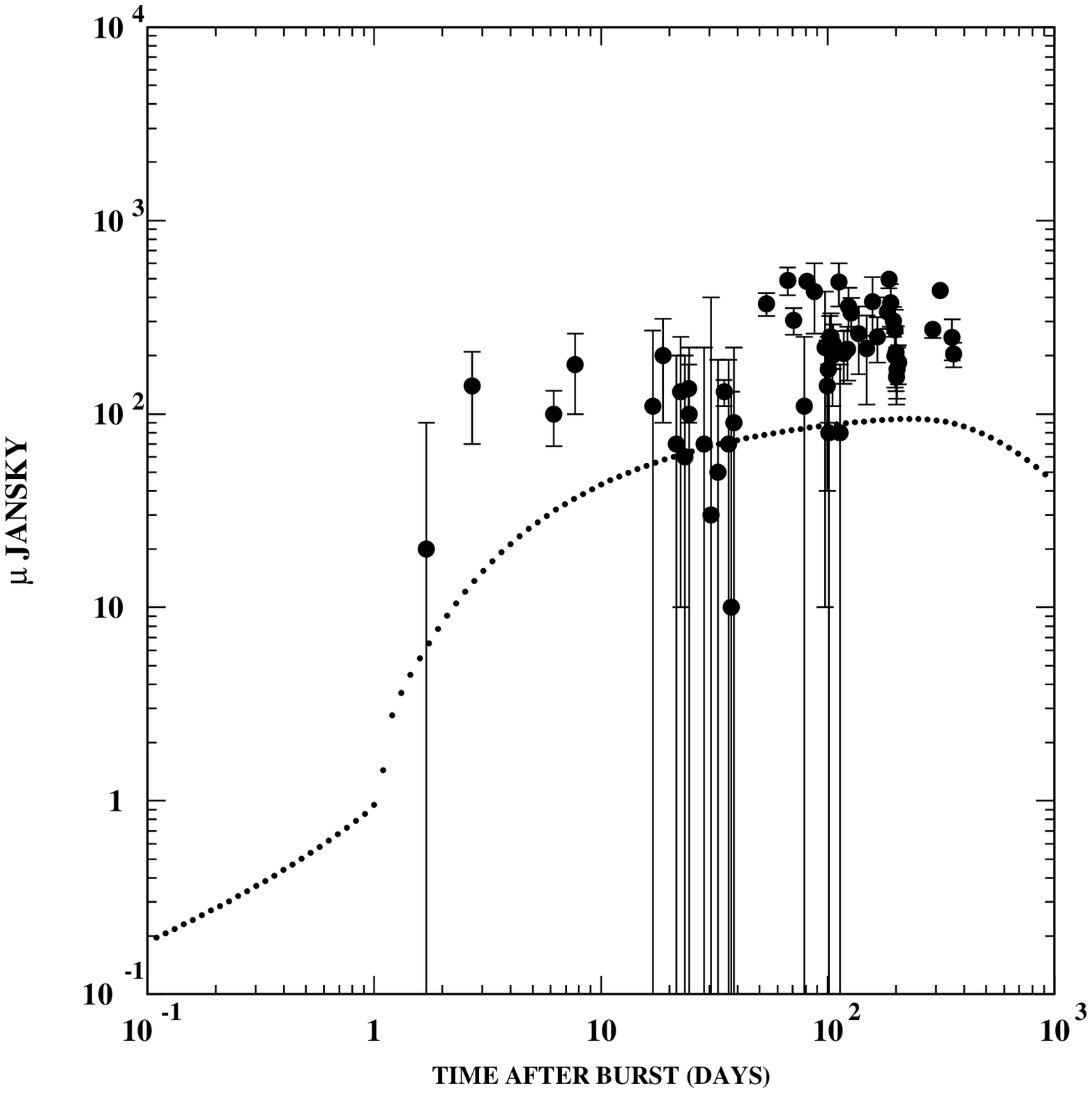, width=8cm} \\ 
%\hskip 1truecm  
\hspace*{.5cm}  
\epsfig{file=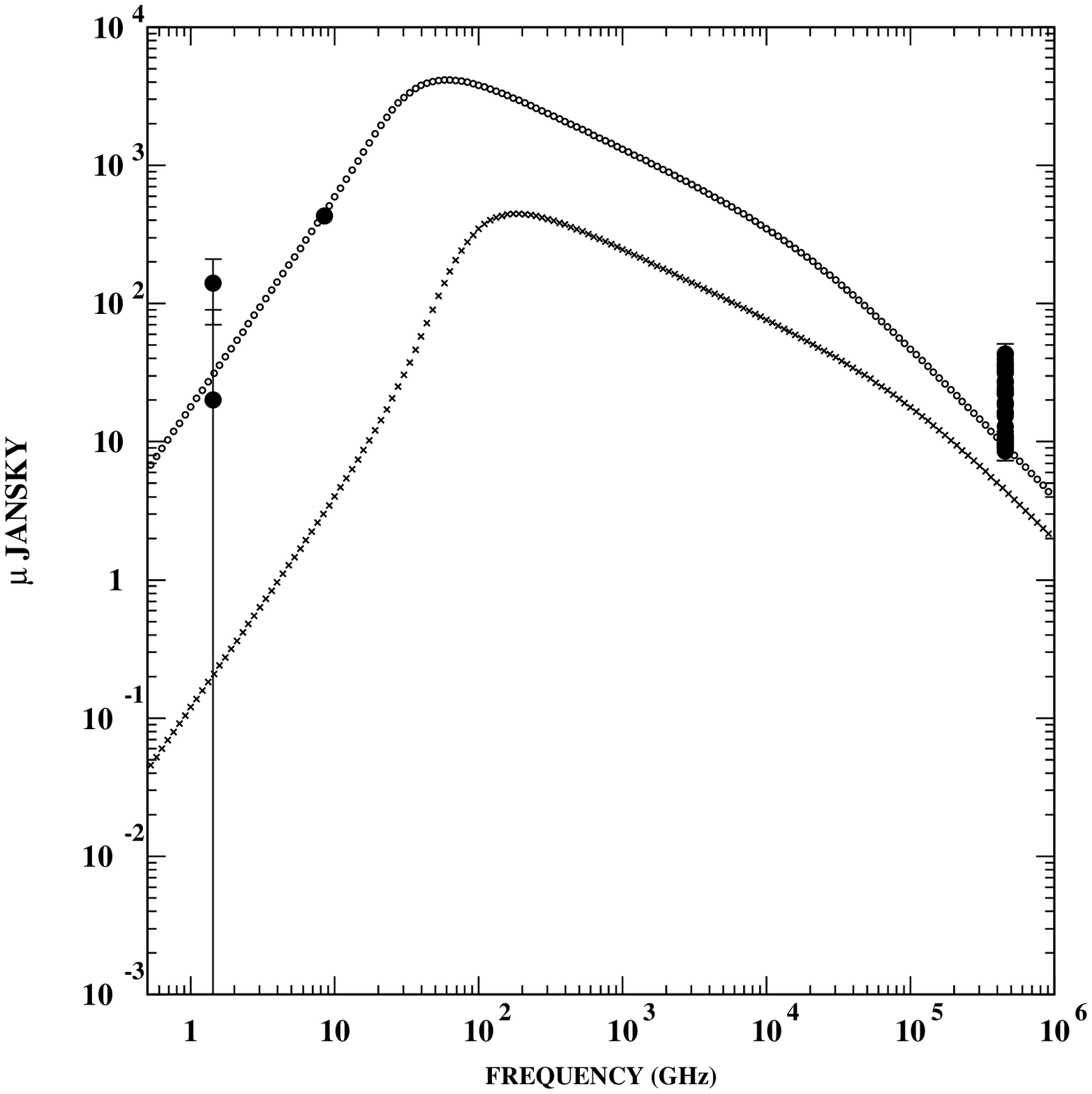, width=8cm} 
\end{tabular}  
\caption{Comparisons between our fitted CB model afterglow, 
Eq.~(\ref{Fnuobser}), and the observed radio afterglow of GRB 970508.
Upper panel: the light curve at 1.43 GHz. 
Lower panel: the spectral behaviour in the time
interval  between 0.12 and 6 
days after burst. The highest peaking curve
corresponds to the earlier time.}
\label{figr050802} 
\end{figure}

\begin{figure}[t]  
\begin{tabular}{cc}  
\hskip 2truecm  
\vspace*{2cm} 
\hspace*{-1.7cm}  
\epsfig{file=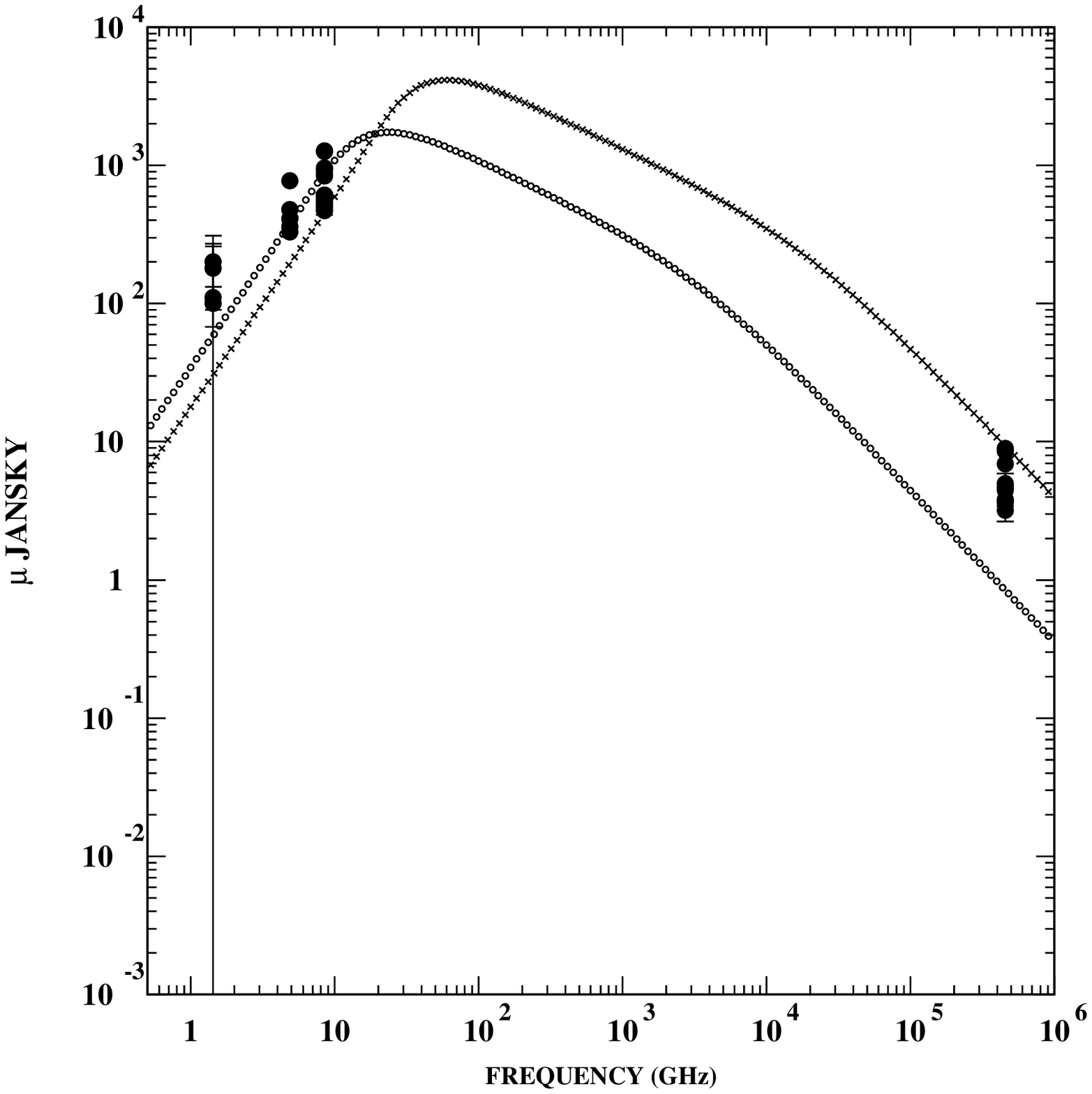, width=8cm} \\ 
%\hskip 1truecm  
\hspace*{.5cm}  
\epsfig{file=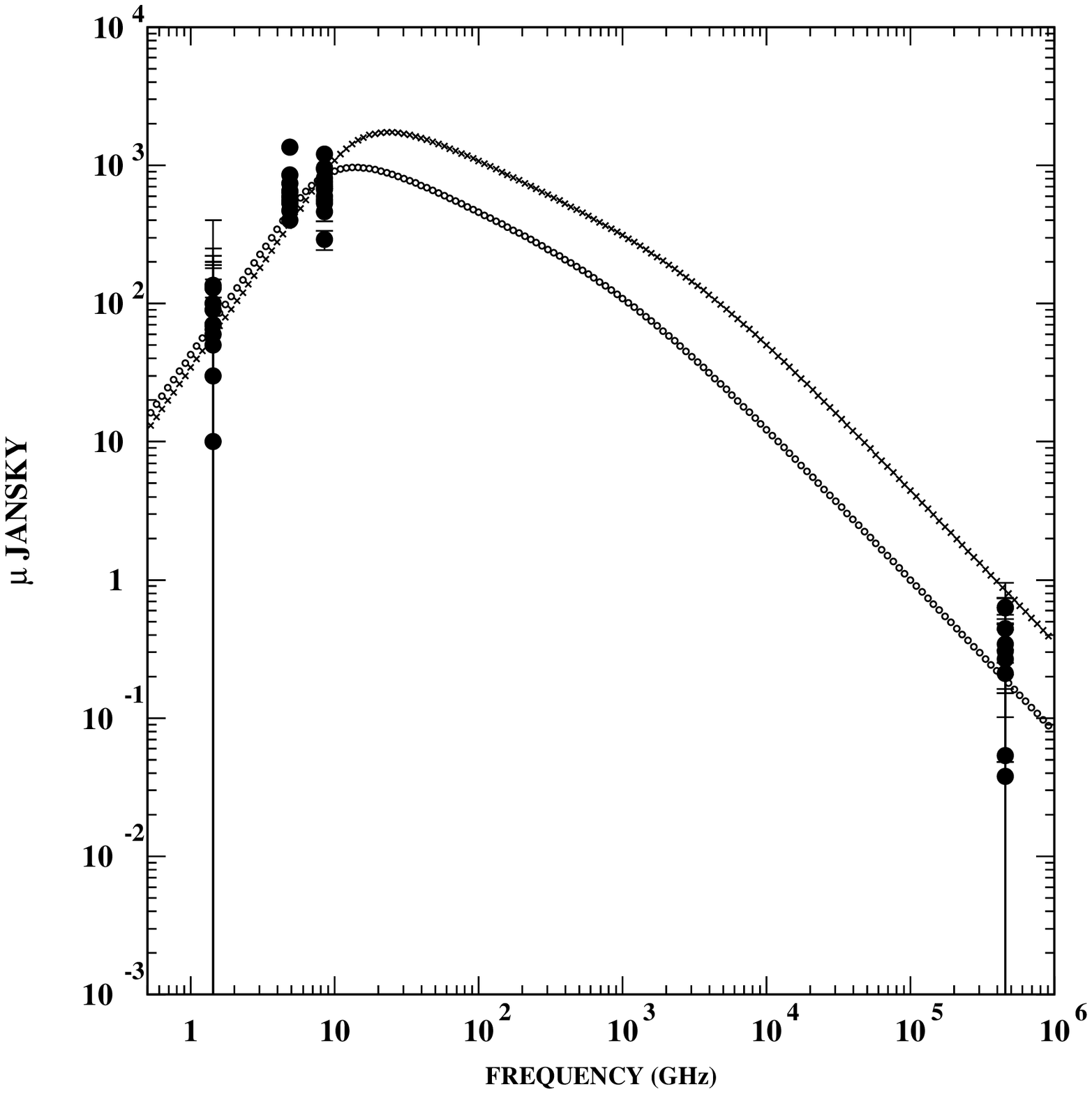, width=8cm} 
\end{tabular}  
\caption{The spectrum of the AG of GRB 970508
from radio to optical frequencies.
Upper panel: in the time interval between 6 and 20 days after burst.
Lower panel: in the time interval between 20 and 40 days after burst.
In both cases the highest peaking curve
corresponds to the earlier time.} 
\label{rad-opt508a} 
\end{figure}

\begin{figure}[t]  
\begin{tabular}{cc}  
\hskip 2truecm  
\vspace*{2cm} 
\hspace*{-1.7cm}  
\epsfig{file=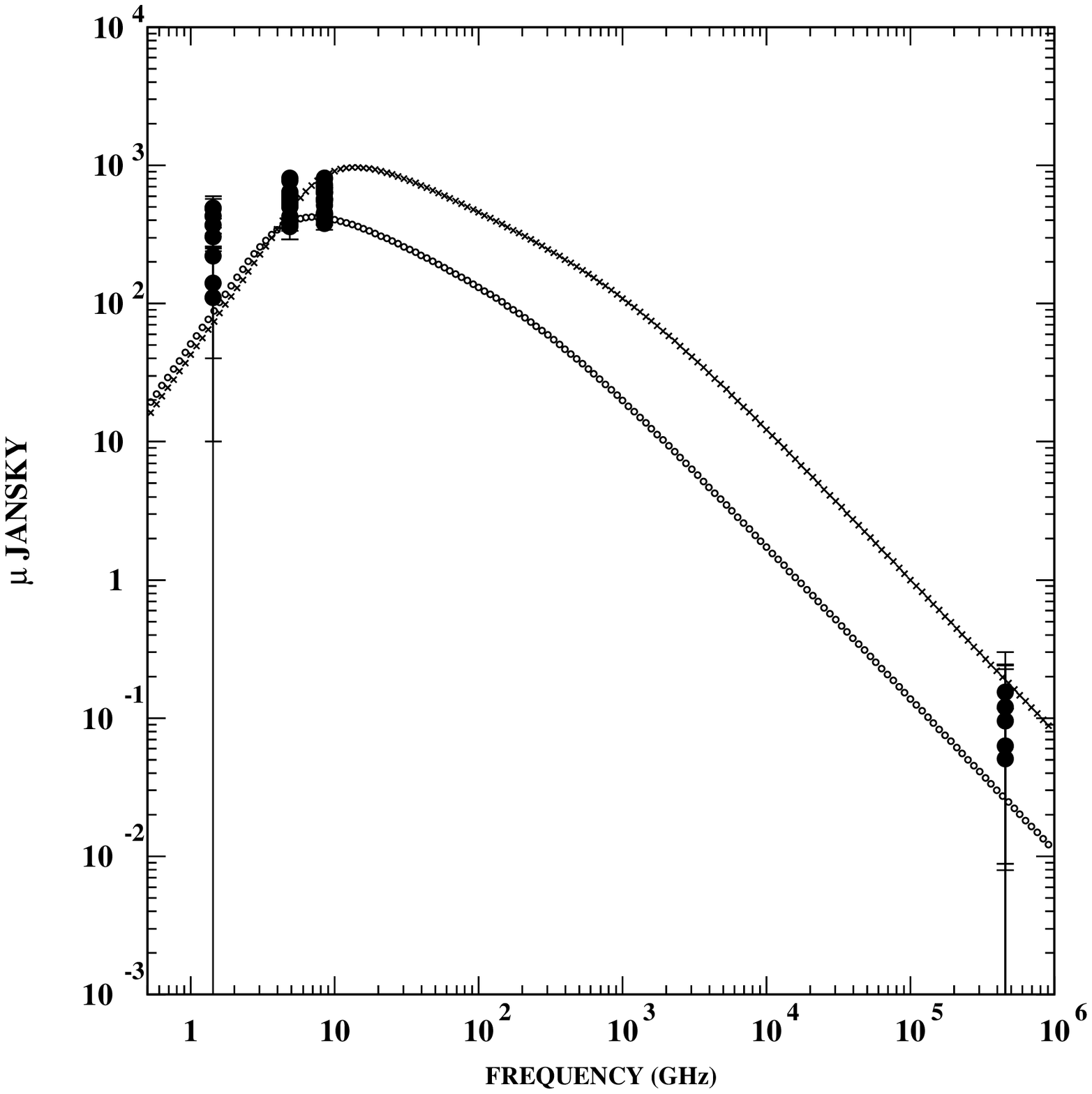, width=8cm} \\ 
%\hskip 1truecm  
\hspace*{.5cm}  
\epsfig{file=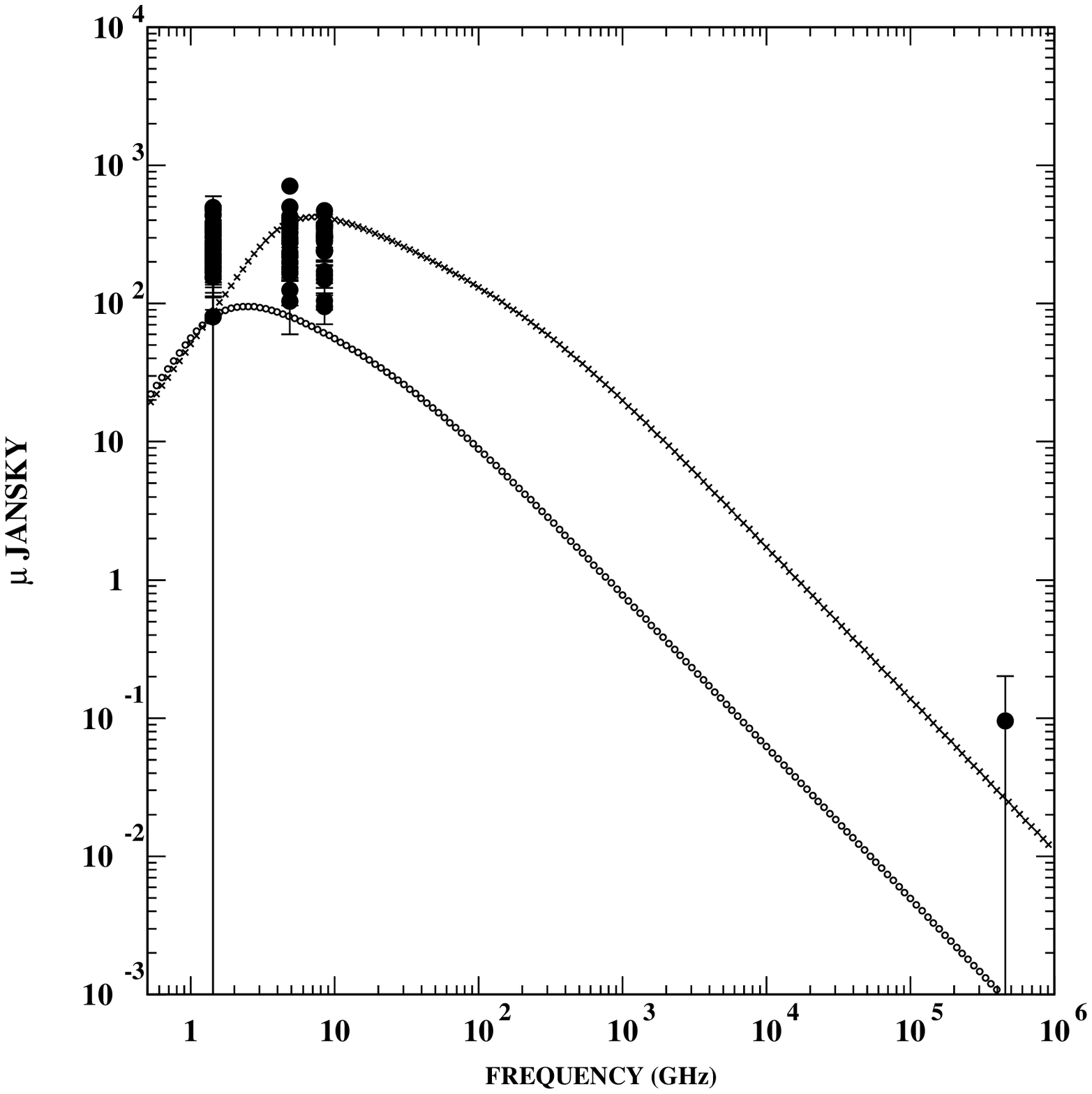, width=8cm} 
\end{tabular}  
\caption{The spectrum of the AG of GRB 970508
from radio to optical frequencies.
Upper panel: in the time interval  between 40 and 100 days after burst.
Lower panel: in the time interval between 100 and 470 days after burst. 
In both cases the highest peaking curve
corresponds to the earlier time.}
\label{rad-opt508b} 
\end{figure}

In Fig.~(\ref{opt508}) we present the comparison between
the measured light curve  for the I,R,V and B  bands after
subtraction of the host galaxy and SN contribution (DDD 2001). 
The theoretical predictions were corrected for the small galactic extinction
in the GRB direction (E(B - V)=0.016, Schlegel et al. 1998) but not for 
extinction in the host
galaxy, since there is no spectral evidence for significant extinction 
there. In Figs.~(\ref{figr050801}) and (\ref{figr050802}) we also present 
the radio light curves at 8.46, 4.86 and 1.43 GHz.
In Figs.~(\ref{figr050802}) to (\ref{rad-opt508b}) we
make the complementary comparison of theory and
observations for the radio to optical spectra,
in five time intervals
extending from 0.12 to 470 days. The results are quite satisfactory.

\subsection{Commentary}

In DDD 2001 we demonstrated that, in the CB model, the spectral index
in the optical to X-ray domain could be extracted from the 
time-dependence of the optical
light curves. The fits resulted in $\rm \alpha=p/2\simeq 1.1$ for
all GRBs of known redshift. This result is in good agreement
with the observed late spectral observations.
We have learned in this section that the CB model also provides an
excellent description of the AG spectra in the broader band that
includes the radio data. Only one new parameter, $\rm \nu_a$, is
involved in the extension to the broader band. And this fitted 
parameter and the injection bend ---at its {\it predicted} frequency and 
time-dependent position---  bring about
the agreement between the different magnitudes and spectral trends
of the radio and optical domains.
  
In some of our fits to broad band spectra, such as the earliest data
on GRBs 000301c,  991216 and 990123
in the upper panels of Figs.~(\ref{rad-opt301}), (\ref{rad-opt216})
and (\ref{rad-opt123a}), respectively, the theoretical curve is
an underestimate of the low-energy spectral intensity.
In other cases, such as GRBs 000926, 991208 and 980425,
the spectral fits are excellent at all times. The lowest frequencies and
earliest times are the most dependent on our simplifications
concerning the GRB geometry, density profile, self-absorption,
cumulation and limb-darkening. We would have been surprised
if these simplifications worked even better than they do, and
the fits do improve if we remove our  approximation
of a fixed spectral index $\rm p=2.2$. But our aim in this paper
is not to obtain spectacularly good fits, but to demonstrate that, 
even in the simplest approximations, the CB model provides
a good description of the broad-band data. The analysis of the
lowest radio frequencies at the earliest times brings forth a
plethora of details that are not of fundamental interest: our
ultimate goal is not to understand these details, but
to investigate what the origin of GRBs actually is.

\section{SN1998bw and GRB 980425} 
 
The time and position of the peculiar gamma ray burst 980425 (Soffita et  
al. 1998) coincided with supernova SN1999bw (Tinney et al. 1998) in the  
spiral galaxy ESO 184-G82,  at a nearby $\rm z=0.0085$ (Tinney et al. 
1998; Sadler et al. 1998; Galama et al. 1998c; Lidman et al. 1998; 
Iwamoto et al. 1998).  Iwamoto et al. (1998) estimated the 
the core collapse of SN1998bw to have happened within 
$-$2 to +7 days 
of GRB 980425. The BeppoSAX Narrow Field Instrument (NFI) located 
10h after burst an X-ray source coincident in position with SN1999bw that  
declined slowly with time between April and November 1998 (Pian et 
al. 2000). A posteriori statistics indicate a very low chance probability  
($\leq 10^{-4}$) of a GRB being so nearly coincident in position.  But  
despite how close  ---if it was associated with SN1999bw--- the progenitor  
of GRB980425 was to us (38 Mpc for  $\rm H=65\, km\,Mpc^{-1}\, s^{-1}$),  
the $\gamma$-ray fluence indicated only $7\times 
10^{48}$ erg equivalent spherical energy release in $\gamma$ rays, much  
smaller than $\sim 3\times 10^{53}$ erg, the mean value for 
 the score of other GRBs with known ---cosmological--- redshifts.
 
\subsection{SN1998bw: the accepted lore}

Like its accompanying GRB,
SN1998bw was also claimed to be a very peculiar radio supernova 
(e.g. Kulkarni et al. 1998). Over the past twenty years approximately two  
dozen  SNe have been detected in the radio: 2 
Type Ib, 5 Type Ic, and the rest Type II. A much larger list of 
more than 100 additional SNe have low radio upper limits (for a 
review see, e.g., Weiler et al. 2000 and references therein). Type Ib/c  
SNe are fairly homogeneous in their radio properties, but 
SN1998bw\footnote{SN1998bw was 
classified initially as  Type Ib (Sadler et al. 1998), then 
Type Ic (Patat and Piemonte 1998), then peculiar Type Ic (Kay et al. 1998),  
then, at an age of 300 to 400 days, again as Type Ib (Patat et al. 
1998).} had a peak 6-cm radio luminosity of $\rm \sim 8\times 10^{28}\, erg  
s^{-1} \, Hz^{-1}$, that is 20 to 40 times brighter than other 
radio Type Ib/c SNe, which fall typically in the range $\rm 1.4-2.6\, \times  
10^{28}\, erg s^{-1} \, Hz^{-1}$. SN1998bw also reached a high radio 
luminosity earlier than any known SN. Simple arguments based on the 
brightness temperature of its radio luminosity (e.g., Readhead 1994) 
required the radiosphere of SN1998bw to have expanded  surprisingly 
fast, at $\rm\geq 200,000\, km\, s^{-1}$, at least during the first few days. 
Its unusually high optical and radio luminosities and  its extraordinarily 
large initial speed of expansion led many authors to conclude that 
SN1999bw was a hypernova (Paczynski 1998) rather than a 
peculiar supernova (e.g., Iwamoto et al. 1998). 

\subsection{The pair SN1998bw/GRB 980425 in the CB model}
 
In Dar and De R\'ujula (2000a) we argued that the only 
peculiarity of SN1998bw was that it was viewed very near its axis. The  
peculiarity of GRB 980425 was its nearness, that allowed for 
its detection at an angle, $\theta\sim 8/\gamma(0)$  
unusually large relative to the other GRBs of known redshift,
for which $\theta\sim 1/\gamma_0$. 
These facts conspired to produce a ``normal'' GRB fluence, and 
resulted in an optical AG dominated by the SN. In DDD 2001,  
we  demonstrated 
that the X-ray AG of this GRB was also ``normal'': it has precisely the  
light curve (in shape and normalization) expected in the CB model
if the X rays are produced by the CBs and {\it not}, 
as the observers assume (Pian et al. 2000), by the supernova.  

In the CB model (Dar and De R\'ujula 2000a), the 
gamma-ray fluence of GRBs
at large viewing angle ($\gamma_0^2\, \theta^2\gg 1$) is
$\propto \delta_0^3\propto \theta^{-6}$. 
The radio AG spectral energy density is proportional to
$\gamma^{1/2}\, \delta^{7/2}$, as implied by 
Eqs.~(\ref{sync}, \ref{Fnucb}, \ref{Fnuobser}), the 
dependence $\rm\nu_b\propto \gamma^3$ and the
relation $\rm\nu_{obs}\propto\nu_{_{CB}}\,\delta$.
As a function of time, the AG peaks when $\gamma\, \theta\sim 1$, 
so that $\gamma\simeq\delta\simeq 1/\theta$ and the
peak value is proportional to $\theta^{-4}$.
Because its proximity and large viewing angle ``conspired'' to make
GRB 980425 appear ``normal'' in gamma rays, its peak radio intensity
should have been enhanced by a factor $\rm \sim
(\theta/mrad)^2$ relative to that of ordinary GRBs. Thus, for $\theta\sim 8.3$
mrad, as estimated for GRB 980425 in Dar and De R\'ujula 2001, its expected
peak radio intensity is $\sim$ 60 times larger than that of
ordinary GRBs. Observationally, it is 50 to 100 times larger.
 
\begin{figure}[t] 
\begin{tabular}{cc} 
\hskip 2truecm 
\vspace*{2cm} 
\hspace*{-1.7cm} 
\epsfig{file=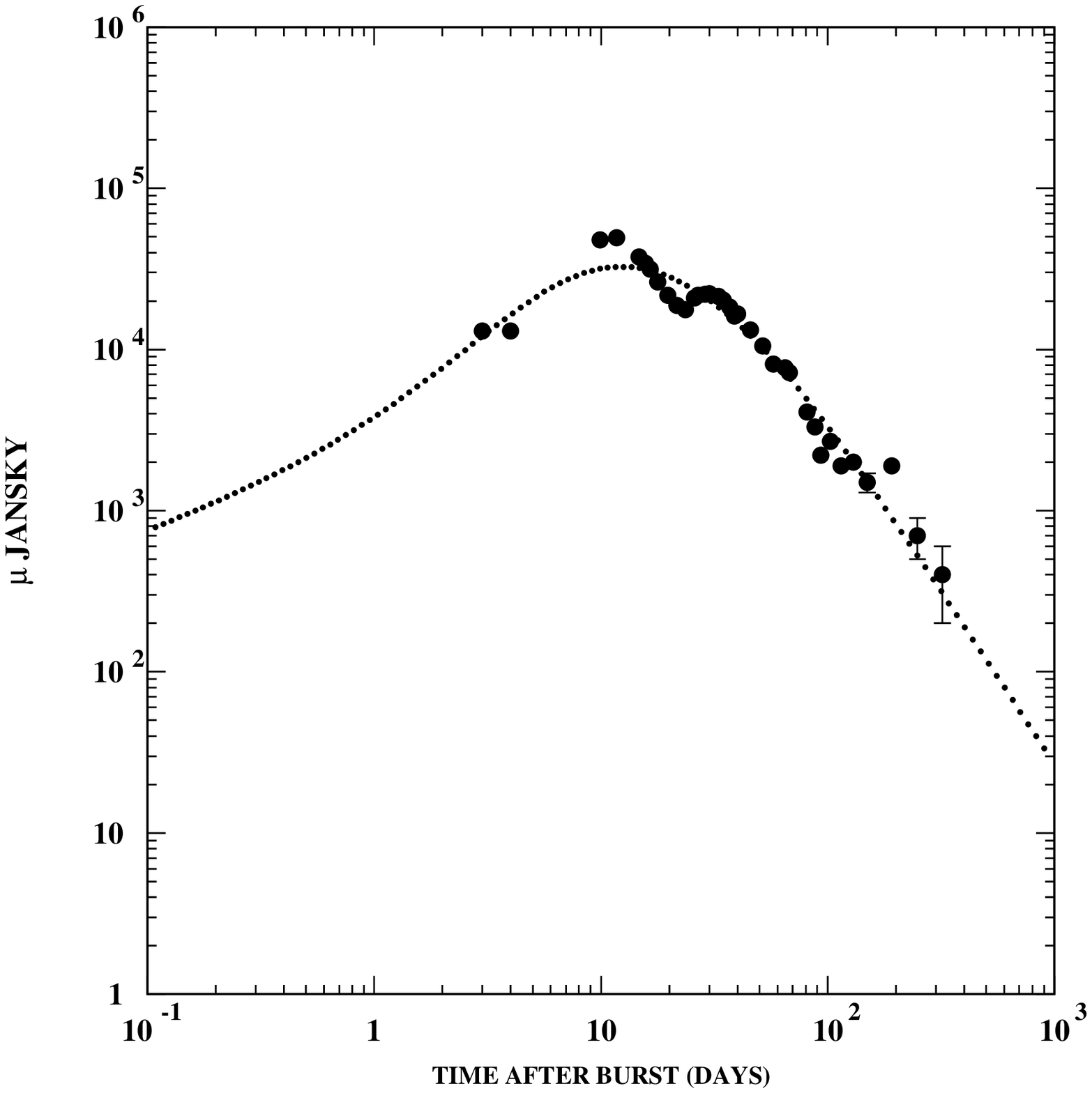, width=8cm} \\ 
%\hskip 1truecm 
\hspace*{.5cm} 
\epsfig{file=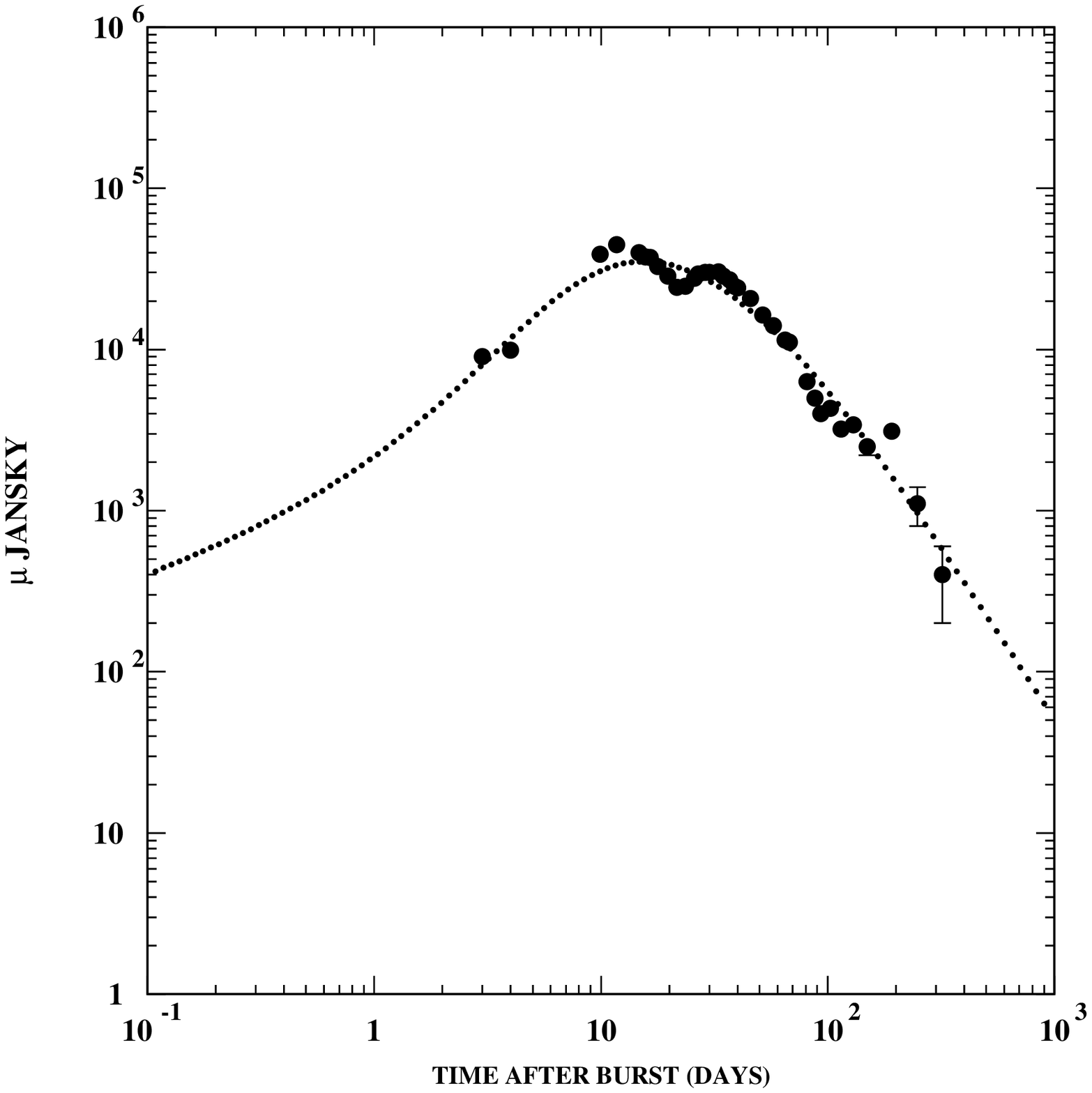, width=8cm} 
\end{tabular} 
\caption{Comparisons between our fitted CB model afterglow of
GRB 980425 at z=0.0085, 
Eq.~(\ref{Fnuobser}), 
and its observed radio afterglow. 
Upper panel: the light curve at 8.64 GHz. 
Lower panel: the light curve at 4.80 GHz.} 
\label{figr042501} 
\end{figure}

In Figs.~(\ref{figr042501}) to (\ref{figr042505}) 
we show our CB model fits to the   
temporal and spectral behaviour of the radio afterglow of GRB 980425.
The fit parameters (in particular the large observation angle $\theta$)
are quite close to the ones  
that explain its GRB fluence (Dar and De R\'ujula 2000a), and its  
X-ray afterglow (DDD 2001). These figures show  
how, in the CB model, the radio AG of GRB 980425 also has a ``normal''  
magnitude and shape.  
That is, once more, {\it if} the radio AG is produced by the CBs and  
{\it not} by the SN, unlike, once again, it is generally assumed (e.g., 
Kulkarni et al. 1998; Li and Chevalier; Weiler et al. 2000). 

\begin{figure}[t] 
\begin{tabular}{cc} 
\hskip 2truecm 
\vspace*{2cm} 
\hspace*{-1.7cm} 
\epsfig{file=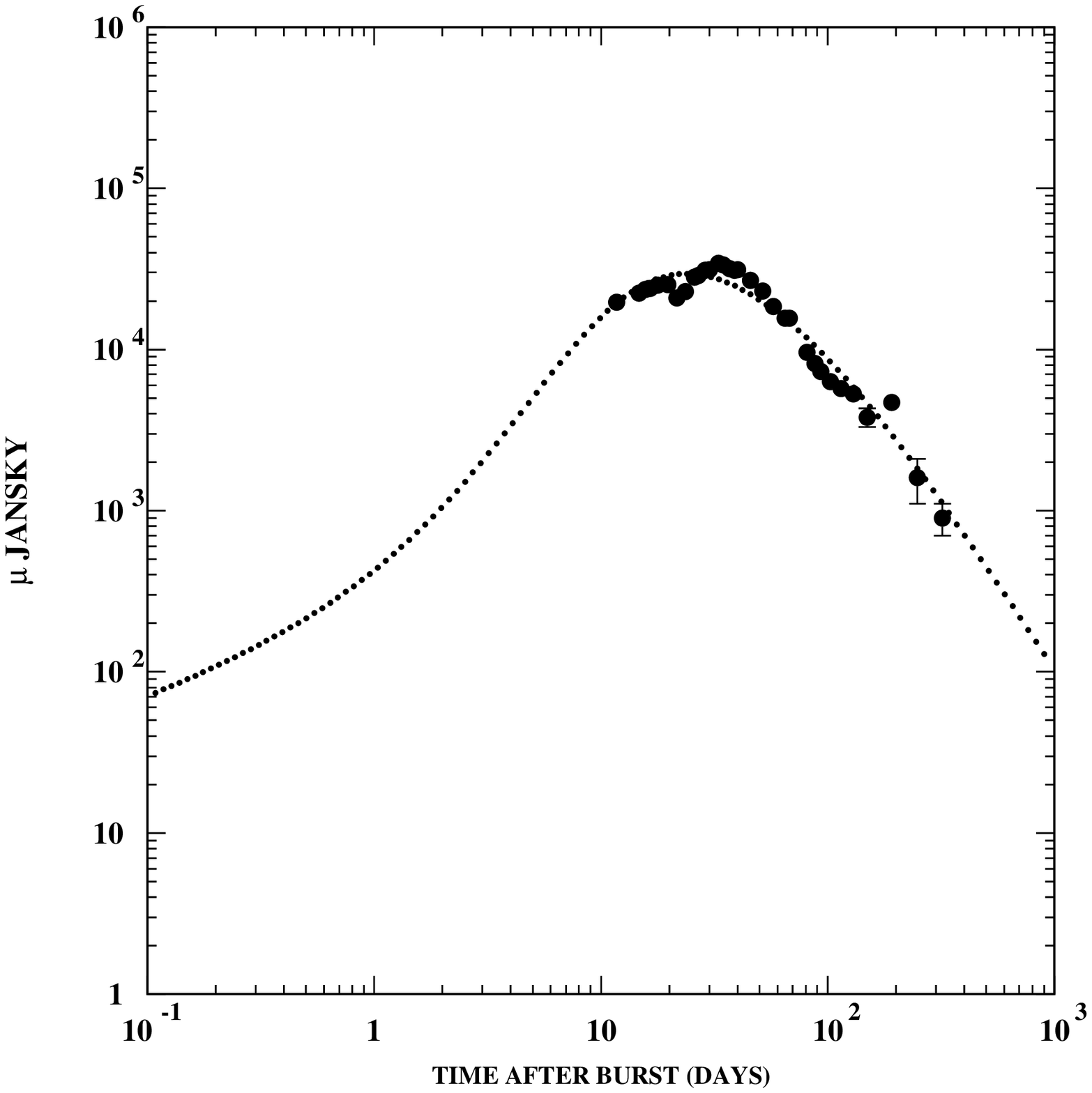, width=8cm} \\ 
%\hskip 1truecm 
\hspace*{.5cm} 
\epsfig{file=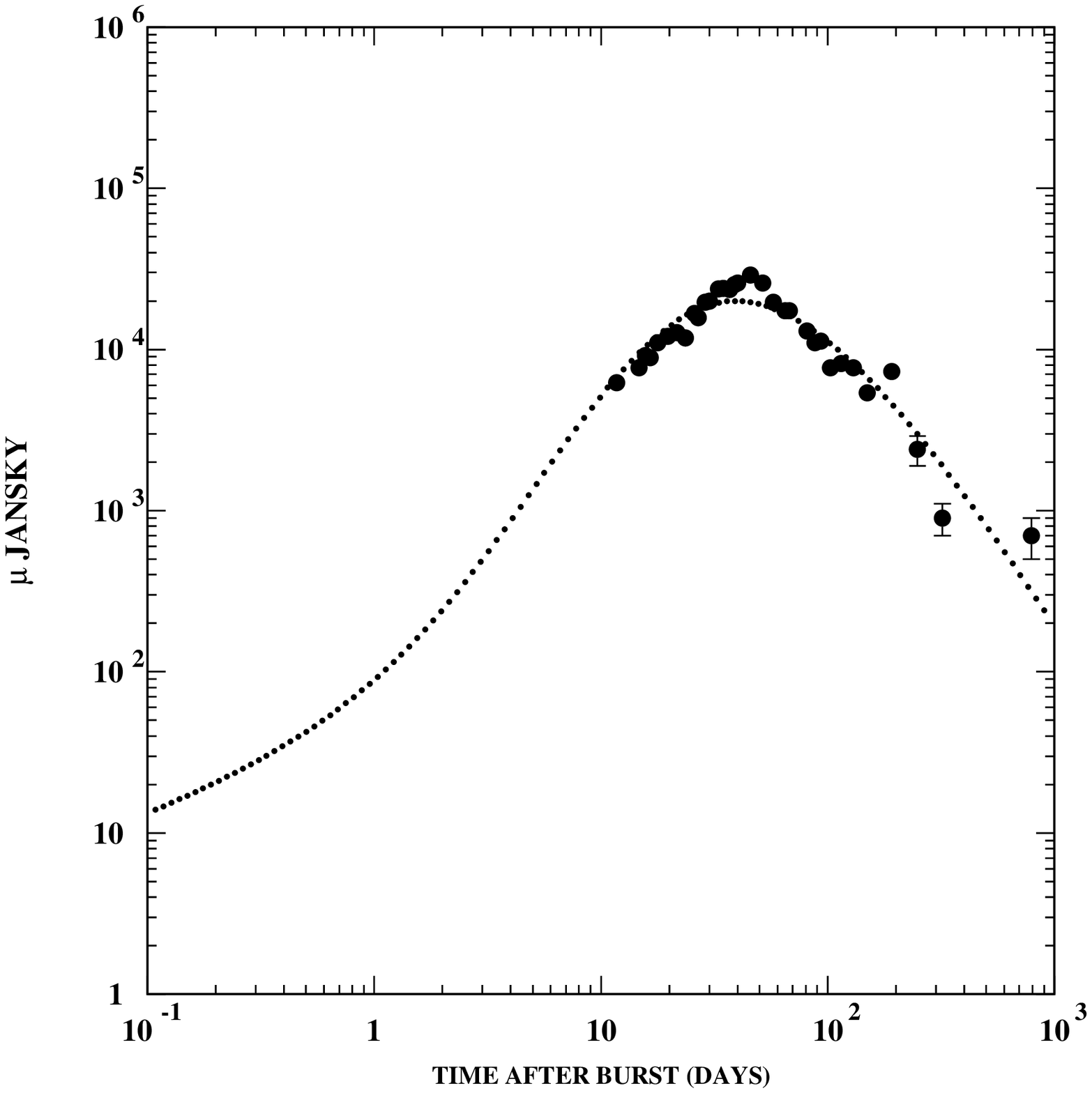, width=8cm} 
\end{tabular} 
\caption{Comparisons between our fitted CB model afterglow, 
Eq.~(\ref{Fnuobser}), 
and the observed radio afterglow of GRB 980425. 
Upper panel: the light curve at 2.49 GHz. 
Lower panel: the light curve at 1.38 GHz.} 
\label{figr042502} 
\end{figure}

In the case of GRB 980425 the relatively large viewing angle
and the subsequently small Doppler factor imply that, at
late times, even the radio frequencies are above the injection bend.
The large $\nu$ behaviour in Figs.~(\ref{figr042504}) and 
(\ref{figr042504}) is $\rm\nu^{-p/2}\sim\nu^{-1.1}$. Also, the late time
trend of the radio light curves in Figs.~(\ref{figr042501}) and
(\ref{figr042502}) approaches the asymptotic 
$\rm t^{2(p+1)/3}\sim t^{-2.1}$.

For GRB 980425 the
radio data are overwhelmingly more abundant and precise than
the X-ray data, and it is interesting to check what the prediction
for the X-ray light curve is, if the input parameters are those
determined in the radio fits. This is done in Fig.~(\ref{X425})
for two values of the electron spectral index $\rm p$. For our
fixed choice, $\rm p=2.2$, the prediction misses the data by a
factor $\sim 20$. There are two excuses for that. First, since GRB
980425 is seen much more ``sideways'' than other GRBs, and its
Doppler factor $\delta$ is much smaller than usual, the cumulation,
illumination and limb-darkening factors play a bigger role than
usual. These factors involve many simplifying assumptions (such as
spherical symmetry) and significantly affect the normalization of
the radio AG, but not that of the X-rays. Second, the extrapolation
from radio to X-rays is over some 10 orders of magnitude in frequency,
and a small change in the spectral photon's slope, $\rm (p - 1)/2$,
entails a very large change in relative magnitude, as can be seen
in Fig.~(\ref{X425}) by comparing the $\rm p=2.2$ and $\rm p=2$
curves.  

The X-ray light-curve of GRB 980245 is essentially 
flat in the time-interval of the first four observed points
(Pian et al. 2000), while the corresponding data
for all other GRBs fall with time much faster. 
The last observational point in Fig.~(\ref{X425}), a preliminary result
from XXM Newton (Pian 2002) and Chandra (Kouveliotou 2002),
falls precisely in the expected  
subsequent fast decline (predicted in DDD 2001) and
definitely not in a naive power-law extrapolation.
The peculiar light curve is a
consequence of the large observing angle (Dar and De R\'ujula 2000a).
For the reasons stated in this paragraph and the preceding one,
we consider the prediction of the X-ray
fluence completely satisfactory.

In DDD 2001, on the basis of the very meager X-ray data, we argued
that the last optically-measured point of the SN1998bw/GRB 980425
pair, at day 778 (Fynbo et al., 2000), was due to the CB's AG and
not to the supernova. Redoing the analysis with the input of the
abundant radio data, we must now revise this conclusion. In
Fig.~(\ref{late425}) we show the result, with inclusion of the
late optical measurement. This point lies more than two orders
of magnitude above the predicted CB's AG: it must be due to the
SN. We do not have an explanation ---specific to the CB-model---
of the fact that this point also lies somewhat above the expectation
based on $\rm ^{56}Co$ decay (Sollerman et al. 2000).

We are claiming that long duration GRBs are associated with a good
fraction of core-collapse SNe.  Yet, SN1998bw was one of the
brightest in its class. The apparent contradiction may be dispelled
by the increasing evidence that SN explosions are fairly asymmetric.
It is quite conceivable that, viewed very close to their ``CB axis''
SNe  appear to be brighter than when observed from other directions.

The conclusion is twofold. GRB 980425 is, {\it in every respect},
normal ($\rm z$ and $\theta$ being chance variables). And, deprived
of very abnormal X-ray and radio outputs ---which are not due to
the supernova, but to its ancillary GRB--- SN1998bw loses most of
its ``peculiarity''.

\subsection{Superluminal motion in SN1998bw/GRB 980425}

The transverse projected velocity in the sky of a CB relative
to its parent SN is, for large $\gamma$ and small $\theta$:
\begin{equation}
\rm
V_{_{CB}}(t)\simeq {\gamma(t)\, \delta(t)\, \theta\over (1+z)}\; c\, ,
\label{supervelocity}
\end{equation}
which, for typical parameters, is extremely superluminal.
The resulting angular separation at time $\rm t$ is:
\begin{equation}
\rm
\Delta \alpha(t)={1\over D_A}\int_0^t\,V_{_{CB}}(t')\;dt'\; ,
\label{Deltaalpha}
\end{equation}
where $\rm D_A=(1+z)^2\, D_L$ is the angular distance
to the SN/CB system. In Fig.~(\ref{superluminal}) we show
$\rm \Delta(t)$ for SN1998bw/GRB 980425 with our 
parameters fit to the corresponding radio data (for our
adopted $\rm H_0$, $\rm D_L\simeq D_A\simeq 38$ Mpc). In 
Dar and De R\'ujula (2000a) we argued that this separation
was sufficient to justify a dedicated effort to search for
a ``binary'' source. It is interesting to discuss what the
situation is with the data currently available.

The most accurate determination of the position in the sky of the
SN1998bw/GRB 980425 system is based on the radio observations made
with the Australian Telescope Compact Array (ATCA, Wieringa et al.
1998). Recall that in the CB model the radio coordinates are those
of the CB (GRB 980425 was a single-pulse GRB, that is, it had a
single dominant CB). In days 3, 4 and 10 the source is reported to
be at (RA 19:35:03.31, Dec -52:50:44.7).  In the subsequent 33
observations, ranging from day 12 to day 790, the  position is (RA
19:35:03.32, Dec -52:50:44.8), some $0.^{\!\!''}18$
away from the original determination,
but not inconsistent with the observational uncertainty of
$0.^{\!\!''}1$. In the penultimate observation at day
320 the source has faded to the point that it is not observable in
2 out of 6 frequencies, and in the last observation at date 790
there is no clear sighting at any frequency.  
The predicted values of $\Delta \alpha$
from Eq.~(\ref{supervelocity}) at some relevant dates are 12, 158,
183 and 292 mas at days 12,  249, 320 and 790, respectively. These
results, the observational error, and the fact that the ATCA
observers were not trying to follow the source's motion imply that
their results are insufficient to claim either that the early change
of position was significant, or that a motion of the CB comparable
to the predicted one is excluded.

Observations of the vicinity of the source of GRB 980425 were made
with the Hubble Space Telescope (HST) at day 778 (Fynbo et al.,
2000), with a tiny astrometric uncertainty of $0.^{\!\!''}018$, and
pointing at ATCA's first reported coordinates. The observations
are compatible with SN1998bw lying at that point, and reveal six
other objects in a ($1.^{\!\!''}0\times 1.^{\!\!''}0$) field centered
there. As a result of our CB model fit to the radio data, as we have
explained, we expect the optical observations to correspond to
SN1998bw, and there it is, at the field's center. We also expect, 
as in Fig.~(\ref{late425}), the CB to be more than two orders of
magnitude fainter:  not observable. It would be nice if this
conclusion was wrong, that is, if the large ``naive'' extrapolation
from radio to optical frequencies in Fig.~(\ref{late425}) was an
underestimate by a considerable factor, which is the case for the
larger extrapolation from optical to X-ray frequencies in
Fig.~(\ref{X425}) (the ``naive'' prediction there is the one labelled
$\rm p=2.2$). In that case, it may be that a subsequent observation
of the same field reveals that one of the closer-by extra sources
has faded away!  Three of these sources are $\sim 0.^{\!\!''}5$
away from the SN, if one of them is the CB, and it is dimming, we
would not excessively mind that this is $\sim 60$\% more distant
than the prediction in Fig.~(\ref{superluminal}), based on a
constant-density approximation for the ISM.

\section{The normalization of afterglows}

The values of the CB model parameters that were fitted to the broad band 
data on GRBs with known redshift ---and are listed in Table III--- are
narrowly distributed around their reference values, except 
for the overall normalization which is much smaller than originally 
anticipated if  $\eta=1$.  
This normalization ``problem'' may point to inaccuracies in the
various hypothesis that we have made. One example 
is the contention that {\it all} of the incident energy of the ISM electrons 
is radiated in the CB. It may well be 
that the moving CB deflects and scatters the ISM  electrons
before they radiate a large fraction of their acquired energy,
as suggested by the results of the numerical simulations of Achterberg et 
al. (2001).   
But abandoning some of our simplifications would be premature. Indeed,
in concluding that in the CB model the normalization of the AGs is more than
one order of magnitude too large, we have used our reference parameters
to compute the expected values. And there is sufficient elasticity
in these  parameters to obtain a consistent overall picture of all GRB 
properties, as we proceed to review and discuss.
 
\section{The parameters of the CB model}

With the current analysis of GRB radio AGs we have completed a 
first round of the study of GRB properties in the CB model, and it
 behooves us
to look back at the various constraints on the relevant parameters.

In Dar and De R\'ujula (2000a) we followed Dar (1998) and Dar and
Plaga (1999) in suggesting that the large peculiar velocities of
neutron stars may be due to a ``natal kick'', induced by a momentum
imbalance in the oppositely-directed jets of CBs accompanying
their birth. On this basis we chose as a reference value
 $\rm E_{CB}\sim 10^{52}$ erg, for a jet with a reference
number of CBs (or prominent peaks in the GRB light-curve)
$\rm n_{_{CB}}\sim 10$. Based on a first
analysis of AG properties, and of GRB $\gamma$-ray fluences and 
individual $\gamma$-ray energies, we set $\gamma_0=10^3$
as a reference value. 

In Dar and De R\'ujula (2000b) we investigated two extreme
models meant to bracket the behaviour of a CB as it
crosses a SN shell, is heated by the collision with
its constituents, and emits observable $\gamma$-rays as
it reaches the shell's transparent outskirts with a 
radius $\rm R_{_{CB}}^{tr}$, proportional to its early
transverse expansion velocity $\rm \beta_{trans}\,c$,
which we assumed to be close to the sound speed in
a relativistic plasma, $\rm\beta_{trans}=1/\sqrt{3}$. In our
``surface'' model, which is no doubt closer to a realistic
description, the energy of the GRB in $\gamma$-rays
is proportional to $\rm n_{_{CB}}\, [R_{CB}^{tr}]^2 \,\gamma_0$.
(Eq.~(45) of Dar and De R\'ujula 2000b).
For the chosen reference parameters, in the surface model,
this prediction overestimates the GRB fluences by about
one order of magnitude. Since the individual $\gamma$-ray
energies corroborate the choice $\gamma_0\sim 10^3$, this
means that $\rm [R_{_{CB}}^{tr}]^2$ (and  $\rm\beta_{trans}$)
are overestimated by roughly one order of 
magnitude\footnote{The reduction of thermal
free-electron surface-density discussed in Sections 6.2 and 6.3
is logically independent from the modifications discussed
here (such as a {\it decrease} in radius) which have to do
with the fluence produced by the small fraction of energetic
radiating electrons.}.

In Dar and De R\'ujula (2001) we analyzed the X-ray ``Fe'' lines
observed in the AGs of some GRBs, which we attributed to
hydrogen recombination in the CBs, with the corresponding 
Lyman-$\alpha$ lines 
boosted by a large factor $\delta/(1+z)\sim 500$. We equated
the total number of photons in the lines to the baryon number
of the jet of CBs, and found agreement with the baryon
number in the jet, $\rm n_{_{CB}}\,N_{CB}$, to within one
order of magnitude. But in the current investigation, 
we have found that the absorption of radio waves keeps the
CBs hot and ionized (Appendices III and V). This means
that our reference value of $\rm N_{CB}$ is likely
to be an overestimate.

In DDD 2001 we proposed a mechanism that would quench the
expansion of a CB in minutes of observer's time, well
after it has exited the SN shell. The CBs reach an asymptotic
radius (Eq.~(16) of DDD 2001):
\begin{equation}
\rm R_{max}^3\simeq 
{3\,N_{CB}\,\beta_{trans}^2\over 2\,\pi\,n_p\,\gamma_0^2}\; .
\label{Rmax}
\end{equation}
On the basis of this calculated radius (for $\rm\beta_{trans}=
1/(3\sqrt{3})$, we found that the normalization of optical
and X-ray AGs agreed with the reference-value expectations.
On the same basis, we find now that the normalization
is overestimated by an order of magnitude. The reason for
the discrepancy is that, in DDD 2001, we effectively placed
the spectral discontinuity at a ``cooling break'' frequency
corresponding to an electron Lorentz factor $\rm \gamma_e\sim 1$,
while we have now argued that the discontinuity should occur
at a higher value $\rm \gamma_e=\gamma(t)$.

Both the GRB fluence and the AG fluence are, in the CB-model,
$\rm F\propto n_{_{CB}}\,R^2$, with $\rm R=R_{_{CB}}^{tr}$ for the
$\gamma$ rays and $\rm R=R_{max}$ for the AG. At a value of
$\rm x_\infty$ fixed by the fit to the AG's temporal behaviour, the AG 
fluence is:
\begin{equation}
\rm
F_{AG}\propto 
n_{_{CB}}\,n_e\,R_{max}^2=
n_{_{CB}}\,N_{CB}\,{1\over \pi\, x_\infty}.
\label{FAG}
\end{equation}

All of the above ``problems'' are solved if we reduce our
``typical'' values of $\rm R_{max}^2$ and $\rm N_{CB}$
by about one order of magnitude, relative to our reference
parameters with, according to Eq.~(\ref{Rmax}), the corresponding 
reduction of the choice of $\rm\beta_{trans}$ by half an order of 
magnitude.

The precise location of the injection bend is
not predictable and a modification by up to
one order of magnitude of its position has
a small effect on the quality of the fits
to observations. An increase of the cooling
break frequency $\rm \nu_b$ implies a
corresponding decrease in AG flux, see Eq. (4),
adding to the uncertainty in the prediction
of the precise overall normalization.

To summarize, the CB model correctly describes, in terms of a very 
limited set of parameters, the properties of GRBs and their
AGs, including their normalizations. This is the case even if
we adhere to all of the detailed assumptions we have made,
even though they are approximations to a no doubt fairly
convoluted physical problem. 

\section{Hyperluminal CBs and radio scintillations} 
 
The radio AGs of GRBs often show temporal variations of a factor of 
two or more on a time scale of hours at early times and on a time scale of  
days at later times; e.g.
GRB 000926 in Figs.~(\ref{figr092602}, \ref{figr092603}), 
GRB 991216 in Fig.~(\ref{figr121602}),
GRB 991208  in Figs.~(\ref{figr120803}, \ref{figr120804}), 
GRB 000418 in Fig.~(\ref{figr041802}),
GRB 990123  in Fig.~(\ref{figr041801}) and
GRB 970508 in Figs.~(\ref{figr050801}, \ref{figr050802}). 

Similar variations have not been seen in the optical and X-ray bands. 
The intensity variations of GRB radio AGs are very reminiscent of 
the ones seen in radio signals from pulsars  in 
our galaxy, interpreted as scintillations due to the 
motion of the line of sight through the refractive, diffractive and   
dispersive ISM of the Galaxy (see, e.g., Lyne and Smith 1982).   
Some very compact  active galactic nuclei also show an intraday 
variability that has been the subject of much debate (e.g., Wagner and  
Witzel 1995 and references therein). At least in one case ---the   
variations in 
the radio intensity of the quasar J1819+3845, the most extremely variable  
AGN known at radio wavelengths (Dennett-Thorpe and de Bruyn 2000)--- it  
was shown unambiguously that the variations are scintillations caused by the  
ISM (Dennet-Thorpe and Bruyn 2002). 
 
The (de)coherence properties in time and frequency 
of the radio scintillations have been used to measure the 
transverse speed of pulsars (e.g., Lyne and Smith 1982). 
Gupta (1995) has demonstrated for a sample of 59 pulsars that their 
transverse speed, $\rm V_{iss}$, measured from their inter-stellar 
scintillations, agrees well with their transverse speed, $\rm V_{pm}$,  
measured from their proper motion (see also Nicastro et al. 2001). 
 
The movement of the line of sight to pulsars is in most cases dominated 
by their proper motion at a transverse velocity $\rm V_{pm}$ 
larger than the turbulent speeds in the ISM, or of the 
sun relative to the ISM, or of the Earth around the sun. 
The mean $\rm V_{pm}$ of Gupta's 59 pulsars is 311 $\rm km\, s^{-1}$ 
and their mean distance is estimated to be 1.96 kpc. 
Their angular speeds are within an order of magnitude of a central value:   
\begin{equation} 
\rm \dot \omega_{ps}\sim 
{\langle V_{pm}\rangle \over \langle D\rangle} 
\simeq 5.1\times 10^{-15}\, rad\, s^{-1}\, . 
\label{vpang} 
\end{equation} 
 
Travelling with a characteristic $\gamma\sim 10^3$ and viewed at typical 
angles $\theta$ of milliradians, CBs have apparent superluminal velocities,
$\rm V_{_{CB}}$ of Eq.~(\ref{supervelocity}), that are 
so high (a few hundred times the speed of light) that they deserve 
to be called {\it hyperluminal}. 
The angular speed in the sky is: 
\begin{equation} 
\rm 
\dot\omega_{_{CB}}(t)= 
{V_{_{CB}}(t)\over D_A}\simeq {\gamma(t)\, \delta(t)\, c\, \theta\over   
(1+z)\, D_A}\, , 
\label{vcbang} 
\end{equation} 
For the reference values $\theta\sim 1$ mrad 
and an initial $\gamma_0\sim\delta_0\sim 10^3$, 
the initial angular speed of a CB at redshift $\rm z=1$ is 
$\rm\dot\omega_{_{CB}}(0)\sim 2.7\times 10^{-15}$ rad s$^{-1}$, 
in the very same range as that of 
Galactic pulsars. 
The CBs' angular velocity $\rm \dot\omega_{_{CB}}(t)$ 
and the resulting (inverse) coherence time of the 
scintillations should decline as  $\rm\gamma(t)\,\delta(t)$ does. 
Both pulsars and CBs are pointlike from the point of view 
of their radio scintillations. Thus, all conditions are met to expect 
pulsar-like scintillations in the radio signals from CBs.

The deviations from a smooth behaviour of the radio signals in the case of
GRB 980425, as can be seen in Figs. (\ref{figr042501}, \ref{figr042502}),
are chromatic, but correlated in time over a much longer period than for
the other GRBs. Because GRB 980425 is so close ($\rm z=0.0085$) and is
viewed at the unusually large angle of $\sim 8$ mrad (Table III and DDD
2001), its apparent angular velocity, Eq.(\ref{vcbang}), is much larger
than for other GRBs. The line of sight to this GRB swept a much bigger
region of galactic ISM than for other GRBs or, for that matter, pulsars.
Thus, we have no independent information on the ISM irregularities causing
scintillations on this large scale. 

The analysis of CB scintillations could result in a measurement of their  
hyperluminal speeds and a decisive test of the cannonball model 
(fireballs do not have relativistic proper motions, firecones stop moving
close to their progenitors and, unlike CBs, have an increasing size that
should rapidly quench their scintillations with time). 
The presently available information (or the current information 
in its published form) is insufficient for us to attempt 
at the moment to extract conclusions about CB hyperluminal velocities 
from the observed scintillation patterns in the radio AG of GRBs.

\section{Conclusions} 

The Cannonball model gives an excellent and 
extremely simple description of all 
measured properties of GRB afterglows, 
including their radio afterglows.

In the CB model, there is an injection bend in the spectrum,
at the predicted time-dependent frequency $\rm \nu_b$ of Eq.~(\ref{nubend}). 
We have shown that the evidence for the correctness of this
prediction is very strong, see Table II and
Figs.~(\ref{index2}) and (\ref{index}).
It is this spectral bend that governs the relative normalization
of the radio and the optical AG, again in agreement with observation,
as shown in all our figures of AG wide-band spectra.

Since we have always set the electron index to its theoretical
value, $\rm p=2.2$, just {\it three intrinsic parameters}
are needed to describe an optical or X-ray AG: 
$\gamma_0$, $\rm x_\infty$ and the normalization; the viewing
angle $\theta$, although it must also be fit, is external to the GRB, like
the redshift and the absorption in the host and in the Galaxy are.
We have shown that, in the CB model, the extension of these results
to the radio domain
requires the introduction of {\it just one extra parameter:} the 
free-free absorption frequency $\rm \nu_a$ of Eq.~(\ref{absfreq}),
and that, in spite of various approximations,
this simplest of descriptions is at the moment entirely satisfactory. 
Notice that what one has to parametrize is a two
dimensional surface: the fluence as a function of frequency and
time. The shape of this surface is that of a relatively simple
``mountain'', various cuts of which at fixed $\rm t$ or $\nu$
are shown if Fig.~(\ref{figCBpreds}). It would be easy, and it may well
be misleading, to overparametrize this rather
featureless surface with more than a few parameters. 

It is instructive to compare, or so Occam would have 
thought, the understanding of wide-band AG spectra in the CB model 
with that in the fireball or firetrumpet models. 
In the latter, the number of intrinsic parameters varies: 
seven [e.g. Berger et al. 2001d], eight [e.g. Yost et al. 2001]
nine [e.g. Yost et al. 2002] and even thirteen [e.g. Galama et al. 2000].
This counting does not include the viewing angle, since
the firetrumpets in these works point precisely at the 
observer\footnote{This serious limitation (DDD 2001) is beginning to
be remedied in the firecone literature, in which the CB-model's
geometry is being ---though with no reference--- ``standardized''
and its consequences (Dar and De R\'ujula 2000a) explained again;
see, e.g.  Rossi, et al. (2001),  
Zhang and Meszaros (2001),
Salmonson and Galama (2001),
Granot J. et al., (2002), Panaitescu and Kumar (2001).}.
Moreover, even before the ``break'' in the time-evolution
---a period during which it is not inconsistent to use the
quasi-spherical self-similar  approximation of Blandford and McKee
(1976) for the expanding material---  the ordering of the ``breaks''
in frequency implies a multiple choice of spectral shapes
and of their evolution (Granot and Sari 2002).

Countrary to established custom, we are not presenting the $\chi^2$
values of our fits, which are generally reasonable and would become
quite good if, again following the consuetudinary path, we artificially
increased the errors to compensate for scintillations in the radio
data and/or uncertainties in attenuation.  The reason is that the
CB model is a very simplified description of a no doubt very
complicated reality (e.g. CBs could be somewhat comet-like, as
opposed to spherical, their inner distributions of density,
ionization, magnetic field and temperature could be non-trivial,
even chaotic, etc, etc).  Even when the physics is much simpler
than in the analysis of radio emissions, and the fits are very good
---as is the case in our description of optical and X-ray AGs in
DDD 2001--- we do not report their quite impressive $\chi^2$
values\footnote{A ``$\chi$-by-eye'' of the figures reporting here
the optical AG light curves should suffice to prove this
statement.}.  We view our ``fits'' as rough descriptions,
rather than true fits.  Under such circumstances, the overintrepretation
of a $\chi^2$ test has every chance of being  misleading, much more
so in models containing many more parameters than the CB model.

For the same reasons, 
and because of the systematic errors in the data, 
the values of the parameters we extract from our fits should not 
be taken entirely at face value, even though the minimization procedure 
---which attributes to the errors a counterfactual purely statistical origin--- 
results in tiny 1 $\sigma$ spreads for the fitted parameters, and in 
$\chi^2 $ values that are in most cases  satisfactory.

In the radio domain, as in every other aspect, the pair SN1998bw/GRB
980425 is particularly fascinating. On the basis of this GRB's
observed fluence and distance, and given the (totally trivial but
all important) dependence of the fluence on observation angle, we
claimed in Dar and De R\'ujula (2000a) that the only peculiarity
of this pair was that it was observed uncharacteristically far from
its axis (for a GRB) and uncharacteristically close to it (for a
SN). In DDD 2001 we proved that the X-ray AG of GRB
980245 was also what is expected in the CB model, depriving the
supernova of its X-ray peculiarity: it did not make the observed
X-rays.  In this paper, by understanding the magnitude, time- and
frequency-dependence of the pair's radio signals ---which were not
emitted, either, by the SN--- we have demonstrated that SN1998bw
was also ``radio normal''. Neither this GRB,  nor its SN have ---in
the CB model--- anything in particular, except the chance occurrences
of the distance and observation angle.  Alas, the unique occasion
to make a fundamental discovery by actually resolving the SN and
the CB, as proposed in Dar and De R\'ujula (2000a), may now be
very difficult, but, as we have explained, not entirely out of the 
question.

By pure coincidence, the apparent angular velocities of galactic
pulsars and cosmological cannonballs are of the same order of
magnitude.  The analysis of radio scintillations, one of the methods
used to measure pulsars velocities, should also be applicable to
the GRB ejecta.  Thus, it ought to be possible to test the CB-model's
prediction of  hyperluminal cannonball velocities.

\section*{Appendix I: The slowdown of a CB} 

We review the functional form of 
the time dependent Lorentz factor $\rm \gamma(t)$, which
is explicit and analytical in a fair approximation (DDD 2001).

In minutes of observer's time, CBs reach a roughly constant radius 
$\rm R_{max}$ and
are parsecs away from their progenitor star,
a domain where a constant-density ISM may be a reasonable approximation.
Relativistic energy-momentum conservation in the  progenitor's      
rest frame results in the equation governing the deceleration
of a CB in the ISM:
\begin{equation}   
\rm  M_{_{CB}}\, d\gamma =-\, m_p\,c^2\, n_p\, \pi\, R_{max}^2\,
                           \gamma^2\, c\, dt_{SN}\, .
\label{conservation}
\end{equation}
Interestingly, the above expression is correct both if the incoming
ISM protons are isotropically reemitted in the CB rest frame,
or if they are ingurgitated by the CB (in the first case, they are
reemitted with average energy $\rm m_p\,c^2\,\gamma^2$ in the
progenitor's frame, in the second, the change in $\gamma$ per
added proton is $\rm d\gamma=-[m_p/M_{_{CB}}]\gamma$).

Use the relation $\rm dt_{SN}=\gamma(t)\, dt$ between the
times  measured in the supernova and CB rest frames,
divide both sides of the Eq.~(\ref{conservation}) by 
$\rm M_{_{CB}}\, \gamma^3 $ and 
integrate to obtain the relation:
\begin{equation}
\rm {1\over \gamma^2(t)} - {1\over \gamma_0^2}\simeq {2\, c\, t\over
x_\infty}\, ,
\label{gammacb}
\end{equation} 
where $\rm t$ is CB time, and:
\begin{equation} 
\rm 
x_\infty\equiv{N_{CB}\over\pi\, R_{max}^2\, n_p}  \, , 
\label{range} 
\end{equation} 
with $\rm N_{CB}\approx M_{_{CB}}/m_p$ the CB's baryon number. 

It is important to know the number of electrons  accumulated
by a CB as its Lorentz factor decreases from $\gamma_0$ to
$\rm\gamma(t)$ (in the
approximation $\rm n_e=n_p$ of a Hydrogenic ISM
this number equals that of scattered or incorporated protons). 
The number rate of accumulation is related to the energy-loss rate of
Eq.~(\ref{conservation}) so that:
\begin{equation}
\rm dN= - \eta\, {M_{_{CB}}\, d\gamma\over m_p\,c^2\,\gamma^2}=
-\eta\, N_{_{CB}}\,{d\gamma\over \gamma^2}\, .
\label{accrate}
\end{equation}
Assuming constant $\eta$, the total number of ISM electrons 
accumulated at a CB time $\rm t$  is then:
\begin{equation}\rm
N(t)=\eta\, N_{_{CB}}\,\left[{1\over \gamma(t)}-{1\over 
\gamma_0}\right]\, . 
\label{acctot}
\end{equation}
 
The time-dependence of $\rm \gamma(t)$ with $\rm t$ the observer's
time is more complicated than Eq.~(\ref{gammacb}): 
 the relation between the two times
($\rm dt_{obs}=dt_{_{CB}}(1+z)/\delta$) introduces a $\gamma$ (or
$\rm t$) dependence via $\delta$. The result for $\rm\gamma(t_{obs})$
is (DDD 2001): 
\begin{eqnarray} 
\rm \gamma&=&\rm\gamma(\gamma_0,\theta,x_\infty;t) 
=\rm {1\over B} \,\left[\theta^2+C\,\theta^4+{1\over C}\right]\nonumber\\   
\rm C&\equiv&\rm 
\left[{2\over B^2+2\,\theta^6+B\,\sqrt{B^2+4\,\theta^6}}\right]^{1/3}   
\nonumber\\ 
\rm B&\equiv&\rm 
{1\over \gamma_0^3}+{3\,\theta^2\over\gamma_0}+ 
{6\,c\, t\over  (1+z)\, x_\infty}\, . 
\label{cubic} 
\end{eqnarray} 
The Lorentz factor of the CB decreases from $\gamma_0$ 
to $\gamma_0/2$ as the CB travels a distance 
$\rm x_\infty/\gamma_0$, whose
reference value is 1.3 kpc, as in Table I. 

\section*{Appendix II: The synchrotron cooling break}

We argue that the conventional synchrotron spectral break occurs at a very
non-relativistic electron energy. The corresponding break in the
radio spectrum is unobservable.

In writing the electron energy-loss rate Eq.~(\ref{Eloss}), we have
assumed that synchrotron radiation, which is quadratic in energy,
dominates (inverse Compton scattering has the same energy dependence,
but it is negligible, since
the magnetic energy density within a CB is much higher than 
the radiation energy density). The general result for the energy loss of
high-energy electrons is of the form:
\begin{equation}
\rm -{dE\over dt}\simeq A_C\,(ln {E\over m_e\,c^2}+a)+A_B\,E+A_S\, E^2\, .
\label{b1}
\end{equation}
The term proportional to $\rm A_C$ describes Coulomb
scattering and ionization losses in the CB, which are negligible at high
energies. The second term represents bremsstrahlung and its
coefficient is (e.g. Shu 1991 )
\begin{equation}
\rm A_B={34.35\,\alpha\, \sigma_T\, c\,\bar{n}_b \over 2\, \pi}\, ,
\label{A31}
\end{equation}
where $\alpha=1/137$ and $\rm \bar n_b$ is the baryon number density in the CB.
Adiabatic losses and electron escape would have the same energy dependence
as bremsstrahlung, but in the AG regime we are discussing the CBs are no longer
expanding and the electron's Larmor radii are so small ---relative to the CB's
radius--- that escape losses should also be negligible.

The spectral index of high energy electrons injected with a power-law 
spectrum steepens by one unit at a ``cooling break'' energy 
$\rm [\beta(E_c)]^2\,E_c=A_B/A_S$.
For  $\gamma \simeq 10^3$ and the reference values of
$\rm \bar n_b=\bar n_e$ and $\rm n_e$,
the synchrotron cooling break is at a subrelativistic energy
($\beta\sim 0.8$). This is in
contrast with the injection bend at the highly relativistic energy
$\rm E_b\simeq \gamma\, m_e\, c^2\,.$
The synchrotron radiation of electrons below the cooling break
is, for the current data, at unobservably low observer's radio frequencies.

\section*{Appendix III: Photoionization and recombination in the CB}

We argue that the synchrotron radiation in a CB is intense 
enough to maintain its plasma partially ionized with ion and 
free electron densities proportional to $\rm \gamma(t)$.

The bound-free cross section for photoionization of atomic hydrogen in its
$\rm n$-th excited state by photons with frequency above the ionization 
threshold,
$\rm \nu_n=3.29\times 10^{15}/n^2\, Hz$, is given by 
$\rm \sigma_\nu(n)=n\,\sigma_1\,\bar{g}_n (\nu/\nu_n)^{-3}$, with
$\rm \sigma_1= {64\, \alpha\, \pi\, a_0^2/(3\, \sqrt{3}})$ $\simeq
7.91\times 10^{-18}$ cm$^2$ ($\rm a_0=0.53\times 10^{-8}$ cm 
is the Bohr radius
and $\rm \bar{g}_n$ is the Gaunt factor 
for photoabsorption by hydrogen).
For the surface flux of photons of Eq.~(\ref{Fnucb}), we obtain 
an ionization rate of the $\rm n$-th level of atomic hydrogen:
\begin{equation} 
\rm R_i(n)=  {\eta\, n_e\, m_e\, c^3 \gamma(t)^2\, n\,  \sigma_1\, 
             (p-2)\over 8\, (p+6)\,(p-1)\, h\, \nu_n}\, 
              \left({\nu_b\over\nu_n}\right)^{(p-2)/ 2}\, .    
\label{ionizationrate1}
\end{equation} 
For the reference values of $\rm n_e$, $\rm p$ and $\gamma=\gamma_0$,
the ionization rate of  
the ground state is $\rm R_i(1)\sim 1.1\times 10^{-2}\,\eta\, s^{-1}$.

The recombination rate per unit volume
of hydrogen in an hydrogenic CB is (Osterbrock
1989):  
\begin{equation} 
\rm R_{rec}\simeq 1.0\times 10^{-12}\,\bar n_e\,
                  \left[{T\over 10^4\,K}\right]^{-0.7}\,s^{-1}\, . 
\label{Recombination}
\end{equation} 
For the reference value of the CB parameters, and T=10,000
K the recombination rate of hydrogen is $\rm R_{rec}\sim 1.1\times
10^{-5}\, s^{-1}$. In quasi equilibrium, the ionization rate per unit
volume, which is proportional to the number density of recombined hydrogen
atoms, must be equal to the recombination rate per unit volume, which is
proportional to the product of the ion and free electron densities in the
CB. Thus, initially the CB is highly ionized.  But, for small values of
$\eta$ and later times when $\rm \gamma(t)$ becomes sufficiently small,
equilibrium between the ionization and recombination rates in the partially
ionized hydrogenic plasma results in $\rm \bar n_i=\bar n_e\propto
\gamma(t)\, .$

\section*{Appendix IV: Synchrotron and Bremsstrahlung Self Absorption}

We argue that the self-attenuation in the CB of the observed radio waves is 
dominated by free-free absorption.

\subsection*{The density of energetic electrons }

In the rest system of a CB, the ISM electrons arrive at a rate
$\rm d\,N_e/dt\simeq n_p\,\gamma\,  \pi\, R_{max}^2\, c\, .$
A fraction $\eta$ of their energy, $\rm E_b=m_e\, c^2\, \gamma \,,$ is 
synchrotron
re-radiated. Let $\rm n_{eff}$ be the density of the emitting
electrons, so that:
\begin{equation}
\rm \eta\, E_b\, {dN_e\over dt}=\int F_{_{CB}}[\nu]\,d\nu
=\int dV\,dE_e\,E_e\,{dn_{eff}\over dE_e}\,{1\over \tau(E_e)}\; ,
\label{neffV}
\end{equation}
with $\rm \tau(E_e)=E_e/(dE_e/dt)$ the  cooling time for electrons
of energy $\rm E_e$,  determined from Eqs.~(\ref{mag}) and
(\ref{Eloss}) to be:
\begin{equation}
\rm
\rm \tau(E_e)=4\,{m_e\over m_p}\,{1\over n_p\,\sigma_{_T}\,c\,\gamma_e^3}\, .
\label{cooltimes}
\end{equation}
For the electron spectrum of Eq.~(\ref{approx1}), the second integral 
in Eq.~(\ref{neffV}) is dominated by energies $\rm E_e\sim E_b$
($\rm \gamma_e\sim \gamma_b=\gamma$). Thus, for a uniform distribution
of electrons in the CB's volume $\rm V$, we obtain:
\begin{eqnarray}
\rm
n_{eff}&\simeq&\rm -\,{\eta\over V}\,{dN_e\over dt}\,\tau(E_b)
\nonumber\\ 
&=&\rm{m_e\over m_p}\,{3\,\eta\over R_{max}\,\sigma_{_T}\,\gamma^2}
\simeq (11\,\eta\, cm^{-3})\left[10^3\over\gamma\right]^2 ,
\label{neff}
\end{eqnarray}
where we have used our reference $\rm R_{max}$.
This density  could be somewhat 
higher if the emitting electrons are concentrated on the CB's front surface.
Note that this density increases with time like $\rm[\gamma(t)]^{-2}$. 

\subsection*{Synchrotron self absorption}
For a power-law distribution, 
$\rm dn_e/d\gamma_e= (p-1)\,n_{eff}\, \gamma_e^{-p}$,  
the correct attenuation coefficient for synchrotron self absorption 
at frequency $\nu$ is
(e.g., Shu 1991, Eqs.~ 19.37,38):
\begin{equation}
\rm \chi_\nu= K_0\, (p-1)\,n_{eff}\,  {B_\perp}^{(p+2)/2}\, \nu^{-(p+4)/2}\, ,
\label{chisyn}
\end{equation}
where: 
\begin{equation}
\rm K_0={c\, r_e\over 4\, \sqrt{3}}\, 
\left ({3\, e\over 2\, \pi\, m_e\, c }\right)^{p+2\over 2}
       \Gamma\left({3\, p+2\over 12}\right) 
       \Gamma\left({3p+22\over 12}\right),
\label{K0}
\end{equation}
with $\rm r_e\, =\, e^2/m_e\, c^2\, .$ 
An  observed frequency $\rm \nu_{obs}$ was emitted at 
$\rm \nu(t)=(1+z)\, \nu_{ob}/\delta(t)$  in 
the CB's rest frame.  As an example, $\rm \nu_{ob}=5$ GHz
from a decelerating CB with
$\rm\delta(t)\simeq \gamma_0/2 \simeq  500$,  if emitted at a typical
$\rm z=1$, corresponds to $\rm \nu=20$ MHz in the CB rest frame. 
For our reference parameters, the synchrotron 
self absorption coefficient in the CB is $\rm\chi_\nu \simeq 2.4\times 
10^{-10}\,\eta\, cm^{-1}$.  Since $\rm n_{eff}\propto \gamma^{-2}$, 
$\rm B\propto \gamma$, 
and, after a few observer's days, $\delta\sim 2\gamma$ and
$\rm\rm\gamma(t)\sim t^{-1/3}$, $\chi_\nu$ decreases 
with time like $\rm \gamma^{(p+1)}\sim t^{-(p+1)/3}\sim t^{-1.1}$.

\subsection*{Bremsstrahlung self absorption} 
The  X-ray AG is dominated first by bremsstrahlung from plasma
electrons and later by synchrotron radiation from the swept up high
energy electrons (DDD 2001). The observed X-ray flux, or the
theoretical UV flux in the CB rest frame, can be used to show that
the CB is partially ionized during the radio AG observations and
that the ionized fraction of the CB plasma is proportional to $\rm
\gamma(t)$ (see Appendix III).  The logarithmic dependence of the
plasma temperature in the Saha equation on the fractional ionization,
keeps the CB's temperature nearly constant during its  AG phase.
Consequently, in the CB rest frame, the free-free attenuation at
a fixed frequency is proportional to $\rm [\gamma(t)]^2$ and the
free-free (bremsstrahlung) absorption coefficient is that of
Eq.~(\ref{chifree}).

The temperature of the partially ionized CB is of ${\cal{O}}(1)$
eV and almost constant during the observed AG.  For $\sim$ 20 MHz
emission from a thermal plasma at such temperature, $\rm g\sim 10$
and for one tenth of the typical bulk CB density, $\rm \bar n_e\sim
10^6\ cm^{-3}$, one obtains from Eq.~(\ref{chifree})
 $\rm\chi_\nu\simeq 4\times 10^{-10}\, cm^{-1}$, which is $\sim
1.7/\eta$ larger than the synchrotron absorption coefficient of
the energetic electrons in the CB (the values of $\eta$ are listed
in Table III).  At a fixed observer frequency, $\rm \nu =(1+z)\,
\nu_{ob}/\delta$, the free-free opacity of the CB decreases roughly
like $\rm \sim\gamma^2\sim t^{-2/3}$ compared with the $\rm \sim
t^{-1.1}$ decline of the synchrotron self opacity.

The conclusion is that free-free absorption is dominant for
as long as the ionization of the CB is considerable.

\section*{Appendix V: Emission of self absorbed radiations}

We argue that the energy of the self absorbed radio waves and 
ionizing photons in the CB is radiated mainly by thermal bremsstrahlung
and line emission from the CB contributing significantly to the 
observed X-ray afterglow.
 
The absorbed radio power is roughly equal to the integrated
emissivity of the CB over all frequencies below $\rm \nu_a$,
defined by Eq.~(\ref{absfreq}).  For the spectrum of Eq.~(\ref{sync}), 
and the normalization of Eq.~(\ref{cbemissivity}), the
absorbed  power is:
\begin{equation}
\rm {dE_{ab}\over dt}\simeq {\eta\, \pi\, R_{max}^2\, n_e\, m_e\, c^3\, 
                       [\gamma(t)]^2\, (p-2)\over (p-1)}\, 
                        \left[{\nu_a\over \nu_b}\right]^{1/2}\, .
\end{equation} 
The CB self absorbed radio energy becomes part of the thermal 
energy of the CB plasma. It is radiated by the plasma
as thermal bremsstrahlung at optical wavelengths (in the CB rest frame).
For our reference parameters, $\rm \nu_a\sim$ 100 MHz and  
$\eta<1$, this absorbed power is smaller than the power
absorbed by photoionization.

The recombination energy is radiated  at a rate 
$\rm \approx R_{rec}\,  x\, N_{cb} \, I $ in the CB rest frame
where $\rm x=\bar n_e/\bar n_b $ is the fraction of ionized hydrogen in 
the CB and $\rm I=13.6\, eV$ is the binding energy of hydrogen in 
its ground state. In the distant observer frame, the observed  radiation
is boosted and collimated by the highly relativistic motion of the CB  
and redshifted by the cosmological expansion to:
\begin{equation} 
\rm {dE_{rec}\over dt} 
\simeq {R_{rec}\, x\, N_{cb}\, I\, (1+z)\, [\delta(t)]^4\over 4\, \pi\, 
D_L^2}\, . 
\label{Erecombination} 
\end{equation} 
For our reference parameters, hydrogen recombination 
produces X-ray 
lines with a total energy flux of $\rm \sim x^2\, \times 10^{-12}\, erg\, 
s^{-1}\,cm^{-2}\, .$ 

Due to their large Doppler shift $\delta$, the 
hydrogen emission lines (and the emission lines from the swept up ISM and
supernova shell material) as well as the CB's  thermal bremsstrahlung, 
are shifted to the observer's X-ray band. They
contribute significantly to the X-ray afterglow and may provide a simple
alternative explanation (Dar and De R\'ujula 2000) to the commonly assumed
Fe-line origin of the X-ray lines observed in the afterglows of GRB
970508: Piro et al. (1998), GRB 970828: Yoshida et al. (1999; 2001), GRB
991216: Piro et al. (2000) and GRB 000214: Antonelli et al. (2000).

\clearpage 
\newpage 

{ \vskip 0.3 true cm
\noindent 
{\bf Table I - Reference parameters} }
\vskip -0.5 true cm 
\begin{table}[h] 
%\vskip 0.1 true cm 
%\huge\bf
\normalsize
\hspace{.0cm} %if you want to center your table act on this argument 
\begin{tabular}{|l|c|c|c|} 
\hline 
%\multicolumn{6}{GRB with redshifts}\\ 
\hline 
Fitted & Value & Definition \\ 
\hline 
$\theta$             & $10^{-3}$          & Observer's viewing angle   \\ 
$\gamma_0$     & $10^3 $             & Lorentz factor at $\rm t=0$  \\ 
$\rm x_\infty$   & 1.3  Mpc            & Deceleration parameter \\ 
%\hline
%$\rm p $            & 2.2                     & Uncooled electron index \\
%\hline
\hline
\hline
Other &  &  \\
\hline
$\delta_0$               & $10^3$              & Doppler factor at $\rm t=0$ \\
$\rm x_\infty/\gamma_0$   & 1.3 kpc & Distance $\rm \gamma_0\to \gamma_0/2$\\ 
$\rm R_{max}$       & $2.2\times10^{14}$  cm         & CB's maximum radius \\ 
$\rm \bar{n}_e$       &$\rm10^{7}\,cm^{-3}$    & CB e number-density \\
$\rm N_{_{CB}}$     & $6\times10^{50}$            & CB's baryon number   \\
\hline 
\hline 
Ambient &  &  \\
\hline
$\rm n_p $      & $\rm 10^{-3}\,cm^{-3}$        & Distant p number-density \\
$\rm n_p^{SN} $      & $\rm1 \,cm^{-3}$       & Close-by p number-density \\
\hline
\hline
\end{tabular} 
\end{table} 
\vskip -0.3 true cm 
\noindent 
{\bf Comments:} The ``Fitted'' parameters are the typical values 
in the fits to optical and X-ray AGs. 
%The spectral index $\rm p$ could be input, rather than fit.
The ``Other'' parameters
are deduced from the fitted ones ($\delta_0$), are calculated 
($\rm R_{max}$),  or are deduced from the rest
($\rm N_{_{CB}}$  and $\rm \bar{n}_e$). ``Ambient''
numbers refer to the ISM, not the CBs.

%\pagebreak 
\vspace{.5  cm} 

\noindent
{\bf
Table II -  The crossing of the bend frequency through the U to I bands
[($\sim$ 10 to 3)  $\times 10^{15}$ Hz]}
%\vskip 0.2 true cm
\begin{table}[h]
%\vskip 0.1 true cm
\hspace{-.1cm} %if you want to center your table act on this argument
\begin{tabular}{|l|c|c|c|c|c|c|l|}
\hline
\hline
GRB   &$\rm\nu_b^0$  & $\rm t_1$ & $\rm\beta(t_1)$ & $\rm t_2$ & $\rm\beta(t_2)$    
\\
\hline
\hline
970508& $3.7$ & 0.1-1.5 & $-0.58\!\pm\! 0.40$ & 12.1 & $-1.12\!\pm\!0.04$ \\
000301c&$5.8$ &1.8 & $-0.90\!\pm\! 0.20$ & 6-8 &$-1.19\!\pm\! 0.15$ \\
000926& $7.3$ &0.9 & $-0.90\!\pm\! 0.18$ & 3.9 &$-1.00\!\pm\! 0.18$ \\
990712& $13$ &0.5-1 & $-0.70\!\pm\! 0.10$ &    &$                 $ \\
991208& $17$ &      & $                 $ & 3.8 &$-1.05\!\pm\! 0.05$ \\
010222& $18$ & 0.20 & $-0.88\!\pm\! 0.10$ & 1-5 &$-1.10\!\pm\! 0.10$ \\
991216& $20$ & 1.67 & $-0.58\!\pm\! 0.08$ &     &$                 $ \\
990510& $27$ & 0.89 & $-0.61\!\pm\! 0.12$ & 3.6 &$-1.29\!\pm\! 0.23$ \\
            &           &          &                                     &        &$-1.11\!\pm\! 0.12$ \\
990123& $45$ & 0.033& $-0.69\!\pm\! 0.10$ & 1-3 &$-0.90\!\pm\! 0.18$ \\
\hline
\end{tabular}
\end{table}
\vskip -0.3 true cm
\noindent
{\bf Comments:} $\rm\nu_b^0$: bend frequency in units of $10^{14}$ Hz.
$\beta$: spectral index.
$\rm t_i$: times after burst in days. For the second entry on
GRB 990510, see the text.\\  
    {\bf References}:\\
{   GRB 970508}: Galama et al.~1998a\\   
{   GRB 990123}: Andersen et al. 1999; Holland et al.~2000\\
{   GRB 990510}: Stanek et al.~1999; Holland et al.~2000; 
Beuermann et al. 1999 \\
{   GRB 990712}: Sahu, et al.~2000\\
{   GRB 991208}: Castro-Tirado et al.~2001\\
{   GRB 991216}: Garnavich et al.~2000a\\
{   GRB 000301c}: Jensen et al. 2001; Rhoads \& Fruchter~2000 \\
{   GRB 000926}: Fynbo et al. 2001; Harrison et al 2001\\
{   GRB 010222}: Stanek et al. 2001; Masetti et al.~2001 \\

\newpage

\noindent
{\bf
Table III -  The Afterglow  Parameters}
%\vskip 0.2 true cm
\begin{table}[h]
%\vskip 0.1 true cm
\hspace{+1.1cm} %if you want to center your table act on this argument
\begin{tabular}{|l|c|c|c|c|c|c|l|}
\hline
\hline
GRB   &$\gamma_0 $ &  $\theta $ &$\rm x_\infty $& $\rm \nu_a$& $\eta$ \\
\hline
000301c       &1061      &2.321    & 0.128    & 552   &  0.025  \\
000926        & 787      &0.235    & 0.083    & 722   &  0.027  \\
991216        & 906      &0.403    & 0.462    &  46   &  0.029  \\
991208        &1034      &0.111    & 1.014    & 103   &  0.011  \\
000418        &1241      &2.061    & 0.332    & 298   &  0.024  \\
990123        &1208      &0.464    & 0.364    &1604   &  0.009  \\
990510        &1009      &0.261    & 0.372    & 107   &  0.015  \\
970508        & 769      & 2.51    & 0.516    & 559   &  0.035  \\
980425        & 495      &7.831    & 0.425    & 102   &  0.007  \\  
\hline
\end{tabular}
\end{table}
\vskip -0.3 true cm
\noindent
{\bf Comments:}
$\gamma_0$: Initial Lorentz factor.
$\theta$: Viewing angle relative to the CB line of motion,
in milliradians.
$\rm x_\infty$: Deceleration parameter in Mpc
($\gamma=\gamma_0/2$ at $\rm x= x_\infty/\gamma_0$).
$\rm \nu_a $: absorption frequency in MHz in the CB rest frame at t=0.
$\eta$: Our best fit normalization divided by the expected normalization  
for the reference parameters that we had chosen in previous works.
 
\vspace{1cm}
{\bf 
\noindent 
Table IV - Frequencies, in GHz, at which the radio AGs  of GRBs  
of known redshift were measured }
\vspace{-.3cm} 
\begin{table}[h] 
\hspace{1cm}  
\begin{tabular}{|l|c|c|c|c|c|l|} 
\hline 
\hline 
 991208 &  991216 & 000301c & 000418 & 000926 \\ 
\hline 
        & 350     & 350    &        &       \\  
        &         & 250    &        &       \\ 
  100   &  100    & 100    &        & 98.48 \\ 
  86.24 &         &        &        &       \\  
  30    &         &        &        &       \\ 
  22.5  &         & 22.5   & 22.46  & 22.5  \\ 
  14.97 & 15      & 15     & 15     & 15    \\ 
  8.46  & 8.46    & 8.46   & 8.46   & 8.46  \\  
  4.86  & 4.86    & 4.86   & 4.86   & 4.86  \\ 
  1.43  & 1.43    & 1.43   &        & 1.43  \\ 
\hline 
 970508 & 980425 & 990123 & 990510 &   \\          
\hline 
        &        &  351   &       &    \\    
        &        &  222   &       &    \\  
        &        &  15    &  13.68&    \\  
 8.46   & 8.64   &  8.46  &   8.65&    \\    
 4.86   & 4.8    &  4.88  &   4.8 &    \\ 
 1.43   & 2.49   &        &       &    \\  
        & 1.38   &  1.38  &       &    \\ 
\hline 
\hline 
\end{tabular} 
\end{table} 
\vskip -0.3 true cm 
\noindent 
{\bf References}:\\ 
GRB 970508: Frail et al. 2000a\\ 
GRB 980425: Kulkarni et al. 1998;\\ 
GRB 990123: Kulkarni et al. 1999b; Galama et al. 1999\\ 
GRB 990510: Harrison et al. 1999\\  
GRB 991208: Galama et al. 2000 \\ 
GRB 991216: Frail et al. 2000b \\  
GRB 000301c: Berger et al 2000\\   
GRB 000418: Berger  et al. 2001a\\ 
GRB 000926: Harrison et al. 2001\\

\clearpage

\clearpage

%\clearpage

\clearpage

\clearpage
 
%\clearpage

\clearpage

%\clearpage

%desde 

\clearpage

\clearpage 

%\clearpage

\clearpage

%\clearpage

\clearpage

\clearpage

\clearpage

\clearpage

%end216

\clearpage

%\clearpage

\clearpage

%\clearpage
 
\clearpage
 
%\clearpage
 
\clearpage

%418starts

\clearpage
 
%418ends

\clearpage

%510starts

\clearpage
 
%510ends

%123starts

%\clearpage
 
%123ends

\clearpage 

\clearpage
 
%\clearpage

\clearpage

%\clearpage

\begin{figure}[t] 
\begin{tabular}{cc} 
\hskip 2truecm 
\vspace*{2cm} 
\hspace*{-1.7cm} 
\epsfig{file=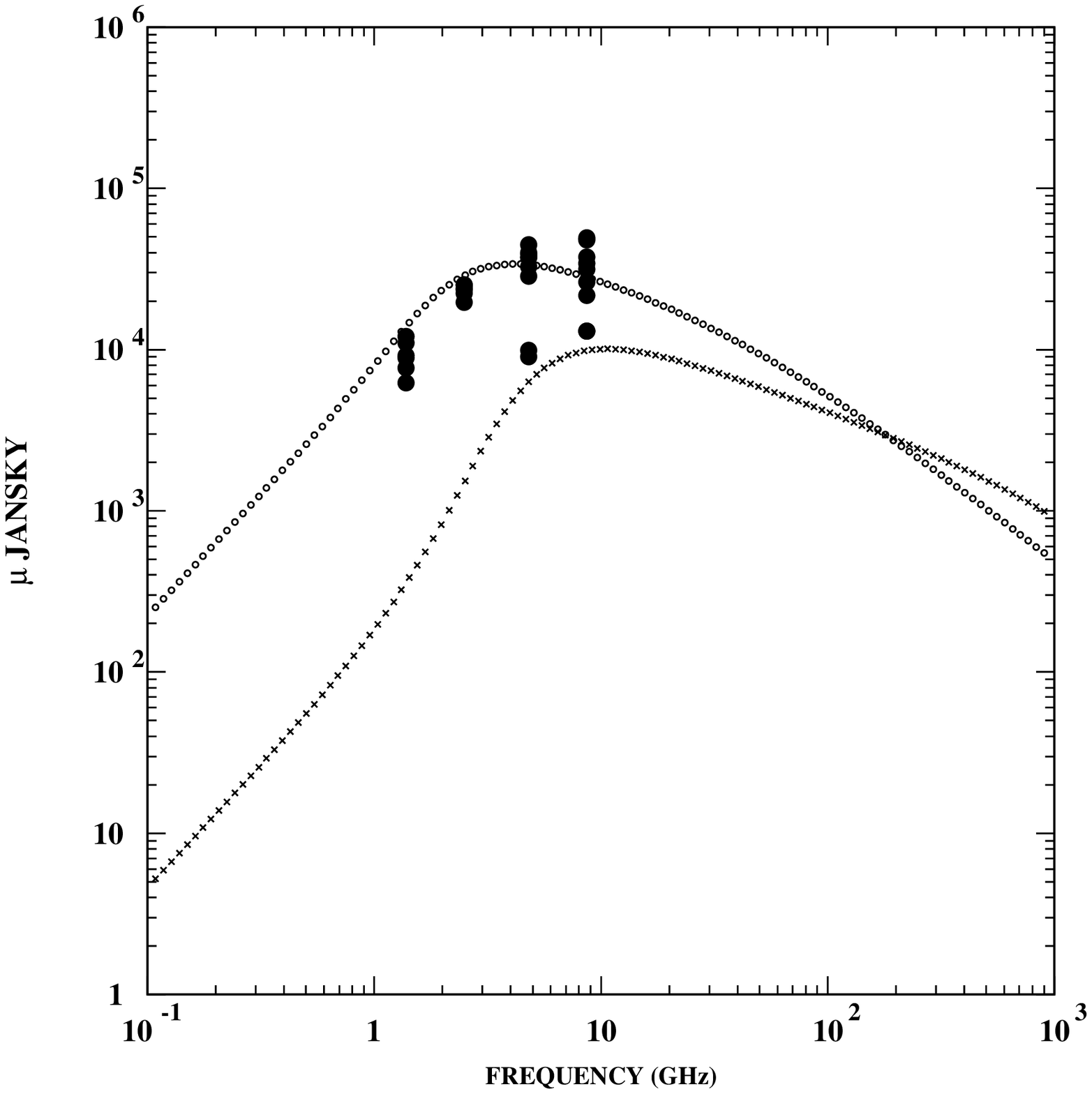, width=8cm} \\ 
%\hskip 1truecm 
\hspace*{.5cm} 
\epsfig{file=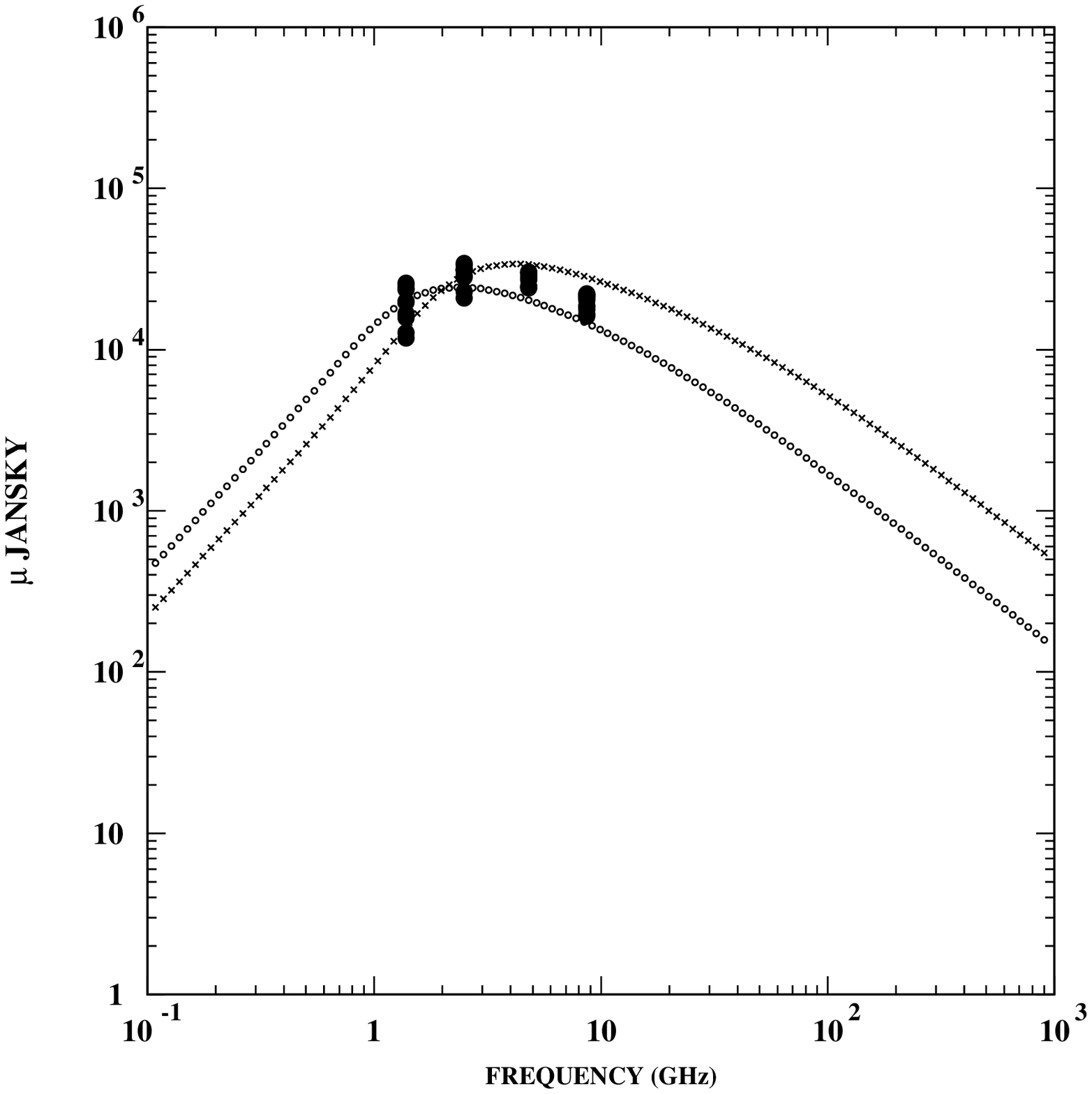, width=8cm} 
\end{tabular} 
\caption{Comparison between the observed spectrum of the AG of GRB 980425 in 
the radio band and the fitted CB model spectrum. 
Upper panel: in the time interval  between 2.5 and 20 days after burst.
Lower panel: in the time interval between 20 and 40 days after burst. 
The highest peaking curve in the upper pannel corresponds to the later 
time and in the lower panel to the earlier time.}
\label{figr042503} 
\end{figure}

\begin{figure}[t] 
\begin{tabular}{cc} 
\hskip 2truecm 
\vspace*{2cm} 
\hspace*{-1.7cm} 
\epsfig{file=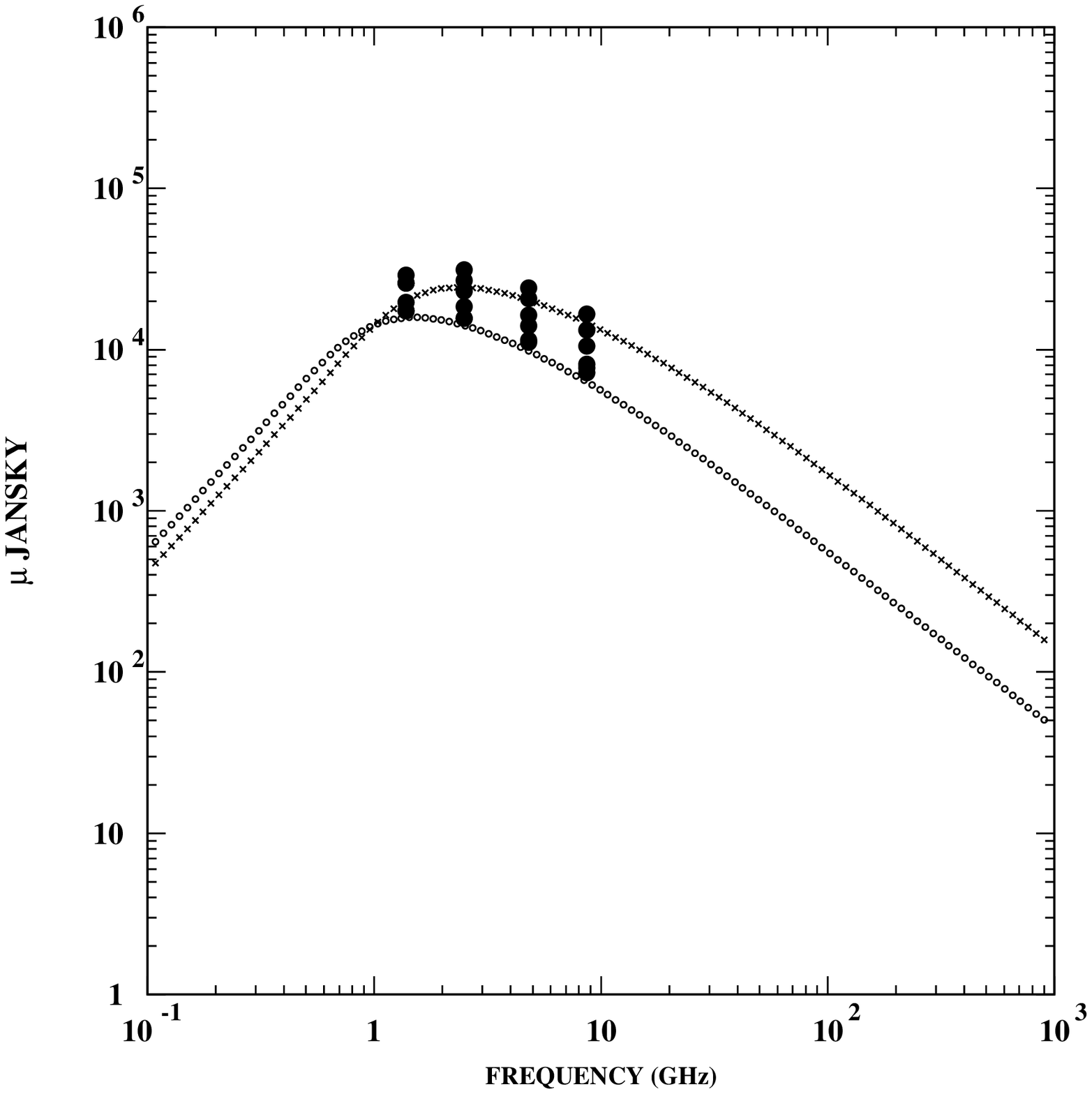, width=8cm} \\ 
%\hskip 1truecm 
\hspace*{.5cm} 
\epsfig{file=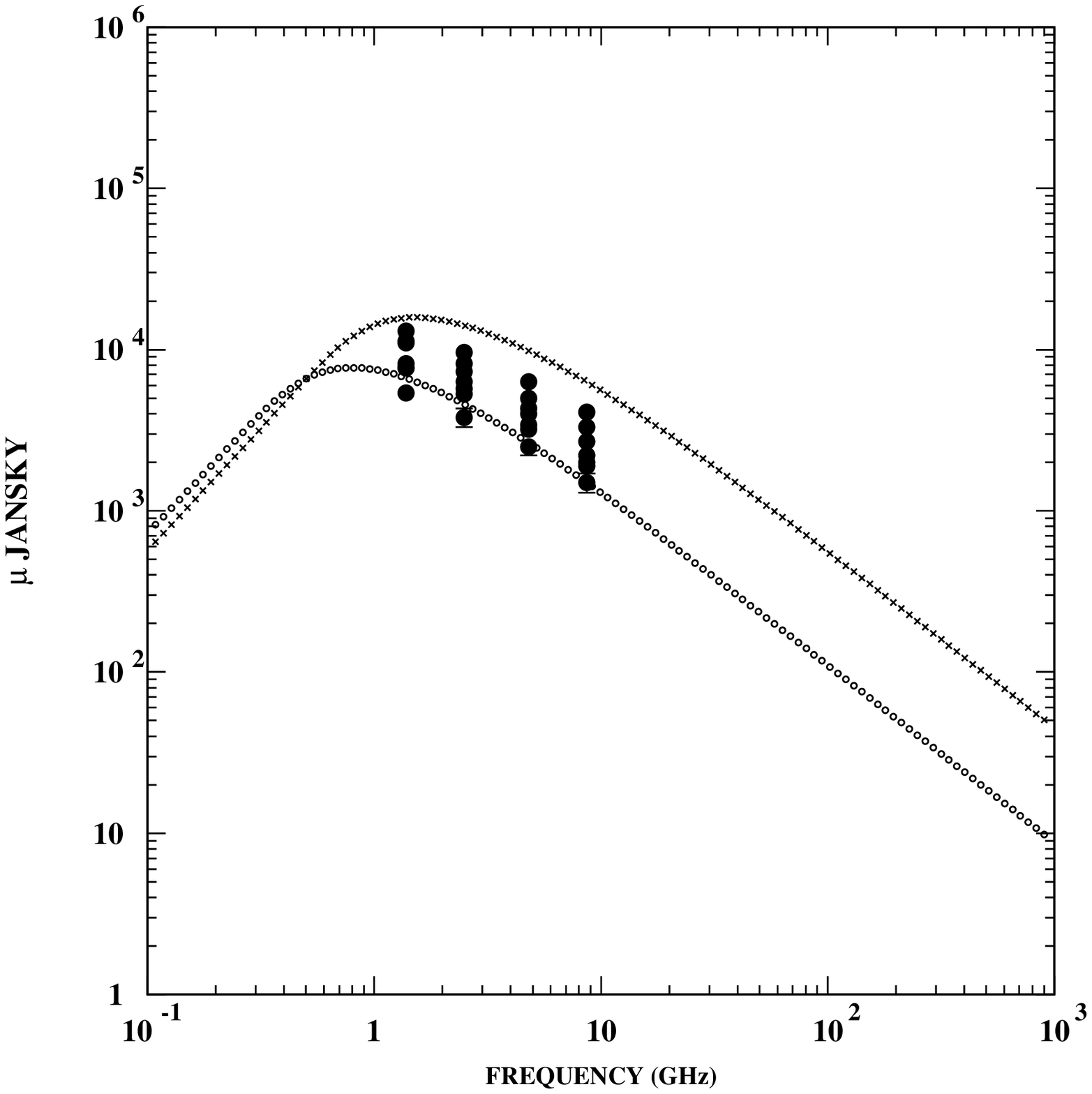, width=8cm} 
\end{tabular} 
\caption{Comparison between the observed spectrum of the AG of GRB 980425 in 
the radio band and the fitted CB model spectrum. 
Upper panel: in the time interval  between 40 and 70 days after burst.
Lower panel: in the time interval between 70 and 150 days after burst. 
In both cases the highest peaking curve
corresponds to the earlier time.}
\label{figr042504} 
\end{figure} 

%\clearpage
 
\begin{figure}[t]  
\hskip 0truecm   
\epsfig{file=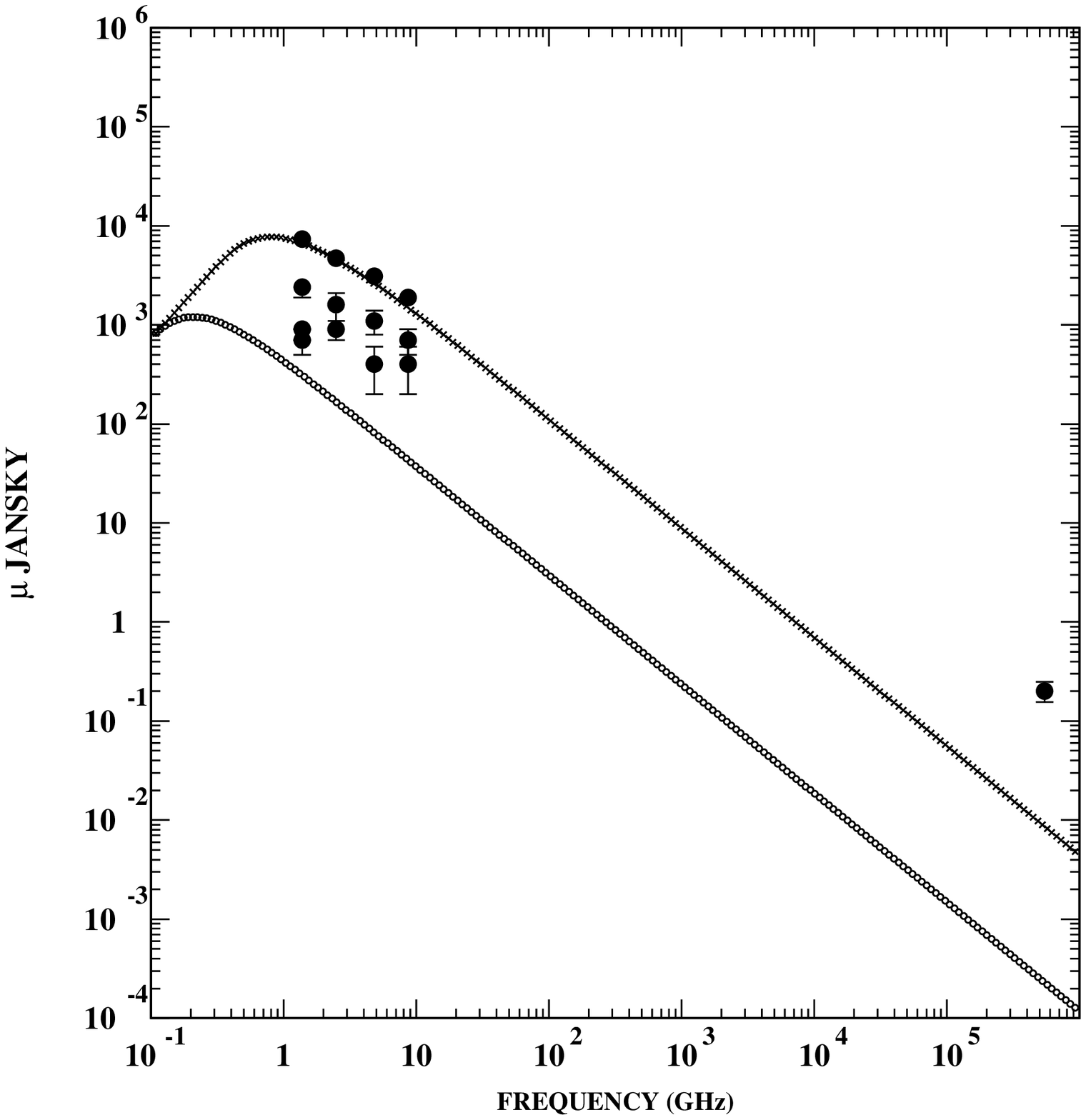, width=8cm}  
\caption{The late spectrum of GRB 980425 in the time interval 
between 150 and 759 days after burst. The highest 
peaking line corresponds to the earlier time. The late isolated
point is the last optical observation, which must correspond
to SN1998bw and not to the AG of the CB of GRB 980425.
}  
\label{late425}  
\end{figure}

\begin{figure}[t]  
\hskip 0truecm   
\vskip 1cm
\epsfig{file=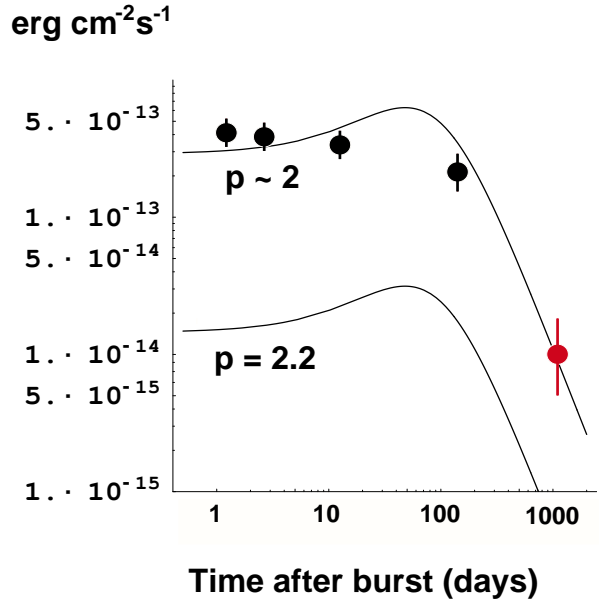, width=8cm}  
\caption{The X-ray afterglow of GRB 980425.}  
\label{X425}  
\end{figure}

\begin{figure}[t]  
\hskip 0truecm   
\epsfig{file=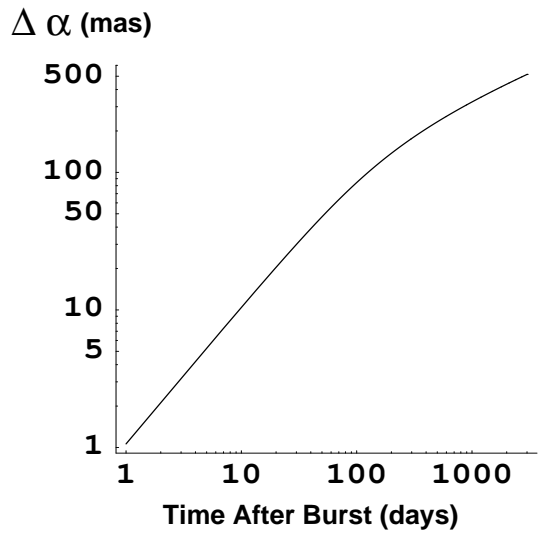, width=8cm}  
\caption{The predicted angular separation of SN1998bw
and GRB 980425, in milliarcseconds, as a function of time.
}  
\label{superluminal}  
\end{figure}

\end{document}